\documentclass[aoas,preprint]{imsart}

\RequirePackage[OT1]{fontenc}
\RequirePackage{amsthm,amsmath}
\RequirePackage[colorlinks,citecolor=blue,urlcolor=blue]{hyperref}

\usepackage[lofdepth,lotdepth]{subfig}
\usepackage{epsfig,subfloat, caption, chicago,keyval}%,, altpage} %chicago,
\usepackage[latin1]{inputenc}
\usepackage{amsmath}
\usepackage{latexsym,graphicx, rotating}
\usepackage{verbatim}
\usepackage{bm}
\usepackage{pdflscape}
\usepackage{array}
\usepackage{pdfpages}

\newcommand{\ai}{^{(AIDS)}}
\newcommand{\na}{^{(non-AIDS)}}
\newcommand{\am}{^{(AIDS\&Mat)}}
\newcommand{\ap}{^{(AIDS\&Preg)}}
\newcommand{\mda}{{D}\am}
\newcommand{\mdap}{{D}\ap}
\newcommand{\dna}{{D}\na}

\newcommand{\apga}{^{(AIDS\&Preg|AIDS)}}
\newcommand{\amgap}{^{(AIDS\&Mat|AIDS\&Preg)}}
\newcommand{\vv}{P\apga}
\newcommand{\uu}{P\amgap}

\newcommand{\md}{{\Phi}}
\newcommand{\mdna}{{\Phi}\na}
\newcommand{\emdna}{\widetilde{\Phi}\na}

\newcommand{\mmr}{{\Psi}} % unknown mmr
\newcommand{\mmmr}{\widehat{\Psi}} % posterior median for mmr
\newcommand{\emmrna}{\widetilde{\Psi}\na} % expected non-aids mmr
\newcommand{\mmrna}{{\Psi}\na} % unknown non-AIDS MMR

\newcommand{\dermmrna}{{\Psi}^{'(non-AIDS)}}
\newcommand{\deremmrna}{\widetilde{\Psi}^{'(non-AIDS)}}

\newcommand{\mult}{\vartheta}

\newcommand{\citep}{\shortcite}

\startlocaldefs
\endlocaldefs

\usepackage{geometry}
\begin{document}
	
\begin{frontmatter}
\title{A Bayesian approach to the global estimation of maternal mortality\protect\thanksref{1}}

\thankstext{1}{The research was funded by research grant R-155-000-146-112 at the National University of Singapore. The views expressed in this paper are those of the authors and do not necessarily reflect the views of the WHO, UNICEF, UNFPA, the World Bank, or the United Nations Population Division.}

\runtitle{Global estimation of maternal mortality}
\begin{aug}
\author{\fnms{Leontine Alkema$^1$, Sanqian Zhang$^2$, Doris Chou$^3$, \\ Alison Gemmill$^4$, Ann-Beth Moller$^3$, Doris Ma Fat$^3$,\\ Lale Say$^3$, Colin Mathers$^3$,  Daniel Hogan$^3$} \snm{} }
\address{$^1$Department of Biostatistics and Epidemiology, School of Public Health and Health \\ Sciences, University of Massachusetts, Amherst; $^2$Department of Statistics, Harvard University; \\ $^3$World Health Organization; $^4$Department of Demography, UC Berkeley.}
\bf{\today}
\end{aug}
		
\runauthor{L. Alkema et al.}
		
\begin{abstract}
	The maternal mortality ratio (MMR) is defined as the number of maternal deaths in a population per 100,000 live births. Country-specific MMR estimates are published on a regular basis by the United Nations Maternal Mortality Estimation Inter-agency Group (UN MMEIG) to track progress in reducing maternal deaths and to evaluate regional and national performance related to Millennium Development Goal (MDG) 5,  which calls for a 75\% reduction in the MMR between 1990 and 2015.
	
	Until 2014, the UN MMEIG used a multilevel regression model for producing estimates for countries without sufficient data from vital registration systems. While this model worked well in the past to assess MMR levels for countries with limited data, it was deemed unsatisfactory for final MDG 5 reporting for countries where longer time series of observations had become available because by construction, estimated trends in the MMR were covariate-driven only and did not necessarily track data-driven trends.
	
	We developed a Bayesian maternal mortality estimation model, which extends upon the UN MMEIG multilevel regression model. The new model assesses data-driven trends through the inclusion of an ARIMA time series model that captures accelerations and decelerations in the rate of change in the MMR. Varying reporting and data quality issues are accounted for in source-specific data models. The revised model provides data-driven estimates of MMR levels and trends and will be used for MDG 5 reporting for all countries.
\end{abstract}


\end{frontmatter}

\clearpage

\section{Introduction}

A maternal death is ``the death of a woman while pregnant or within 42 days of termination of pregnancy, irrespective of the duration and site of the pregnancy, from any cause related to or aggravated by the pregnancy or its management but not from accidental or incidental causes'', as defined in International Statistical Classification of Diseases and Related Health Problems, Tenth Revision (ICD-10) \shortcite{icd}. 
In this definition, which we adhere to in this paper, a maternal death may be due to a direct obstetric cause (resulting from complications during pregnancy, delivery, and postpartum period) or an indirect cause, which refers to maternal deaths that result from existing diseases or diseases that developed during pregnancy which were aggravated by physiological effects of pregnancy.

The number of maternal deaths in a population is the product of two factors: the risk of mortality associated with a single pregnancy, and the number of pregnancies or births that are experienced by women of reproductive age. The maternal mortality ratio (MMR) is defined as the number of maternal deaths in a population per 100,000 live births; thus, it depicts the risk of a maternal death relative to the number of live births. 

Country-specific estimates of the MMR are used to track progress in reducing maternal mortality and to evaluate regional and national performance related to Millennium Development Goal (MDG) 5, which calls for a 75\% reduction in the MMR between 1990 and 2015. Estimates are constructed and published by the UN Maternal Mortality Estimation Inter-agency Group (UN MMEIG); the latest set of estimates for 1990--2013 was published in May 2014 (WHO et al 2014).  \nocite{whomm2014} Monitoring country-level trends in maternal mortality is challenging for many developing countries because they lack high quality vital registration (VR) data systems, which enable countries to record deaths, and causes of death, that occur in their populations.  Instead,  intermittent national surveys are the main source of data for estimating the MMR in developing countries, and these surveys have various data quality issues, including small sample sizes and reliance on respondent recall of past events. Given the data limitations for estimating the MMR for countries  without high quality vital registration systems, the MMEIG developed a statistical model to obtain maternal mortality estimates. In the UN MMEIG 2014 publication, a multilevel regression model formed the basis of the maternal mortality estimates for countries  without high quality vital registration systems, whereby country-specific estimates followed from country-specific random intercepts and country-specific covariates, with global regression coefficients \shortcite{Wilmoth2012}. While that model has proven to work well to assess MMR levels for countries with limited data, its main limitation was that the model specification resulted in country-specific trend estimates that were determined by the country-specific covariates only, as opposed to trends in a country's maternal mortality data. Hence, the resulting country-specific MMR estimates did not necessarily reflect trends in a country's maternal mortality data. To assess countries' achievement regarding progress since 1990, an improved model was deemed necessary to produce estimates in the final year of MDG 5 reporting. 

In this paper, we present a Bayesian maternal mortality estimation model, referred to as the BMat model, which extends upon the UN MMEIG multilevel regression model to improve upon its main limitation: BMat combines the rate of change implied by the UN MMEIG multilevel regression model with an ARIMA time series model to capture data-driven changes in country-specific MMRs. In BMat varying data quality is accounted for in source-specific data models, which reduces bias while also facilitating the inclusion of multiple data sources for a given country. The revised model provides data-driven estimates of MMR levels and trends, is reasonably well calibrated and can be used for all countries, regardless of data availability and quality. The BMat model has been accepted by the UN MMEIG for producing the MMR estimates for MDG 5 reporting in September 2015.

This paper is organized as follows: we first provide an overview of the data sources that are available for measuring maternal mortality and summarize the approach to estimating the MMR that was used by the MMEIG in 2014. We then describe the BMat model and highlight the main differences with the MMEIG 2014 model. In the results section, we present estimates of the MMR for both models to illustrate the main types of differences that arise. Lastly, we present findings related to the performance of the new approach in validation exercises and conclude with a discussion of limitations and future research areas.

\section{Data: UN MMEIG maternal mortality database}
The UN MMEIG 2014 database was used in this paper as input to the BMat model to obtain results which are directly comparable to the MMEIG 2014 estimates produced with the previous model. The UN MMEIG 2015 database, and corresponding BMat estimates, will be available in September 2015.

\subsection{Maternal mortality data sources}
The UN MMEIG 2014 database contains information related to maternal mortality from vital registration (VR) systems, special inquiries, surveillance systems, household surveys and censuses. For all data sources which collect information related to maternal deaths as well as all-cause deaths to women aged 15-49, the observed proportion of maternal deaths (PM) among all-cause deaths was  taken as the preferred summary for use in estimating maternal mortality. The PM is generally preferred over observed maternal deaths or other summary outcomes because it is less affected by underreporting of the all-cause deaths: potential underreporting of all-cause deaths would affect the numerator and the denominator of the PM proportionately as long as causes of death are not underreported differentially. Therefore, in processing maternal mortality-related data, observed PMs took priority over observed maternal deaths. For all observed PMs, corresponding PM-based observations of the MMR were obtained by multiplying the PM by UN estimates for the ratio of all-cause deaths to births to obtain the ratio of maternal deaths relative to births. 

National vital registration systems record the number of deaths to women of reproductive ages, as well as the cause associated with each death using ICD coding. Based on the all-cause and maternal deaths, the PM can be constructed. An illustration of observed PMs from VR systems, based on annual numbers of deaths, is given in Figure~\ref{fig-countries} for Japan and El Salvador in the first column, with the corresponding PM-based observations of the MMR added in the second column. Under ideal circumstances, when all deaths are captured and all causes are accurately classified, VR systems provide perfect information on the number of maternal deaths within the country. However, even if routine registration of deaths is in place, maternal deaths may be reported incorrectly if deaths are unregistered or misclassified, where misclassification of deaths refers to incorrect coding in vital registration systems, due either to error in the medical certification of cause of death or error in applying the correct ICD code. In addition to misclassification issues, the VR-based PM may also differ from the true PM because of the inclusion of \textit{late} maternal deaths, which are deaths that take place after 42 days but within one year of termination of pregnancy (usually a small fraction of maternal and late maternal deaths combined). While our interest lies in maternal deaths, late maternal deaths were included in VR-based PMs in the MMEIG 2014 estimation round to maintain consistency between ICD-9 and ICD-10 coding of maternal deaths -- the exclusion of late maternal deaths is not possible based on ICD-9 coding.\footnote{For the MMEIG 2015 estimates, late maternal deaths were excluded from the ICD-10 coded observations. See Alkema et al. (forthcoming).\nocite{alkemaetalforthcoming}}

To provide a more rigorous assessment of maternal deaths, specialized studies are carried out to investigate whether there are non-reported or misclassified maternal deaths and, in some instances, whether there are unregistered deaths. The result of such a study is  illustrated in Figure 1 for Japan in 2005, with the observed PM plotted in purple. This investigation of the classification of deaths (the accuracy of the reported causes of deaths) in 2005 found that the VR system in Japan underreported maternal deaths by 35\% for that year due to misclassification and/or underreporting of maternal deaths. % note: incompl  = yes in this study
Similarly, for El Salvador, a study around 2006 suggested that the PM was about twice as high as compared to the VR-based PM. The high levels of misclassification in VR systems as discussed for Japan and El Salvador are common and typically on the order of 50\% \shortcite{Wilmoth2012}, and need to be accounted for when constructing estimates of maternal mortality.

\begin{figure}[htbp]
	\begin{center}
\includegraphics[width = 1.1\textwidth]{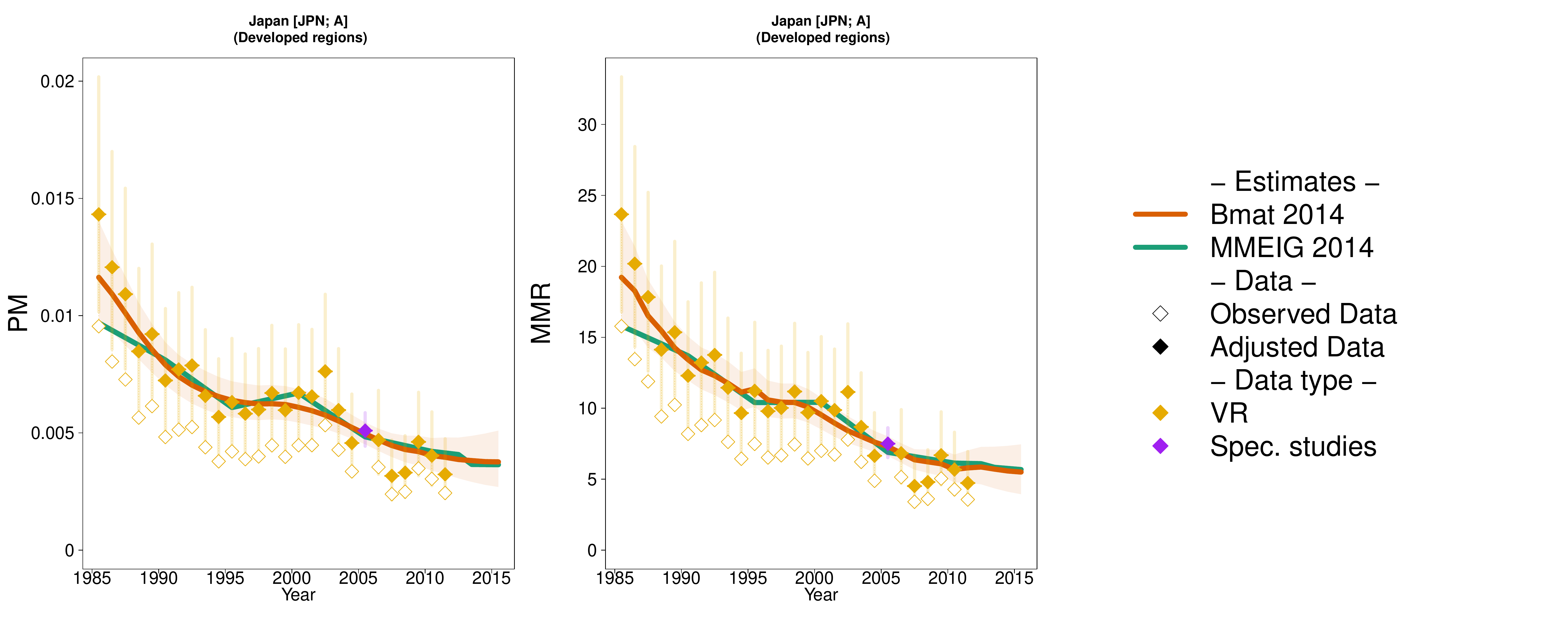} \\
\includegraphics[width = 1.1\textwidth]{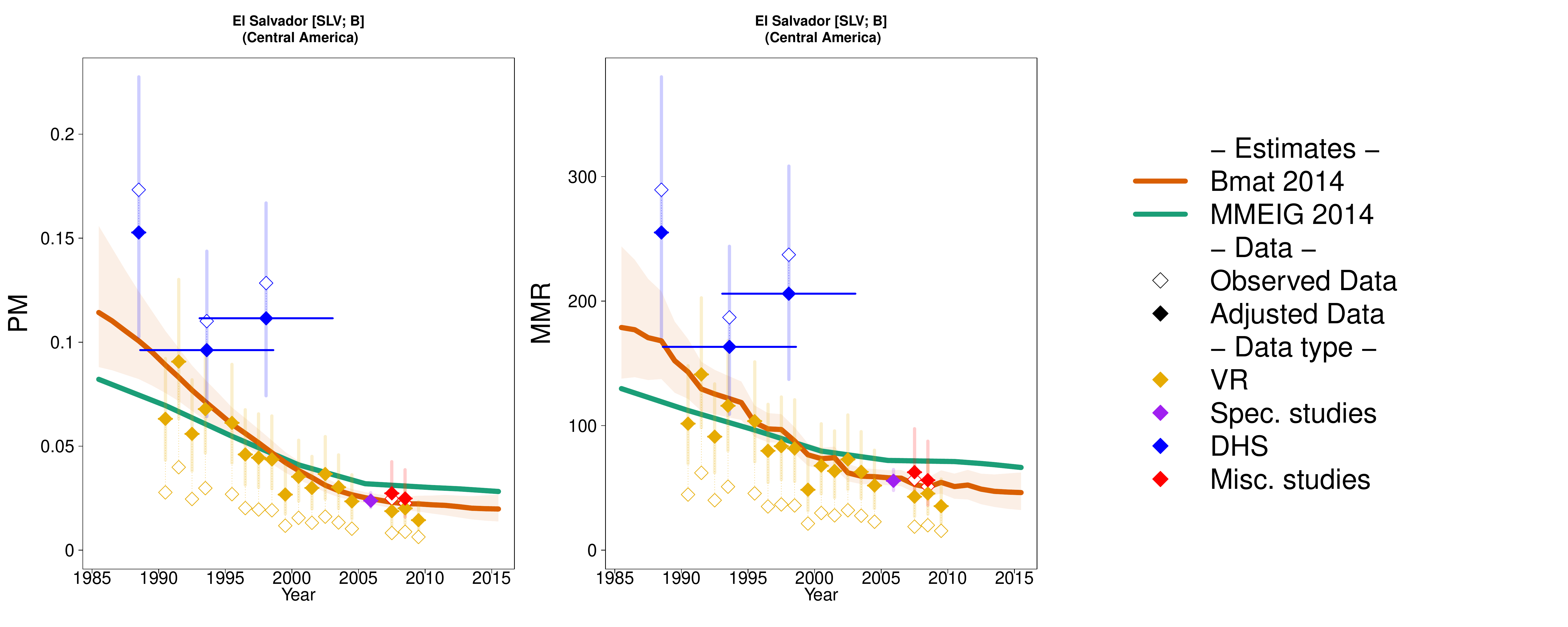} \\
\includegraphics[width = 1.1\textwidth]{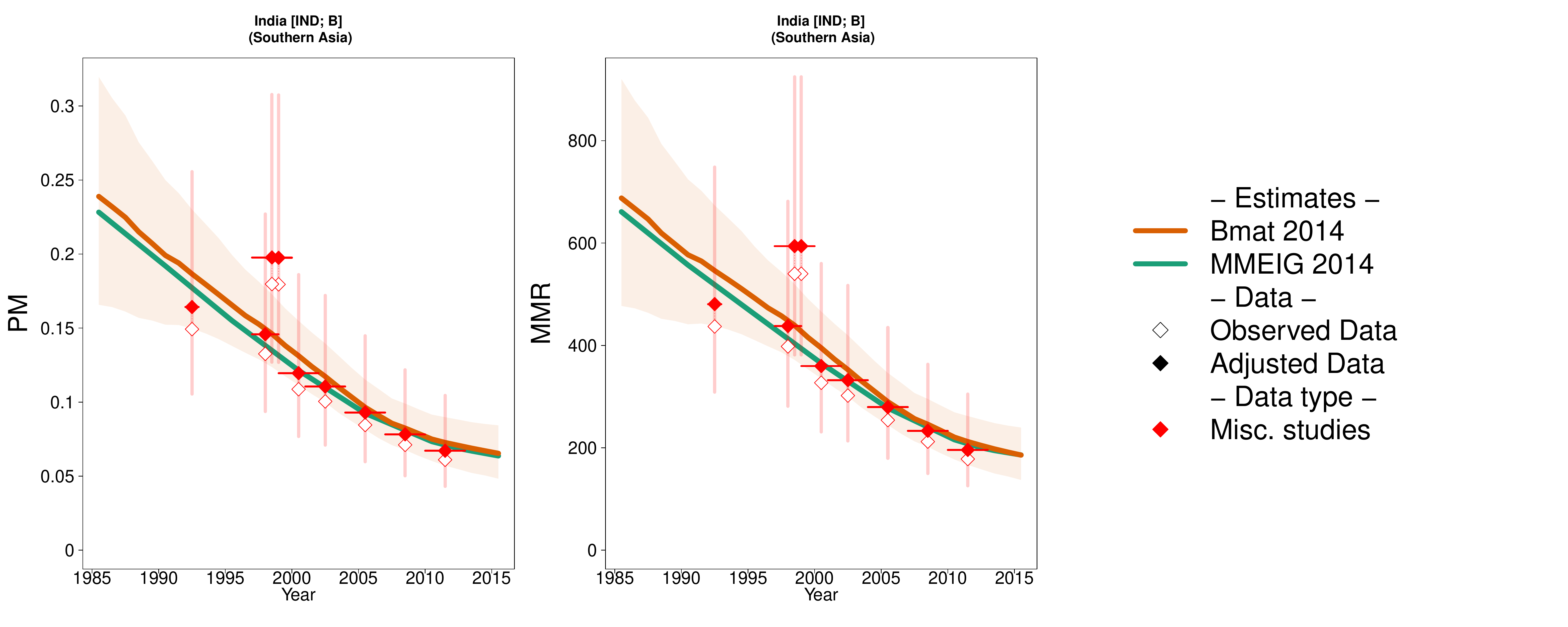} \\
	\end{center}
	Figure 1 continues on the next page
\end{figure}

\begin{figure}[htbp]
	\begin{center}
\includegraphics[width = 1.1\textwidth]{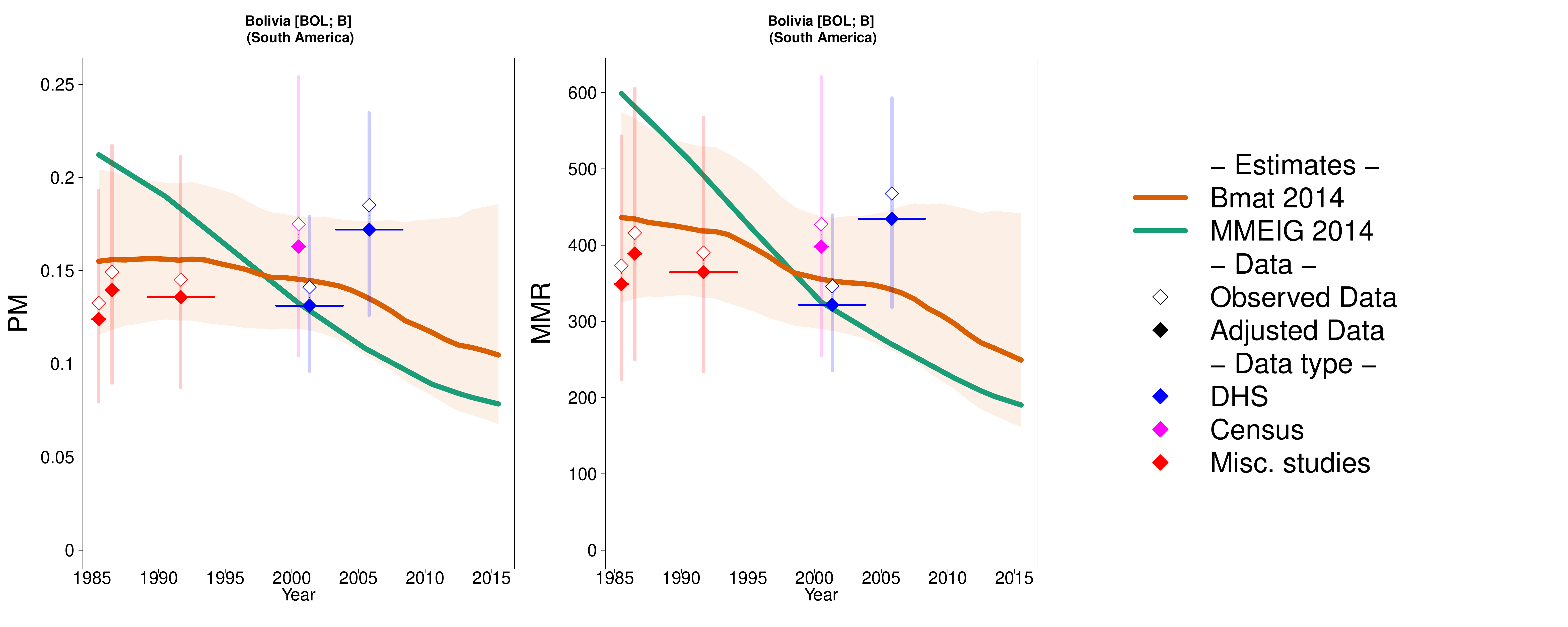} \\
\includegraphics[width = 1.1\textwidth]{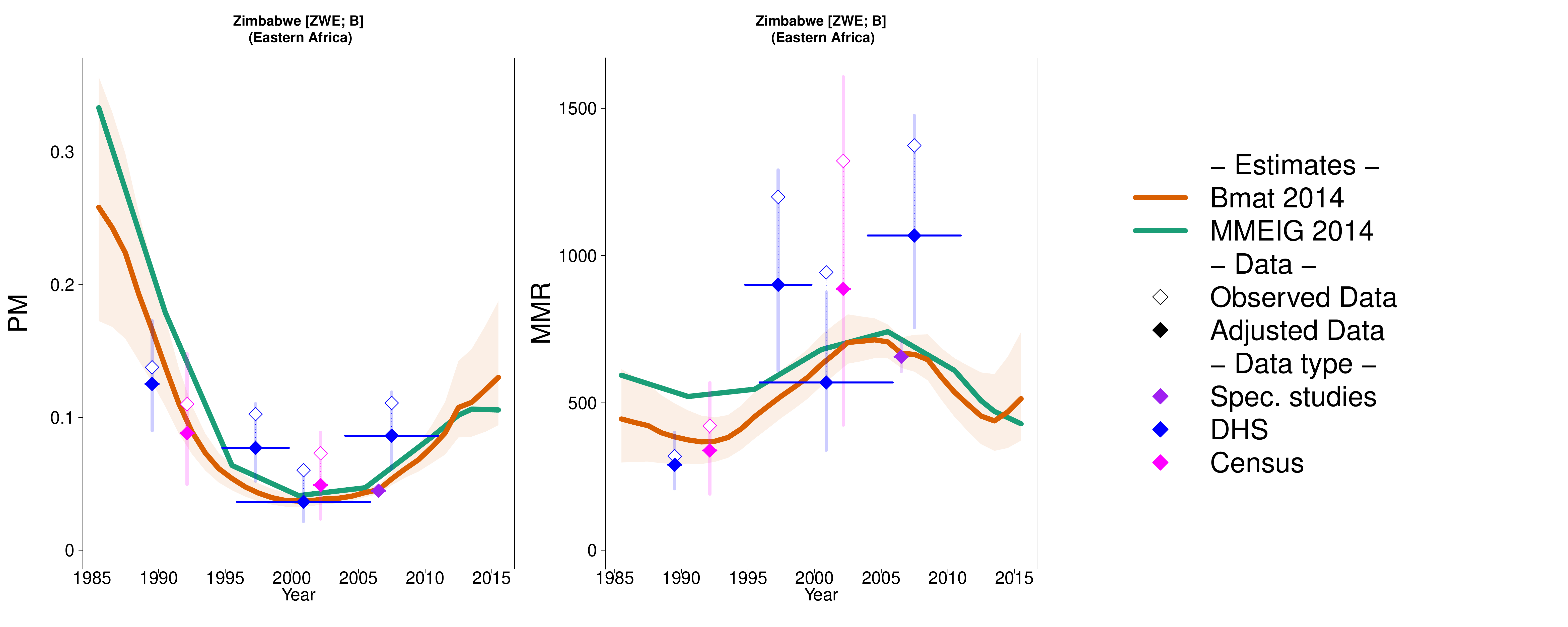} \\
\includegraphics[width = 1.1\textwidth]{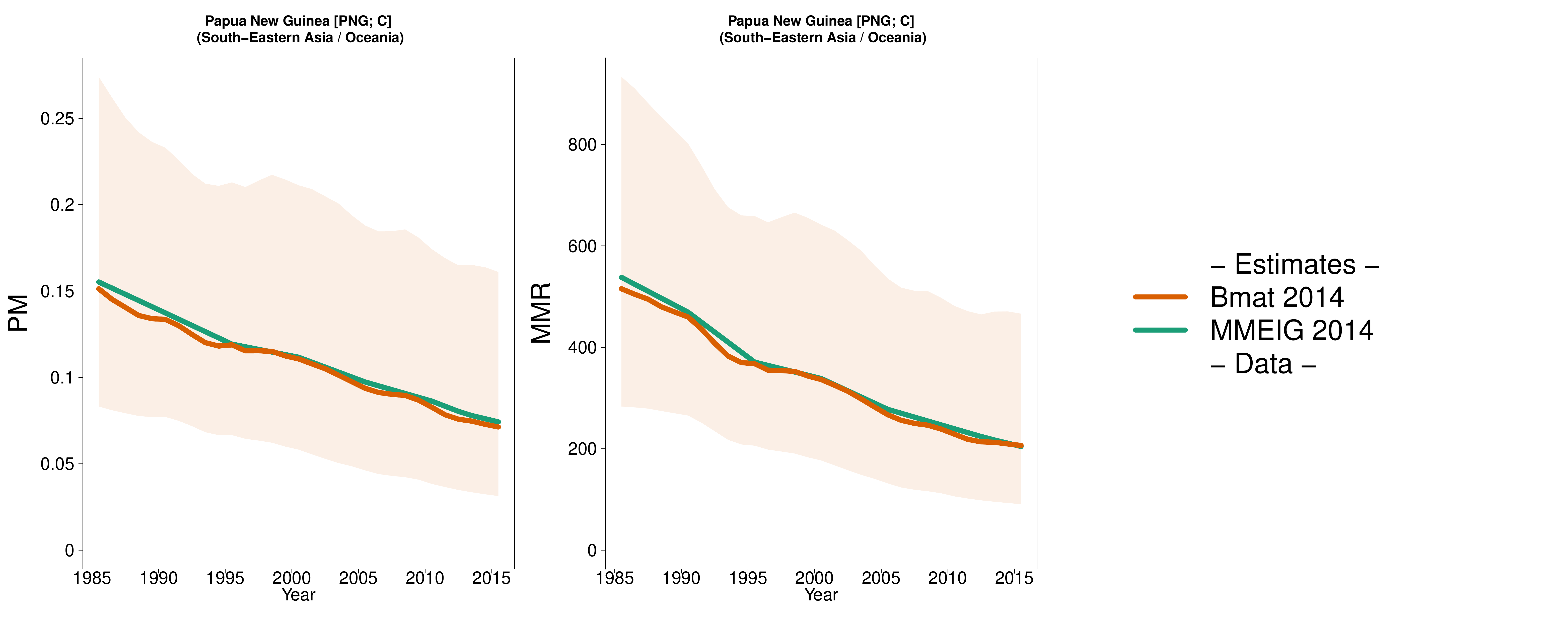} \\
				\caption{{\bf Data series and estimates of the PM (proportion of all-cause deaths that are maternal) and the MMR (number of maternal deaths per 100,000 live births) for Japan, El Salvador,  India, Bolivia, Zimbabwe and Papua New Guinea.} BMat estimates are illustrated by the solid red lines and 80\% CIs are shown by the red shaded areas. The UN MMEIG 2014 estimates are illustrated with the green lines. Reported (unadjusted) and adjusted observations are displayed. The vertical line with each adjusted observation indicates the approximate 80\% confidence interval for the PM or MMR associated with that observation, based on point estimates for reporting adjustments and total error variance. } \label{fig-countries}
	\end{center}
\end{figure}

In many countries, a vital registration system is either not in place or is not nationally representative. For estimating maternal mortality, other available data sources include data from surveillance systems, surveys, and censuses. For example in India, a Sample Registration System is in place to monitor mortality, where a representative sample of households are monitored for vital events such as births and deaths. In this system, information on cause-specific mortality is collected through Verbal Autopsy (VA) whereby trained physicians (and/or software) categorize deaths by cause based on symptoms associated with recorded deaths, which are provided by household members of the deceased. Data from these systems can be used to calculate PMs for use in modeled estimates (see Figure 1). 

Household surveys and censuses that collect information on deaths may collect information related to \textit{pregnancy-related} deaths, if, for a death to a woman of reproductive age, respondents are asked whether the deceased women died during pregnancy or shortly thereafter. Pregnancy-related deaths are the combination of maternal deaths and deaths during pregnancy that are due to incidental and accidental causes unrelated to the pregnancy. If the survey or census also collects data   on deaths occurring among women of reproductive age, one can compute the pregnancy-related PM, which is the ratio of pregnancy-related deaths to all-cause deaths. The Demographic Health Survey (DHS) program  commonly collects this information, as do censuses. Example observations of pregnancy-related PMs (and PM-derived MMRs) from censuses are illustrated for Bolivia and Zimbabwe in Figure 1, and DHS observations are available for the same countries, as well as for El Salvador.  In El Salvador, the DHS data points are much higher than the VR data points, which may be explained by the inclusion of pregnancy-related deaths in the DHS records, misclassification in the VR, or some other reason(s) including sampling errors and non-sampling errors (e.g. due to how the questionnaire was administered in the DHS, which may be substantial). Finally, some studies provide an observed (pregnancy-related) MMR as opposed to an observed (pregnancy-related) PM, if information on births and maternal deaths is collected. For such observations, the observed (pregnancy-related) PMs are obtained by multiplying the observed MMR by the ratio of UN estimates of births to all-cause deaths.

\clearpage
\subsection{Summary of data availability and data quality issues}
The overview of data available by source from 1985 to 2013 (from the the 2014 MMEIG maternal mortality database) is given in Table~\ref{tb-databysource}, summarized in terms of observations (the number of annual records for VR data and the number of studies for other data sources) and country-years of information. The majority of observations and observed country-years are from VR systems. The other three source types, which are specialized studies, miscellaneous studies reporting on maternal mortality and miscellaneous studies reporting on pregnancy-related mortality (including DHSs), each provide around 5 to 7\% of the total number of observations. The number of country-years provided by miscellaneous studies reporting on pregnancy-related mortality is much larger at 30\% because the majority of observation from this source type refer to multi-year periods. 

Data availability varies greatly across countries, as illustrated for the selected countries in Figure 1: while data are available for the countries discussed so far, no observations are available for Papua New Guinea.
The overview of data availability by country is given in Table~\ref{tb-dataoverview}. Of the 183 countries for which the UN MMEIG constructs MMR estimates, no data are available for 20 countries, and for 22 countries, only one data point is available.

\begin{table}[htbp]
		\caption{Overview of data availability by study type based on the 2014 MMEIG database. Observations refer to the number of annual records for VR data and the number of studies for other data sources.}\label{tb-databysource}
	\begin{tabular}{|l||c|c|c|c|}\hline
		Study type & VR & Specialized & Misc. studies - & Misc. studies -  \\ 
		& & studies & maternal & pregn.related \\ 
		\hline
		
		\#  observations (\% out of 2216 total) & 1824 (82\%) & 113 (5\%) & 122 (6\%) & 157 (7\%)\\
		\hline
		\#  country-years(\% out of 3152 total) & 1824 (58\%) & 239 (8\%) & 144 (5\%) & 945 (30\%) \\ 
		\hline
	\end{tabular}

\end{table}

\begin{table}[htbp]
		\caption{Overview of data availability by country based on the 2014 MMEIG database. Observations refer to the number of annual records for VR data and the number of studies for other data sources. }\label{tb-dataoverview}
	\begin{tabular}{|l||c|c|c|c|c|}\hline
		Total number of observations & 0 & 1 & 2 & 3-5 & $>$5 \\ 
		\hline
		
		Number of countries (\% out of 183 countries ) & 20 (11\%) & 22 (12\%) & 14 (8\%) & 30 (16\%) & 97 (53\%) \\ 
		\hline
	\end{tabular}

\end{table}

Estimating maternal mortality is challenging because of limited data availability for many countries, and moreover, because of reporting and data quality issues for the observations that \textit{are} available. Firstly, many data inputs differ systematically from our outcome of interest, as explained in the previous section and summarized in Table~\ref{tab-DATASOURCESandADJ}. Information regarding the extent of systematic over- or underreporting of maternal mortality is limited. Secondly, observations tend to be subject to substantial random error, including stochastic or sampling error because maternal deaths are generally rare events, as well as random errors introduced in the data collection and processing procedure. This is illustrated in Figure~1 for the observations in the selected countries: the adjusted observations are based on a transformation of the reported observation to account for reporting issues mentioned in Table~\ref{tab-DATASOURCESandADJ}, and the vertical lines indicate the approximate 80\% confidence interval for the true PM and MMR based on uncertainty due to stochastic, sampling and/or random errors (explained further in Section~\ref{sec-methods}). The confidence intervals indicate that the uncertainty associated with the observations is large. 

\clearpage 
\begin{table}[htbp]
\begin{center}
	\caption{Overview of data sources that are included in the maternal mortality model and the reporting issues that are associated with the reported outcomes for each source type and have been addressed in the MMR model.}
	\label{tab-DATASOURCESandADJ}   
\begin{tabular}{@{}|p{0.3\textwidth}|p{0.6\textwidth}|}
\hline
\textbf{Data source type}  &\textbf{Reporting issues that are addressed in MMR modeling}\\ \hline
Vital registration (VR)	&	\begin{itemize}
	\item Misclassification of maternal deaths; 
	\item Inclusion of late maternal deaths
\end{itemize}
\\ \hline
Specialized studies	 & \begin{itemize}
	\item None
\end{itemize}\\	 \hline
Other miscellaneous data sources reporting on maternal mortality	&
\begin{itemize}
	\item Underreporting of maternal deaths
\end{itemize}
		\\ \hline
Other miscellaneous data sources  reporting on pregnancy-related mortality& 
\begin{itemize}
	\item Underreporting of pregnancy-related deaths; 
	\item Over-reporting of maternal deaths due to the inclusion of pregnancy-related deaths that are not maternal
\end{itemize}
\\ \hline

\end{tabular}

\end{center}
\end{table}

%\clearpage

\section{Constructing maternal mortality estimates}\label{sec-methods}
Given the limited data availability and issues associated with the reporting of maternal mortality, modeling of maternal mortality is required for producing estimates based on the available data sources. In this section, we discuss the BMat modeling approach that we developed for estimating the MMR and MDG 5 reporting. The goal was to construct estimates based on all available data and additional information on data quality issues. The new modeling approach extends upon the MMEIG 2014 estimation approach. Hence, we first summarize the the MMEIG 2014 estimation approach to introduce the main ideas that form the basis of the new model. Based on discussions of limitations of the 2014 approach that we seek to improve upon, we introduce and explain in detail the BMat model. We start with general notation and notes related to the database and data preprocessing that are applicable to both the MMEIG 2014 and BMat approach.

\paragraph{Notation} In the model description in this section, lowercase Greek letters refer to unknown parameters and uppercase Greek letters to variables which are functions of unknown parameters (modeled estimates). Roman letters refer to variables that are known or fixed, including data (lowercase) and estimates provided by other UN sources or the literature (upper case). 

$\md_{c,t_1, t_2}$ denotes the main quantity to be estimated, which is the number of maternal deaths for country $c$ for any period $[t_1, t_2)$, with $\md_{c,t} \equiv \md_{c,t, t+1}$, the number of maternal deaths in country $c$ in calendar year $t$. $\mmr_{c,t}$ denotes the final outcome of interest, the MMR for the respective country-year. 

UN country-year estimates for births, all-cause deaths and AIDS deaths to women of reproductive ages are denoted by $B, D$ and $D^{(AIDS)}$ respectively with subscripts consistent with subscripts for $\md$, e.g., $D_{c,t_1, t_2}$ refers to the number of all-cause deaths in country $c$ in period $[t_1, t_2)$ and $D_{c,t} \equiv D_{c,t, t+1}$. The UN estimates for births are taken from the UN World Population Prospects \shortcite{wpp2012}. All-cause mortality information is provided by WHO \shortcite{wholt} and the AIDS deaths are obtained from UNAIDS \shortcite{UNAIDS2013}.

Observations were combined across countries and indexed by $i = 1, 2, \hdots, N$; $c[i]$ refers to the country of the observation, $h[i]$ to the exact start date of the observation period, $e[i]$ to its exact end date, and $t[i]$ refers to the calendar year of the midpoint of the observation period. 
When a single subscript $i$ is used with modeled or UN estimates of demographic variables, the variable represents the country-period of the $i$-th observation, e.g., $\md_i \equiv \md_{c[i],h[i], e[i]}$ and $D_i \equiv D_{c[i],h[i], e[i]}$. 

\paragraph{Maternal mortality database and data preprocessing} 
Data were obtained from the MMEIG maternal mortality database. Recorded information for the $i$-th record in the database includes information about the study population and study period, as well as the definition of the deaths (maternal or pregnancy-related) and any available information on the number of maternal or  pregnancy-related deaths $m_i$, all-cause deaths $d_i$, the births associated with the sample population of women of reproductive ages for the study period $b_i$, and/or direct observations of the (maternal or pregnancy-related) PM $y_i = m_i/d_i$ or MMR $m_i/b_i$. 
All records in the database are processed to obtain an observed PM $y_i$ for each record $i$. For most records, the PM was calculated directly from the available information on the number of maternal or  pregnancy-related deaths $m_i$ and all-cause deaths $d_i$; $y_i = m_i/d_i$. An exception to this rule was made for specialized studies (in particular, inquiries) that include an investigation into all-cause deaths in addition to maternal deaths. For such studies, the observed number of maternal deaths $m_i$ is considered to be the best available information on maternal mortality and the PM data input associated with the study is given by $y_i = m_i/D_i$. For observations where the PM cannot be calculated directly, UN estimates of the number of births and deaths in the study period in the country were used to obtain the corresponding PM. For example, if record $i$ reported only the MMR $m_i/b_i$, the corresponding PM used as a data input was obtained as follows: $y_i = m_i/b_i\cdot B_i/D_i$, where $B_i$ and $D_i$ are the UN estimates for births and all-cause deaths in the study period in the country. 

\subsection{MMEIG 2014 model}\label{sec-who}
\subsubsection{Summary}
The MMEIG maternal mortality estimation methods and data sources used are described in detail elsewhere \citep{Wilmoth2012,whomm2014}. Here we summarize the approach, focusing on the main set-up which is the starting point of the BMat model.

In the UN MMEIG 2014 approach, estimates were published for 5-year periods. For constructing estimates, countries were classified into groups A, B and C based on data availability. For countries in group A, high-quality VR records were available for a sufficient number of years such that the VR-based PM data, adjusted for misclassification issues as introduced in the previous section, could be used directly to construct maternal mortality estimates. 

For countries in group C, no data were available while for countries in group B, data were available but deemed insufficient w.r.t. availability and/or associated biases and error variances to be used directly to obtain estimates. For countries in group B and C, a model was used to obtain estimates, in which maternal deaths were modeled as the sum of non-AIDS maternal deaths $\mdna_{c,t}$ and AIDS maternal deaths $\mda_{c,t}$:
\begin{eqnarray*}
\md_{c,t} &=& \mdna_{c,t} + \mda_{c,t}.
\end{eqnarray*}
AIDS maternal deaths are deaths of HIV positive women who die because of the aggravating effect of pregnancy on HIV and are thus considered as indirect maternal deaths. The non-AIDS maternal deaths refer to maternal deaths due to direct or non-HIV/AIDS-related indirect causes. Given the substantial impact of the HIV/AIDS epidemic on mortality in many countries, in particular in sub-Saharan Africa, the number of AIDS maternal deaths was modeled separately to be able to capture the trends in maternal mortality associated with the epidemic. 

\subsubsection{Modeling AIDS maternal deaths}\label{sec-aidsmat}
The AIDS maternal deaths $D\am_{c,t}$ were modeled as follows
\citep{Wilmoth2012}:
\begin{eqnarray}
D\am_{c,t} &=& D\ai_{c,t}\vv_{c,t}\uu, 
\end{eqnarray}
where number of AIDS deaths $D\ai_{c,t}$ is obtained from UNAIDS \shortcite{UNAIDS2013},\\ $\vv_{c,t}$ refers to  the proportion of AIDS deaths that occurs during the maternal risk period, and finally $\uu$ is the proportion of AIDS deaths among women during the maternal risk period that qualify as maternal because of some causal relationship with the pregnancy, delivery or postpartum period. The proportion of AIDS deaths that occur during the maternal risk period $\vv_{c,t}$ was obtained from the general fertility rate $x_{c,t}^{(GFR)}$ (obtained from UN estimates), 
the relative risk $R$ of dying from AIDS for a pregnant versus non-pregnant woman  %R=k
and the average woman-years lived in the maternal risk period per live birth $F$: %=c, %(set equal to 1 year, including the 9 month gestation, plus 42 days postpartum, and an additional 1.5 months to account for pregnancies not ending in a live birth); 
\begin{eqnarray*}
	\vv_{c,t} &=& \frac{F\cdot R \cdot x_{c,t}^{(GFR)}}{1+F\cdot (R-1)\cdot x_{c,t}^{(GFR)}}.
\end{eqnarray*}
Values were assigned to the unknown parameters based on a combination of literature review and expert opinion (WHO et al. 2014), with $R=0.3$ ($R$ is less than one because HIV-positive women who become pregnant are generally of better health than the general population of non-pregnant HIV-positive women), $\uu=0.3$ and $F=1$. 

\subsubsection{Modeling non-AIDS maternal deaths}
To obtain estimates of non-AIDS maternal deaths, a multilevel regression model was developed.  After considering various model specifications, the MMEIG 2014 model used as a dependent variable the non-AIDS PM 
$\mdna_{c,t}/\dna_{c,t}$, which is the proportion of non-AIDS maternal deaths among the total number of non-AIDS deaths of women of reproductive ages $\dna_{c,t}$. The mean response in the multilevel model is modeled with three covariates with fixed coefficients and a random country intercept: 
\begin{eqnarray}\label{eq-whoregr}
\log\left(\mdna_{c,t}/D\na_{c,t}\right)&=&  \alpha_{c} - \beta_1 \log(x_{c,t}^{(GDP)}) +\beta_2 \log(x_{c,t}^{(GFR)}) -\beta_3 x_{c,t}^{(SAB)},
\end{eqnarray}
with covariates GDP $x_{c,t}^{(GDP)}$, the general fertility rate $x_{c,t}^{(GFR)}$, and the percentage of births with a skilled attendant present $x_{c,t}^{(SAB)}$ (WHO et al. 2014), and random country intercept $\alpha_c$. Country intercepts are modeled hierarchically:
\begin{eqnarray}
\alpha_c|\alpha_{r[c]}, \sigma_{country}^2 &\sim& N(\alpha_{r[c]}, \sigma_{country}^2), \hspace{0.1cm} \alpha_r \sim N(\alpha_{world}, \sigma_{region}^2),
\end{eqnarray}
with regional and global intercepts $\alpha_{r}$ and $\alpha_{world}$  and across-country and across-region variances $\sigma_{country}^2$ and $\sigma_{region}^2$. Of interest to highlight here is that the rate of change in the non-AIDS MMR is determined by rates of change in covariates and global regression coefficients.

In the MMEIG 2014 approach, the multilevel model was fitted to all available data worldwide (including from group A) using the following data model:
\begin{eqnarray*}
	\log(y_i^*)|\mdna_{i},\sigma^2  &\sim& N(\log(\mdna_{i}/D\na_i), \sigma^2).
\end{eqnarray*}
where $\sigma^2$ is the unknown error variance, assumed to be the same for all observations, and $y_i^*$ refers to the $i$-th adjusted non-AIDS PM observation in the data set. $y_i^*$ is a transformed version of the original observation $y_i$, $y_i^* = f(y_i)$. The transformation $f(\cdot)$ depends on the data source and characteristics and includes, where appropriate, (i) the addition of maternal deaths that are missing because of underreporting, (ii) the removal the AIDS deaths from the numerator and the denominator, and (iii) the removal of pregnancy-related deaths which are not maternal. The types of adjustments involved are discussed in more detail in Section~\ref{sec-datamodel}, as part of the BMat model specification. In the MMEIG 2014 approach, point estimates of the MMR were obtained based on pre-fixed settings of the adjustment parameters and the uncertainty assessment was based on a monte carlo procedure whereby model estimates were obtained based on different combinations of adjustment parameters.

\subsubsection{MMEIG 2014 model limitations}

MMEIG 2014 estimates are shown in Figure 1 for the selected countries. The exact adjusted data points (the $y_i^*$'s) are not shown in Figure 1 to improve figure legibility but are similar to the adjusted version of the raw data points that are shown in the figure. In Figure 1, Japan is the only group A country. For this country, the MMEIG 2014 MMR estimates are given by the period-based adjusted VR data. Papua New Guinea does not have any maternal mortality data and is in group C; its modeled estimates are based on covariates, the regional intercept and the inputs to the AIDS maternal deaths model. The other countries in the figure are in group B, including El Salvador, because data availability and/or quality was not sufficient to use the data directly for estimating the MMR. Estimates for all group B countries were obtained from the multilevel model and the AIDS model. For El Salvador, India and Bolivia, the number of AIDS maternal deaths are negligible and the MMEIG 2014 estimates are given by the multilevel regression model, which suggests decreasing MMRs in all three countries. In Zimbabwe, the number of maternal AIDS deaths is not negligible and included in the MMEIG estimates. As a result, the MMEIG estimates indicate an increase around the year 2000 which is due to the addition of AIDS indirect deaths. 

The comparison of adjusted data and MMR estimates for the countries in Figure 1 highlights the main limitations of the MMEIG 2014 approach. Firstly, the rate of change in the (non-AIDS) MMR for countries in group B follows from Eq.~\ref{eq-whoregr} and is thus determined by rates of change in covariates and global regression coefficients, which can result in inaccurate estimates of trends. Among the example countries in group B, we find that the estimates represent the observed trend in India well but are less reflective of the observed trends in adjusted data points for El Salvador, Bolivia and Zimbabwe. The estimate for the MMR in the 1990s is higher than the level suggested by the data in Bolivia and Zimbabwe, and lower than the data suggest in El Salvador. Secondly, when fitting the multilevel model, all adjusted observations are treated equally regardless of data quality and uncertainty associated with the observation. This is not appropriate given the great variation in error variance associated with the different observations. The uncertainty is illustrated in Figure 1 in the form of 1-single-observation-based confidence intervals for the unknown MMR for the study period (explained in more detail in Section~\ref{sec-commu}). In El Salvador, a recent specialized study provides more information about the MMR than any of the survey-based entries, yet all observations are weighted equally and the model estimates are not able to capture the level suggested by the high-quality study. 

In summary, the MMEIG 2014 estimation approach for group B does not provide sufficient insights into observed trends in countries with longer time series of observations because estimated trends are covariate-driven and not informed by trends in country-specific data, and because all (adjusted) observations are treated equally in model fitting. 

\subsection{The Bayesian maternal estimation model (BMat)}

\subsubsection{Summary}
We developed a revised maternal mortality estimation approach which improves upon the MMEIG modeling approach. The main set-up is summarized as follows:
\begin{itemize}
	\item[(I)] Maternal deaths are modeled for each country-year as the sum of non-AIDS and AIDS maternal deaths, $\md_{c,t} = \mdna_{c,t}  + \mda_{c,t}$, with AIDS maternal deaths given by the MMEIG 2014 approach, as explained in Section~\ref{sec-aidsmat}.
	
	\item[(II)] The number of non-AIDS maternal deaths $\mdna_{c,t}  = \emdna_{c,t}\cdot \mult_{c,t},$ which is the product of the  expected number of non-AIDS maternal deaths $\emdna_{c,t}$,given by the MMEIG 2014 multilevel regression model (Eq.~\ref{eq-whoregr}), and a country-year-specific  multiplier $\mult_{c,t}$, which is modeled with an ARIMA time series model.
	
	\item[(III)]  Data models (the likelihood function used) take  into account varying data quality in the form of reporting issues and varying stochastic and sampling variance. A usability measure of VR data is introduced to distinguish between VR data of varying quality. 
	
\end{itemize}

The revised model is fitted in a Bayesian framework and referred to as BMat (a Bayesian maternal mortality estimation model). BMat is able to track high quality data very closely, to handle countries that move from survey-based data sources in earlier time periods to newly scaled up VR in later time periods, and to combine information from data and covariates for countries with limited data while producing covariate-driven estimates for countries without data. The flexibility of the BMat model eliminates the need for grouping countries: one model is used for all countries, regardless of data sources available.

The model for the non-AIDS maternal deaths (point II), and the database and data models used (point III) are explained in more detail in the remainder of this section. The complete maternal mortality model is given in the appendix.

\subsubsection{Modeling non-AIDS maternal deaths}
The non-AIDS maternal deaths are modeled as follows for all countries:
\begin{eqnarray}\label{eq-mult}
\mdna_{c,t}  &=& \emdna_{c,t}\cdot \mult_{c,t},
\end{eqnarray}
where $\emdna_{c,t}$ refers to the ``expected'' non-AIDS deaths and $\mult_{c,t}$ is a country-year-specific multiplier. The expected non-AIDS deaths $\emdna_{c,t}$ are obtained through the MMEIG 2014 multilevel model (Eq.~\ref{eq-whoregr}); we assume that the regression model for the non-AIDS PM provides the expected non-AIDS maternal deaths:
\begin{eqnarray*}
	log(\emdna_{c,t})&=& \log(D\na_{c,t}) +  \alpha_{c} - \beta_1 \log(x_{c,t}^{(GDP)}) +\beta_2 \log(x_{c,t}^{(GFR)}) -\beta_3 x_{c,t}^{(SAB)},
\end{eqnarray*}
% % decide about minuses or not!
with country-specific intercept $\alpha_c$ and covariates GDP, GFR and SAB (minus signs are added to the GDP and SAB term such that we expect all $\beta$'s to be positive). 

Country-year-specific multiplier $\mult_{c,t}$ in Eq.~\ref{eq-mult} allows for data-driven deviations from the regression-model-implied levels and trends in non-AIDS deaths. For ease of interpretation and to motivate the modeling choice for the multiplier, we divide both sides of Eq.~\ref{eq-mult} by the number of births to express equality in terms of the non-AIDS MMR $\mmrna_{c,t}$, and then obtain the annualized continuous rate of reduction (ARR) in the non-AIDS MMR:
\begin{eqnarray}
\mmrna_{c,t}  &=& \emmrna_{c,t}\cdot \mult_{c,t}, \nonumber \\
\Rightarrow  \dermmrna_{c,t} &=& \deremmrna_{c,t}+ \mult'_{c,t}, \label{eq-arr}
\end{eqnarray}
where $\emmrna_{c,t} = \emdna_{c,t}/B_{c,t}$ refers to the expected non-AIDS MMR, and the addition of $'$ denotes the ARR of the indicator, i.e. $\dermmrna_{c,t} = -\log(\mmrna_{c,t+1}/\mmrna_{c,t})$ and  $\mult'_{c,t} = -\log(\mult_{c,t+1}/\mult_{c,t})$. Based on Eq.~\ref{eq-arr}, we can interpret $\mult'_{c,t}$ as a distortion term that is added to the ARR of the non-AIDS MMR. Hence $\mult'_{c,t}$ reflects the difference between the observed and covariate-based expected ARR. If a country goes through a period of scale-up of maternal care which results in faster declines than expected based on the covariates, $\mult'_{c,t}>0$, while 
$\mult'_{c,t}<0$ during periods where investments in maternal care are lacking and rates of decline are lower than expected. Given that the covariates capture the main expected change in the non-AIDS MMR, we expect that the distortions $\mult'_{c,t}$ fluctuate around zero and may be positively autocorrelated. This motivates the choice of a stationary autoregressive moving average process (ARMA) process for $\mult'_{c,t}$. We assume that $\mult'_{c,t}$ follows an ARMA(1,1) process, which is specified as follows:
\begin{eqnarray*}
	\mult'_{c,t} &=& \phi \mult'_{c,t-1} - \theta \epsilon_{c,t-1} + \epsilon_{c,t},\\
	\epsilon_{c,t} &\sim& N(0, \sigma_{c}^2),
\end{eqnarray*}
with autoregressive parameter $\phi$, moving average parameter $\theta$ and the variance of the innovation terms given by $\sigma_{c}^2$. The variance of the innovation terms is country-specific to capture variability across countries w.r.t. fluctuations in their rates of change. The variances are estimated using a hierarchical model to aid the estimation for countries with limited data, as follows:
\begin{eqnarray*}
%\sqrt{{\gamma}_{0,c}}&=&\sqrt{\gamma_0}\cdot (1+\lambda_c),
\sigma_{c}&=&\sigma^{(\epsilon)}\cdot (1+\lambda_c),\\
\lambda_c &\sim& TN_{(-1,2)}(0,\sigma_{\lambda}^2),
\end{eqnarray*}
where $\sigma^{(\epsilon)}$ represents the most likely estimate of the standard deviation of the innovation terms for $\mult'$, $\lambda_c$ the country-specific multiplier of the standard deviation. $TN_{(A,B)}(a, b^2)$ represents a truncated normal distribution
with mean $a$ and variance $b^2$, truncated to lie between $A$ and $B$, such that $\lambda_c$ is a draw from a truncated normal distribution, truncated between \\  -1 and 2, and $\sigma_{\lambda}^2$ represents the across-country variability in the multipliers of the global standard deviation. Priors for the ARMA parameters are chosen to guarantee stationarity and causality of the process and moreover, to allow for positive autocorrelation at any time lag.
The initial conditions for $\mult'_{c,t}$ follow from the stationary distribution of the process. This distribution and the priors are given in the appendix. 

An ARMA(1,1) model for $\mult'_{c,t}$ implies that $\log(\mult_{c,t})$ is an autoregressive integrated moving average ARIMA(1,1,1) process with an unspecified level. We fix $\mult_{c,1990}=1$ to make the model in Eq.~\ref{eq-mult} identifiable. With this choice, $\mmrna_{c,1990}= \emmrna_{c,1990}$ such that the regression model determines the level of the non-AIDS MMR in the year 1990. Sensitivity analyses whereby the base year was changed to a different year, including 2000 and 2010, suggested that the choice has little effect on the point estimates but may change the uncertainty assessment for countries with no or very limited data, with uncertainty increasing for years further removed from the base year. Given the demand for data-driven estimates, in particular so for the most recent years, we decided to fix the base year in 1990 as opposed to a more recent year.

\begin{figure}[htbp]
	\begin{center}
		\includegraphics[width=1\textwidth]{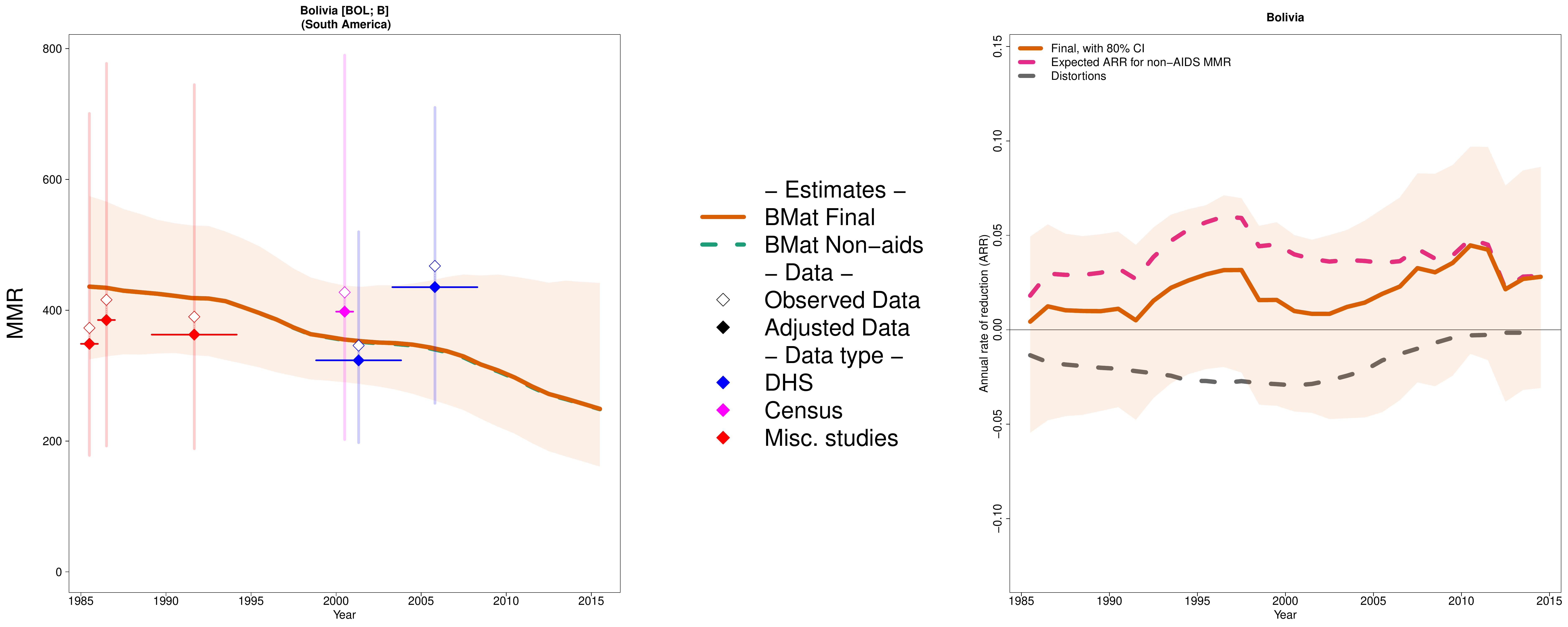}
		\caption{\textbf{Illustration of BMat model set-up for Bolivia.} Left: MMR (maternal deaths per 100,000 live births) data and estimates (see Figure 1 for more details). Right: Annual rate of reduction in the non-AIDS MMR. } \label{fig-bang}
	\end{center}
\end{figure}

The model set-up is illustrated in Figure~\ref{fig-bang} for Bolivia. AIDS MMR deaths are negligible in Bolivia thus the MMR is approximately equal to the non-AIDS MMR. The covariate-based ARR estimates for the non-AIDS MMR (pink dashed line) follow from the multilevel regression model and suggest that the MMR decreased from 1985 to 2015. However, the data points suggest that the MMR declined less rapidly in the 1980s and 1990s than estimated by the covariates. Hence, in the BMat model, the estimate for the ARR distortion term $\mult'_{c,t}$ is negative for that period and the estimate for the (final) ARR is lower than the covariate-based estimate to capture the slower decline. This results in final BMat MMR estimates (red) that capture the trend as suggested by the data. When projecting past the most recent study period, the ARR distortion $\mult'_{c,t}$ converges to zero, hence the estimated final ARR converges to the rate of change implied by the regression model. In other words, when projecting beyond a country's data, the rate of change will converge towards the covariate-implied rate of change, as seen in the projected ARR for Bolivia.

%\clearpage

\subsubsection{VR data: usability, classification and inclusion criteria, and pre-processing }\label{sec-vrpreprocess}
The data inclusion criteria in BMat follow the MMEIG 2014 inclusion criteria except for VR data. For VR data, we consider the \textit{usability} of the observations for model inclusion, as well as in the VR data model. For VR observation $i$, its usability $u_{i}$ is defined as the fraction of all-cause deaths in the country-year for which causes have been assessed in the VR. It is the product of the completeness of the VR and the percentage of deaths with a well-defined cause: 
$$u_{i}= P_{i}^{(complete)}\cdot (1-p_{i}^{(ill)}),$$ where $p^{(ill)}_{i}$ refers to the proportion of VR deaths with ill-defined causes (as reported) and $P_{i}^{(complete)}$ refers to the estimated completeness of the VR, the proportion of all-cause deaths which were captured in the VR: $P_{i}^{(complete)} = \min\{1,d_{i}/D_{i}\}$, where $d_{i}$ refers to the number of all-cause deaths observed in the VR system in the respective country-year. 

Based on the assessment of usability and VR data availability, we categorize VR data as type I, II or III. A VR observation for country $c$ in year $t$ is classified to be of type I if its usability $u_{i}>80$\% and if the observation belongs to a continuous string (with no more than 1 year gap in between) of at least 3 VR observations in the respective country with usability above 60\%. With this classification, type I VR data is considered to be of high VR  quality because of its high usability and because of its occurrence in a window of high usability VR data, suggesting that the vital registration system is well-established. A VR observation is classified to be of type II if its usability $u_{i}>60$\% and if the observation belongs to a continuous string (with no more than 1 year gap in between) of at least 3 VR observations with usability above 60\%. While VR observations from type II may provide information about maternal mortality, they are considered less reliable that type I observations because of the low usability of the VR data. Finally, type III refers to miscellaneous data from vital registration and mortality reporting systems for which we generally cannot assess data quality. All remaining VR observations are excluded from the analysis. VR data are also excluded in periods with specialized studies to avoid double-counting of information because specialized studies are often conducted as extensions of existing VR data systems. 

In some included VR country-years, the observed number of maternal deaths $m_i$ is zero. Such a data entry cannot be used directly because the $\log(y_i)$ is not defined. Instead, the country-year is combined with surrounding years; the PMs and reference years for the observations for surrounding years are recalculated to include the all-cause deaths of the year with zero maternal deaths, and the country-year with zero maternal deaths is excluded. 

\subsubsection{Data models}\label{sec-datamodel}

For $y_i$, the observed proportion of maternal or pregnancy-related deaths among all-cause deaths, we assume that for all data sources
\begin{eqnarray}
\log(y_{i})|\Gamma_i, \gamma_i &\sim& N\left(\log\left(\frac{\Gamma_{i}/\gamma_i}{D_i}\right), \sigma_i^2\right),\label{eq-datamodel}
\end{eqnarray}
where $\Gamma_{i}$ refers to the true maternal or pregnancy-related deaths for study-period $i$, $\gamma_i$ the underreporting parameter, % (the expected number of true maternal (or pregnancy-related) deaths associated with each observed maternal (or pregnancy-related) deaths), 
$D_i$ to the all-cause deaths and $\sigma_i^2$ to the error variance of the $i$-th observation. $\gamma_i$ and $\sigma_i^2$ will be discussed for each source type separately. The expression for $\Gamma_{i}$ depends on whether information on maternal or pregnancy-related deaths was collected:
\begin{eqnarray*}
	\Gamma_{i} &=& 
	\left\{	\begin{array}{ll}
		\md_i,& \text{definition $i$ is maternal},\\
		\mdna_i/\omega_{c[i]}+\mdap_i,& \text{definition $i$ is pregnancy-related}.
	\end{array}
	\right.
\end{eqnarray*}
The pregnancy-related AIDS deaths $\mdap_i = D\ai_{c[i],t[i]}\vv_{c[i],t[i]}$ were discussed in Section~\ref{sec-aidsmat}. Parameter $\omega_{c}$ is the definition-adjustment parameter for country $c$, which refers to the proportion of pregnancy-related non-AIDS deaths that are maternal in country $c$. Past UN analyses suggested that around 90\% of non-AIDS  pregnancy-related death can be classified as maternal deaths in countries in sub-Saharan Africa, compared to around 85\% in countries in other regions \shortcite{Wilmoth2012}. Similar to the MMEIG 2014 adjustment, this information was used in the BMat model to specify a prior distribution on $\omega_c$'s:
\begin{eqnarray*}
\omega_c&\sim& 
\left\{ \begin{array}{ll}
	TN_{(0,1)}(0.9,0.05^2), &\text{ for sub-Saharan Africa},\\
	TN_{(0,1)}(0.85,0.05^2), &\text{ for other regions}.
\end{array} \right.
\end{eqnarray*}

\paragraph{Specialized studies} For specialized studies, underreporting parameter $\gamma_i = 1$ because no underreporting is expected. Error variance  $\sigma_i^2 = s_{i}^2$, which is the stochastic variance associated with the $i$-th inquiry. This variance is approximated using a Binomial sampling model for the total number of maternal deaths %(out of the studies envelope of all-cause deaths)
and the delta method, with the maximum outcome set at 0.5.  

\paragraph{VR data}
The VR type I, II and III categorization and usability measures were used in the analysis of the VR underreporting adjustment $\gamma_i \def \gamma_{c[i],t[i]}$. When developing the BMat model, we explored various approaches to model this adjustment factor, e.g., based on time series models as described in \shortcite{chao2014informative}. While we plan to further explore such approaches in future research, for the current BMat model, we adhered to the main steps taken in the MMEIG 2014 approach to obtain VR adjustments and calculated an initial VR underreporting adjustment $g_{c,t}$ for all VR observations. These initial adjustment were used directly for type I data. We extended the MMEIG 2014 approach by (i) allowing for the possibility of larger adjustments for type II and III VR data, and (ii) incorporating additional uncertainty related to the adjustment in the total error variance for VR data of all types. 

The initial adjustments were obtained as follows (based on the MMEIG approach unless noted otherwise): for any country $c$ with type I and/or II data, as well as data from specialized studies, an initial estimate $g_{c,t}$ is calculated from the available information for all years $t$ while for all other country-years, the initial estimate $g_{c,t} = 1.5$, which corresponds to the specification of underreporting used in the MMEIG 2014 estimates. For countries with specialized studies, $g_{c,t}$ during periods with such studies was given by the ratio of the PM in the specialized study to the VR-based for the study period. 
Linear interpolation was performed for years in-between observed VR multipliers. For forward (or backward) extrapolation in the MMEIG 2014 approach, the VR multiplier was assumed to increase or decrease linearly from the most recent (and oldest) $g_{c,t}$ to 1.5 in 5 years. In BMat, the VR multiplier was kept constant  at the level of the most (or least) recent observed VR multiplier to avoid the introduction of spurious trends when lacking information on true trends in misclassification. The only exception was made in back-extrapolations if the least recent VR multiplier value was below 1.5, in which case the MMEIG approach of an extrapolation to 1.5 in 5 years was used, such that the addition of a recent specialized study that suggests misclassification lower than 1.5 does not reduce the MMR for the entire past for countries without additional studies. If the least recent study suggests an adjustment level which is higher than 1.5, then this level is used throughout in backward extrapolations because it is deemed less likely that the misclassification was better in the past as compared to more recent years. 

For all country-years of type I, the final adjustment is equal to the initial adjustment, $\gamma_{c,t} = g_{c,t}$. For lower-quality VR data from type II and III, $\gamma_{c,t}$ is assigned a prior distribution to allow for the possibility that the adjustment is greater than the initial estimate. The lower bound for $\gamma_{c,t}$ is given by the initial estimate $g_{c,t}$ and its upper bound $g^{(upper)}_{c,t}$ is given by
\begin{eqnarray*}
	g^{(upper)}_{c,t}&=&
	\left\{
	\begin{array}{ll}
		g_{c,t} + (3-g_{c,t})\cdot(0.8 - u_{c,t})/0.2, &\text{ for VR-II with $g_{c,t}<3$},\\
		3, & \text{ otherwise},
	\end{array}\right.
\end{eqnarray*}
such that  $g^{(upper)}_{c,t}$ decreases from 3 to $g_{c,t}$ as $u_{c,t}$ increases from 60 to 80\% (from lowest quality type II to type I). The prior $p(\gamma_{c,t})$ was set up such that the possibility of a greater outcome decreases as the quality (usability) of type II VR data improves, using a mixture of a uniform distribution with support on ($g_{c,t}, g^{(upper)}_{c,t})$ and point mass on $g_{c,t}$:
\begin{eqnarray*}
	p(\gamma_{c,t}) &=&  \Pr (\gamma_{c,t} \neq g_{c,t})\cdot \frac{\bm{1}_{(g_{c,t}, g_{c,t}^{(upper)})}(\gamma_{c,t})}{g_{c,t}^{(upper)}-g_{c,t}} +\Pr (\gamma_{c,t} = g_{c,t})\cdot \delta(\gamma_{c,t} - g_{c,t}),
\end{eqnarray*}
where $\delta(\cdot)$ denotes the Dirac delta function and $\Pr(\gamma_{c,t} = g_{c,t}) = \frac{g_{c,t}-\min\{1, g_{c,t}-0.5\}}{g_{c,t}^{(upper)}-\min\{1, g_{c,t}-0.5\}}$, such that point mass $P(\gamma_{c,t} = g_{c,t})$ increases to 1 as $g^{(upper)}_{c,t}$ decreases to $g_{c,t}$ with improved usability. 

For all VR country-years, total error variance $s_i^2$ in Eq.~\ref{eq-datamodel} includes includes the stochastic variability in the VR-based PM as well as the uncertainty in the VR misclassification adjustment $\gamma_{c[i],t[i]}$ (beyond the additional uncertainty in the point estimate of the adjustment for type II and III data). The total variance is obtained through a monte carlo simulation of VR-deaths $m_{i}^{(h)}$ for draws $h=1,2, \hdots, H$, with
\begin{eqnarray*}
	\log(\tilde{\gamma}_{c,t}^{(h)}) &\sim& N(\log(g_{c,t}), 0.25^2),\\
	m_{i}^{(h)} &\sim&  Bin(d_i, y_i/g_{c,t}\cdot \tilde{\gamma}^{(h)}_{c,t}), 
\end{eqnarray*}
in which we assume that the true adjustment factor $\tilde{\gamma}_{c,t}^{(h)}$ is roughly within 60 to 160\% of $g_{c,t}$, and we obtain a draw $m_{i}^{(h)}$ based on estimated true PM $y_i/g_{c,t}$ and adjustment $\tilde{\gamma}_{c,t}^{(h)}$. The inclusion of the additional uncertainty associated with the VR adjustment $\gamma$ is motivated by the great variability in observed misclassification adjustments \shortcite{chao2014informative}. 
The error variance for the log(PM) is approximated with the delta method and a maximum of 0.5 for the total error $\sigma_{i}$ is used for type I data. 

\paragraph{Data from other sources}
For data from other sources, in line with MMEIG 2014, the underreporting parameter $\gamma_i = 1.1$ to account for 10\% underreporting \shortcite{Wilmoth2012}. The error variance 
\begin{eqnarray*}
\sigma_i^2 &=& s_i^2 + \sigma_{(DHS)}^2\cdot \bm{1}_{\text{DHS}}(i) + \sigma_{(not-DHS)}^2\cdot (1-\bm{1}_{\text{DHS}}(i)),
\end{eqnarray*}
with sampling error variance $s_i^2$ and a non-sampling error variance term $\sigma^2_{(DHS)}$ for observations from DHS with $\bm{1}_{DHS}(i)= 1$, and  $\sigma^2_{(not-DHS)}$ for other sources. Diffuse priors are used for the non-sampling variance parameters. The sampling error $s_i$ was calculated for all data entries with available micro data using the jackknife repeated replication procedure, which is the standard method used by the DHS to generate standard errors for complex statistics such as mortality and fertility rates \shortcite{dhsreport2014}. %. Please see pages 293-294 
For other sources where sampling errors are not available, we set the sampling error equal to the maximum of 0.25 and the maximum sampling error observed within the country (if any). 

\subsubsection{Computation}\label{sec-compu}
A Markov Chain Monte Carlo (MCMC) algorithm was employed to sample from the posterior distribution of the parameters with the use of the software \texttt{JAGS} \shortcite{plummer2003}. Six parallel chains were run with a total of 65,000 iterations in each chain. Of these, the first 5,000 iterations in each chain were discarded as burn-in and every 120th iteration after was retained. The resulting chains contained 3,000 samples each. Standard diagnostic checks (using trace plots and the Gelman and Rubin diagnostic \shortcite{gelmanrubin1992}) were used to check convergence. 

\subsubsection{Communicating uncertainty}\label{sec-commu}
Estimates of relevant quantities are given by the posterior medians while 80\% credible intervals (CIs) were constructed from the 10th and 90th percentiles of the posterior samples. Given the inherent uncertainty in MMR estimates, 80\% CIs are presented by the UN MMEIG instead of the more conventional 95\% ones. 

To communicate what information one observation $y_i$ provides about $\md_i$, and thus about the true PM or MMR, based on Eq.(\ref{eq-datamodel}), we first constructed approximate 80\% confidence intervals for $\Gamma_{i}/(\gamma_iD_i)$, given by $\exp\left(\log(y_i) \pm z_{0.9}\hat{\sigma}_i\right)$, where $\hat{\sigma}_i^2$ refers to calculated total variance for observations from VR and specialized studies, and the combination of sampling error and the posterior median of the non-sampling error variance for observations from other sources. We then transformed these confidence intervals to the required reporting scale (PM or MMR) using posterior median estimates $\hat{\gamma}_i$ and $\hat{\omega}_c$ of adjustment parameters $\gamma_i$ and $\omega_c$. 

%\clearpage
\subsubsection{Model validation}\label{sec-val}

Model performance was assessed through two out-of-sample validation exercises. In the first exercise, 20\% of the observations were left out at random. In the second exercise, we left out approximately 20\% of the most recent data, here 
corresponding to all data in or after 2007: fitting the BMat model to this training set resulted in point estimates and CIs that would have been constructed in 2007 based on the proposed method. To validate model performance, we calculated various validation measures based on the left-out observations. The considered measures were based on prediction errors for the MMR (as opposed to the PM to improve interpretation of findings), where an error refers to the difference between the left-out observation and the median of its posterior predictive distribution based on the training set, as well as coverage of 80\% prediction intervals (to quantify the calibration of the prediction intervals). For the second validation exercise, as in similar past studies on the estimation of child mortality \shortcite{alkema2014global}, we also compared the estimates and CIs for the MMR obtained from the training set to the estimates from the full data set to check the predictive performance of the model.

For the left-out observations, errors in the observed MMR are defined as $e_i = (y_i - \tilde{y}_i)D_i/B_i$, where $\tilde{y}_i$ denotes the posterior median of the predictive distribution for a left-out observed PM $y_i$ based on the training set, and relative errors are given by $e_i/(\tilde{y}_i D_i /B_i)$. Coverage is given by $1/N \sum 1[y_i \geq  l_i] \cdot 1[ y_i \leq r_i]$, where $N$ denotes the total number of left-out observations considered and $l_i$ and $r_i$ the lower and upper bounds of the 80\% predictions intervals for the $i$-th observation. ``Updated'' estimates, denoted by $\mmmr_{c,t}$ for country $c$ in year $t$, refer to the median MMR estimates obtained from the full data set. The error in the estimate based on the training sample is defined as $\hat{e}_{c,t} = \mmmr_{c,t} - \mmmr_{c,t}^{(2007)}$, where $\mmmr_{c,t}^{(2007)}$ refers to the posterior median estimate based on the training sample as constructed in 2007, while relative error is defined as $\hat{e}_{c,t}/\mmmr_{c,t} \cdot 100$. Coverage was calculated in a similar matter as for the left-out observations, based on the lower and upper bound of the $80$\% CIs for the MMR obtained from the training set. Results are reported for $t=2007$.

\section{Results}

\subsection{BMat MMR estimates}
The BMat estimates are shown for the selected countries in Figure 1. MMEIG 2014 estimates are shown in the same figure. 

BMat estimates are generally consistent with the MMEIG 2014 estimates for former group A countries but small differences exist. In MMEIG 2014, estimates for group A countries are given by the adjusted VR data for 5-year periods. The BMat model estimates may differ because BMat estimates are constructed for 1-year periods, the BMat model allows for smoothing of the stochastic fluctuations in individual data points, and in the BMat model, the VR adjustment is kept constant in forward and in (most) backwards extrapolations (as opposed to the convergence to 1.5 as used in the MMEIG 2014 estimates). In addition, BMat takes into account different levels of uncertainty. Data from specialized studies tend to have a much smaller error variance compared to VR data. Hence, in the model, BMat estimates tend to be close to observations from specialized studies and estimates tend to be more certain for periods with such studies. For Japan, the only group A country in Figure 1, these differences result in slightly updated MMR estimates but differences since 1990 are small.

Former Group B countries are those countries with lower quality VR data and/or data from other sources. In MMEIG 2014, these data points informed only the  estimation of the level of maternal mortality in the country but not the trend; trend estimates were driven by the covariates. By introducing the country-year-specific ARIMA(1,1,1) multiplier, BMat estimates can capture data-driven trends, as explained for Bolivia in the previous section. For the other group B countries in Figure 1, the BMat and MMEIG 2014 estimates are similar for India but differ for El Salvador and Zimbabwe, where BMat estimates deviate from the covariate-based estimates where data are available and in disagreement with the covariate-based trend. For El Salvador, the BMat estimates capture the MMR level as indicated by the specialized study around 2006. The VR data in El Salvador is of type II, with initial adjustment given by 2.3 and upper bound $g_{c,t}^{(upper)}$ greater than 2.3 because usability varies between 60 and 65\% during the study period. The BMat estimates differ from the covariate-based MMEIG 2014 estimates in the earlier period because of the adjusted VR data as well as data from another source (DHS). For Zimbabwe, the modeled estimates for 2006 correspond to the MMR as recorded in the specialized study. The estimates for earlier years differ from the MMEIG estimates because data suggest lower MMR levels.

For former group C countries, with no data available, the BMat model continues to produce covariate-based estimates which are similar to the MMEIG 2014 estimates, as illustrated in Figure 1 for Papua New Guinea.

PM and MMR data and estimates for all countries are given in supplementary figure S1. 

\clearpage
\subsection{Validation results}
Validation results are presented separately for developed and developing countries (according to the categorization used for MDG reporting) because results may differ between those two groups. Model validation results based on the left-out observations are shown in Tables~\ref{tab-errors-obs}. Median errors and median relative errors are generally close to zero, except when leaving out recent data in developing countries: the median error in the MMR is 6.5 deaths per 100,000 live births, corresponding to a relative error of 15.2\%. This suggests that projections may be somewhat optimistic and overestimate the MMR decline in developing countries. Coverage of 80\% prediction intervals is reasonable and between 80 and 90\% for both exercises and for both developing as well as developed countries. The greatest asymmetry regarding the percentage of left-out observations above or below the prediction intervals is observed when leaving out recent data in developed countries, with 11.6\% of the observations falling above, and 4.7\% falling below their respective intervals. 

The comparison between estimates based on the training and full data set for the year 2007 is given in Table \ref{tab-errors-est}. Errors, which here are the differences in the MMR estimates for the year 2007 between full and training data set, are reasonably small with median absolute relative errors just above 10\% for both developed as well as developing countries. Coverage of the 80\% CIs based on the training set is greater than 80\%, as expected and desired, indicating that the addition of more recent data does not tend to move the MMR estimates outside previously constructed CIs.

\begin{table}[htbp]
\begin{center}
\caption{\textbf{Validation results based on left-out observations, for developed and developing countries.} The outcome measures are: median error (ME), absolute error (MAE), relative error (MRE) and absolute relative error (MARE) for the MMR (per 100,000 live births), as well as the \% of left-out observations below and above the 80\% prediction interval (PI) based on the training set. Results for exercise II refer to the most recent left-out observation in each country.}
\label{tab-errors-obs}
\begin{tabular}{@{}|l|c|cc|cc|cc|@{}}
\hline
&\# of left-out &  \multicolumn{2}{c|}{error in MMR} & \multicolumn{2}{c|}{relative error (\%)}	 &  \multicolumn{2}{c|}{outside 80\% PI} \\ 	
& observations  & ME &	MAE &   MRE & MARE & 	\% Below & 	\% Above \\ 	\hline
\multicolumn{8}{|l|}{Exercise I: appr. 20\% of observations were excluded at random} \\\hline
Developed countries &  187 & 0.4 & 1.9 & 3.5 & 23.7 & 9.1 & 5.9 \\ 
Developing countries & 248 & 1.6 & 8.3 & 2.3 & 16.8 & 4.4 & 6.9 \\ 
\hline
\multicolumn{8}{|l|}{Exercise II: all observations in and after 2007 were excluded}\\\hline
Developed countries &  43   & 0.2 & 1.5 & 2.5 & 30.0 & 11.6 & 4.7\\
Developing countries & 80& 6.5 & 17.1 & 15.2 & 31.0 & 7.5 & 11.2\\
\hline
\end{tabular}

\end{center}
\end{table}

\begin{table}[htbp]
	\caption{{\bf Validation results for MMR estimates, for the validation exercise II whereby all observations in and after 2007 were left out, for developed and developing countries.} 
		Outcome measures are reported for the MMR (per 100,000 live births) in 2007 for countries for which observations were left out and given by the median error (ME), absolute error (MAE), relative error (MRE) and absolute relative error (MARE), as well as the \% of countries for which MMR estimates in 2007 are below and above the 80\% CIs based on the training set.}
	\label{tab-errors-est}   
	\begin{center}
		\begin{tabular}{@{}|l|c|cc|cc|cc|@{}}
			\hline
			& &  \multicolumn{2}{c|}{error in MMR} & \multicolumn{2}{c|}{relative error (\%)}	 &  \multicolumn{2}{c|}{outside 80\% CI} \\ 	
			& \# of countries & ME &	MAE &   MRE & MARE & 	\% Below & 	\% Above \\ 	\hline
			
			Developed countries & 43 & 0.2 & 1.0 & 2.8 & 12.2 & 9.3 & 7.0 \\ 
			Developing countries & 80 & 3.3 & 9.5 & 4.7 & 11.9 & 2.5 & 6.2 \\ 
			\hline
		\end{tabular}
	\end{center}
\end{table}
%\clearpage

\section{Discussion}
Estimating maternal mortality is challenging because of limited maternal mortality data availability, especially for recent years for many developing countries, and moreover, the substantial uncertainty surrounding observations due to reporting issues and random errors associated with the observations. We have described the BMat model for estimating the MMR for all countries in the world. This Bayesian model extends the approach previously used by the UN MMEIG to better capture trends in, and uncertainty around, country data: it combines the rate of change implied by a multilevel regression model with a flexible time series model to capture data-driven changes in country-specific MMRs. As compared to the past MMEIG approach, the new modeling approach eliminates the need  for grouping countries based on data availability: one model is used for all countries, regardless of data sources available. In addition, BMat includes a data model to adjust for systematic and random errors associated with different types of data sources, such that observations are adjusted and weighted appropriately when constructing MMR estimates.

Model validation exercises suggested that the BMat model is reasonably well calibrated but indicated, based on out-of-sample projections for more recent periods, that recent maternal mortality declines may be overestimated in developing countries. An investigation into the regression model used to estimate and project the systematic change in the non-AIDS MMR may result in the selection of an alternative model with improved performance. 
The calibration of the current model was found to be satisfactory with respect to coverage of prediction intervals for left-out observations and confidence intervals for MMR estimates, and differences in MMR points estimates between full and training data set were small.

With the development of BMat, we illustrated how a regression model can be extended through the addition of an ARIMA process to provide a more flexible model set-up to capture levels and trends in the outcome of interest, as indicated by the data, while still providing regression-based results for populations and/or time periods for which limited or no information is available. 
At the same time, we showed how to improve upon simplified assumptions regarding observational errors, i.e., the use of error variances that are equal across observations in the MMEIG 2014 model, by incorporating appropriate data models.
The change, from a relatively standard (frequentist) regression model to a Bayesian model that combines regression functions with time series models and includes appropriate data models, may be appropriate for the modeling of other global health or demographic indicators, for which regression models are currently being used. 

Important limitations to our study are related to maternal mortality reporting issues, the estimation of AIDS maternal deaths, and the dependency of the estimation of maternal mortality on the estimation of other demographic indicators. 
The estimation of underreporting in surveys and miscellaneous data sources, as well as the estimation of the proportion of pregnancy-related deaths that are maternal, are hindered by limited availability of data which allow for a detailed analysis. Additional information is needed to overcome these current data limitations. Regarding the estimation of pregnancy-related deaths which are not maternal, future analyses may be possible based on an increasing number of studies that report both pregnancy-related as well as maternal mortality information. Further analysis of the misclassification of maternal deaths in vital registration systems would aid the estimation of maternal mortality in developed countries with well-functioning VR systems, as well as in countries in which VR systems have improved in recent years. The second limitation of our study is due to the limited information on AIDS maternal mortality \shortcite{Wilmoth2012}, which complicates the reconstruction of trends in maternal mortality in countries with generalized HIV/AIDS epidemics. This limitation is of lesser concern for more recent years in which the contribution of AIDS maternal deaths to the overall number of maternal deaths has decreased. Lastly, because of the dependency of the maternal mortality estimation on the estimation of all-cause deaths to women of reproductive ages as well as the number of births, the challenges and limitations that apply to the estimation of these demographic indicators are also applicable to maternal mortality estimation. We did not include the uncertainty surrounding these demographic indicators into the uncertainty assessment for maternal mortality because such uncertainty assessments are generally not available. Future research work regarding the assessment of uncertainty in these demographic indicators 
may result in the reporting of uncertainty intervals for a wider range of demographic indicators and allow for a more complete uncertainty assessment for maternal mortality. 

Given the ending of the MDGs this year, discussions about a new set of goals have been ongoing and will result in the approval of the Sustainable Development Goals (SDGs) later this year (see {http://sustainabledevelopment2015.org/}). Regarding maternal mortality, the SDGs will include the goal to reduce the global MMR to 70 deaths per 100,000 births by 2030, and in addition, to reduce the MMR in each country to be lower than twice the global MMR, thus 140, by the same year \shortcite{epmm}. Continued monitoring of levels of and changes in maternal mortality will be necessary to evaluate progress and aid resource allocation and priority setting. Given the great uncertainty surrounding maternal mortality indicators, greater emphasis should be placed on the communication of uncertainty intervals to avoid confusion, e.g. when point estimates of the MMR in the baseline year for SDG assessment are updated in light of new data, as described for under-five mortality monitoring \shortcite{oestergaard&2013}. Despite the challenges related to the measurement of maternal mortality and the limitations of MMR estimates, the importance of monitoring maternal survival necessitates the continued production of estimates and justifies a continued effort to further improve the validity of maternal mortality estimates.

\section*{APPENDIX}
The BMat model for the MMR $\mmr_{c,t} = \md_{c,t}/B_{c,t}$ is specified as follows: 
\begin{eqnarray*}
\md_{c,t} &=& \mdna_{c,t} + \mda_{c,t},\\
\mdna_{c,t} &=& \mmrna_{c,t}\cdot B_{c,t},\\
\mmrna_{c,t}  &=& \emmrna_{c,t}\cdot \mult_{c,t}.
\end{eqnarray*}

The expected non-AIDS MMR $\emmrna_{c,t}$ is obtained using a multilevel regression model for the expected non-AIDS maternal deaths $\emdna_{c,t}$:
\begin{eqnarray*}
\emmrna_{c,t} &=& \emdna_{c,t}/B_{c,t},\\
log(\emdna_{c,t})&=& \log(D\na_{c,t}) +  \alpha_{c} - \beta_1 \log(x_{c,t}^{(GDP)}) +\beta_2 \log(x_{c,t}^{(GFR)}) -\beta_3 x_{c,t}^{(SAB)},\\
\alpha_c|\alpha_{r[c]}, \sigma_{country}^2 &\sim& N(\alpha_{r[c]}, \sigma_{country}^2),\\ \alpha_r|\alpha_{world}, \sigma_{region}  &\sim& N(\alpha_{world}, \sigma_{region}^2),\\
\alpha_{world} &\sim& N(log(0.001), 100),\\
\sigma_{\alpha, country} &\sim& U(0,5),\\
\sigma_{\alpha, region} &\sim& U(0,5),\\
\beta_h &\sim& N(0.5, 1000), \text{ for } h=1,2,3.
\end{eqnarray*}

The log-transformed multiplier $\mult_{c,t}$ is modeled with an ARIMA(1,1,1) model with $\mult_{c,1990} =1$ and for $t=1985, 1986, \hdots, 2014$:
\begin{eqnarray*}
\log(\mult_{c,t+1}) &=& \log(\mult_{c,t}) - \mult'_{c,t},\\
\mult'_{c,t} &=& -\log(\mult_{c,t+1}/\mult_{c,t}),\\
 &=& \phi \mult'_{c,t-1} - \theta \epsilon_{c,t-1} + \epsilon_{c,t},\\
\epsilon_{c,t} &\sim& N(0, \sigma_{c}^2).
\end{eqnarray*}
Initial conditions follow from the stationary distribution for $\mult'_{c,t}$:
\begin{eqnarray*}
\mult'_{c,1985} &\sim& N(0, {\gamma}_{0,c}),\\
\epsilon_{c,1985} &\sim& N(\sigma_{c}^2/{\gamma}_{0,c}\cdot \mult'_{c,1985}, \hspace{0.1cm} \sigma_{c}^2(1-\sigma_{c}^2/{\gamma}_{0,c})),
\end{eqnarray*}
where $ {\gamma}_{0,c}$ refers to the stationary variance of $\mult'_{c,t}$, ${\gamma}_{0,c} = \frac{1-2\phi\theta + \theta^2}{1-\phi^2}\cdot
\sigma_{c}^2$. Priors for $\phi$ and $\theta$ and the hierarchical distribution for ${\gamma}_{0,c}$ are as follows:
\begin{eqnarray*}
\phi &\sim& U(0,1),\\
\theta &\sim& U(-1,0),\\
\sqrt{\gamma_{0,c}}&=&\sqrt{\gamma_0}\cdot (1+\lambda_c),\\
\sqrt{\gamma_0} &\sim& U(0,0.025),\\ 
\lambda_c &\sim& TN_{(-1,2)}(0,\sigma_{\lambda}^2),\\
\sigma_{\lambda} &\sim& U(0,2),
\end{eqnarray*}
where the upper bound for $\sqrt{\gamma_{0,c}}$ was based on the assumption that for country-periods without any data, we expect the true ARR to be roughly within $\pm 0.05$ of the covariate-based ARR. 
Model constraints were included such that $\emdna_{c,t} < D\na_{c,t}$ and $ \md_{c,t} < D_{c,t}$. 

Data models for observations $i=1,2,\hdots, N$ are given by:
\begin{eqnarray*}
\log(y_{i})|\Gamma_i, \gamma_i &\sim& N\left(\log\left(\frac{\Gamma_{i}/\gamma_i}{D_i}\right), \sigma_i^2\right),\\
\Gamma_{i} &=& 
\left\{	\begin{array}{ll}
	\md_i,& \text{definition $i$ is maternal},\\
	\mdna_i/\omega_{c[i]}+\mdap_i,& \text{definition $i$ is pregnancy-related},
\end{array}	\right.\\
\omega_c&\sim& 
\left\{ \begin{array}{ll}
TN_{(0,1)}(0.9,0.05^2), &\text{ for sub-Saharan Africa},\\
TN_{(0,1)}(0.85,0.05^2), &\text{ for other regions}.
\end{array} \right.
\end{eqnarray*}
The reporting adjustment parameter $\gamma_i=1$ for specialized studies and 1.1 for miscellaneous studies. For VR observations with $\gamma_i \equiv \gamma_{c[i],t[i]}$, country-year adjustment $\gamma_{c,t} = g_{c,t}$ for type I data while for VR data of types I and II, it is assigned a prior distribution $p(\gamma_{c,t})$:
\begin{eqnarray*}
	p(\gamma_{c,t}) &=&  \Pr (\gamma_{c,t} \neq g_{c,t})\cdot \frac{\bm{1}_{(g_{c,t}, g_{c,t}^{(upper)})}(\gamma_{c,t}) }{g_{c,t}^{(upper)}-g_{c,t}} +\Pr (\gamma_{c,t} = g_{c,t})\cdot \delta(\gamma_{c,t} - g_{c,t}),\\
	g^{(upper)}_{c,t}&=&
	\left\{
	\begin{array}{ll}
		g_{c,t} + (3-g_{c,t})\cdot(0.8 - u_{c,t})/0.2, &\text{ for VR-II with $g_{c,t}<3$},\\
		3, & \text{ otherwise}.
	\end{array}\right. 
\end{eqnarray*}	
Finally, the total error variance $\sigma_i^2$ is given by stochastic variance for VR data and specialized studies, and estimated as follows for other sources:
\begin{eqnarray*}
\sigma_i^2 &=& s_i^2 + \sigma_{(DHS)}^2\cdot \bm{1}_{\text{DHS}}(i) + \sigma_{(not-DHS)}^2\cdot (1-\bm{1}_{\text{DHS}}(i)),\\
\sigma_{(DHS)} &\sim& U(0.1, 0.5),\\
\sigma_{(not-DHS)} &\sim& U(0.1, 0.5).
\end{eqnarray*}
where $s_i^2$ denotes sampling error variance.

\paragraph{Acknowledgments}
The authors are very grateful to all members of the (Technical Advisory Group of the) United Nations Maternal Mortality Estimation Inter-agency Group for discussions and comments which have greatly improved this work. Additional thanks to Bilal Barakat, Jeff Eaton and Emily Peterson for specific comments and suggestions related to the BMat model and earlier versions of this manuscript. We also thank the numerous survey participants and the staff involved in the collection and publication of the data that we analyzed.

\bibliographystyle{chicago}
\bibliography{bibmaternal20151110}

\begin{thebibliography}{}

\bibitem[\protect\citeauthoryear{Alkema, Chou, Hogan, Zhang, Moller, Gemmill,
  Fat, Boerma, Temmerman, Mathers, and Say}{Alkema
  et~al.}{ming}]{alkemaetalforthcoming}
Alkema, L., D.~Chou, D.~Hogan, S.~Zhang, A.-B. Moller, A.~Gemmill, D.~M. Fat,
  T.~Boerma, M.~Temmerman, C.~Mathers, and L.~Say (forthcoming).
\newblock Global, regional and national levels and trends in maternal mortality
  between 1990 and 2015, with scenario-based projections to 2030: a systematic
  analysis by the {UN} {M}aternal {M}ortality {E}stimation {I}nter-{A}gency
  {G}roup.
\newblock {\em The Lancet\/}.

\bibitem[\protect\citeauthoryear{Alkema, New, et~al.}{Alkema
  et~al.}{2014}]{alkema2014global}
Alkema, L., J.~R. New, et~al. (2014).
\newblock {Global estimation of child mortality using a Bayesian B-spline
  bias-reduction model}.
\newblock {\em The Annals of Applied Statistics\/}~{\em 8\/}(4), 2122--2149.

\bibitem[\protect\citeauthoryear{Chao and Alkema}{Chao and
  Alkema}{2014}]{chao2014informative}
Chao, F. and L.~Alkema (2014).
\newblock {How Informative are Vital Registration Data for Estimating Maternal
  Mortality? A Bayesian Analysis of WHO Adjustment Data and Parameters}.
\newblock {\em Statistics and Public Policy\/}~{\em 1\/}(1), 6--18.

\bibitem[\protect\citeauthoryear{Gelman and Rubin}{Gelman and
  Rubin}{1992}]{gelmanrubin1992}
Gelman, A. and D.~Rubin (1992).
\newblock Inference from iterative simulation using multiple sequences.
\newblock {\em Statistical Science\/}~{\em 7}, 457--511.

\bibitem[\protect\citeauthoryear{{ICF International Inc.}}{{ICF International
  Inc.}}{2014}]{dhsreport2014}
{ICF International Inc.} (2014).
\newblock {Guidelines for the MEASURE DHS Phase III Main Survey Report}.
\newblock Technical report.
\newblock http://dhsprogram.com/pubs/pdf/DHSM6/
  Final\_Report\_Tab\_Plan\_24Oct2014\_DHSM6.pdf.

\bibitem[\protect\citeauthoryear{Oestergaard, Alkema, and Lawn}{Oestergaard
  et~al.}{2013}]{oestergaard&2013}
Oestergaard, M.~Z., L.~Alkema, and J.~E. Lawn (2013).
\newblock Millennium {D}evelopment {G}oals national targets are moving targets
  and the results will not be known until well after the deadline of 2015.
\newblock {\em International Journal of Epidemiology\/}~{\em 42\/}(3),
  645--647.

\bibitem[\protect\citeauthoryear{Plummer}{Plummer}{2003}]{plummer2003}
Plummer, M. (2003).
\newblock {JAGS}: {A} {P}rogram for {A}nalysis of {B}ayesian {G}raphical
  {M}odels {U}sing {G}ibbs {S}ampling.
\newblock In {\em Proceedings of the 3rd International Workshop on Distributed
  Statistical Computing (DSC 2003), March 20-22, Vienna, Austria. ISSN
  1609-395X. http://mcmc-jags.sourceforge.net/}.

\bibitem[\protect\citeauthoryear{UNAIDS}{UNAIDS}{2013}]{UNAIDS2013}
UNAIDS (2013).
\newblock {\em Global report: UNAIDS report on the global AIDS epidemic 2013}.
\newblock Geneva: Joint United Nations Programme on HIV/AIDS.

\bibitem[\protect\citeauthoryear{{United Nations {P}opulation
  {D}ivision}}{{United Nations {P}opulation {D}ivision}}{2013}]{wpp2012}
{United Nations {P}opulation {D}ivision} (2013).
\newblock {\em {W}orld {P}opulation {P}rospects. {T}he 2012 {R}evision}.
\newblock {U}nited {N}ations publication.

\bibitem[\protect\citeauthoryear{{WHO, UNICEF, UNFPA and The World Bank}}{{WHO,
  UNICEF, UNFPA and The World Bank}}{2014}]{whomm2014}
{WHO, UNICEF, UNFPA and The World Bank} (2014).
\newblock Trends in maternal mortality 1990-2013: Estimates developed by {WHO,
  UNICEF, UNFPA} and {T}he {W}orld {B}ank.
\newblock {ISBN} ISBN 978 92 4 150722 6.

\bibitem[\protect\citeauthoryear{Wilmoth, Mizoguchi, Oestergaard, Say, Mathers,
  Zureick-Brown, Inoue, and Chou}{Wilmoth et~al.}{2012}]{Wilmoth2012}
Wilmoth, J.~R., N.~Mizoguchi, M.~Z. Oestergaard, L.~Say, C.~D. Mathers,
  S.~Zureick-Brown, M.~Inoue, and D.~Chou (2012).
\newblock {A New Method for Deriving Global Estimates of Maternal Mortality}.
\newblock {\em Statistics, Politics, and Policy\/}~{\em 3\/}(2).

\bibitem[\protect\citeauthoryear{{World Health Organization}}{{World Health
  Organization}}{2010}]{icd}
{World Health Organization} (2010).
\newblock {\em International statistical classification of diseases and related
  health problems, tenth revision: Instruction manual.}
\newblock Geneva: World Health Organization.

\bibitem[\protect\citeauthoryear{{{W}orld {H}ealth {O}rganization}}{{{W}orld
  {H}ealth {O}rganization}}{2014}]{wholt}
{{W}orld {H}ealth {O}rganization} (2014).
\newblock {\em Life tables for {WHO} Member States 1990- 2012}.
\newblock Geneva: {W}orld {H}ealth {O}rganization.

\bibitem[\protect\citeauthoryear{{World Health Organization}}{{World Health
  Organization}}{2015}]{epmm}
{World Health Organization} (2015).
\newblock {Strategies toward ending preventable maternal mortality (EPMM)}.
\newblock Technical report.
\newblock
  http://apps.who.int/iris/bitstream/10665/153544/1/9789241508483\_eng.pdf?ua=1.

\end{thebibliography}

%\clearpage
	\begin{supplement}[id=suppA]
		\sname{Figure S1}
		\stitle{Data series and estimates of the PM (proportion of all-cause deaths that are maternal) and the MMR (number of maternal deaths per 100,000 live births) for  for 183 countries. }
		%\slink[doi]{...} ...-AOASXXXXSUPP}
		%\sdatatype{"Supplementary figure S1".pdf} 
		\sdescription{BMat estimates are illustrated by the solid red lines and 80\% CIs are shown by the red shaded areas. The UN MMEIG 2014 estimates are illustrated with the green lines. Reported (unadjusted) and adjusted observations are displayed. The vertical line with each adjusted observation indicates the approximate 80\% confidence interval for the PM or MMR associated with that observation, based on point estimates for reporting adjustments and total error variance. Estimates are shown for the period 1990-2015, data before 1990 were used in model fitting. \\
			Note: these estimates are obtained by fitting the model to 2014 MMEIG data base. Therefore, the BMat 2014 estimates presented here differ from the BMat and MMEIG 2015 estimates, which will be based on more recent data (Alkema et al. (forthcoming))\nocite{alkemaetalforthcoming}.}
	\end{supplement}
	


\section*{Supplementary material}

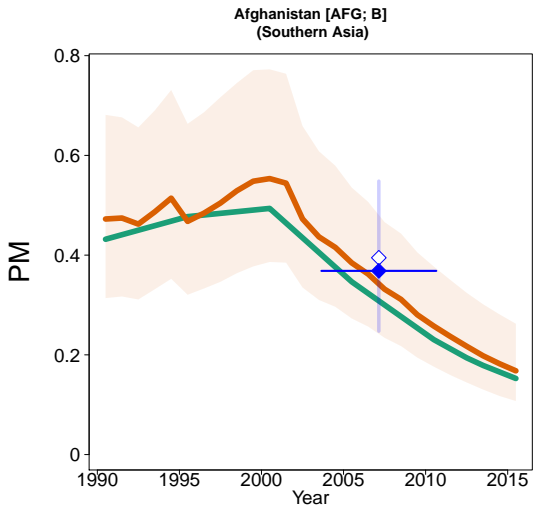

Afghanistan [AFG; B]
(Southern Asia)

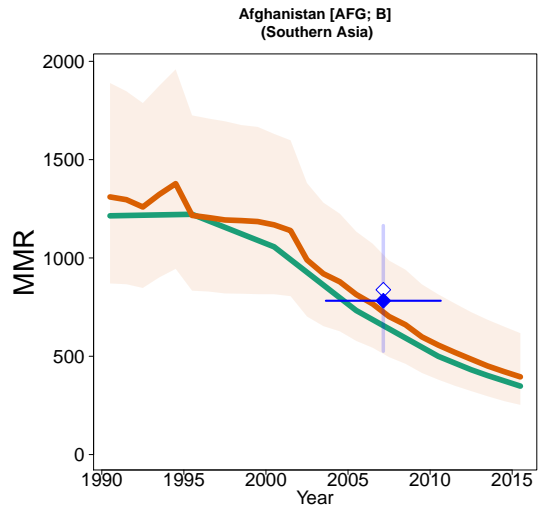

Afghanistan [AFG; B]
(Southern Asia)

– Estimates –
Bmat 2014
MMEIG 2014
– Data –
◇ Observed Data
◆ Adjusted Data
– Data type –
◆ DHS

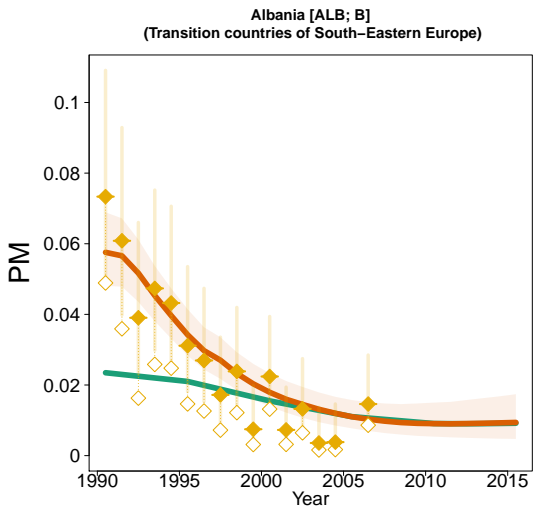

Albania [ALB; B]
(Transition countries of South–Eastern Europe)

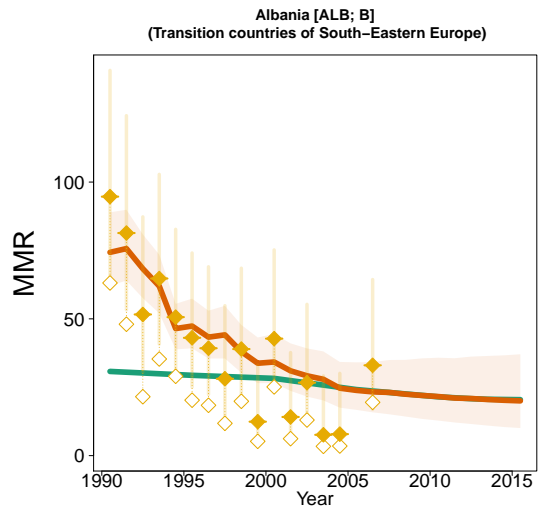

Albania [ALB; B]
(Transition countries of South–Eastern Europe)

– Estimates –
Bmat 2014
MMEIG 2014
– Data –
◇ Observed Data
◆ Adjusted Data
– Data type –
◆ VR

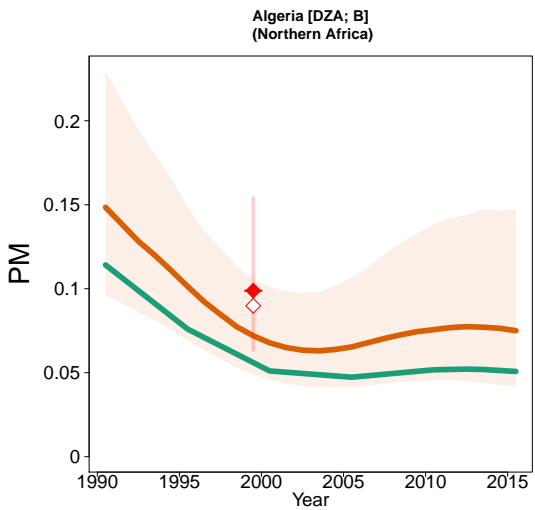

Algeria [DZA; B]
(Northern Africa)

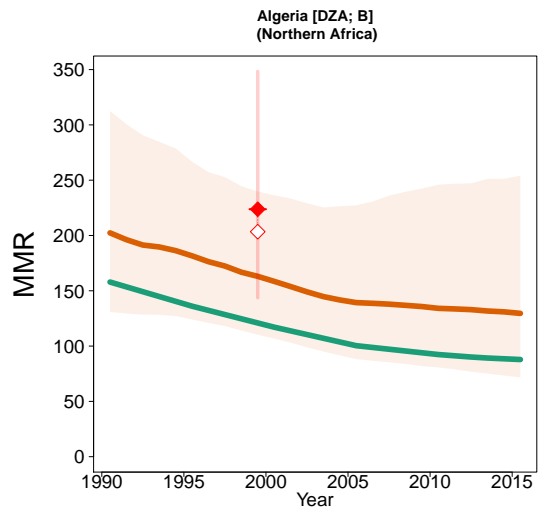

Algeria [DZA; B]
(Northern Africa)

– Estimates –
Bmat 2014
MMEIG 2014
– Data –
◇ Observed Data
◆ Adjusted Data
– Data type –
◆ Misc. studies

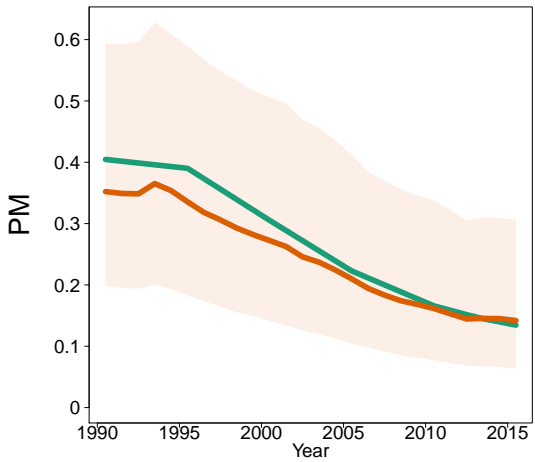

**Angola [AGO; C]**
**(Middle Africa)**

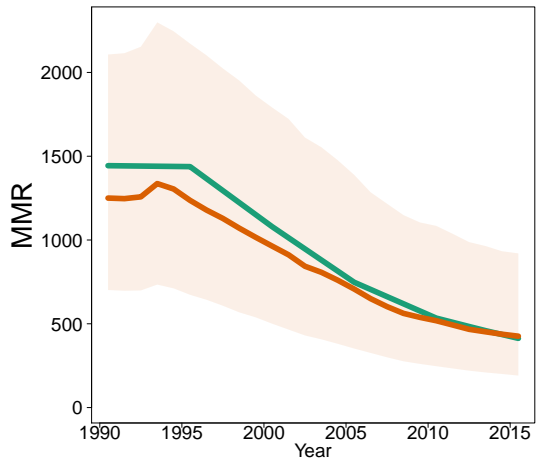

**Angola [AGO; C]**
**(Middle Africa)**

– Estimates –
Bmat 2014
MMEIG 2014
– Data –

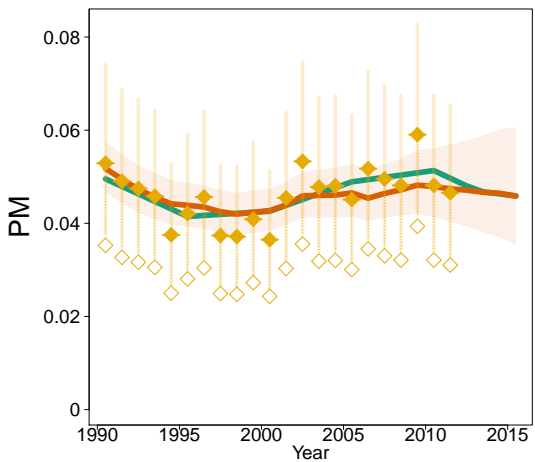

**Argentina [ARG; A]**
**(South America)**

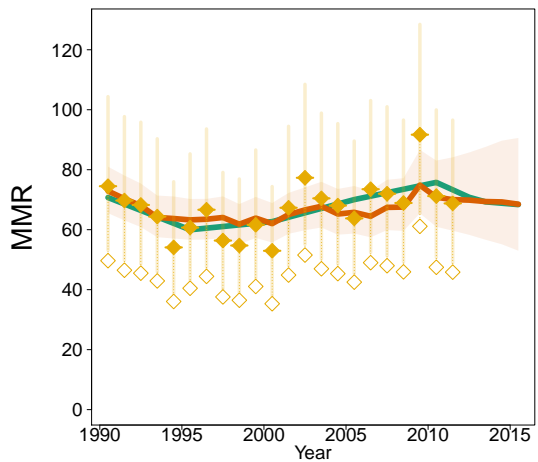

**Argentina [ARG; A]**
**(South America)**

– Estimates –
Bmat 2014
MMEIG 2014
– Data –
◇ Observed Data
◆ Adjusted Data
– Data type –
◆ VR

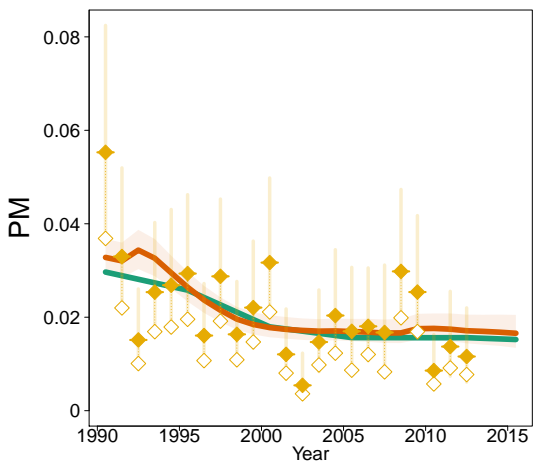

**Armenia [ARM; B]**
**(Western Asia)**

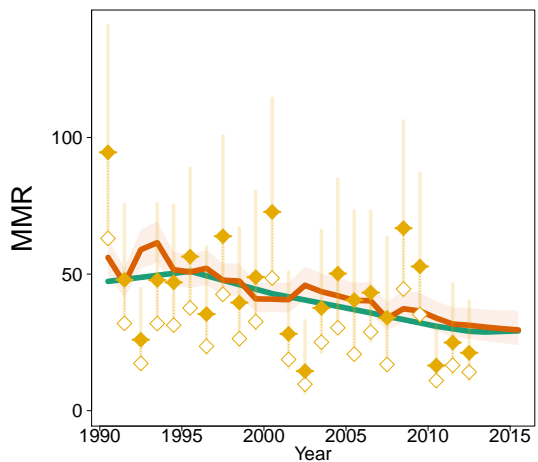

**Armenia [ARM; B]**
**(Western Asia)**

– Estimates –
Bmat 2014
MMEIG 2014
– Data –
◇ Observed Data
◆ Adjusted Data
– Data type –
◆ VR

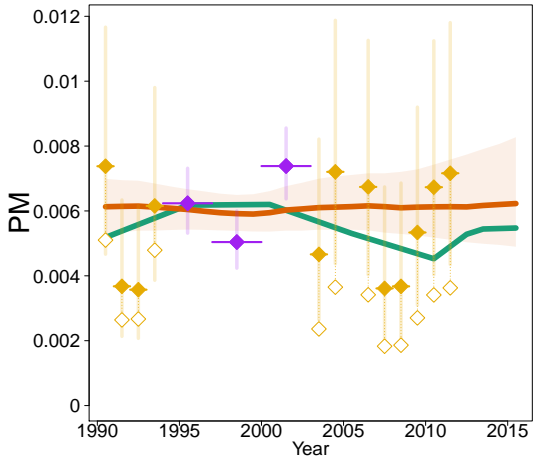

**Australia [AUS; A]**
**(Developed regions)**

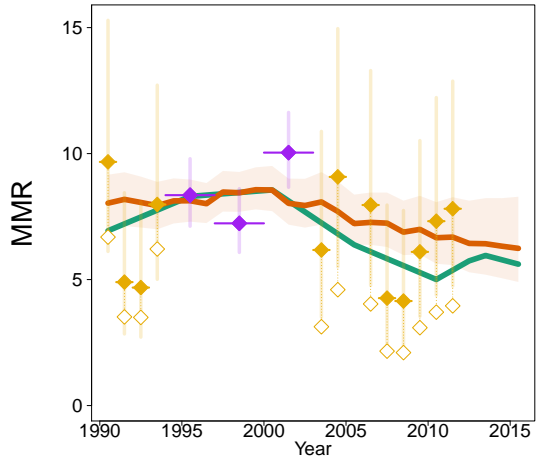

**Australia [AUS; A]**
**(Developed regions)**

– Estimates –
Bmat 2014
MMEIG 2014
– Data –
◇ Observed Data
◆ Adjusted Data
– Data type –
◆ VR
◆ Spec. studies

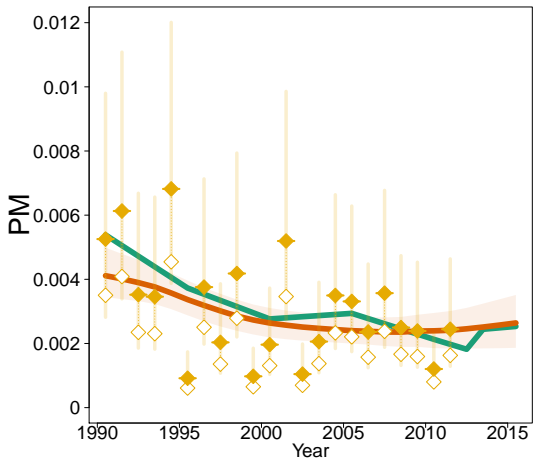

**Austria [AUT; A]**
**(Developed regions)**

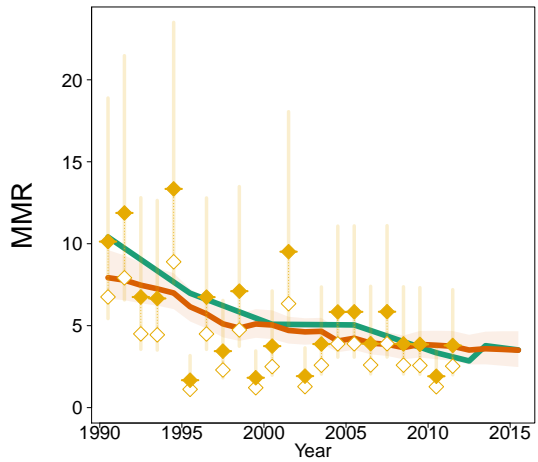

**Austria [AUT; A]**
**(Developed regions)**

– Estimates –
Bmat 2014
MMEIG 2014
– Data –
◇ Observed Data
◆ Adjusted Data
– Data type –
◆ VR

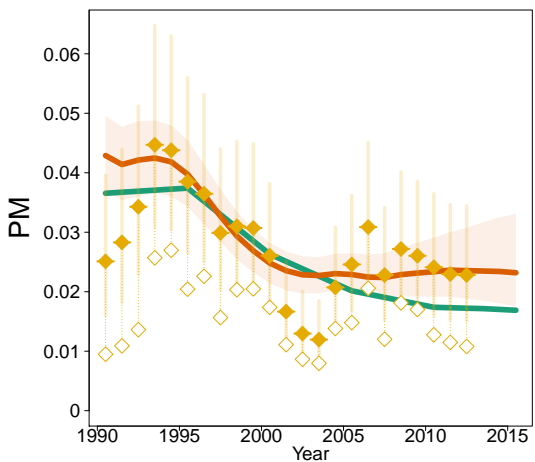

**Azerbaijan [AZE; B]**
**(Western Asia)**

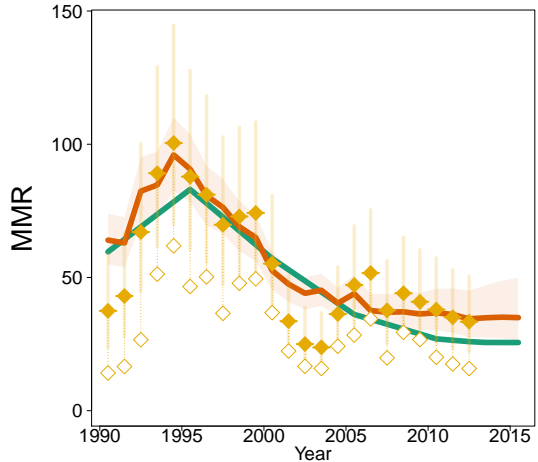

**Azerbaijan [AZE; B]**
**(Western Asia)**

– Estimates –
Bmat 2014
MMEIG 2014
– Data –
◇ Observed Data
◆ Adjusted Data
– Data type –
◆ VR

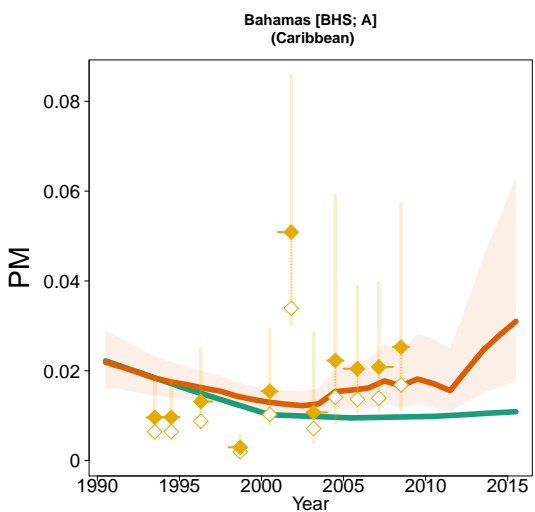

**Bahamas [BHS; A]**
**(Caribbean)**

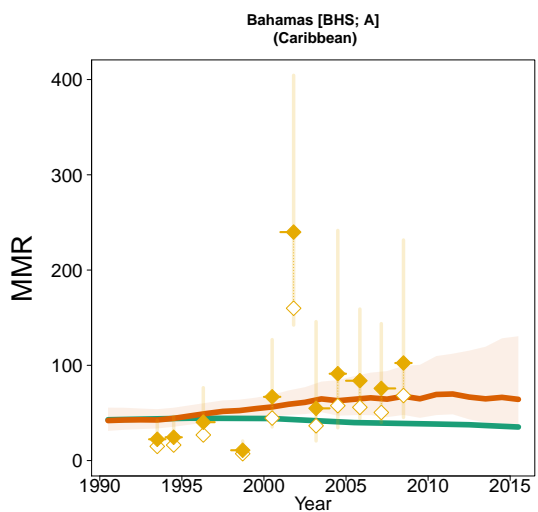

**Bahamas [BHS; A]**
**(Caribbean)**

– Estimates –
Bmat 2014
MMEIG 2014
– Data –
◇ Observed Data
◆ Adjusted Data
– Data type –
◆ VR

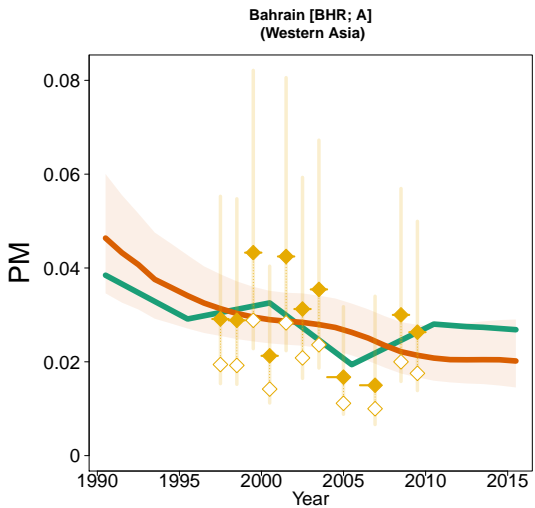

**Bahrain [BHR; A]**
**(Western Asia)**

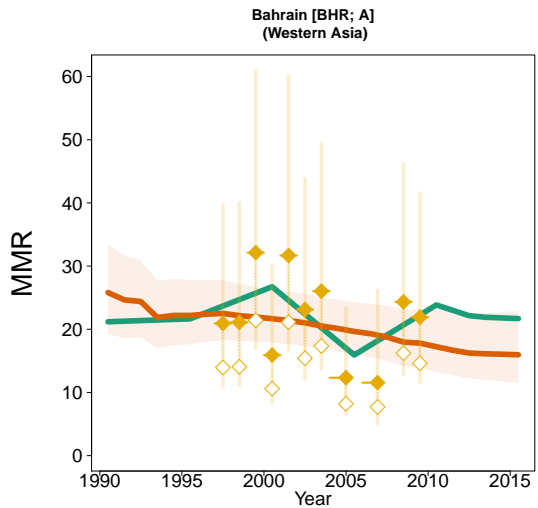

**Bahrain [BHR; A]**
**(Western Asia)**

– Estimates –
Bmat 2014
MMEIG 2014
– Data –
◇ Observed Data
◆ Adjusted Data
– Data type –
◆ VR

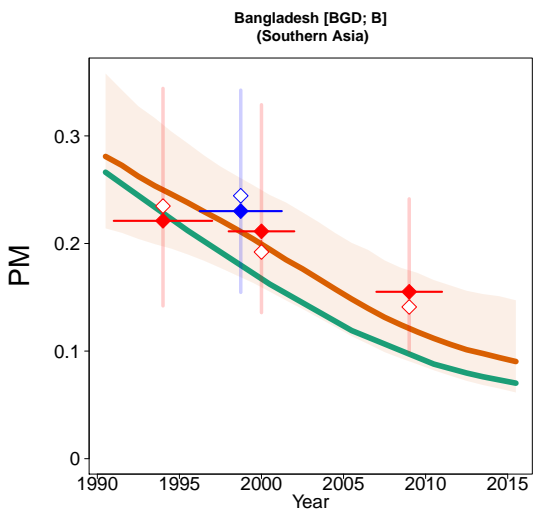

**Bangladesh [BGD; B]**
**(Southern Asia)**

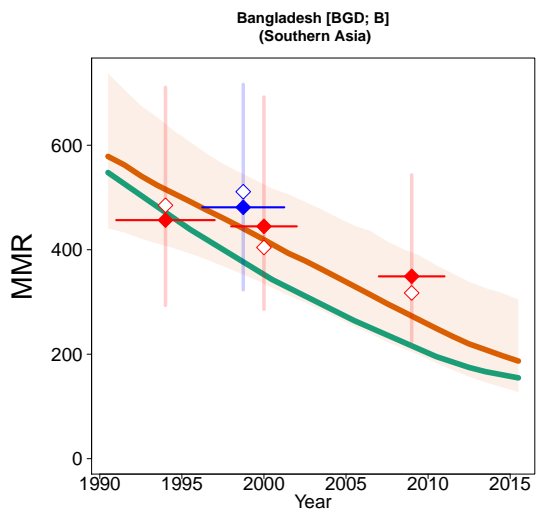

**Bangladesh [BGD; B]**
**(Southern Asia)**

– Estimates –
Bmat 2014
MMEIG 2014
– Data –
◇ Observed Data
◆ Adjusted Data
– Data type –
◆ DHS
◆ Misc. studies

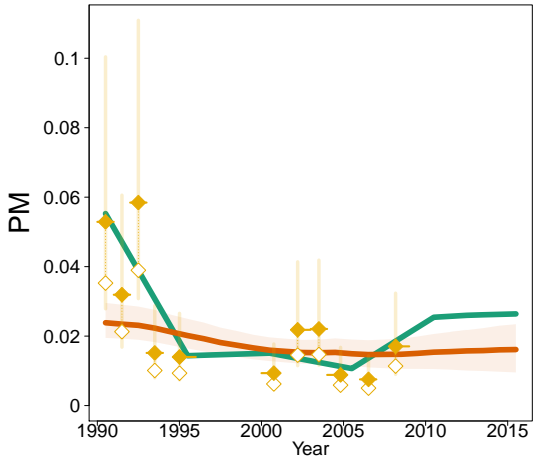

**Barbados [BRB; A]**
**(Caribbean)**

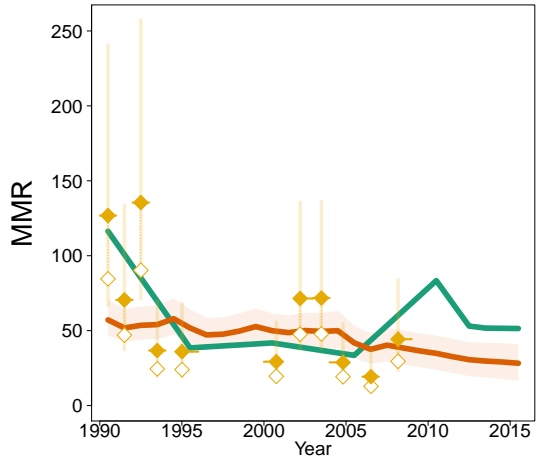

**Barbados [BRB; A]**
**(Caribbean)**

– Estimates –
Bmat 2014
MMEIG 2014
– Data –
◇ Observed Data
◆ Adjusted Data
– Data type –
◆ VR

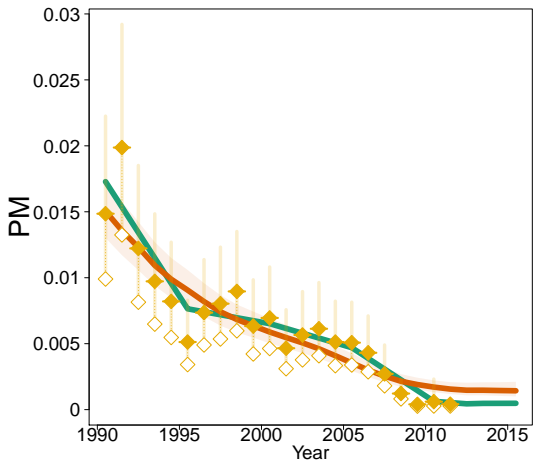

**Belarus [BLR; A]**
**(Developed regions)**

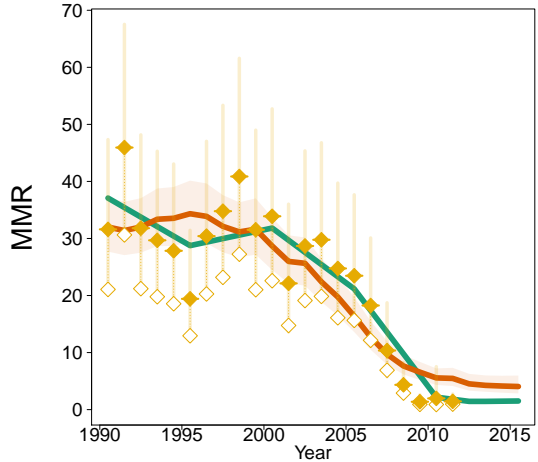

**Belarus [BLR; A]**
**(Developed regions)**

– Estimates –
Bmat 2014
MMEIG 2014
– Data –
◇ Observed Data
◆ Adjusted Data
– Data type –
◆ VR

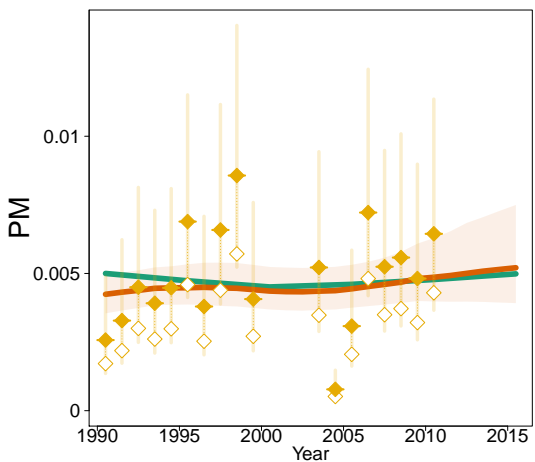

**Belgium [BEL; A]**
**(Developed regions)**

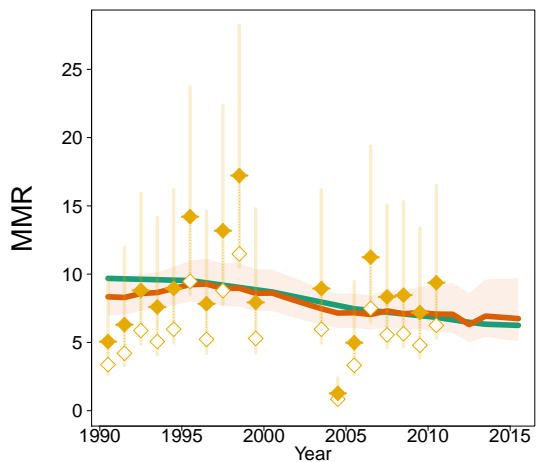

**Belgium [BEL; A]**
**(Developed regions)**

– Estimates –
Bmat 2014
MMEIG 2014
– Data –
◇ Observed Data
◆ Adjusted Data
– Data type –
◆ VR

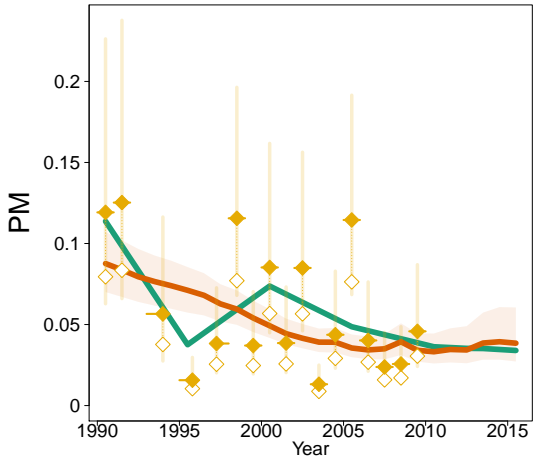

**Belize [BLZ; A]**
**(Central America)**

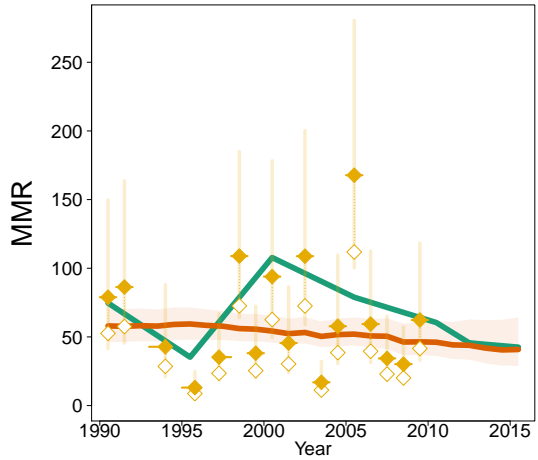

**Belize [BLZ; A]**
**(Central America)**

– Estimates –
**Bmat 2014**
**MMEIG 2014**
– Data –
◇ Observed Data
◆ Adjusted Data
– Data type –
◆ VR

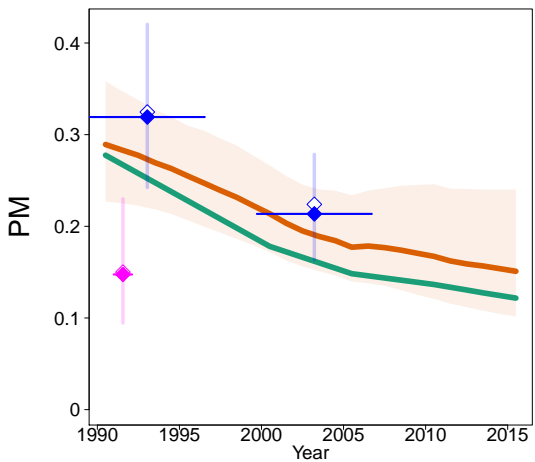

**Benin [BEN; B]**
**(Western Africa)**

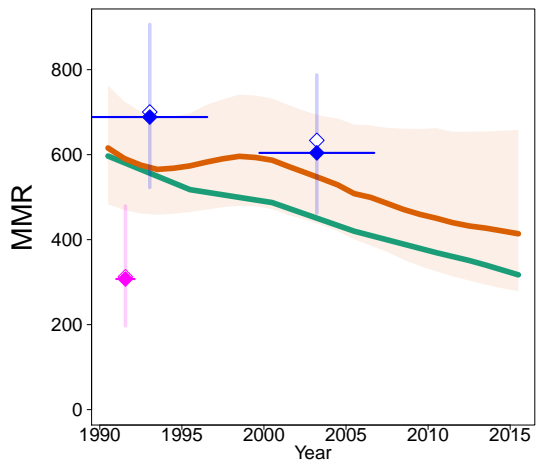

**Benin [BEN; B]**
**(Western Africa)**

– Estimates –
**Bmat 2014**
**MMEIG 2014**
– Data –
◇ Observed Data
◆ Adjusted Data
– Data type –
◆ DHS
◆ Census

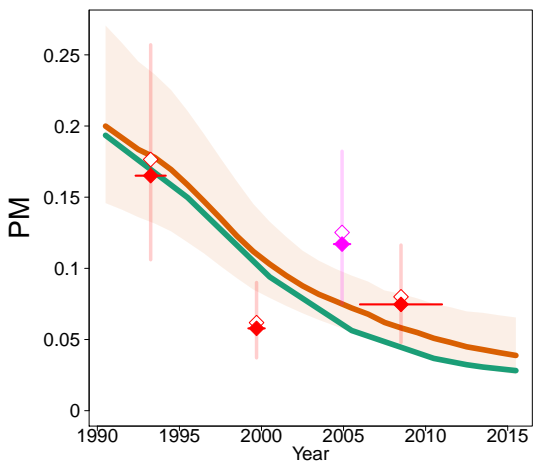

**Bhutan [BTN; B]**
**(Southern Asia)**

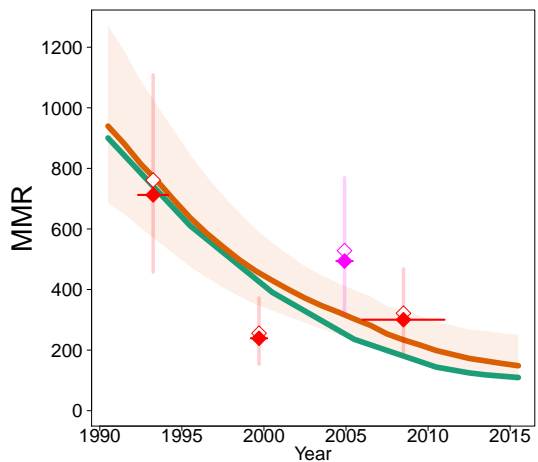

**Bhutan [BTN; B]**
**(Southern Asia)**

– Estimates –
**Bmat 2014**
**MMEIG 2014**
– Data –
◇ Observed Data
◆ Adjusted Data
– Data type –
◆ Census
◆ Misc. studies

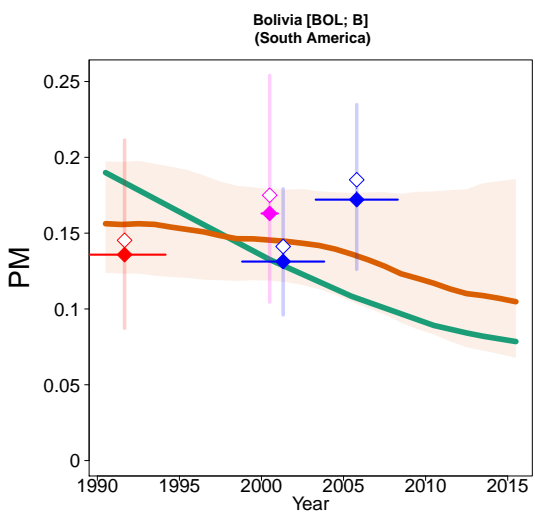

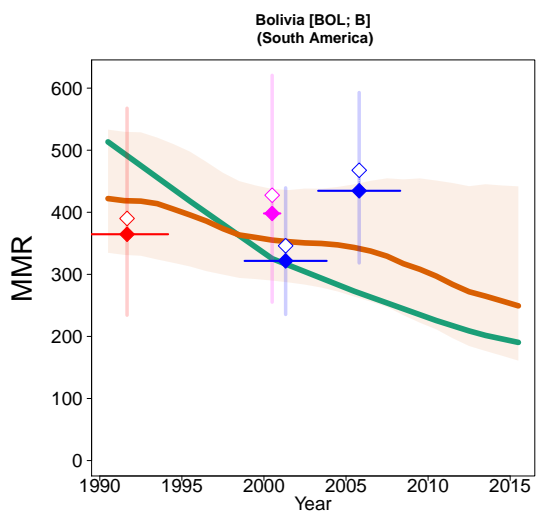

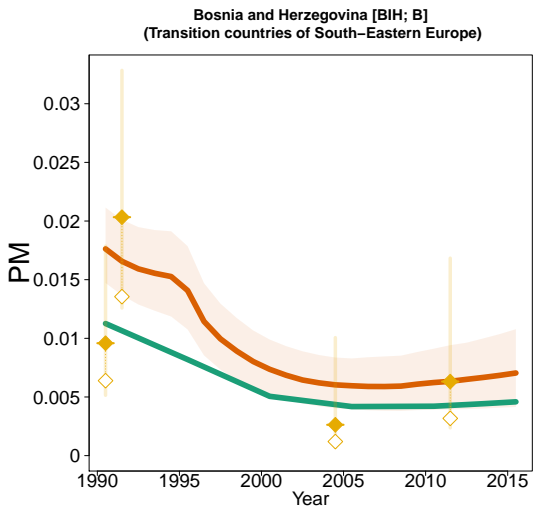

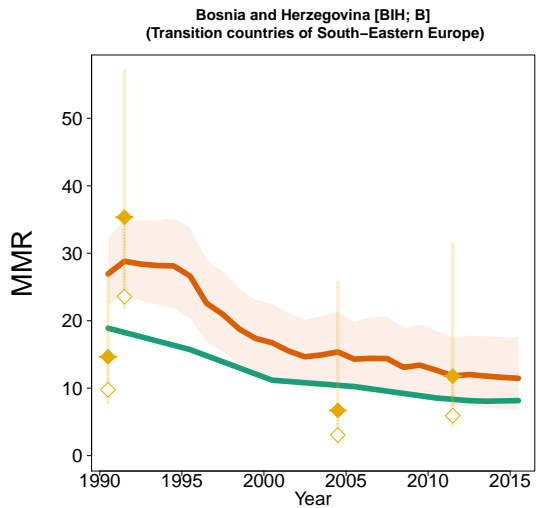

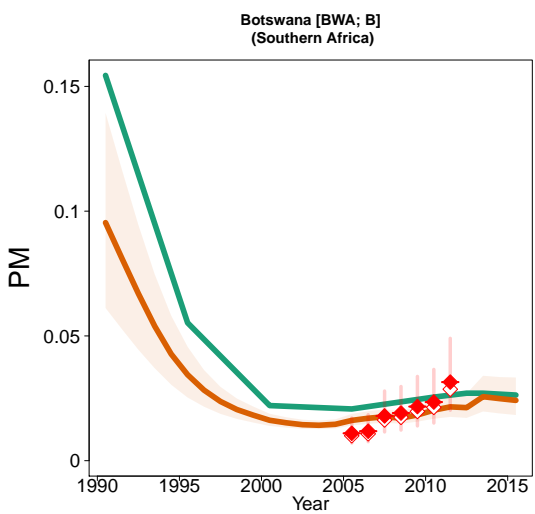

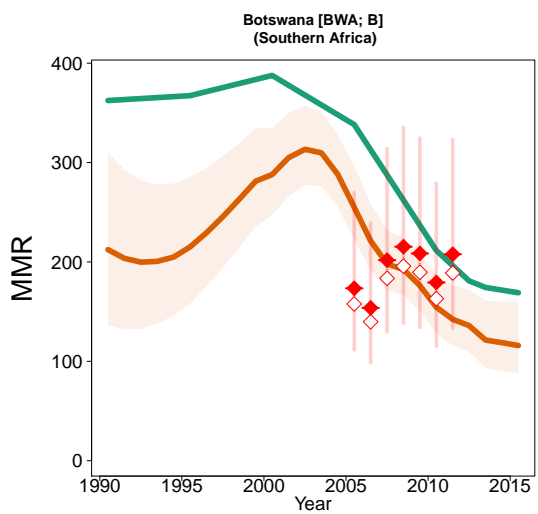

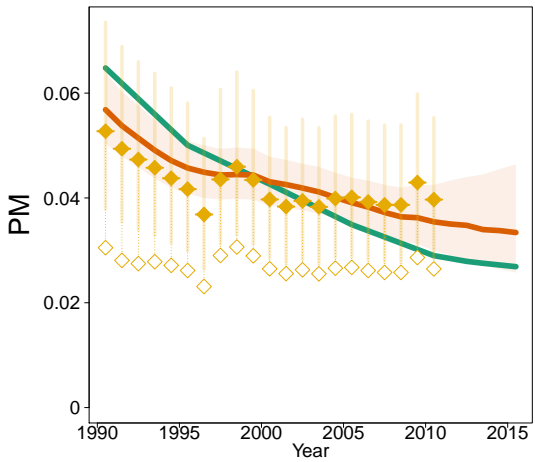

**Brazil [BRA; B]**
**(South America)**

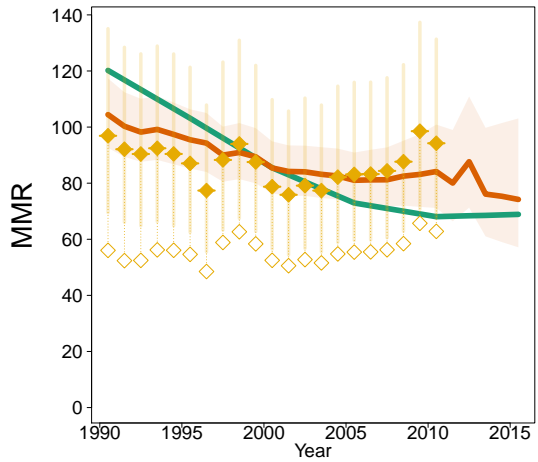

**Brazil [BRA; B]**
**(South America)**

– Estimates –
Bmat 2014
MMEIG 2014
– Data –
Observed Data
Adjusted Data
– Data type –
VR

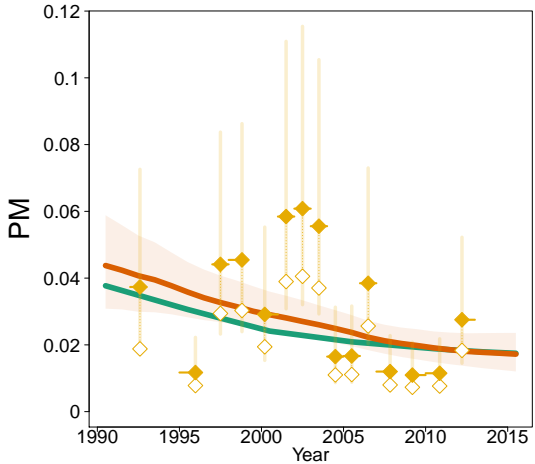

**Brunei Darussalam [BRN; B]**
**(South-Eastern Asia / Oceania)**

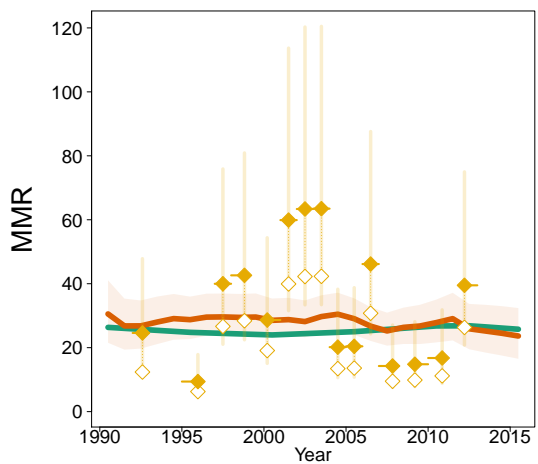

**Brunei Darussalam [BRN; B]**
**(South-Eastern Asia / Oceania)**

– Estimates –
Bmat 2014
MMEIG 2014
– Data –
Observed Data
Adjusted Data
– Data type –
VR

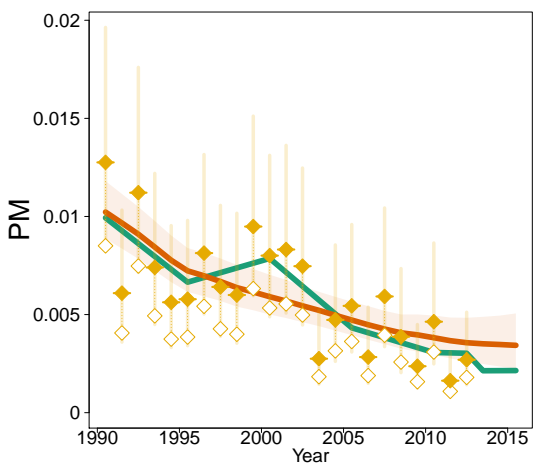

**Bulgaria [BGR; A]**
**(Developed regions)**

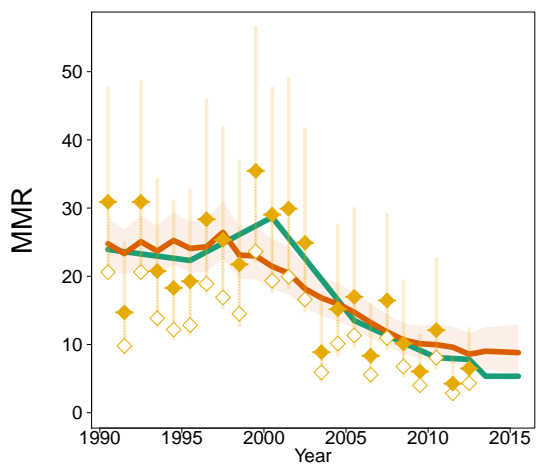

**Bulgaria [BGR; A]**
**(Developed regions)**

– Estimates –
Bmat 2014
MMEIG 2014
– Data –
Observed Data
Adjusted Data
– Data type –
VR

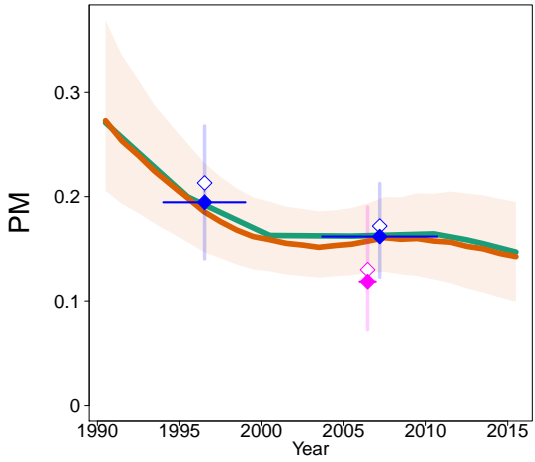

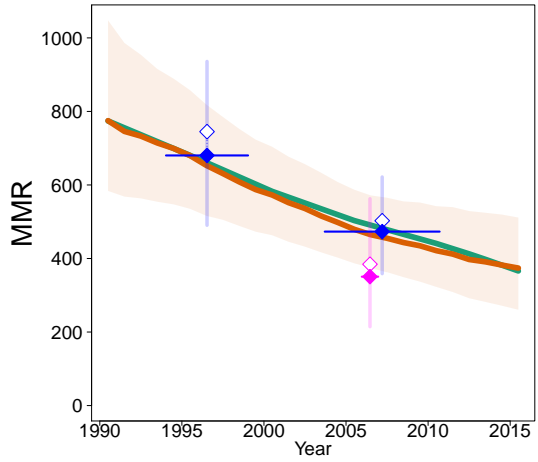

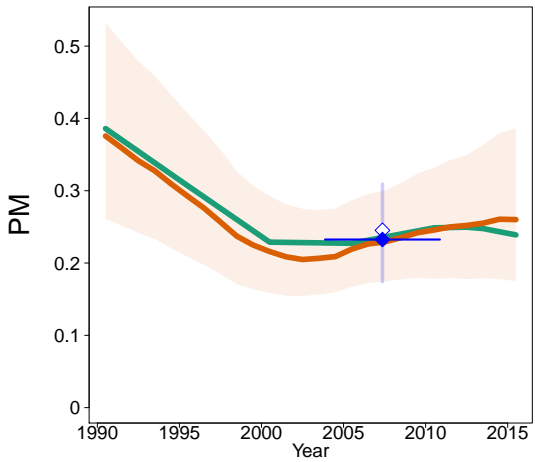

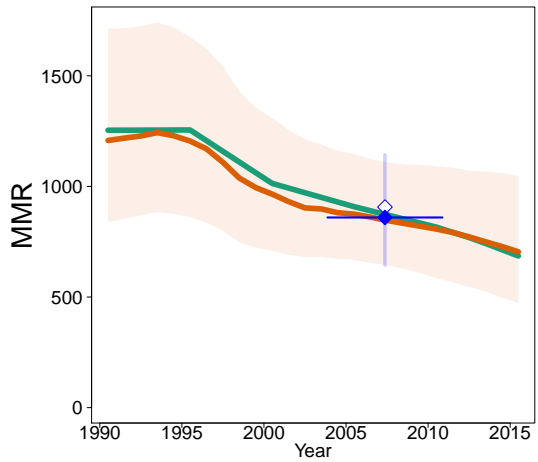

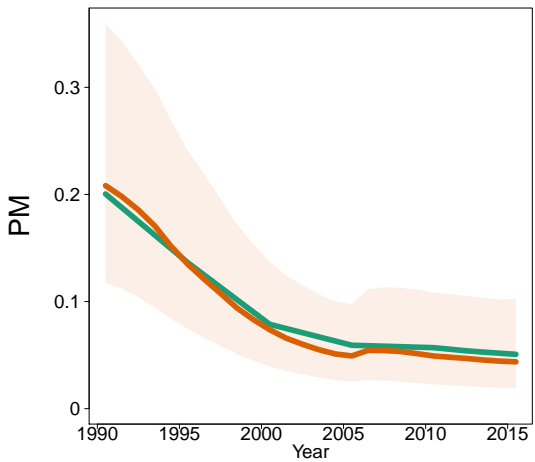

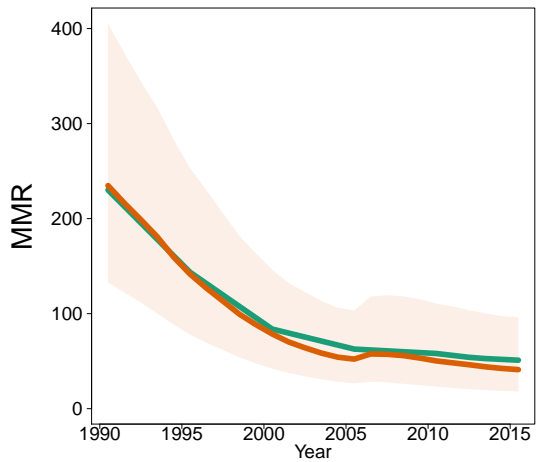

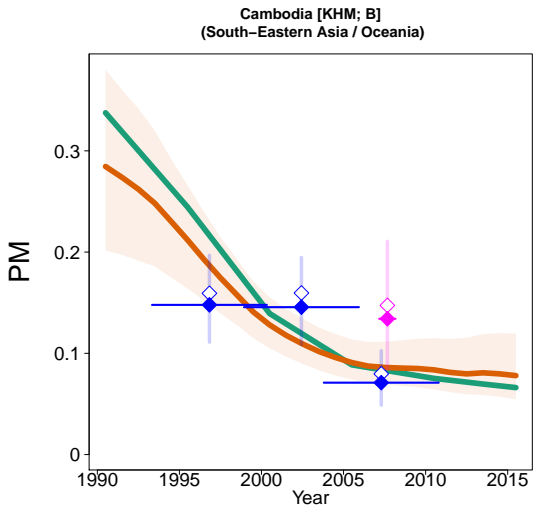

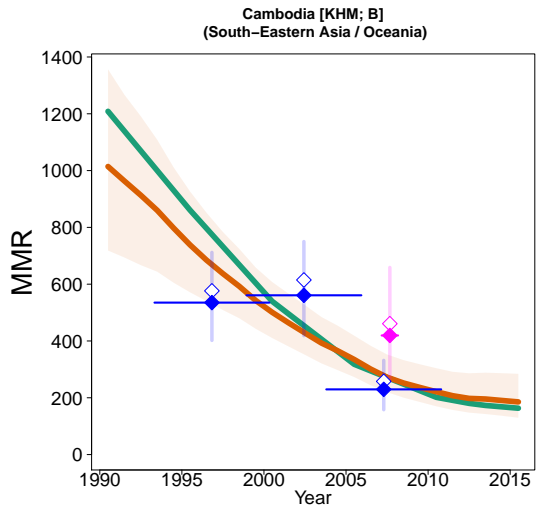

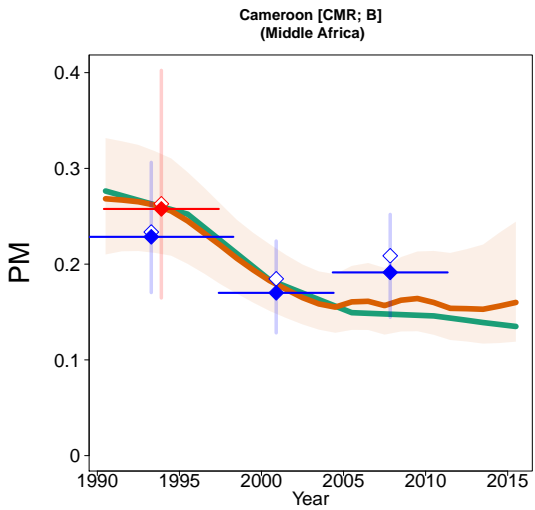

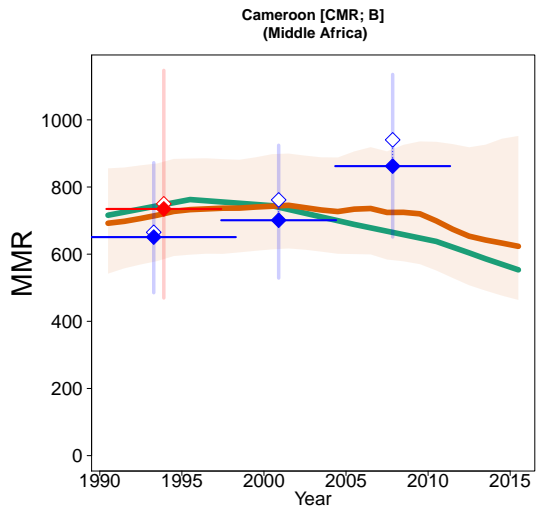

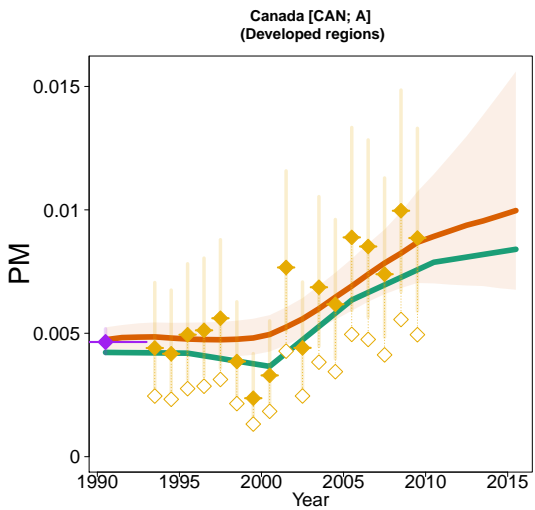

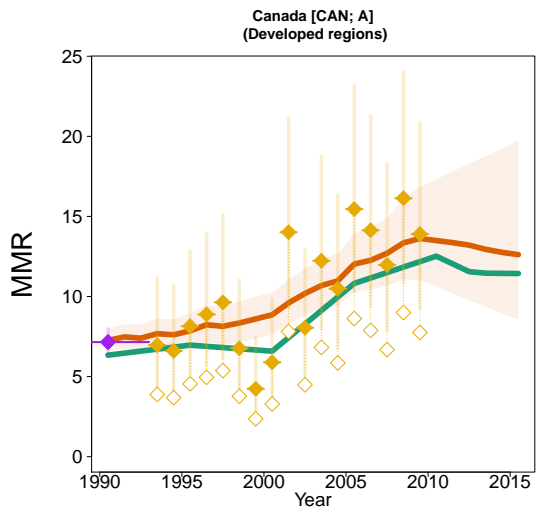

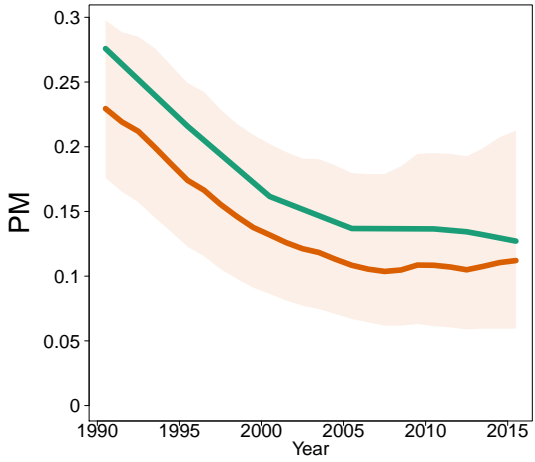

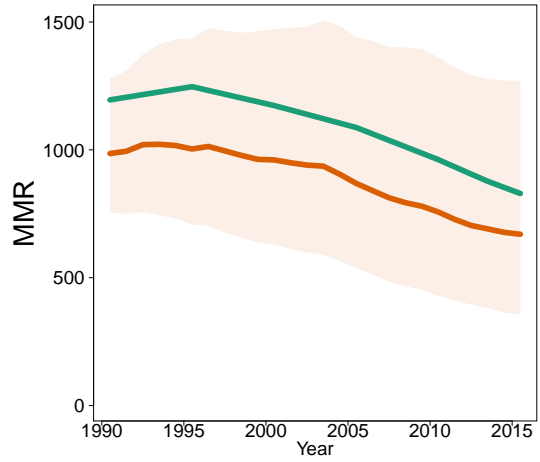

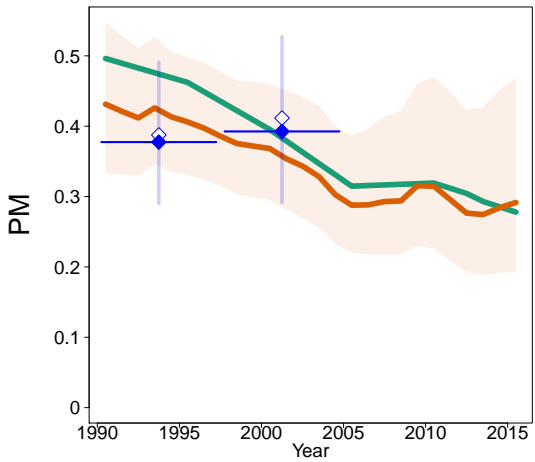

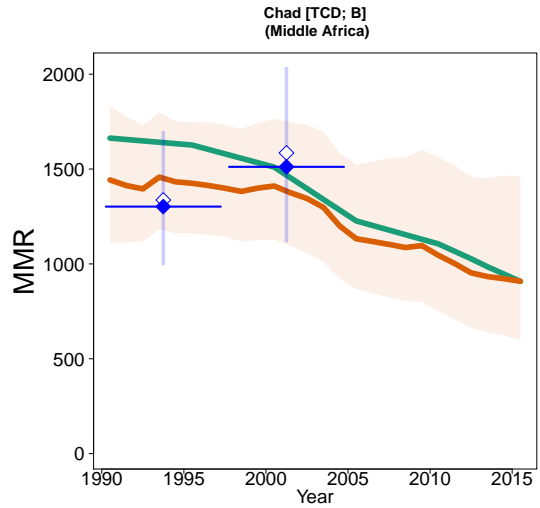

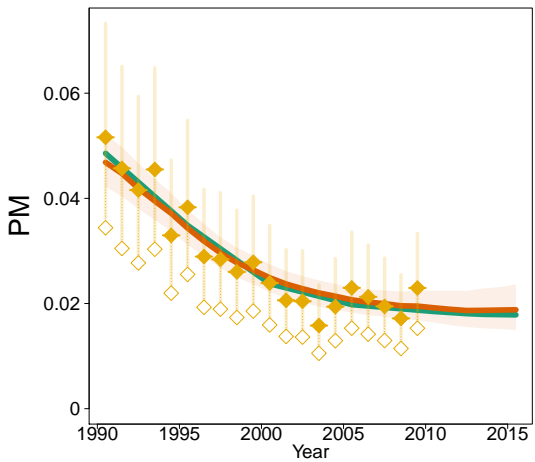

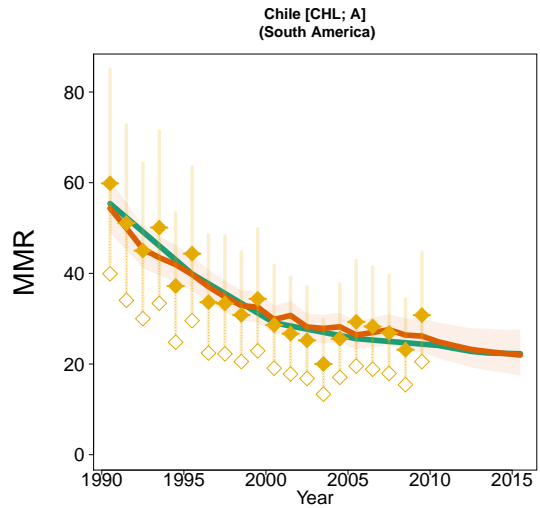

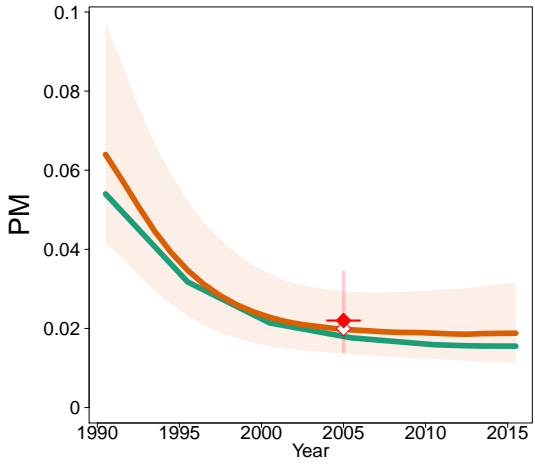

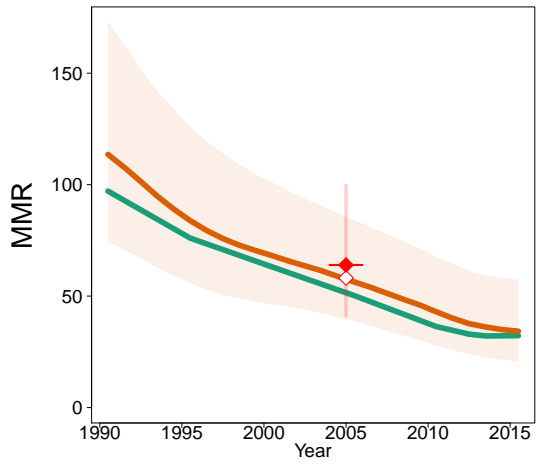

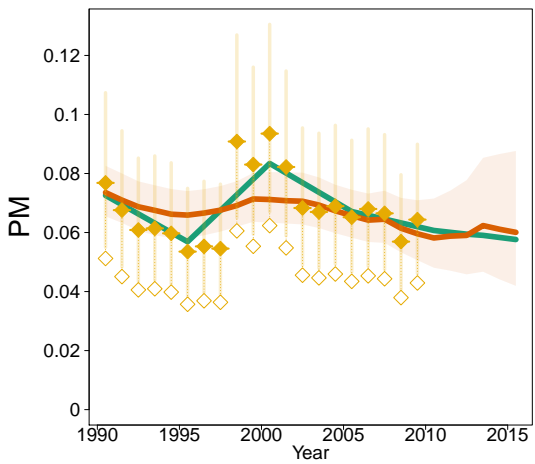

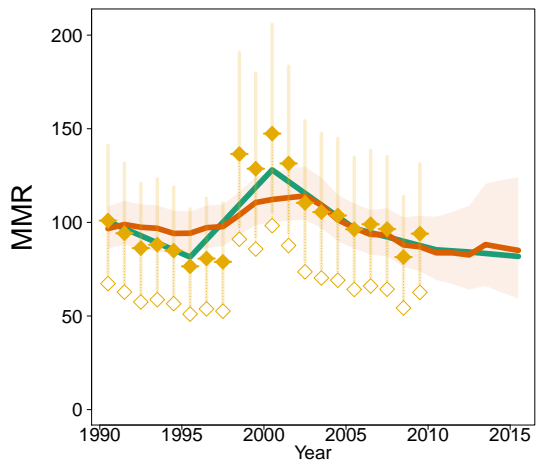

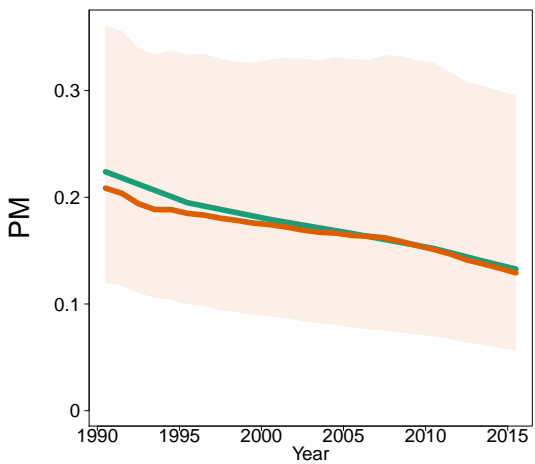

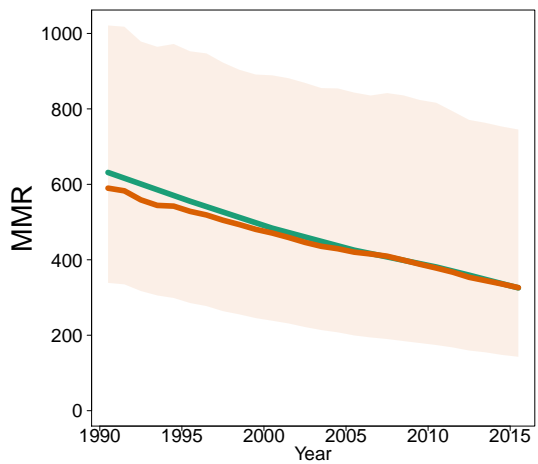

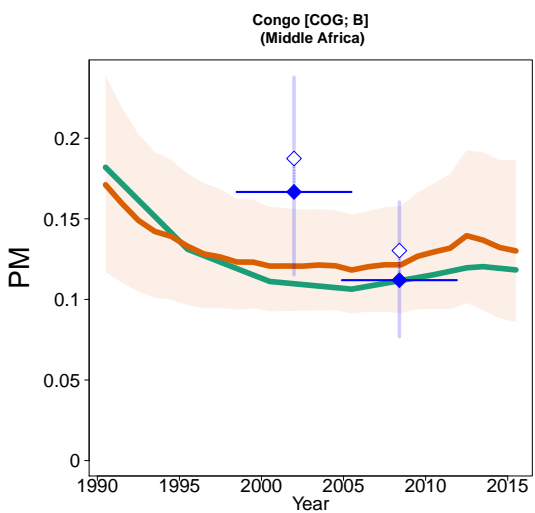

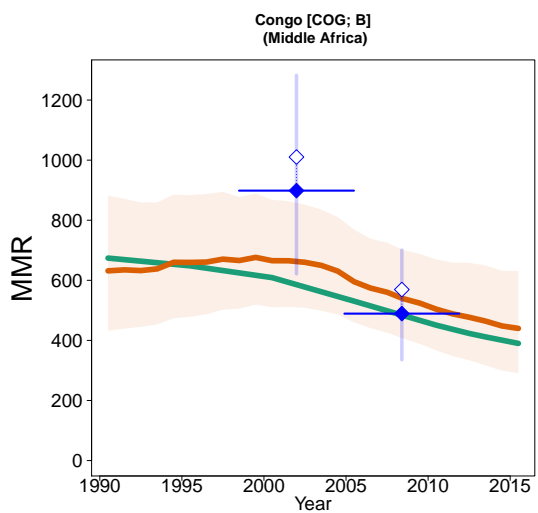

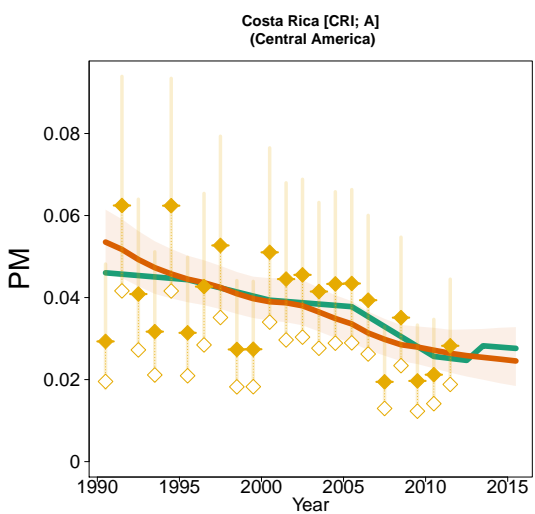

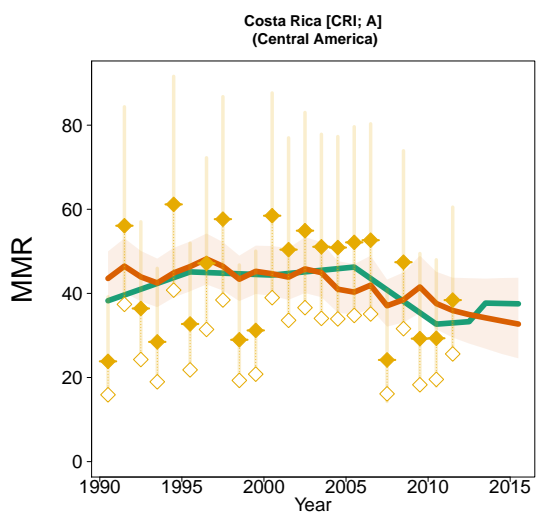

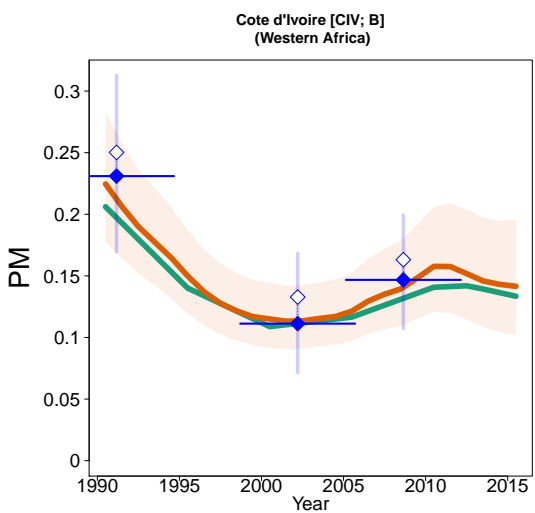

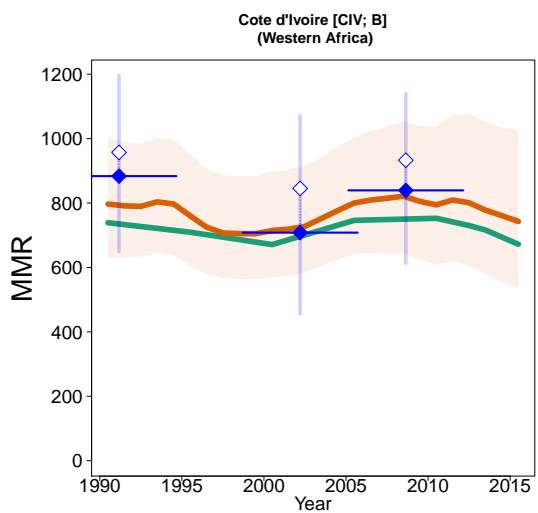

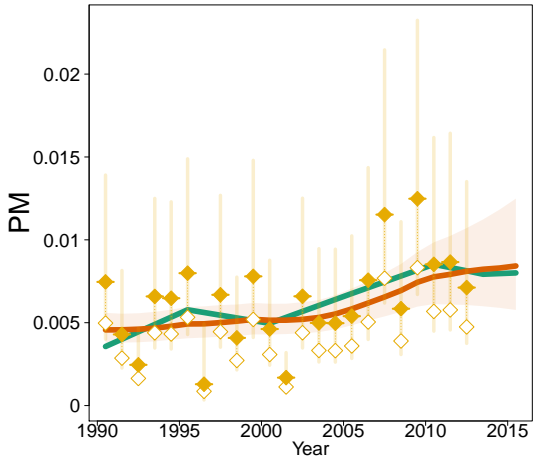

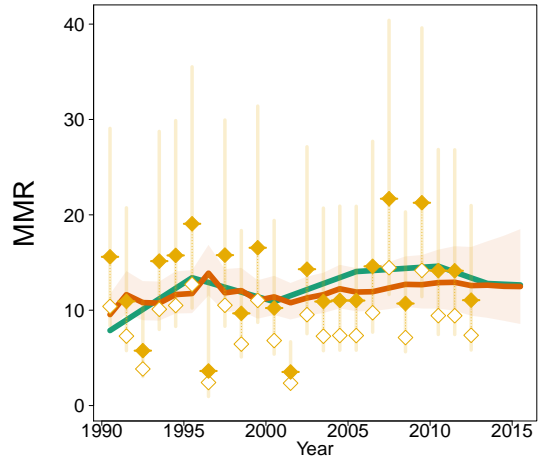

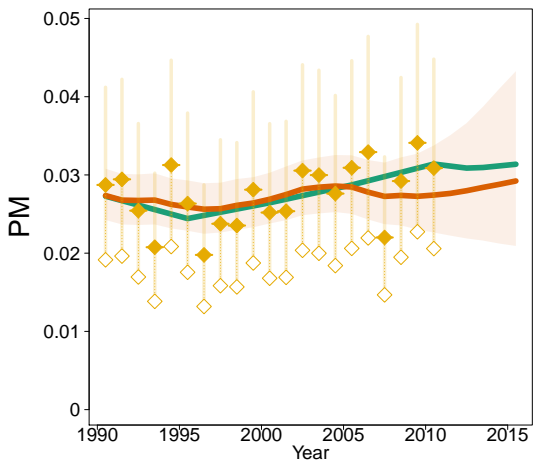

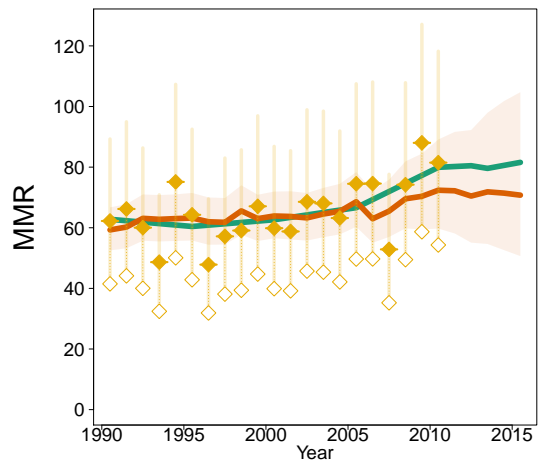

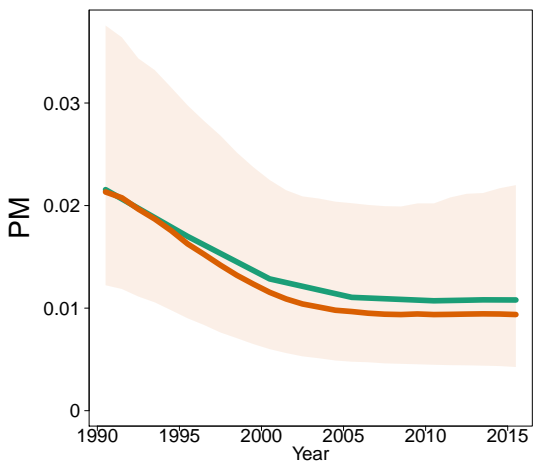

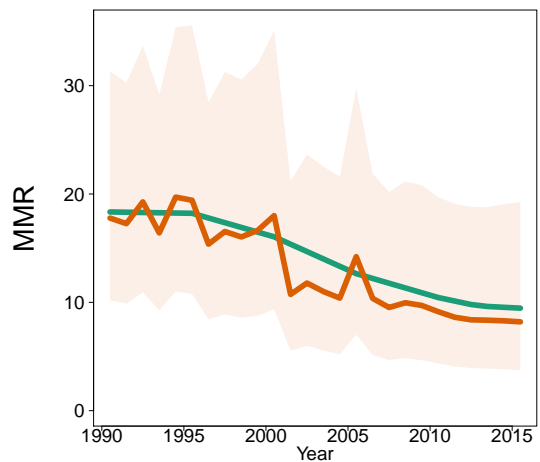

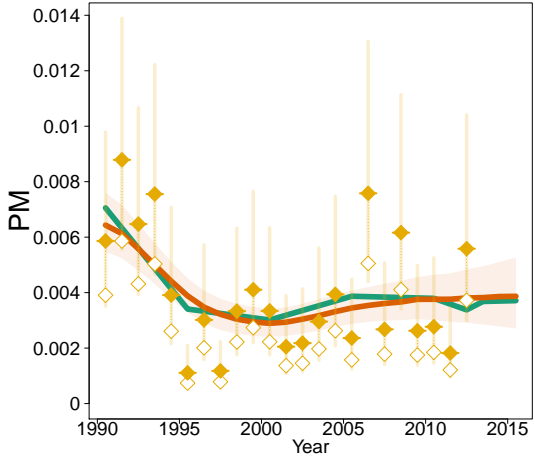

**Czech Republic [CZE; A]**
**(Developed regions)**

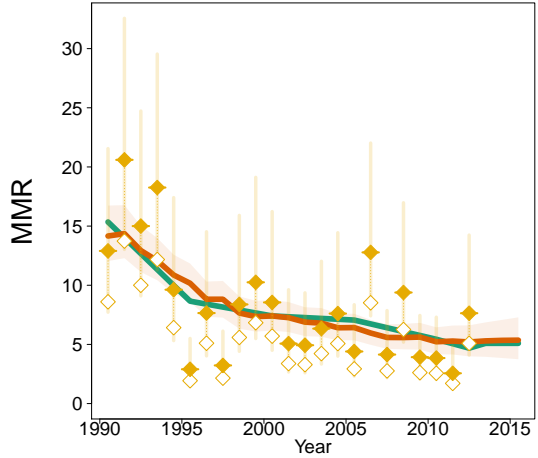

**Czech Republic [CZE; A]**
**(Developed regions)**

– Estimates –
Bmat 2014
MMEIG 2014
– Data –
◇ Observed Data
◆ Adjusted Data
– Data type –
◆ VR

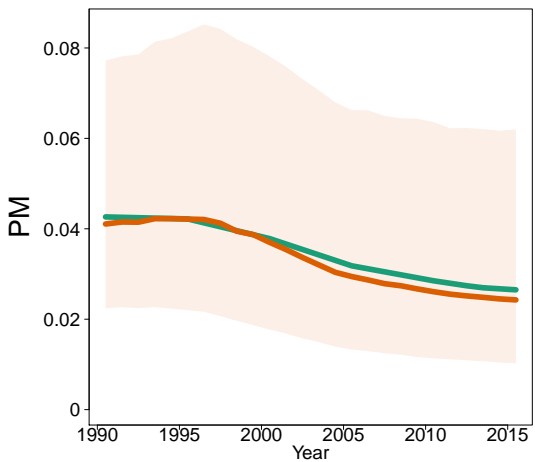

**Democratic People's Republic of Korea [PRK; C]**
**(Eastern Asia)**

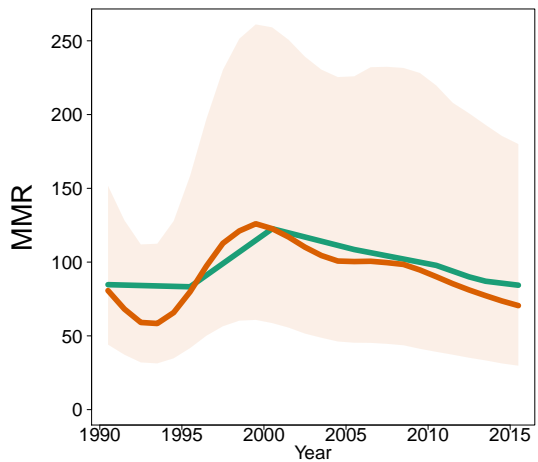

**Democratic People's Republic of Korea [PRK; C]**
**(Eastern Asia)**

– Estimates –
Bmat 2014
MMEIG 2014
– Data –

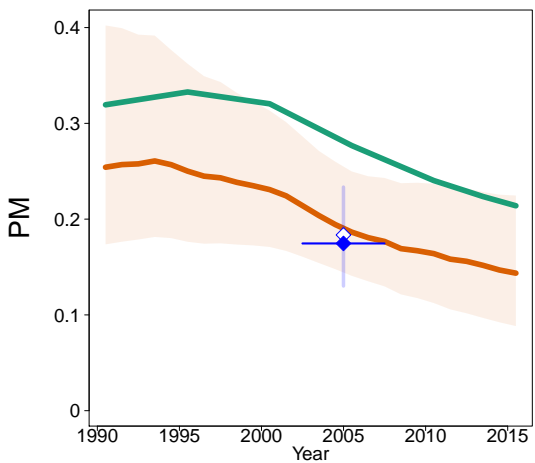

**Democratic Republic of the Congo [COD; B]**
**(Middle Africa)**

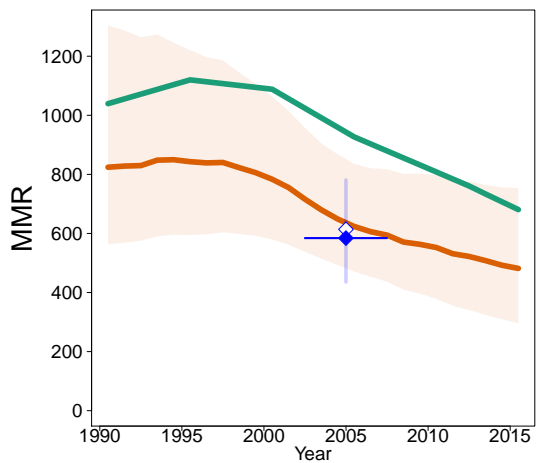

**Democratic Republic of the Congo [COD; B]**
**(Middle Africa)**

– Estimates –
Bmat 2014
MMEIG 2014
– Data –
◇ Observed Data
◆ Adjusted Data
– Data type –
◆ DHS

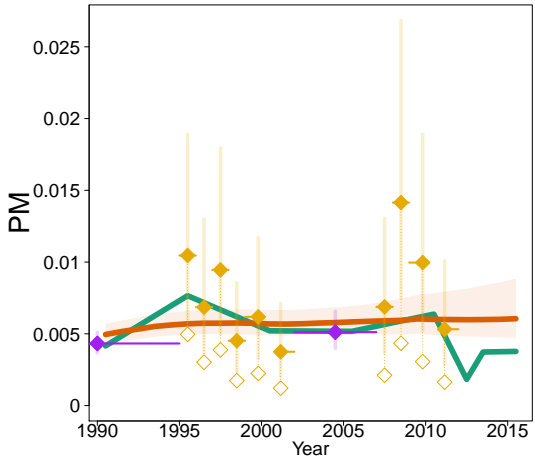

**Denmark [DNK; A]**
**(Developed regions)**

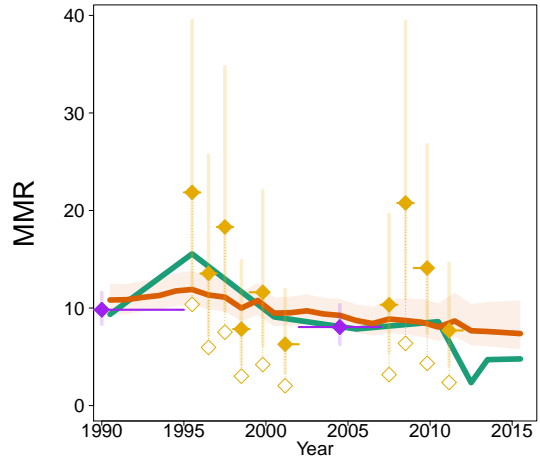

**Denmark [DNK; A]**
**(Developed regions)**

– Estimates –
Bmat 2014
MMEIG 2014
– Data –
◇ Observed Data
◆ Adjusted Data
– Data type –
◆ VR
◆ Spec. studies

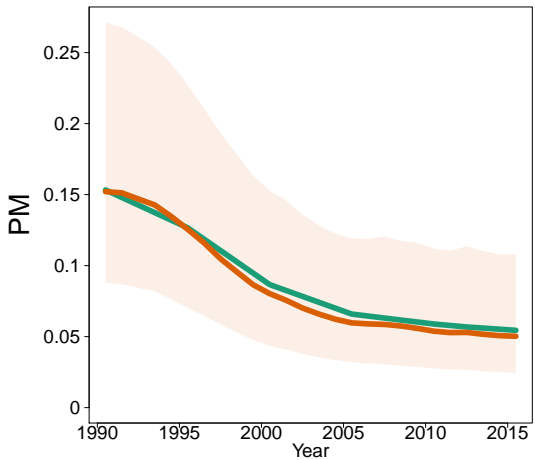

**Djibouti [DJI; C]**
**(Eastern Africa)**

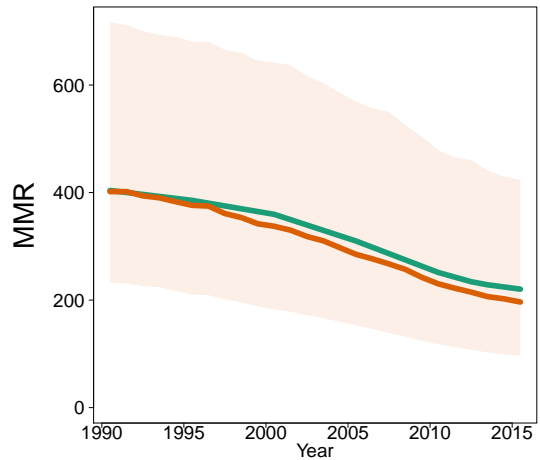

**Djibouti [DJI; C]**
**(Eastern Africa)**

– Estimates –
Bmat 2014
MMEIG 2014
– Data –

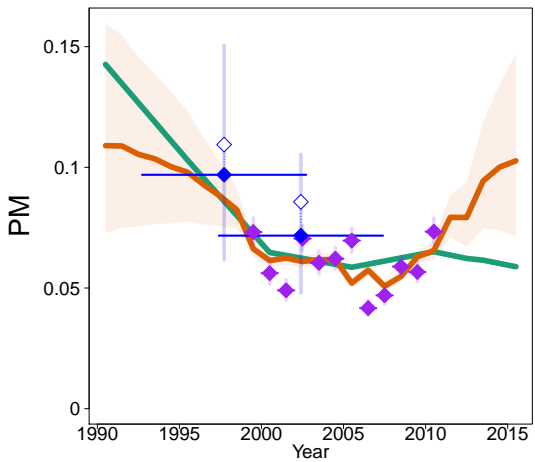

**Dominican Republic [DOM; B]**
**(Caribbean)**

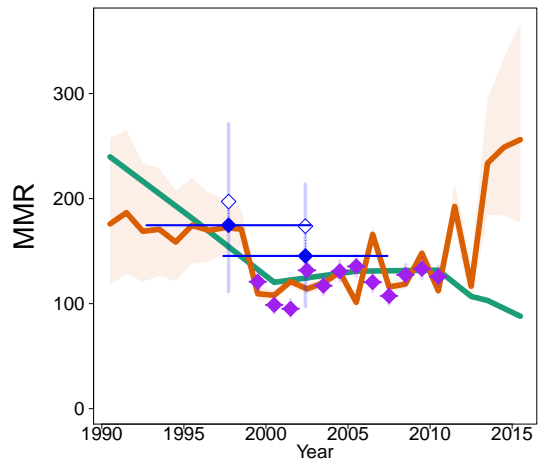

**Dominican Republic [DOM; B]**
**(Caribbean)**

– Estimates –
Bmat 2014
MMEIG 2014
– Data –
◇ Observed Data
◆ Adjusted Data
– Data type –
◆ Spec. studies
◆ DHS

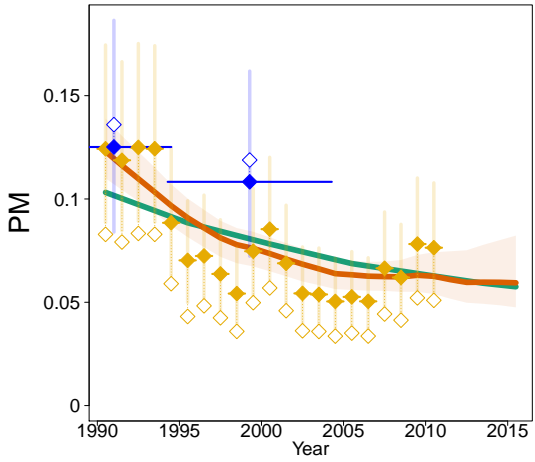

**Ecuador [ECU; B]**
**(South America)**

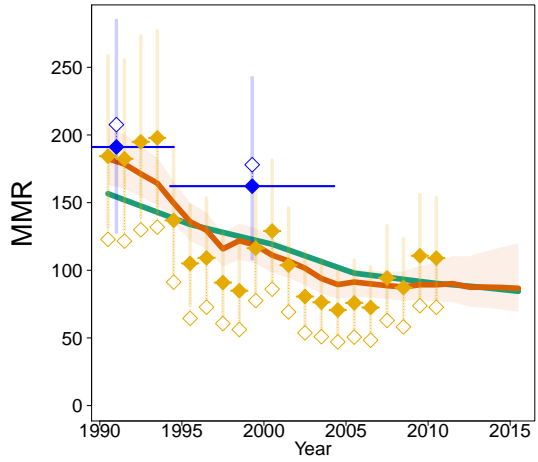

**Ecuador [ECU; B]**
**(South America)**

– Estimates –
Bmat 2014
MMEIG 2014
– Data –
Observed Data
Adjusted Data
– Data type –
VR
DHS

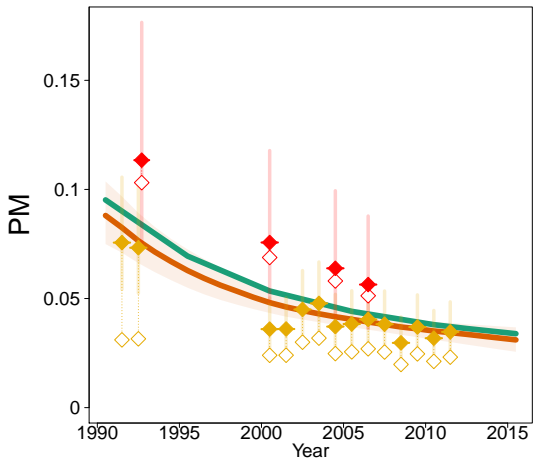

**Egypt [EGY; B]**
**(Northern Africa)**

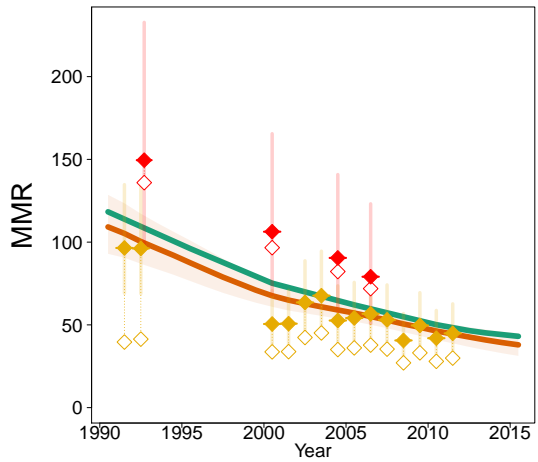

**Egypt [EGY; B]**
**(Northern Africa)**

– Estimates –
Bmat 2014
MMEIG 2014
– Data –
Observed Data
Adjusted Data
– Data type –
VR
Misc. studies

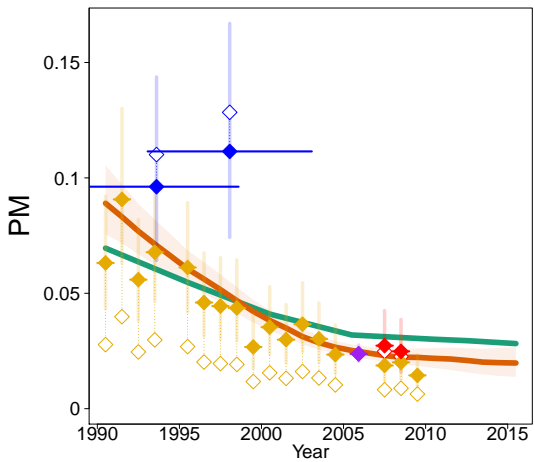

**El Salvador [SLV; B]**
**(Central America)**

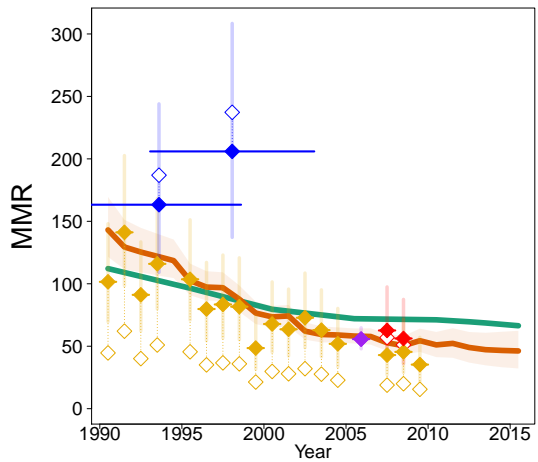

**El Salvador [SLV; B]**
**(Central America)**

– Estimates –
Bmat 2014
MMEIG 2014
– Data –
Observed Data
Adjusted Data
– Data type –
VR
Spec. studies
DHS
Misc. studies

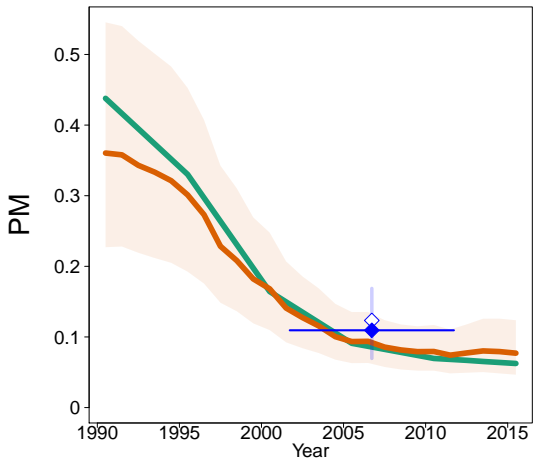

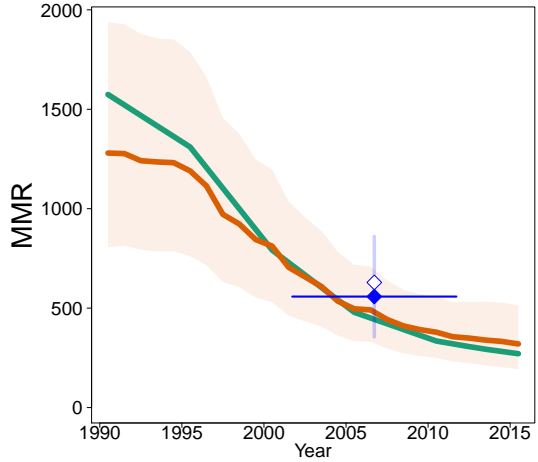

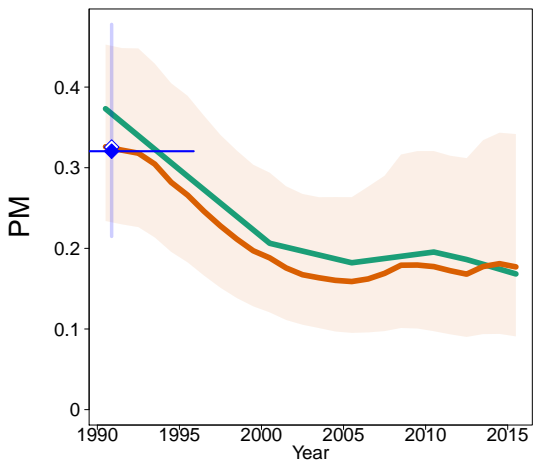

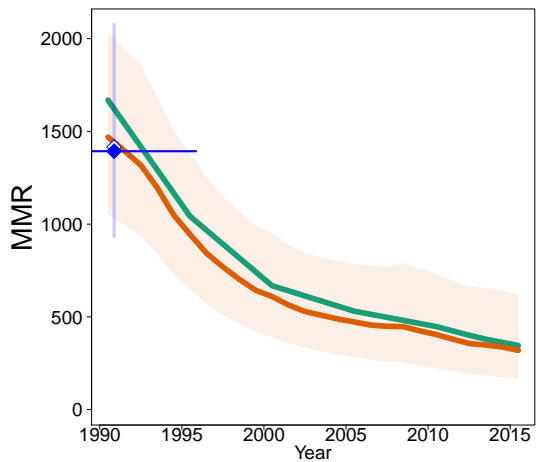

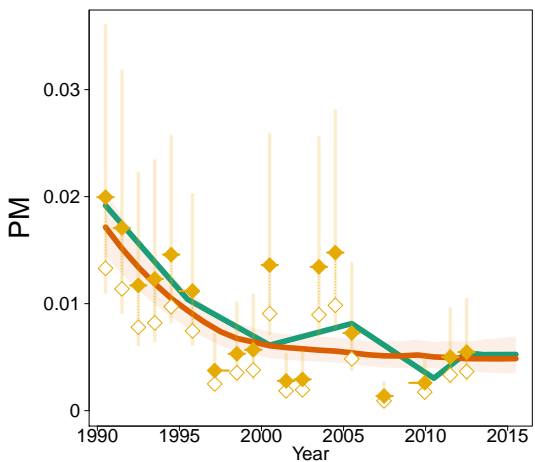

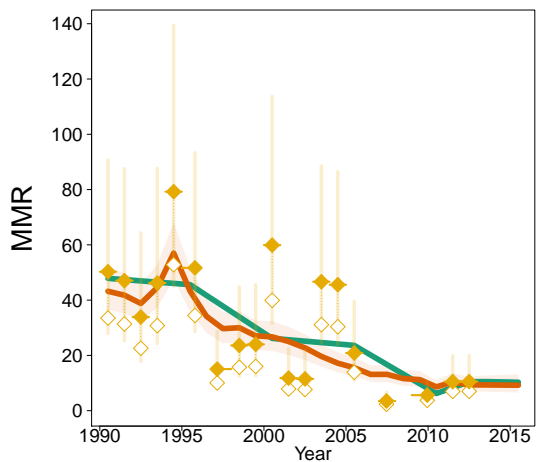

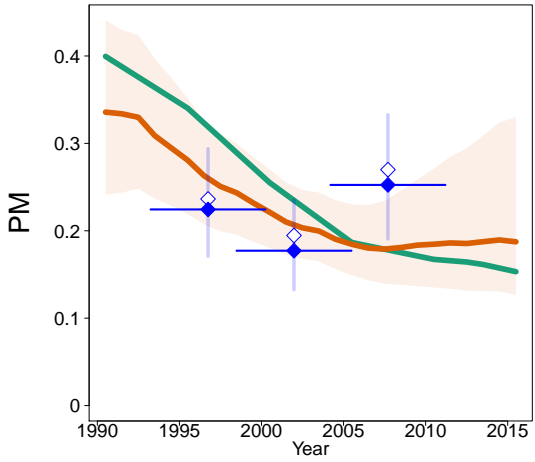

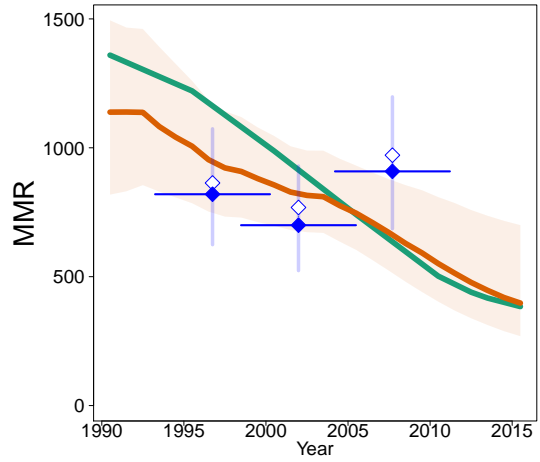

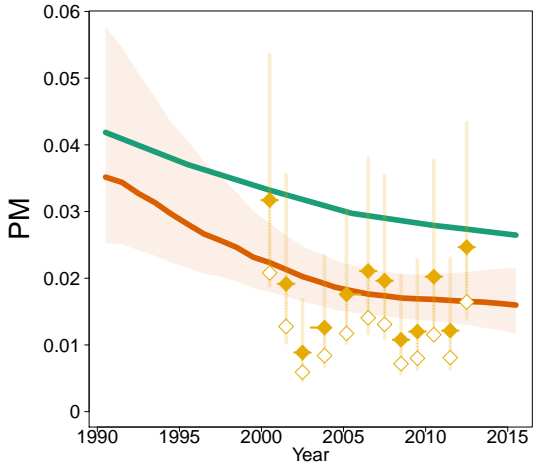

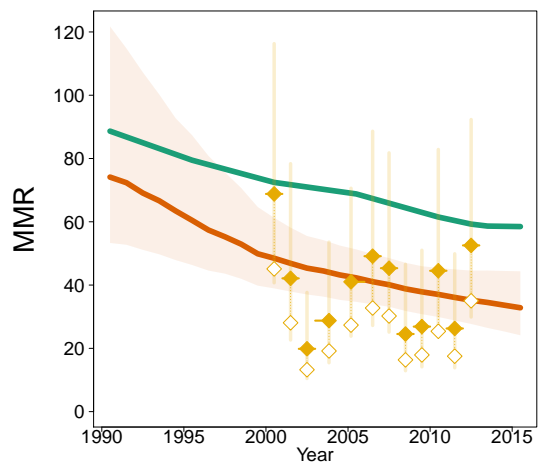

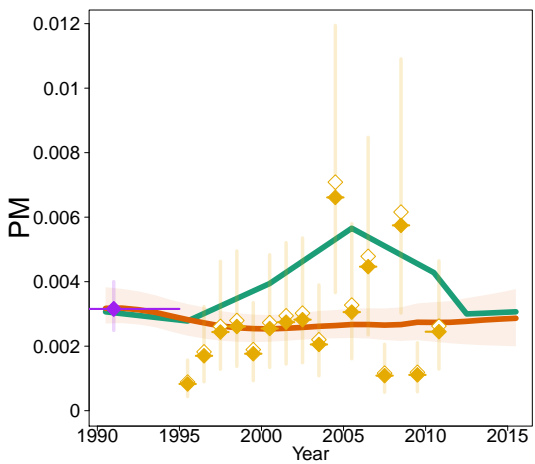

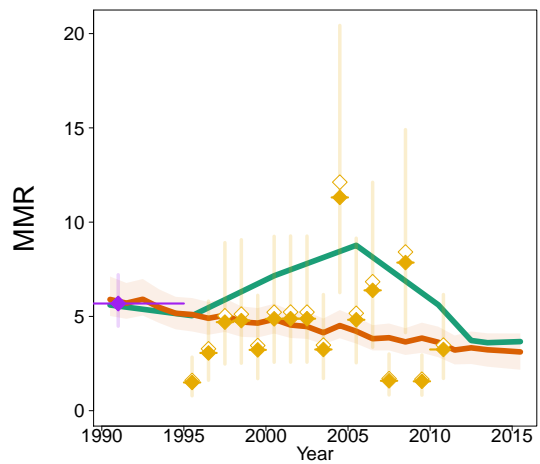

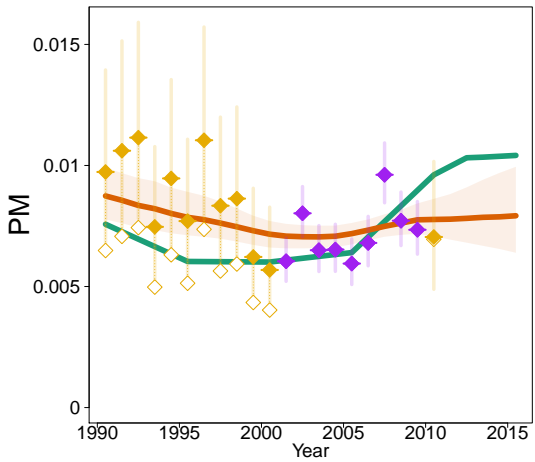

**France [FRA; A]**
**(Developed regions)**

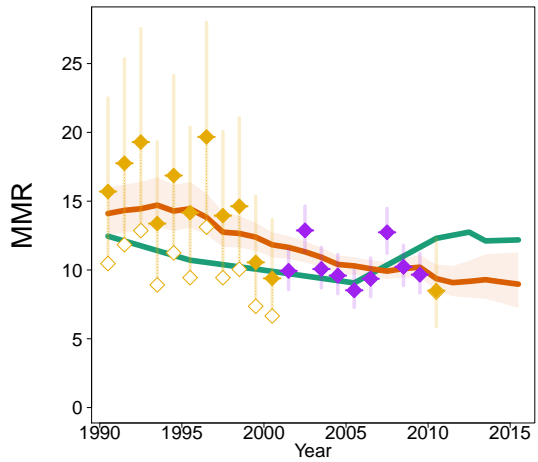

**France [FRA; A]**
**(Developed regions)**

– Estimates –
— Bmat 2014
— MMEIG 2014
– Data –
◇ Observed Data
◆ Adjusted Data
– Data type –
◆ VR
◆ Spec. studies

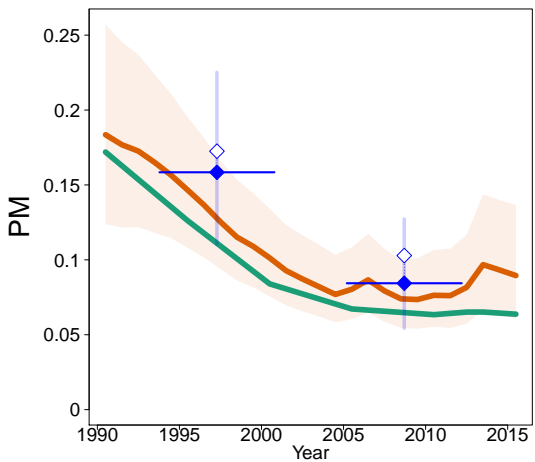

**Gabon [GAB; B]**
**(Middle Africa)**

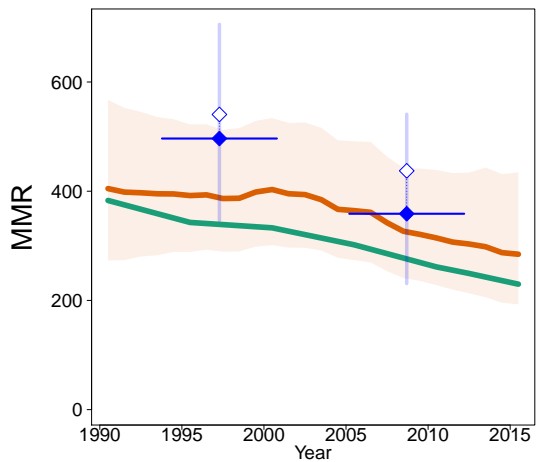

**Gabon [GAB; B]**
**(Middle Africa)**

– Estimates –
— Bmat 2014
— MMEIG 2014
– Data –
◇ Observed Data
◆ Adjusted Data
– Data type –
◆ DHS

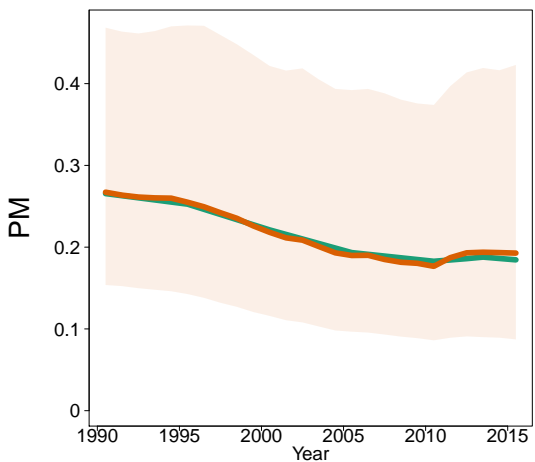

**Gambia [GMB; C]**
**(Western Africa)**

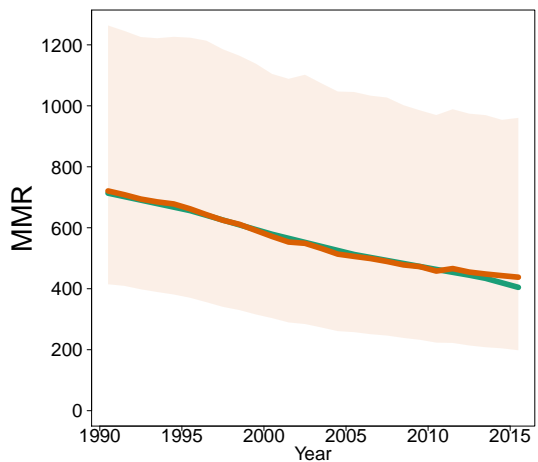

**Gambia [GMB; C]**
**(Western Africa)**

– Estimates –
— Bmat 2014
— MMEIG 2014
– Data –

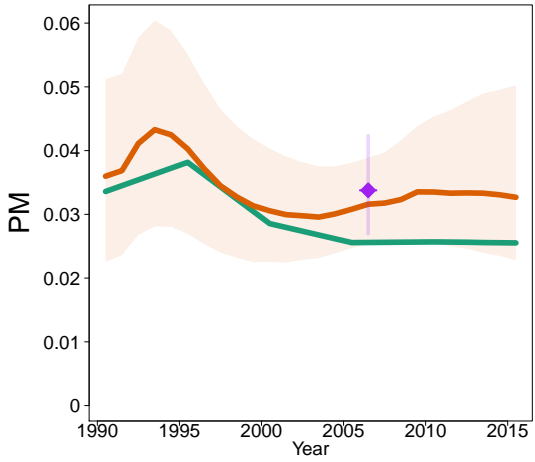

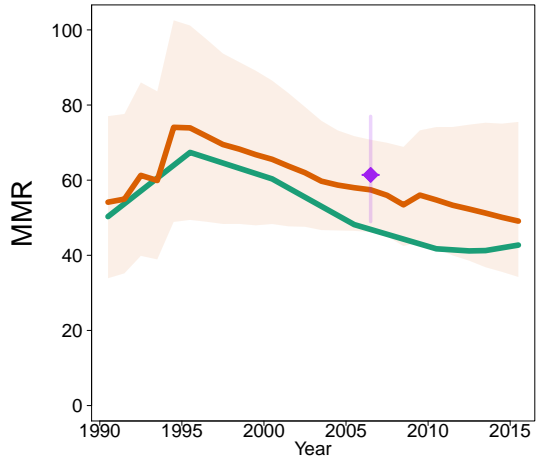

– Estimates –
Bmat 2014
MMEIG 2014
– Data –
◇ Observed Data
◆ Adjusted Data
– Data type –
◆ Spec. studies

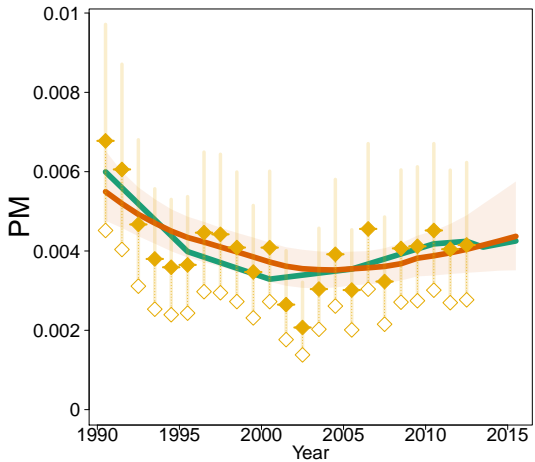

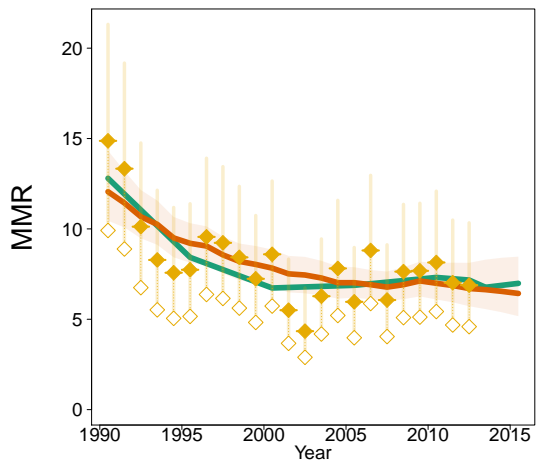

– Estimates –
Bmat 2014
MMEIG 2014
– Data –
◇ Observed Data
◆ Adjusted Data
– Data type –
◆ VR

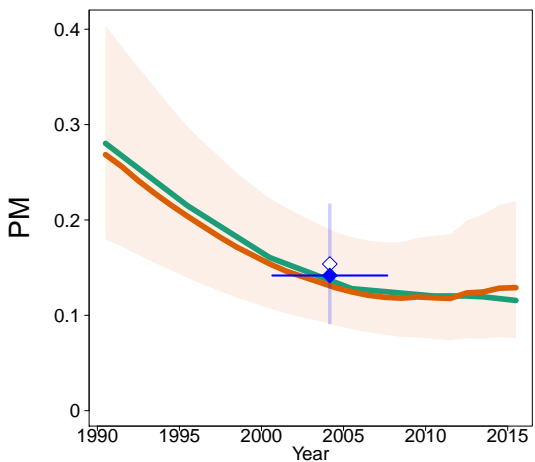

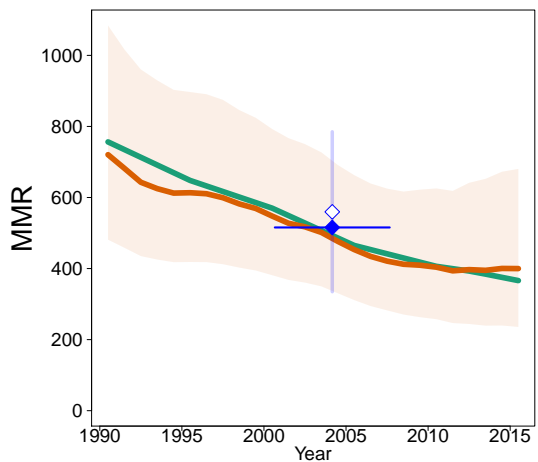

– Estimates –
Bmat 2014
MMEIG 2014
– Data –
◇ Observed Data
◆ Adjusted Data
– Data type –
◆ DHS

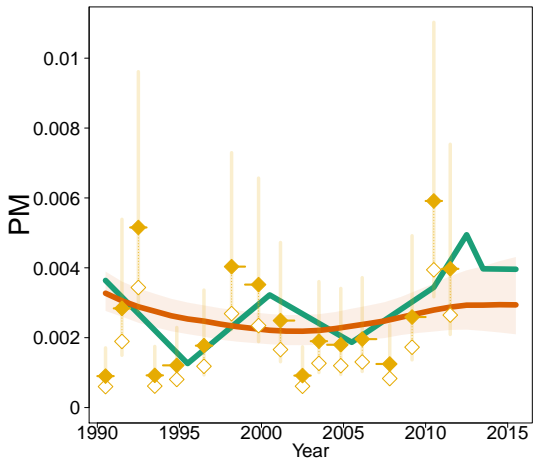

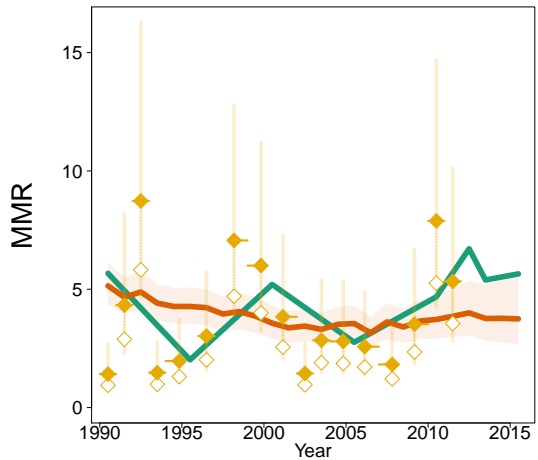

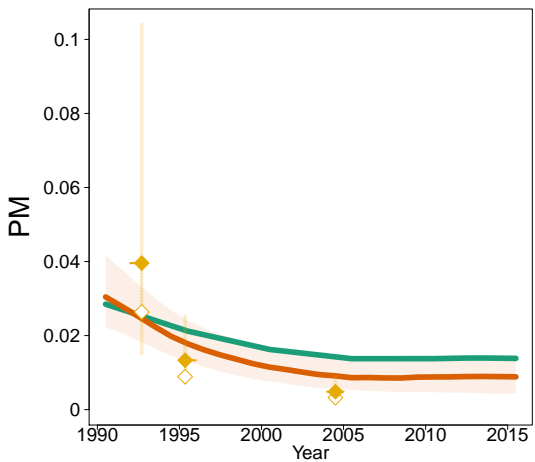

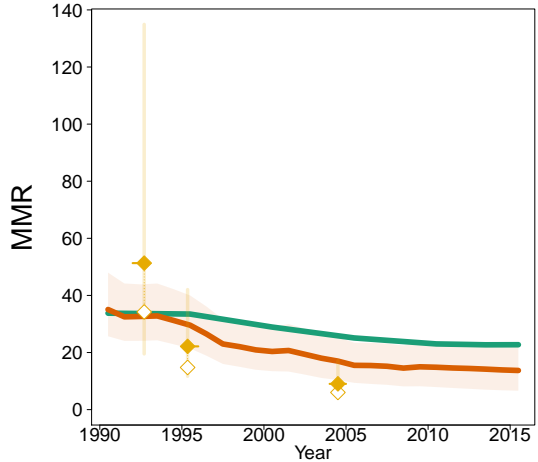

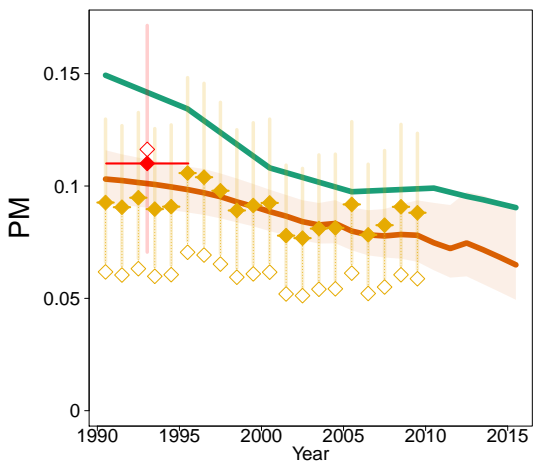

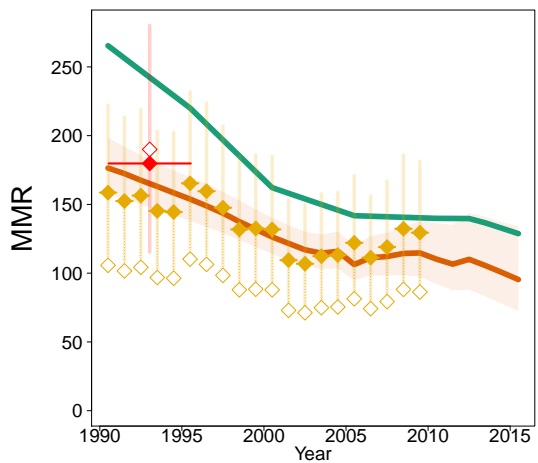

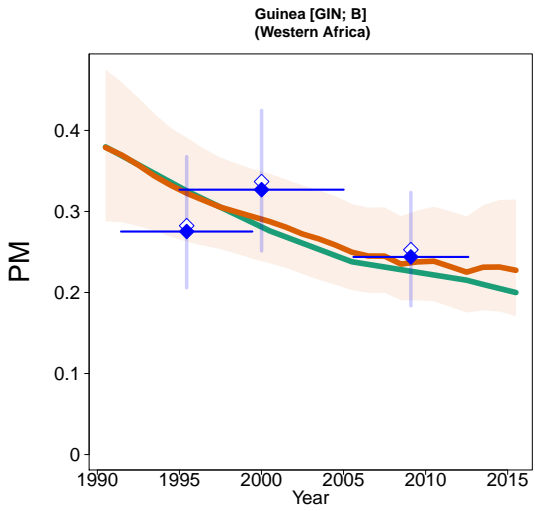

**Guinea [GIN; B]**
**(Western Africa)**

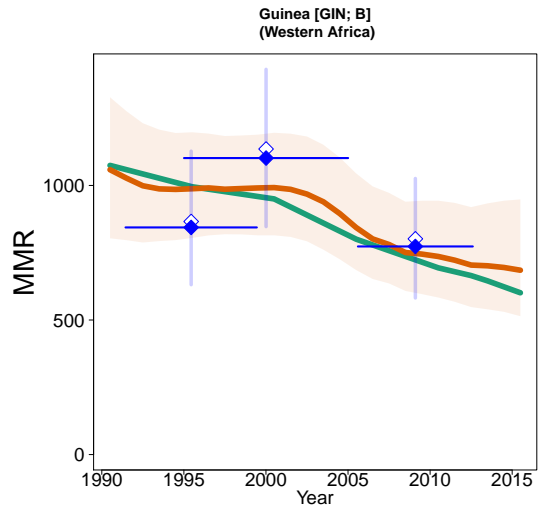

**Guinea [GIN; B]**
**(Western Africa)**

– Estimates –
Bmat 2014
MMEIG 2014
– Data –
◇ Observed Data
◆ Adjusted Data
– Data type –
◆ DHS

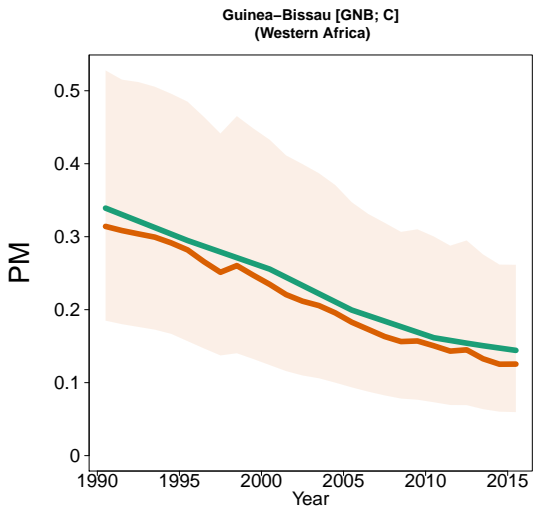

**Guinea–Bissau [GNB; C]**
**(Western Africa)**

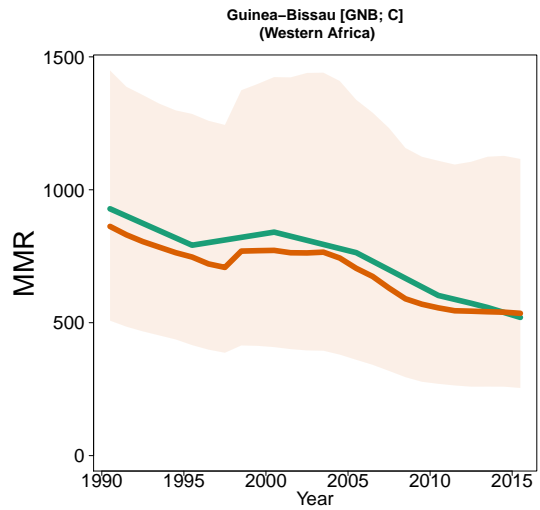

**Guinea–Bissau [GNB; C]**
**(Western Africa)**

– Estimates –
Bmat 2014
MMEIG 2014
– Data –

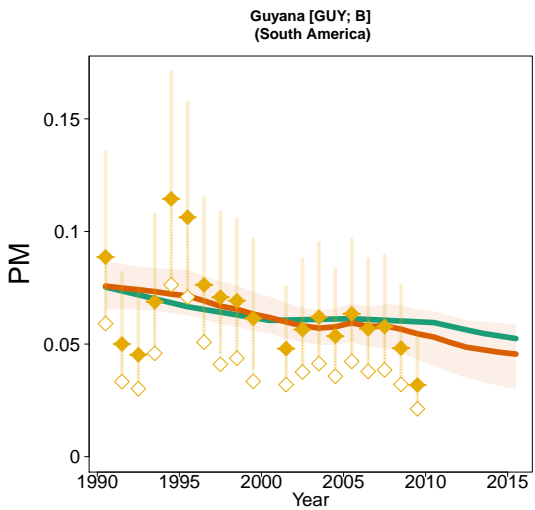

**Guyana [GUY; B]**
**(South America)**

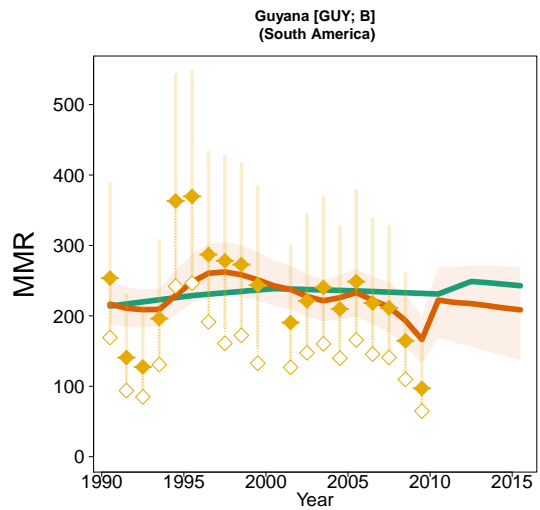

**Guyana [GUY; B]**
**(South America)**

– Estimates –
Bmat 2014
MMEIG 2014
– Data –
◇ Observed Data
◆ Adjusted Data
– Data type –
◆ VR

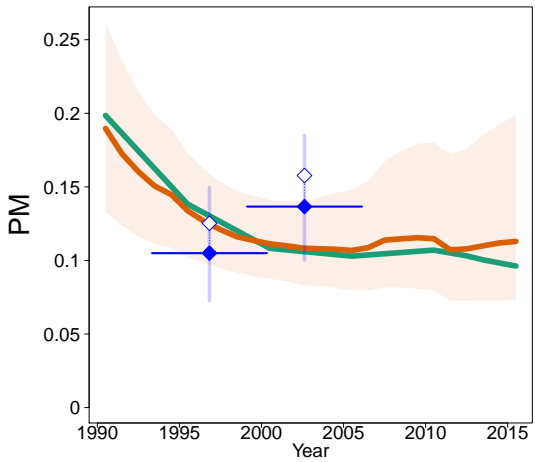

**Haiti [HTI; B]**
**(Caribbean)**

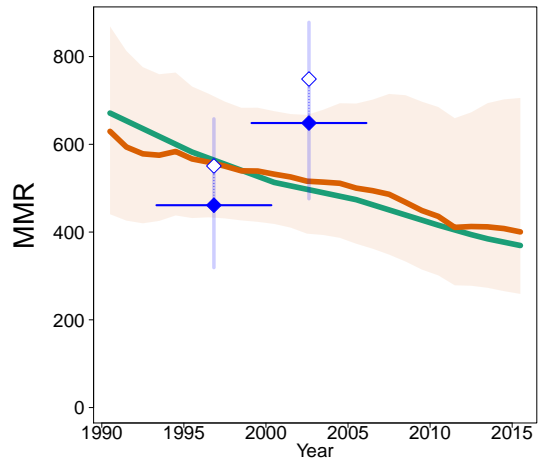

**Haiti [HTI; B]**
**(Caribbean)**

– Estimates –
Bmat 2014
MMEIG 2014
– Data –
◇ Observed Data
◆ Adjusted Data
– Data type –
◆ DHS

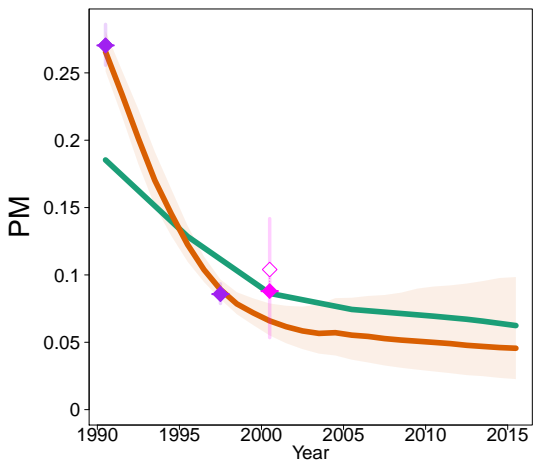

**Honduras [HND; B]**
**(Central America)**

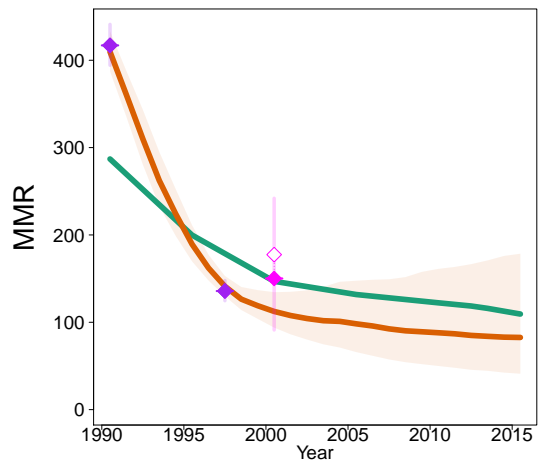

**Honduras [HND; B]**
**(Central America)**

– Estimates –
Bmat 2014
MMEIG 2014
– Data –
◇ Observed Data
◆ Adjusted Data
– Data type –
◆ Spec. studies
◆ Census

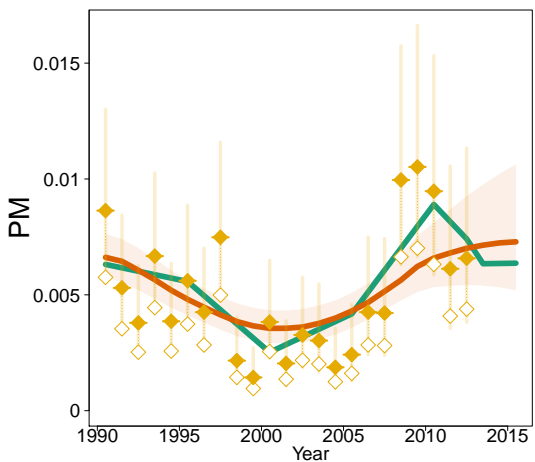

**Hungary [HUN; A]**
**(Developed regions)**

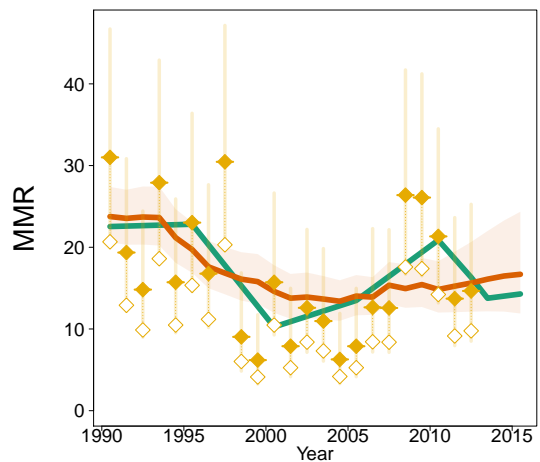

**Hungary [HUN; A]**
**(Developed regions)**

– Estimates –
Bmat 2014
MMEIG 2014
– Data –
◇ Observed Data
◆ Adjusted Data
– Data type –
◆ VR

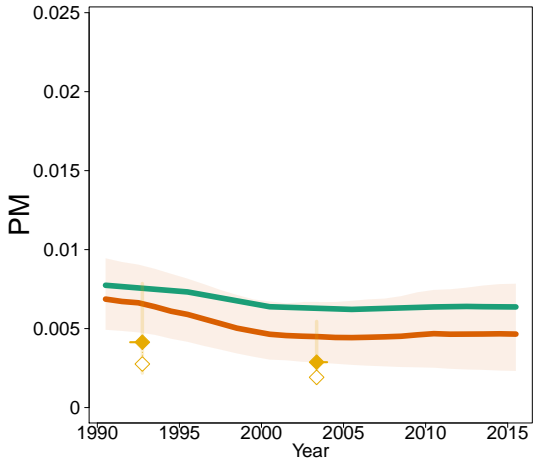

**Iceland [ISL; A]**
**(Developed regions)**

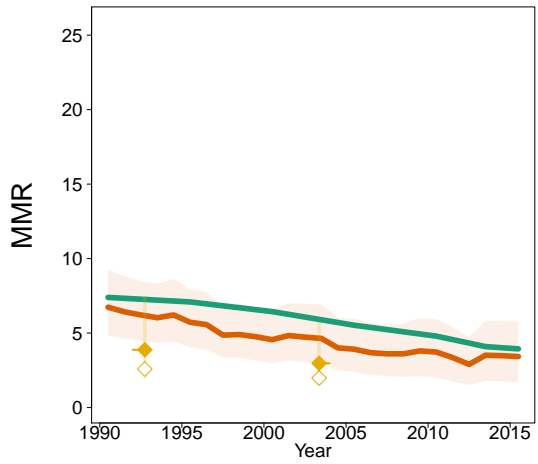

**Iceland [ISL; A]**
**(Developed regions)**

– Estimates –
Bmat 2014
MMEIG 2014
– Data –
◇ Observed Data
◆ Adjusted Data
– Data type –
◆ VR

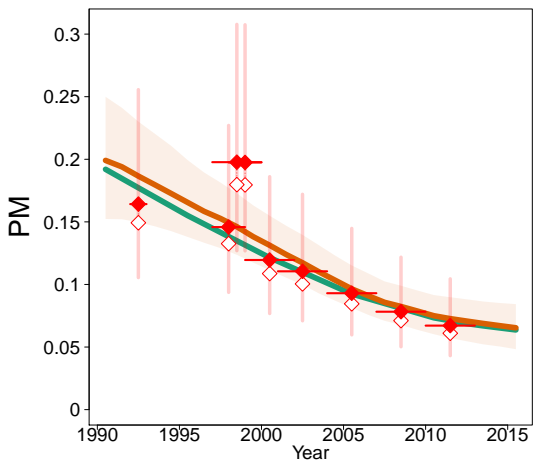

**India [IND; B]**
**(Southern Asia)**

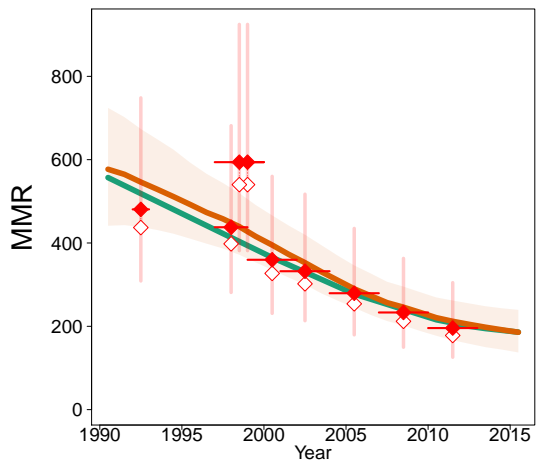

**India [IND; B]**
**(Southern Asia)**

– Estimates –
Bmat 2014
MMEIG 2014
– Data –
◇ Observed Data
◆ Adjusted Data
– Data type –
◆ Misc. studies

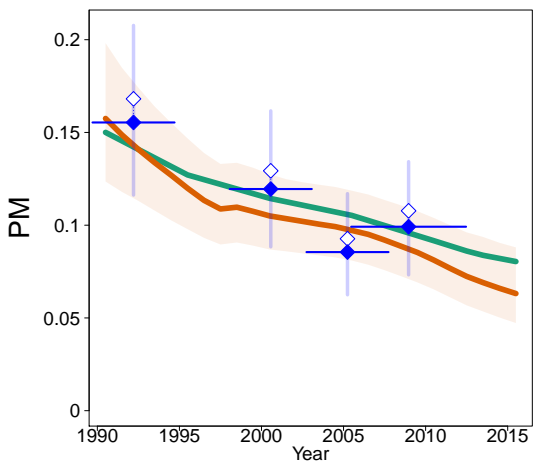

**Indonesia [IDN; B]**
**(South–Eastern Asia / Oceania)**

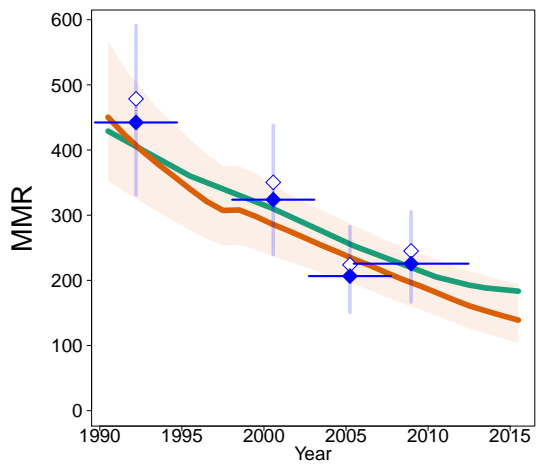

**Indonesia [IDN; B]**
**(South–Eastern Asia / Oceania)**

– Estimates –
Bmat 2014
MMEIG 2014
– Data –
◇ Observed Data
◆ Adjusted Data
– Data type –
◆ DHS

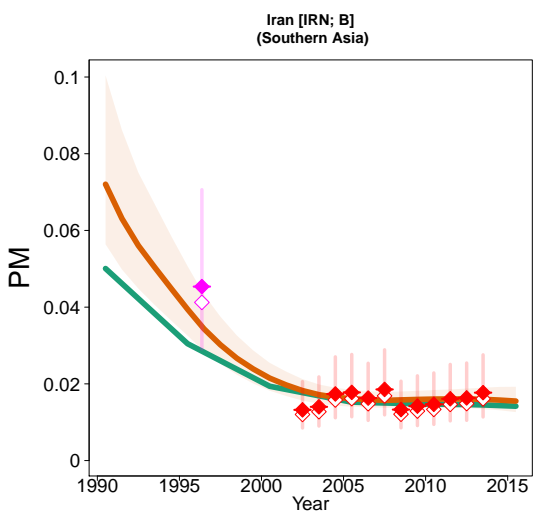

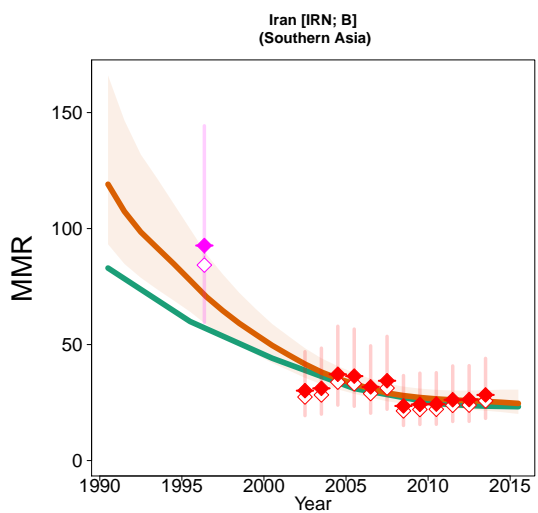

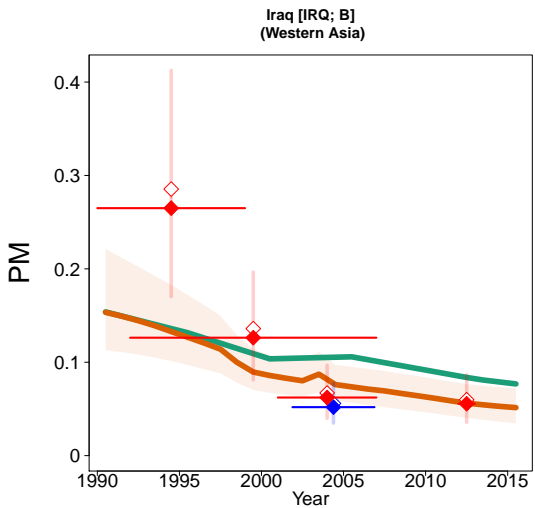

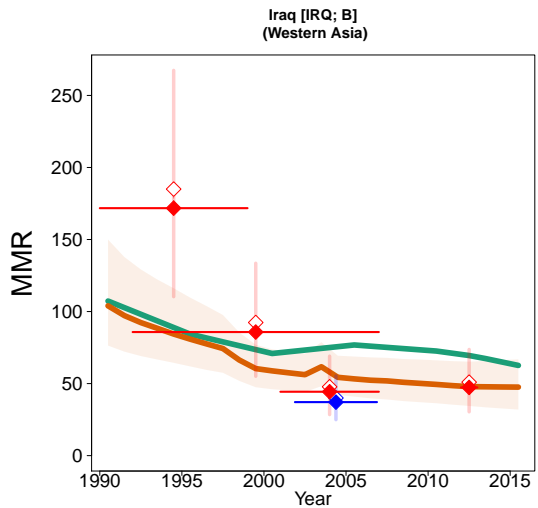

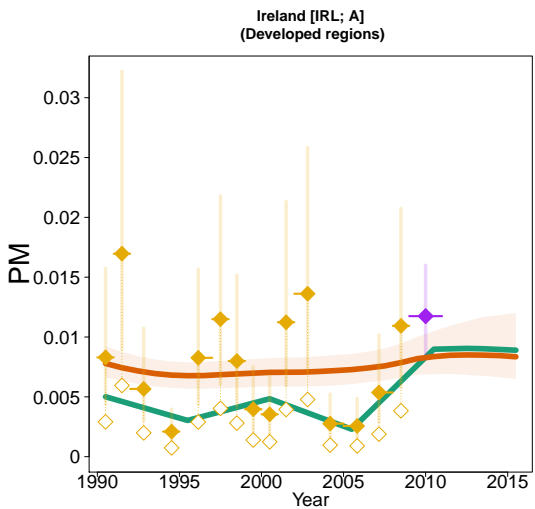

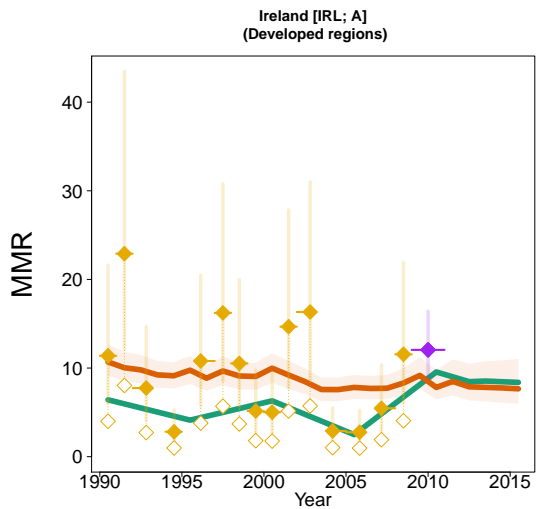

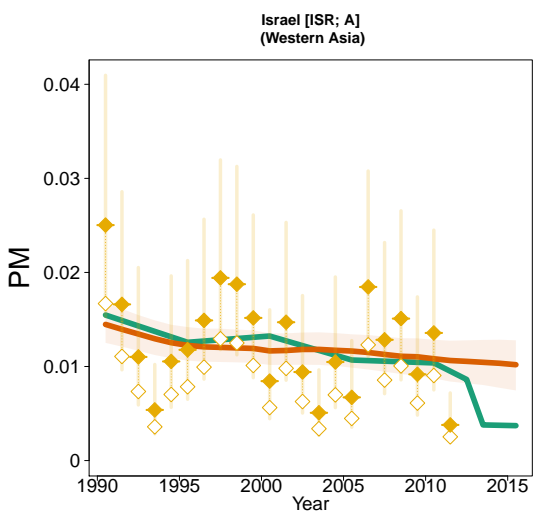

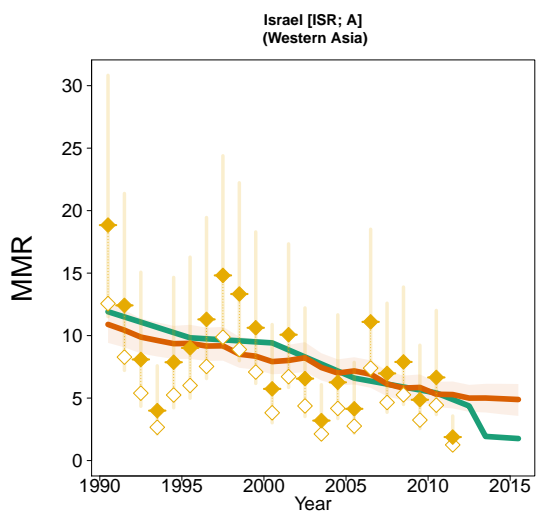

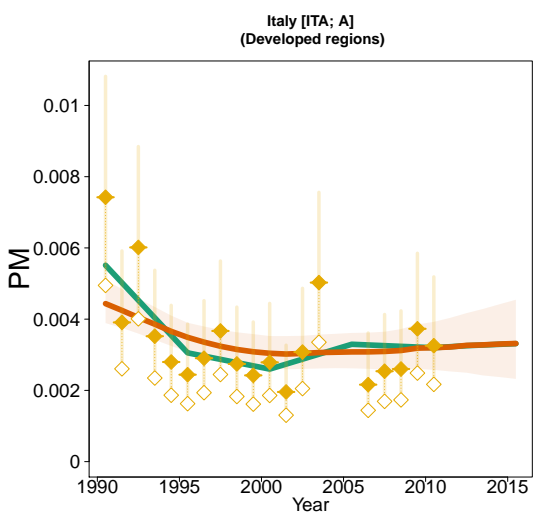

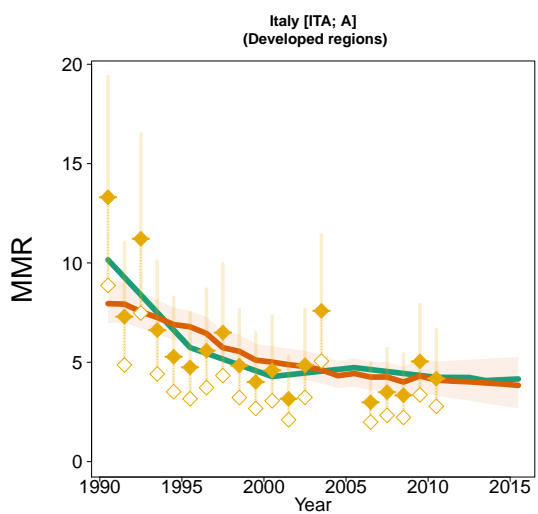

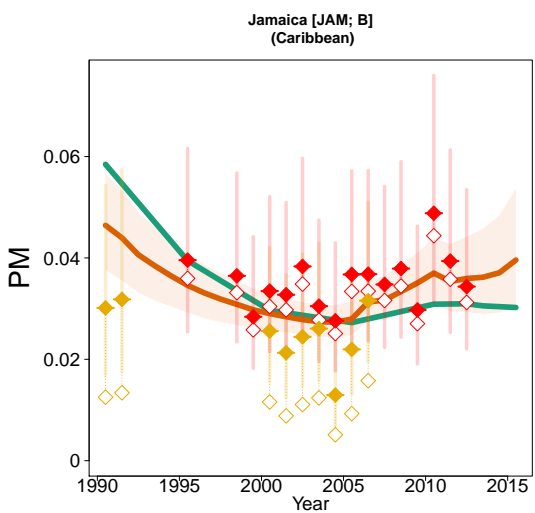

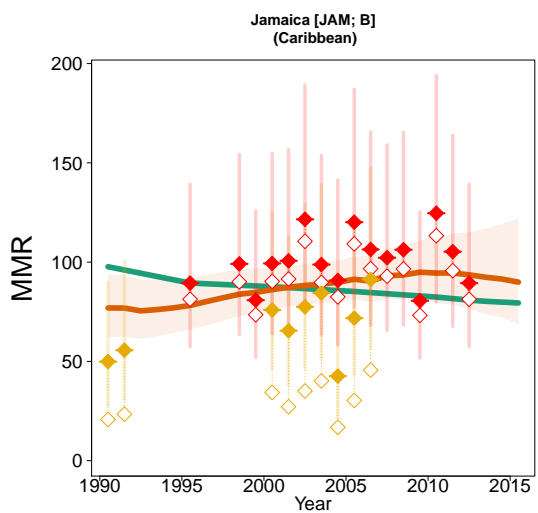

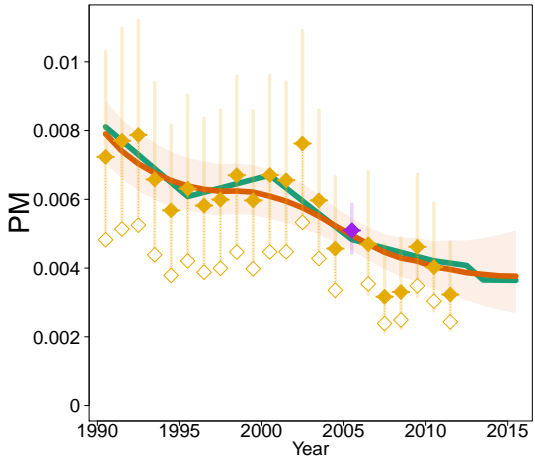

**Japan [JPN; A]**
**(Developed regions)**

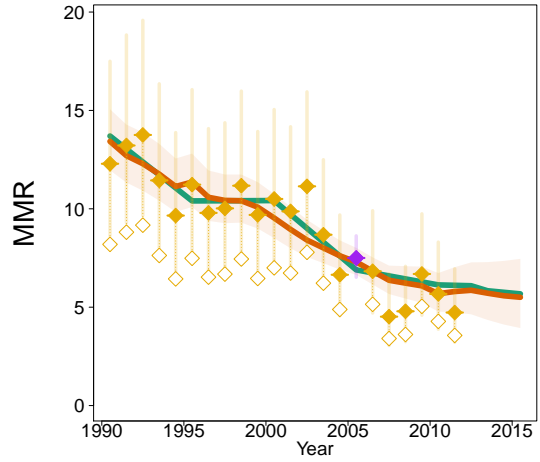

**Japan [JPN; A]**
**(Developed regions)**

– Estimates –
— Bmat 2014
— MMEIG 2014
– Data –
◇ Observed Data
◆ Adjusted Data
– Data type –
◆ VR
◆ Spec. studies

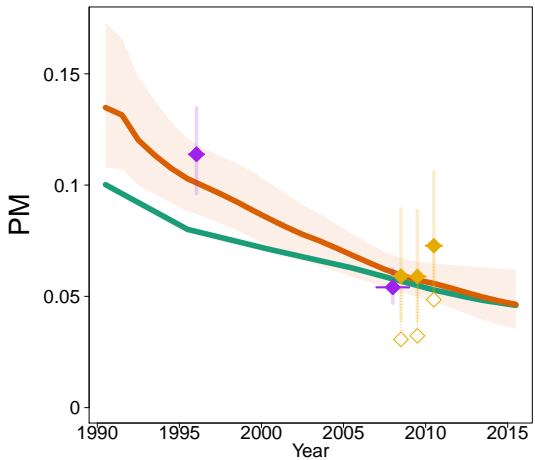

**Jordan [JOR; B]**
**(Western Asia)**

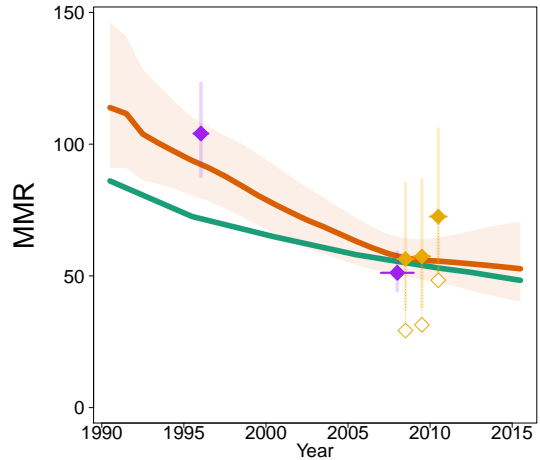

**Jordan [JOR; B]**
**(Western Asia)**

– Estimates –
— Bmat 2014
— MMEIG 2014
– Data –
◇ Observed Data
◆ Adjusted Data
– Data type –
◆ VR
◆ Spec. studies

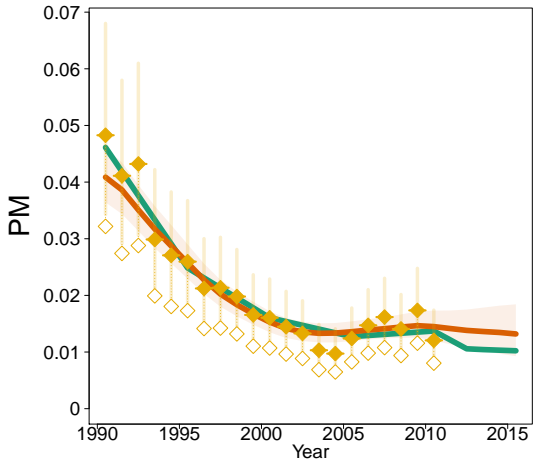

**Kazakhstan [KAZ; A]**
**(Central Asia)**

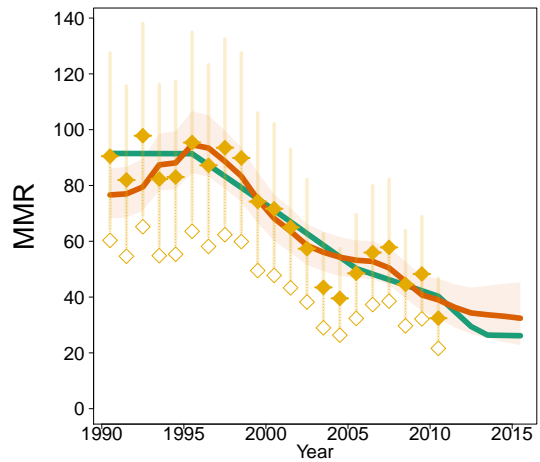

**Kazakhstan [KAZ; A]**
**(Central Asia)**

– Estimates –
— Bmat 2014
— MMEIG 2014
– Data –
◇ Observed Data
◆ Adjusted Data
– Data type –
◆ VR

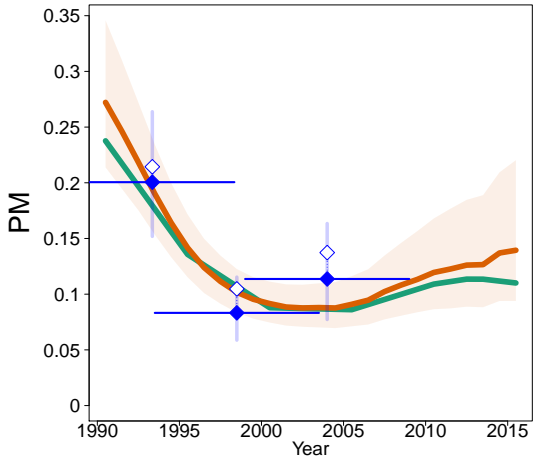

**Kenya [KEN; B]**
**(Eastern Africa)**

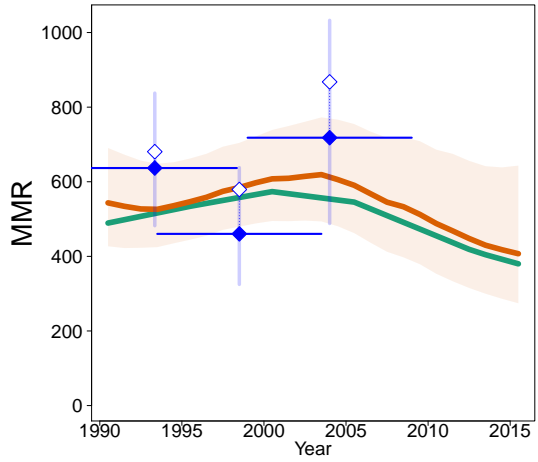

**Kenya [KEN; B]**
**(Eastern Africa)**

– Estimates –
Bmat 2014
MMEIG 2014
– Data –
Observed Data
Adjusted Data
– Data type –
DHS

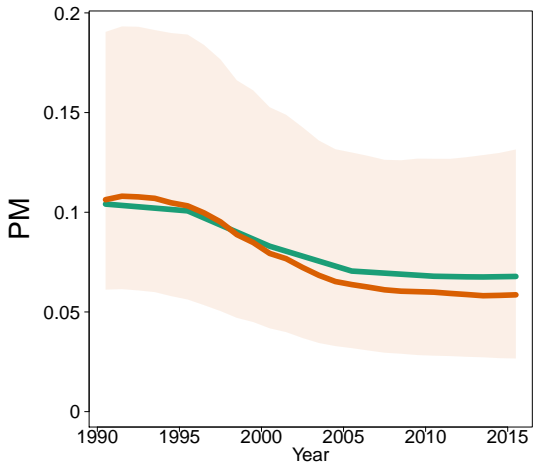

**Kiribati [KIR; C]**
**(South–Eastern Asia / Oceania)**

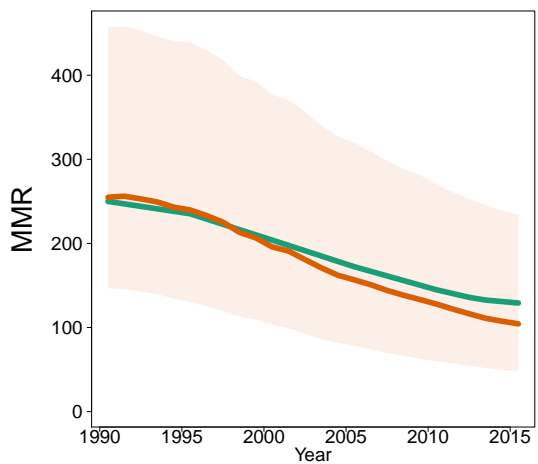

**Kiribati [KIR; C]**
**(South–Eastern Asia / Oceania)**

– Estimates –
Bmat 2014
MMEIG 2014
– Data –
Observed Data
Adjusted Data
– Data type –

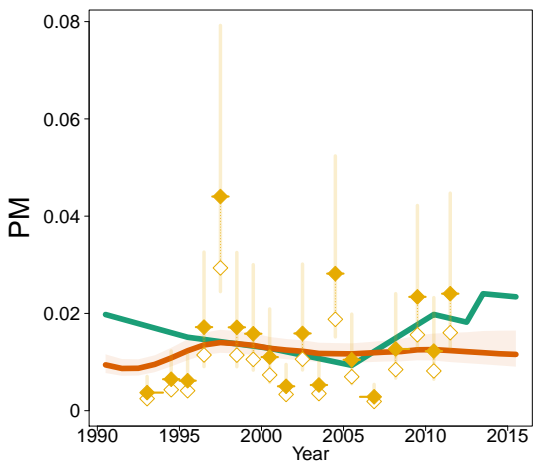

**Kuwait [KWT; A]**
**(Western Asia)**

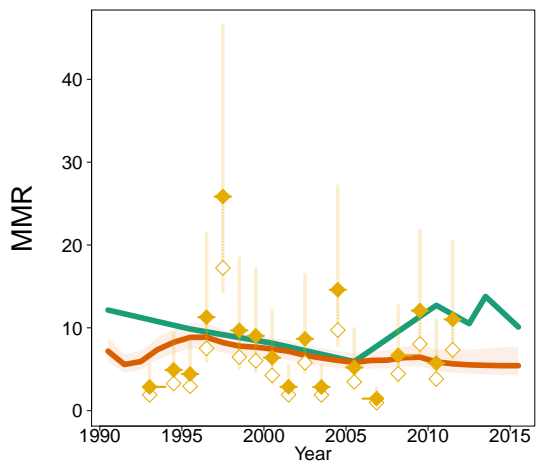

**Kuwait [KWT; A]**
**(Western Asia)**

– Estimates –
Bmat 2014
MMEIG 2014
– Data –
Observed Data
Adjusted Data
– Data type –
VR

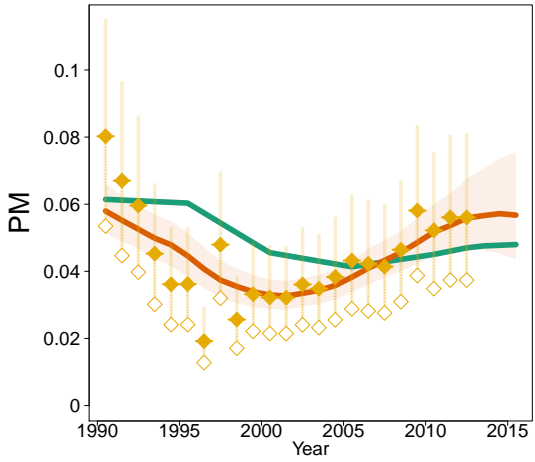

**Kyrgyzstan [KGZ; B]**
**(Central Asia)**

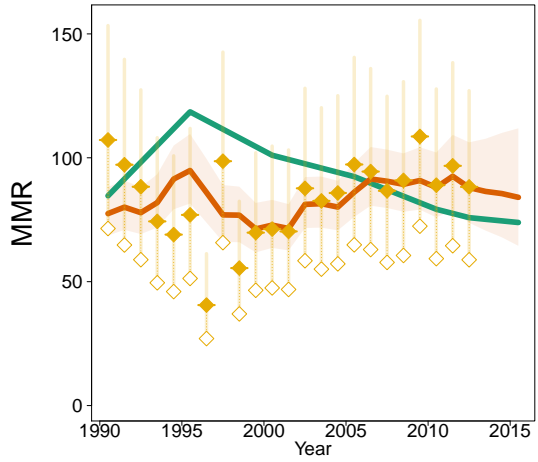

**Kyrgyzstan [KGZ; B]**
**(Central Asia)**

– Estimates –
Bmat 2014
MMEIG 2014
– Data –
◇ Observed Data
◆ Adjusted Data
– Data type –
◆ VR

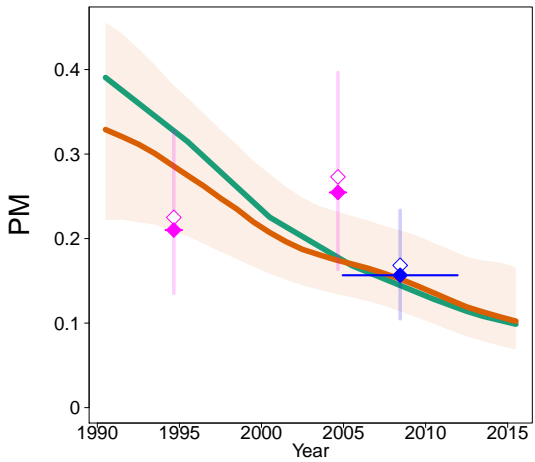

**Lao [LAO; B]**
**(South−Eastern Asia / Oceania)**

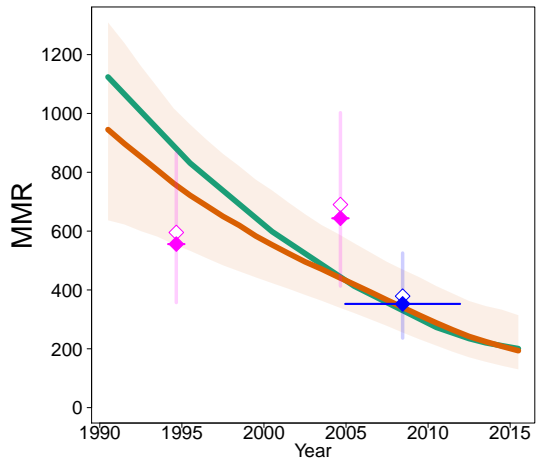

**Lao [LAO; B]**
**(South−Eastern Asia / Oceania)**

– Estimates –
Bmat 2014
MMEIG 2014
– Data –
◇ Observed Data
◆ Adjusted Data
– Data type –
◆ DHS
◆ Census

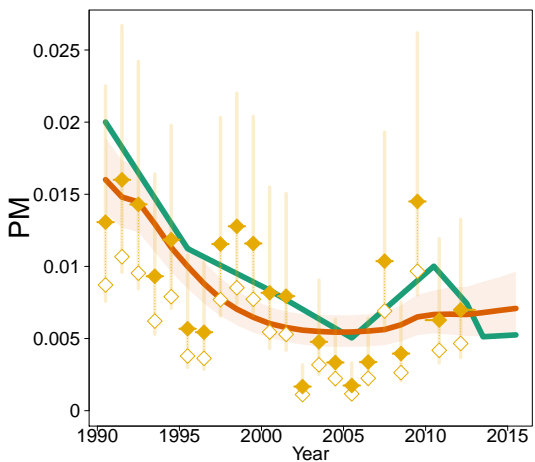

**Latvia [LVA; A]**
**(Developed regions)**

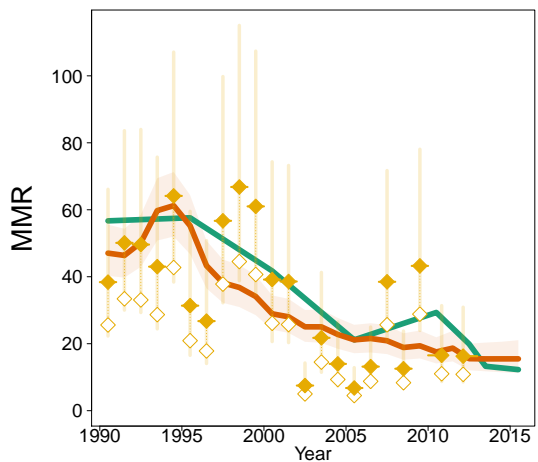

**Latvia [LVA; A]**
**(Developed regions)**

– Estimates –
Bmat 2014
MMEIG 2014
– Data –
◇ Observed Data
◆ Adjusted Data
– Data type –
◆ VR

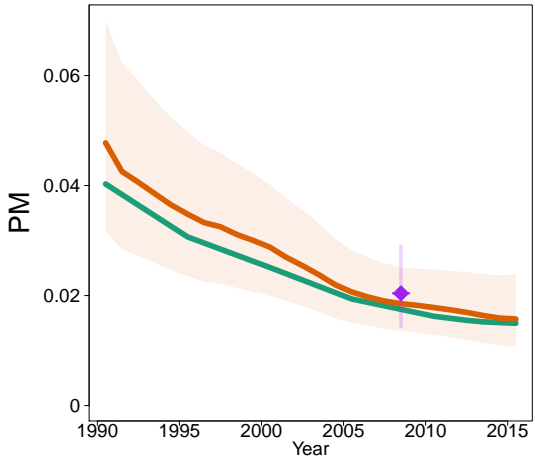

**Lebanon [LBN; B]**
**(Western Asia)**

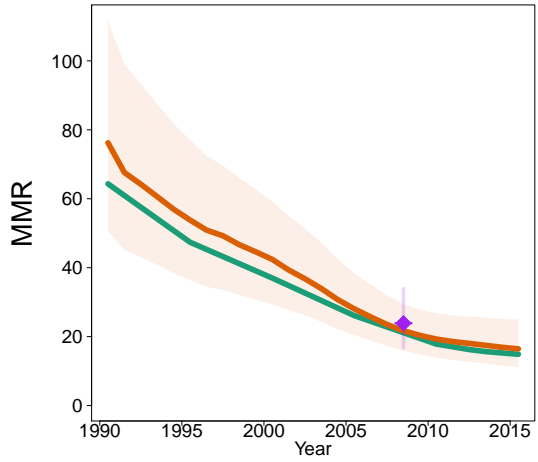

**Lebanon [LBN; B]**
**(Western Asia)**

– Estimates –
Bmat 2014
MMEIG 2014
– Data –
◇ Observed Data
◆ Adjusted Data
– Data type –
◆ Spec. studies

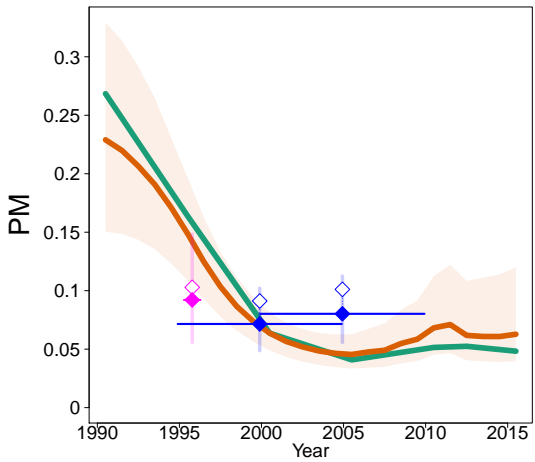

**Lesotho [LSO; B]**
**(Southern Africa)**

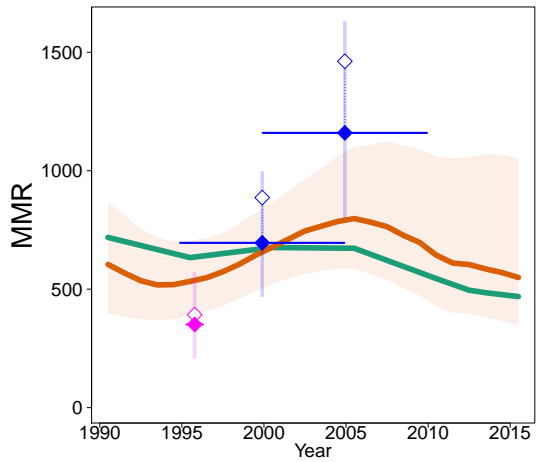

**Lesotho [LSO; B]**
**(Southern Africa)**

– Estimates –
Bmat 2014
MMEIG 2014
– Data –
◇ Observed Data
◆ Adjusted Data
– Data type –
◆ DHS
◆ Census

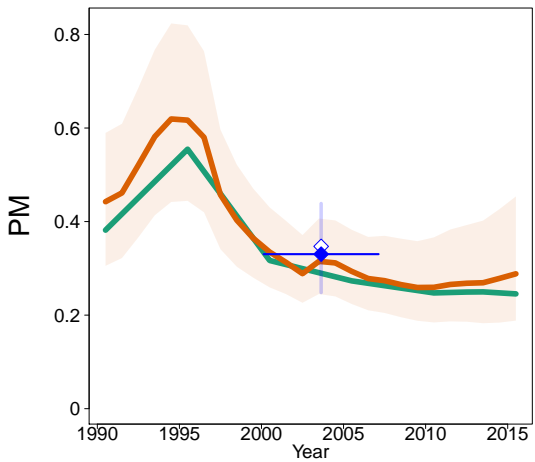

**Liberia [LBR; B]**
**(Western Africa)**

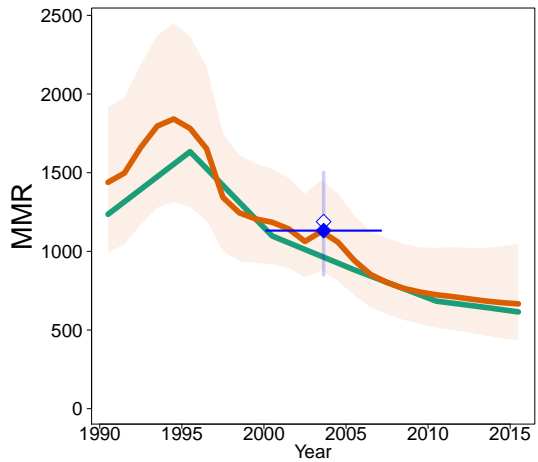

**Liberia [LBR; B]**
**(Western Africa)**

– Estimates –
Bmat 2014
MMEIG 2014
– Data –
◇ Observed Data
◆ Adjusted Data
– Data type –
◆ DHS

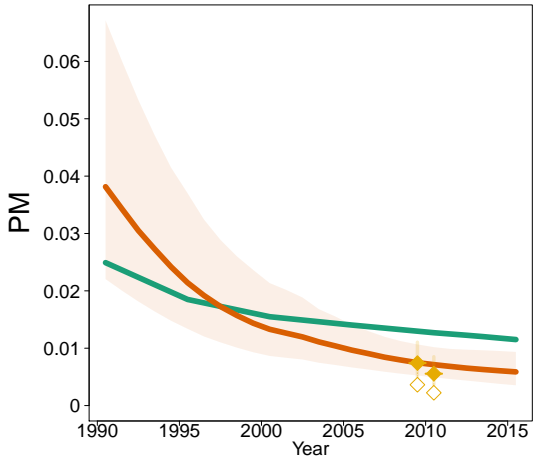

**Libya [LBY; B]**
**(Northern Africa)**

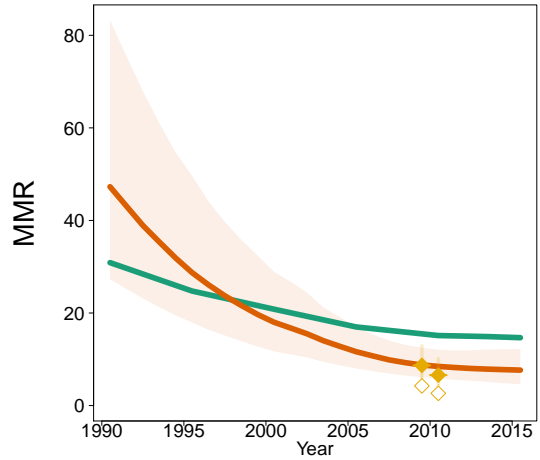

**Libya [LBY; B]**
**(Northern Africa)**

– Estimates –
Bmat 2014
MMEIG 2014
– Data –
◇ Observed Data
◆ Adjusted Data
– Data type –
◆ VR

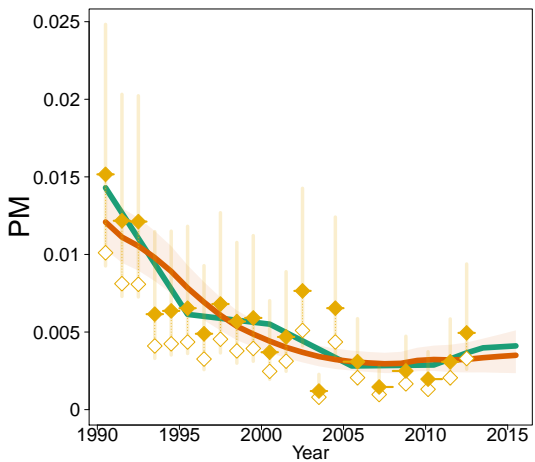

**Lithuania [LTU; A]**
**(Developed regions)**

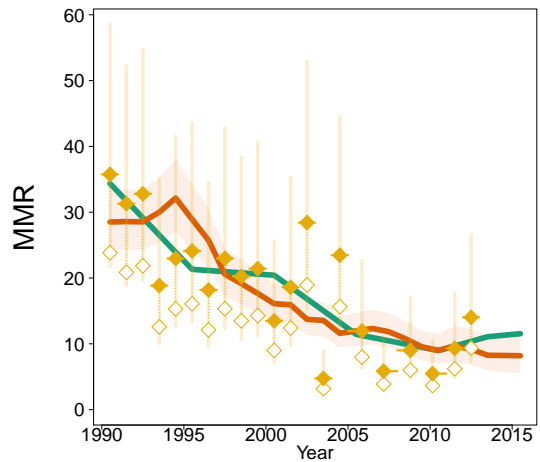

**Lithuania [LTU; A]**
**(Developed regions)**

– Estimates –
Bmat 2014
MMEIG 2014
– Data –
◇ Observed Data
◆ Adjusted Data
– Data type –
◆ VR

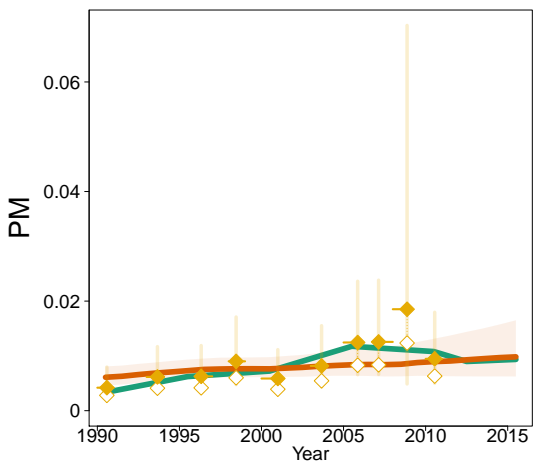

**Luxembourg [LUX; A]**
**(Developed regions)**

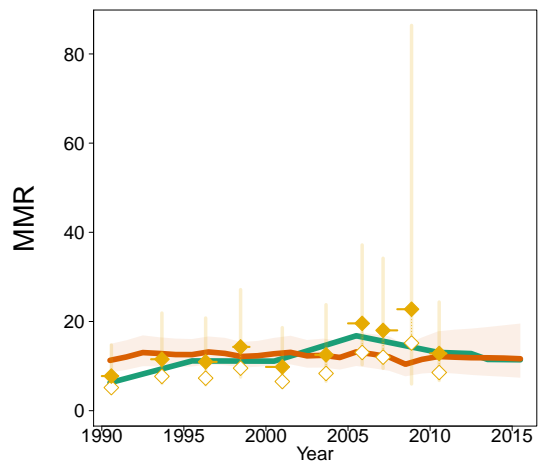

**Luxembourg [LUX; A]**
**(Developed regions)**

– Estimates –
Bmat 2014
MMEIG 2014
– Data –
◇ Observed Data
◆ Adjusted Data
– Data type –
◆ VR

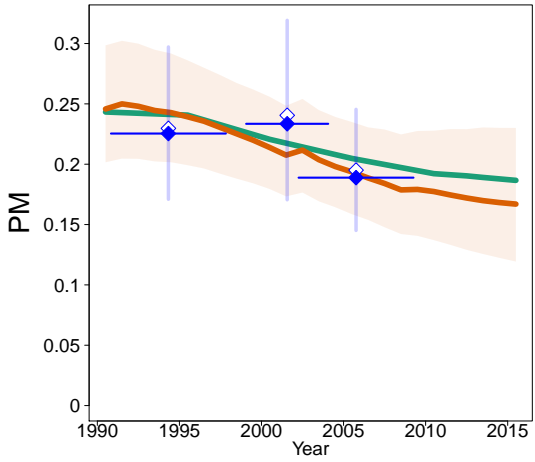

**Madagascar [MDG; B]**
**(Eastern Africa)**

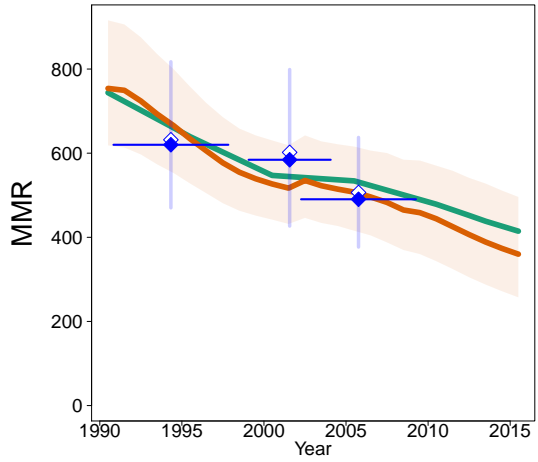

**Madagascar [MDG; B]**
**(Eastern Africa)**

– Estimates –
Bmat 2014
MMEIG 2014
– Data –
◇ Observed Data
◆ Adjusted Data
– Data type –
◆ DHS

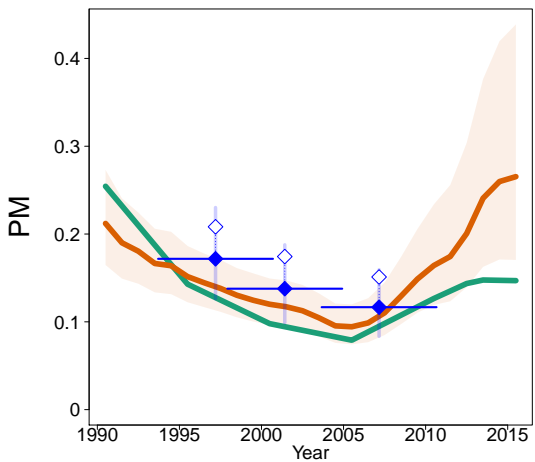

**Malawi [MWI; B]**
**(Eastern Africa)**

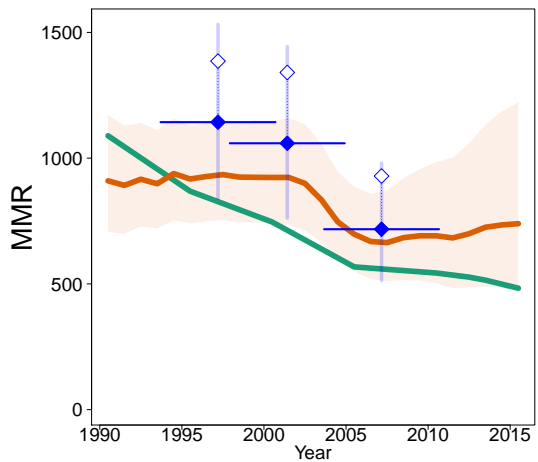

**Malawi [MWI; B]**
**(Eastern Africa)**

– Estimates –
Bmat 2014
MMEIG 2014
– Data –
◇ Observed Data
◆ Adjusted Data
– Data type –
◆ DHS

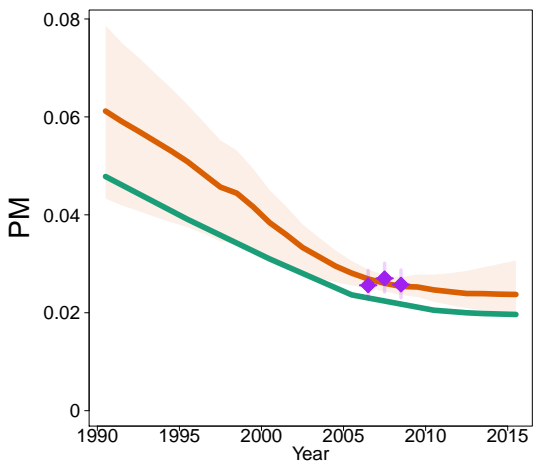

**Malaysia [MYS; B]**
**(South−Eastern Asia / Oceania)**

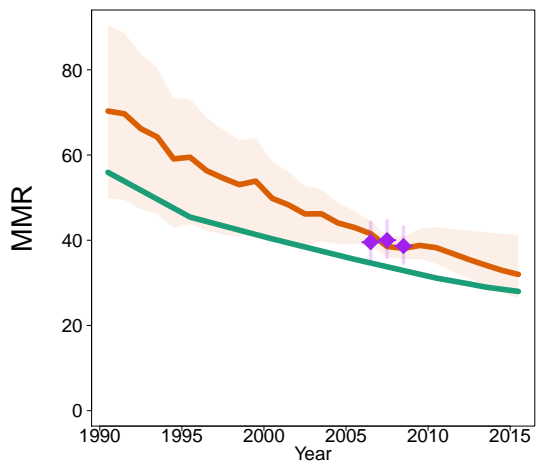

**Malaysia [MYS; B]**
**(South−Eastern Asia / Oceania)**

– Estimates –
Bmat 2014
MMEIG 2014
– Data –
◇ Observed Data
◆ Adjusted Data
– Data type –
◆ Spec. studies

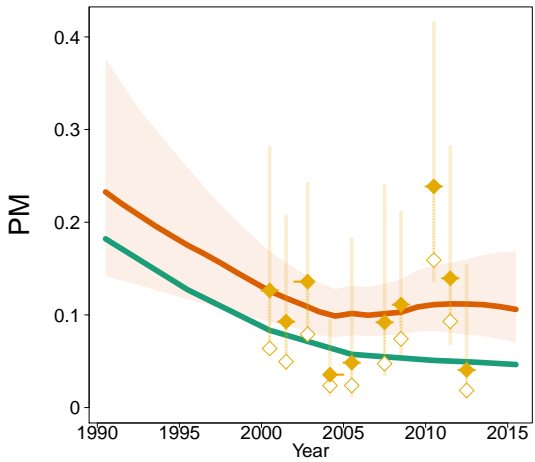

**Maldives [MDV; B]**
**(Southern Asia)**

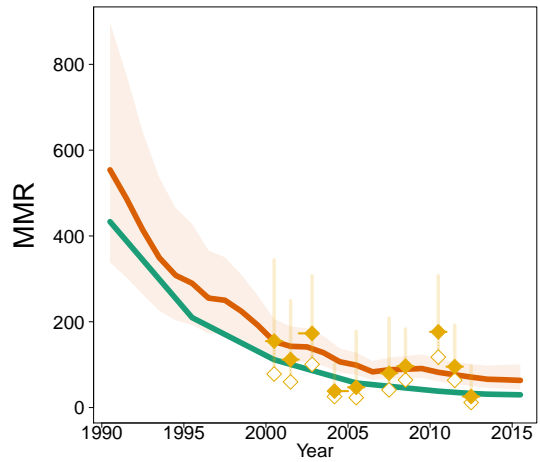

**Maldives [MDV; B]**
**(Southern Asia)**

– Estimates –
Bmat 2014
MMEIG 2014
– Data –
◇ Observed Data
◆ Adjusted Data
– Data type –
◆ VR

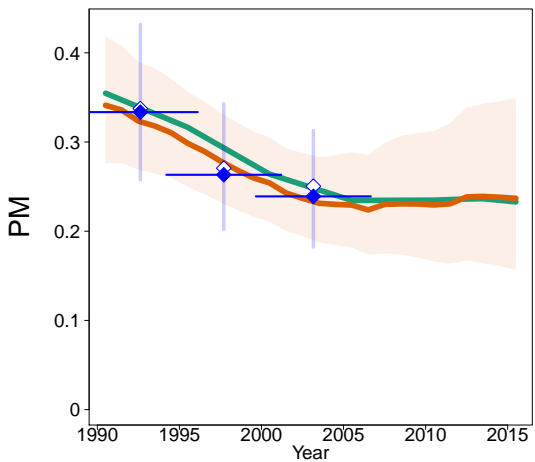

**Mali [MLI; B]**
**(Western Africa)**

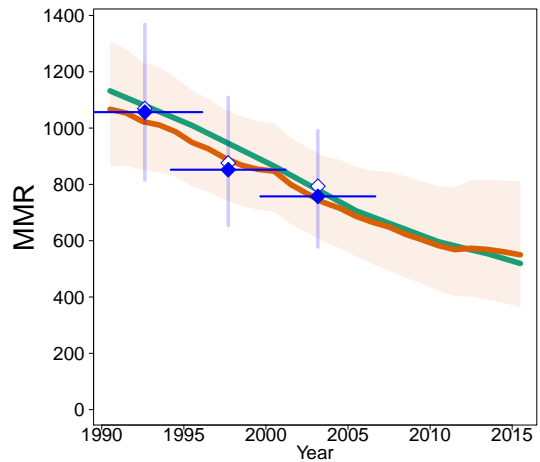

**Mali [MLI; B]**
**(Western Africa)**

– Estimates –
Bmat 2014
MMEIG 2014
– Data –
◇ Observed Data
◆ Adjusted Data
– Data type –
◆ DHS

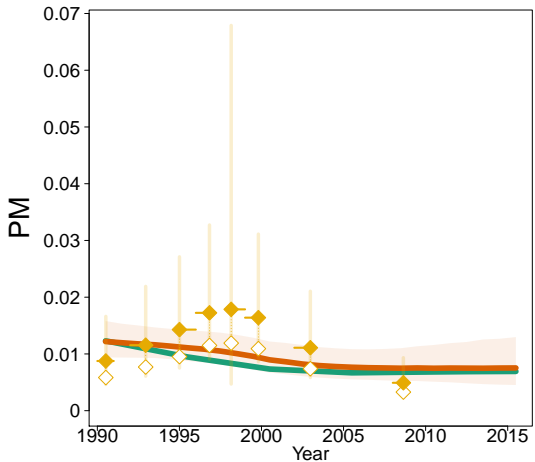

**Malta [MLT; A]**
**(Developed regions)**

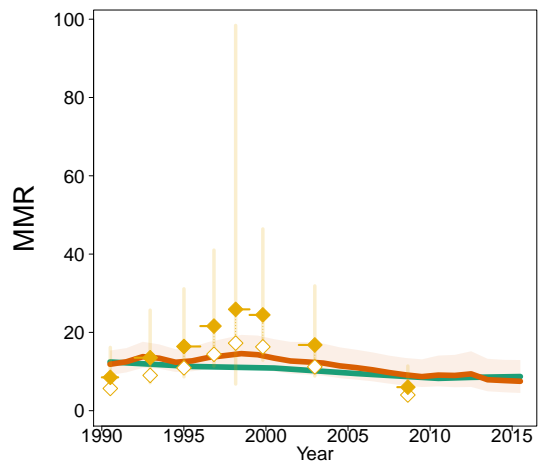

**Malta [MLT; A]**
**(Developed regions)**

– Estimates –
Bmat 2014
MMEIG 2014
– Data –
◇ Observed Data
◆ Adjusted Data
– Data type –
◆ VR

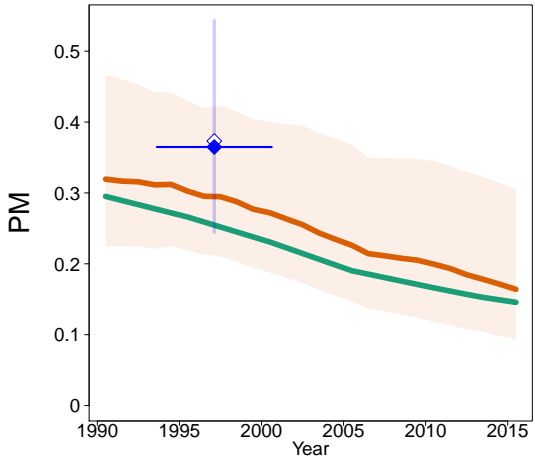

**Mauritania [MRT; B]**
**(Western Africa)**

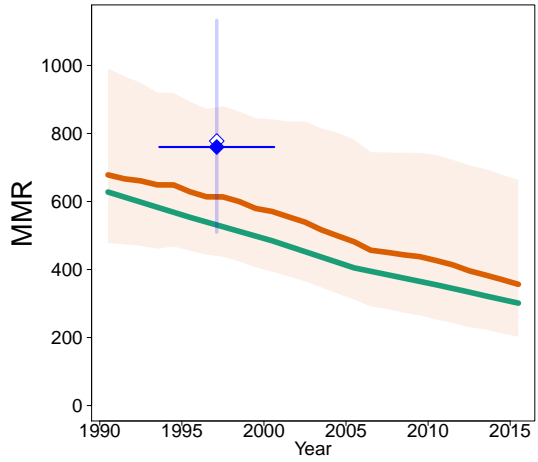

**Mauritania [MRT; B]**
**(Western Africa)**

– Estimates –
Bmat 2014
MMEIG 2014
– Data –
◇ Observed Data
◆ Adjusted Data
– Data type –
◆ DHS

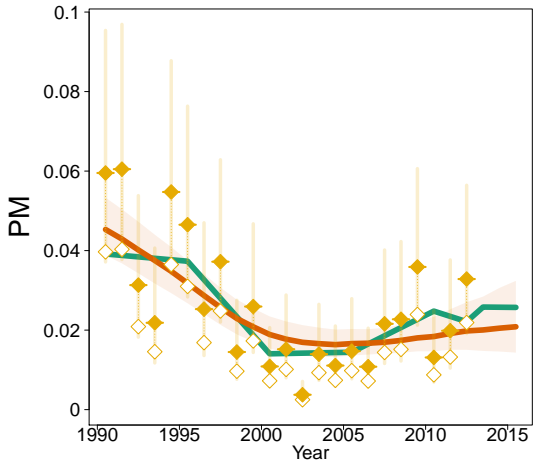

**Mauritius [MUS; A]**
**(Eastern Africa)**

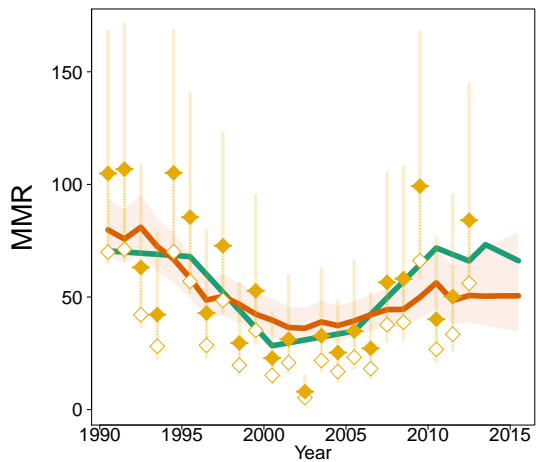

**Mauritius [MUS; A]**
**(Eastern Africa)**

– Estimates –
Bmat 2014
MMEIG 2014
– Data –
◇ Observed Data
◆ Adjusted Data
– Data type –
◆ VR

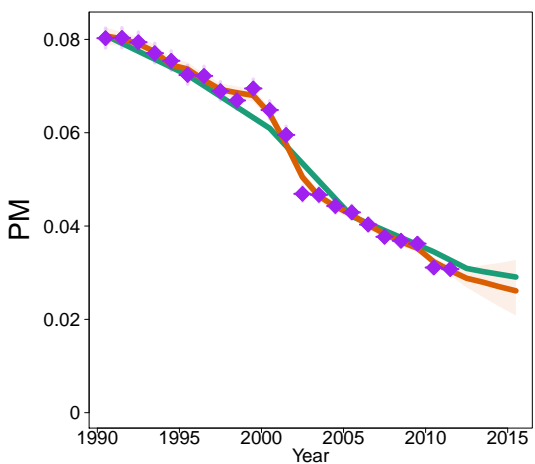

**Mexico [MEX; A]**
**(Central America)**

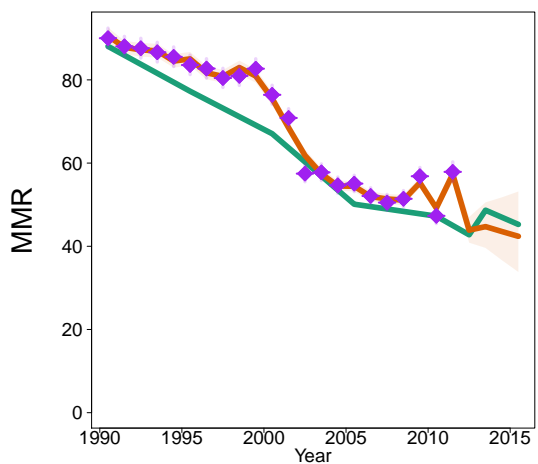

**Mexico [MEX; A]**
**(Central America)**

– Estimates –
Bmat 2014
MMEIG 2014
– Data –
◇ Observed Data
◆ Adjusted Data
– Data type –
◆ Spec. studies

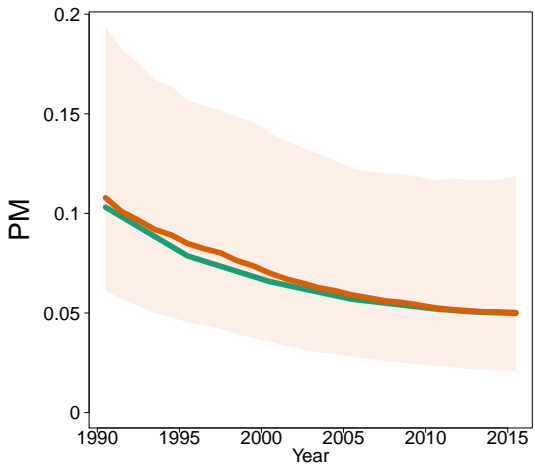

**Micronesia [FSM; C]**
**(South−Eastern Asia / Oceania)**

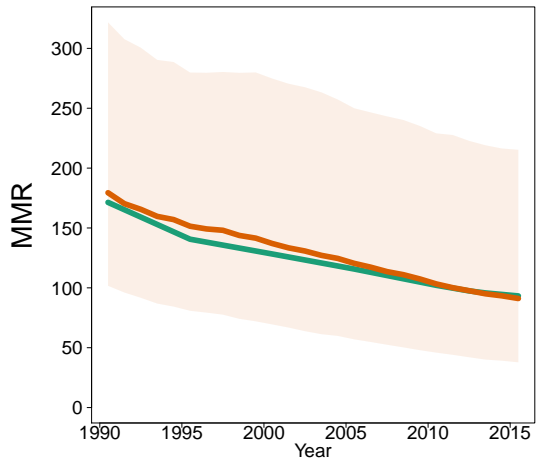

**Micronesia [FSM; C]**
**(South−Eastern Asia / Oceania)**

– Estimates –
Bmat 2014
MMEIG 2014
– Data –

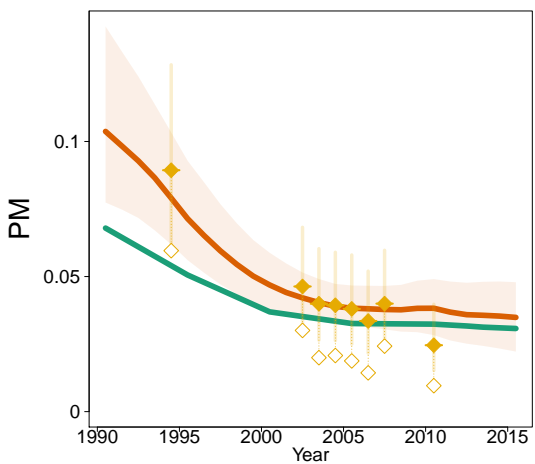

**Mongolia [MNG; B]**
**(Eastern Asia)**

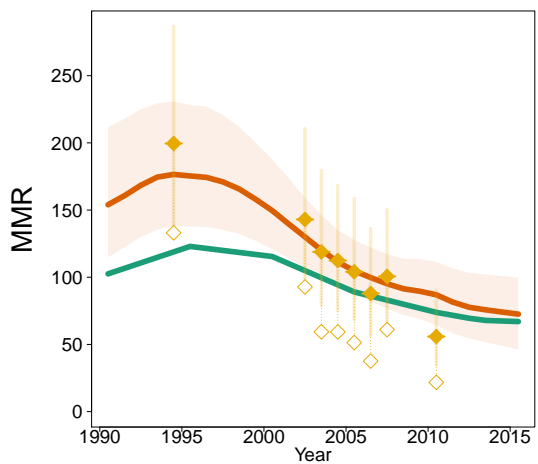

**Mongolia [MNG; B]**
**(Eastern Asia)**

– Estimates –
Bmat 2014
MMEIG 2014
– Data –
◇ Observed Data
◆ Adjusted Data
– Data type –
◆ VR

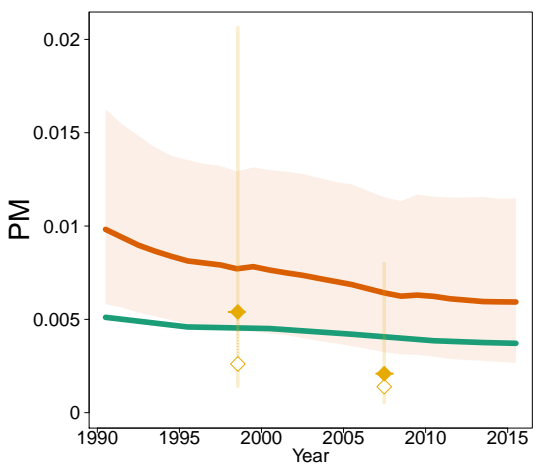

**Montenegro [MNE; B]**
**(Transition countries of South−Eastern Europe)**

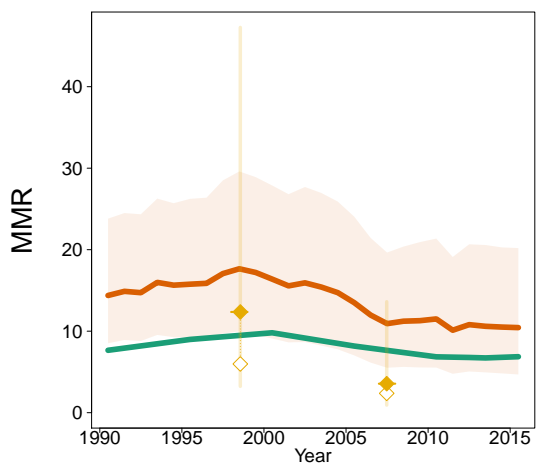

**Montenegro [MNE; B]**
**(Transition countries of South−Eastern Europe)**

– Estimates –
Bmat 2014
MMEIG 2014
– Data –
◇ Observed Data
◆ Adjusted Data
– Data type –
◆ VR

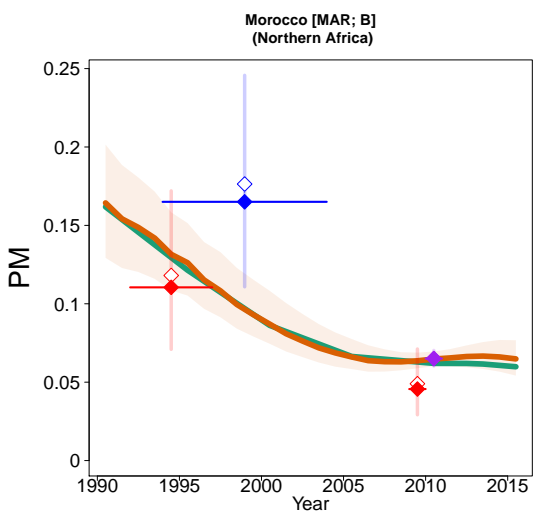

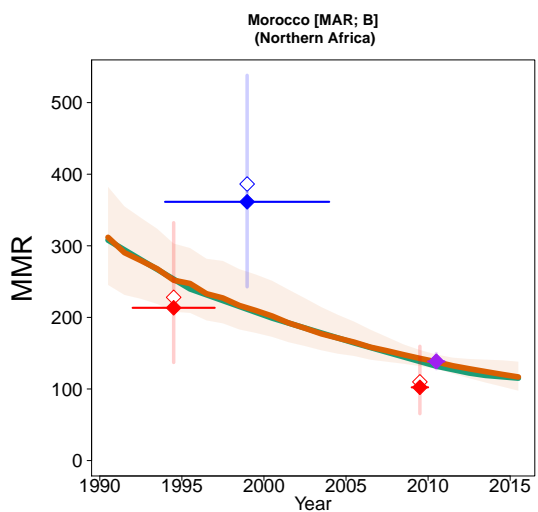

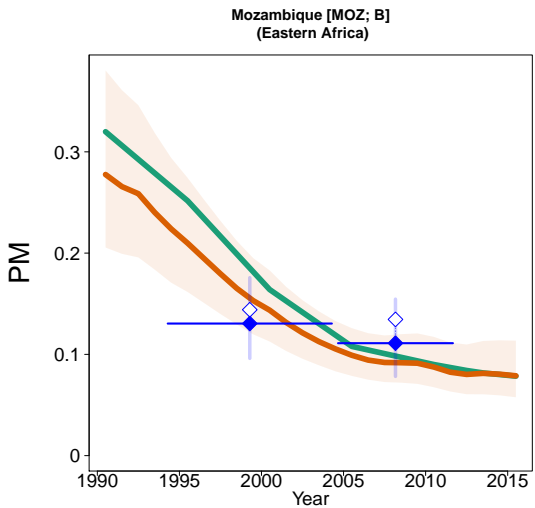

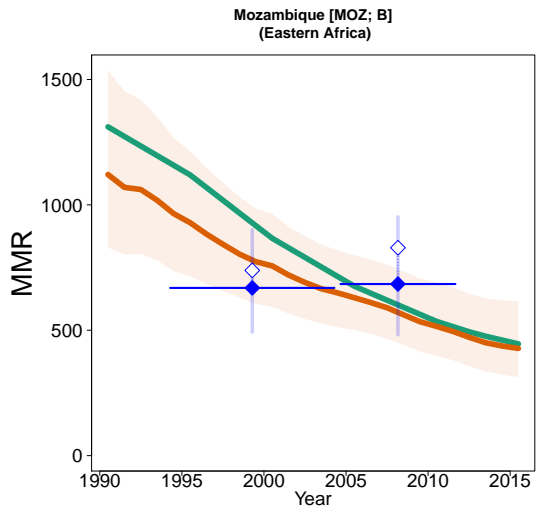

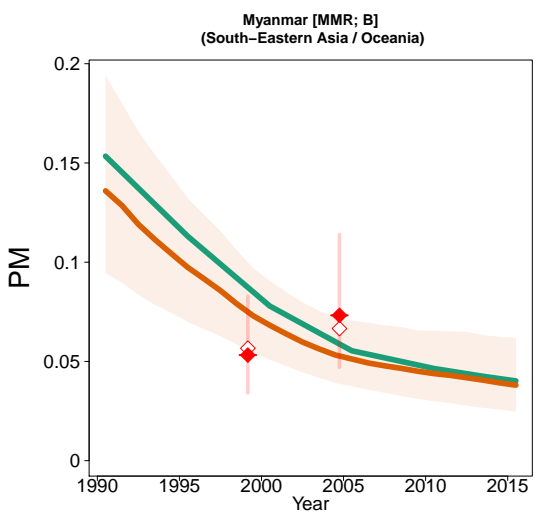

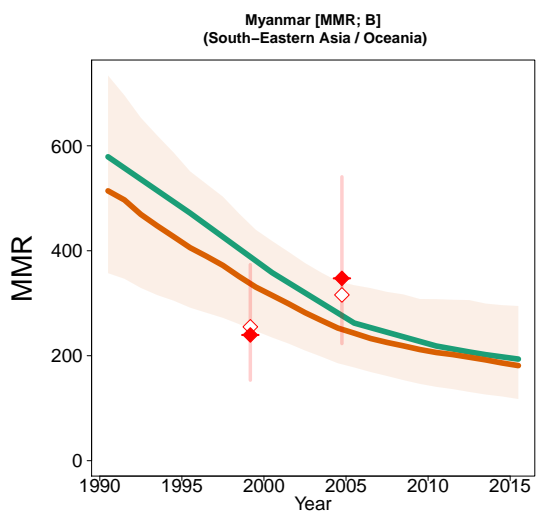

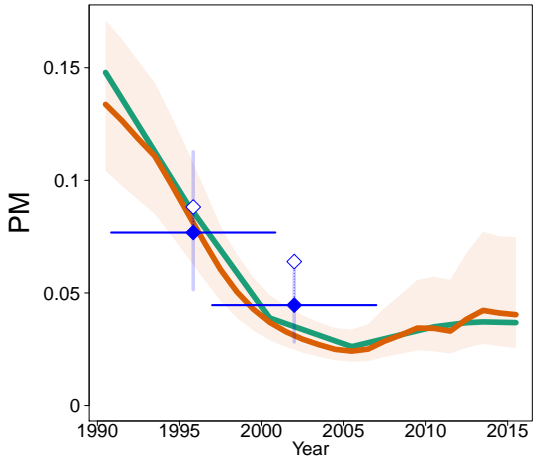

Namibia [NAM; B]
(Southern Africa)

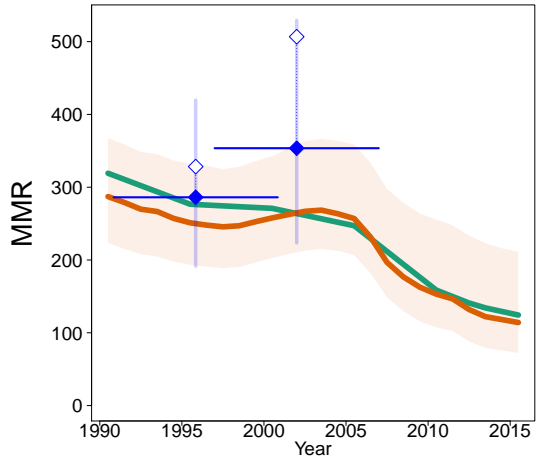

Namibia [NAM; B]
(Southern Africa)

– Estimates –
Bmat 2014
MMEIG 2014
– Data –
Observed Data
Adjusted Data
– Data type –
DHS

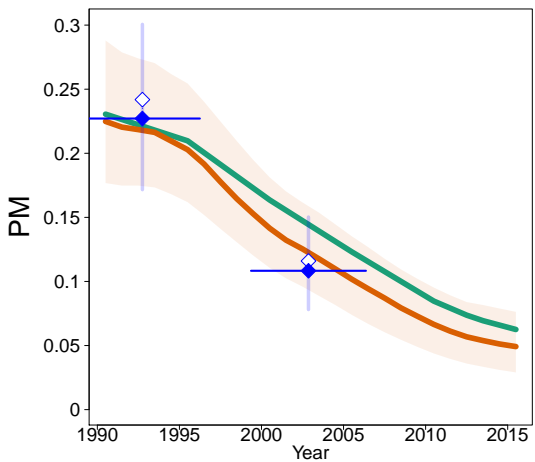

Nepal [NPL; B]
(Southern Asia)

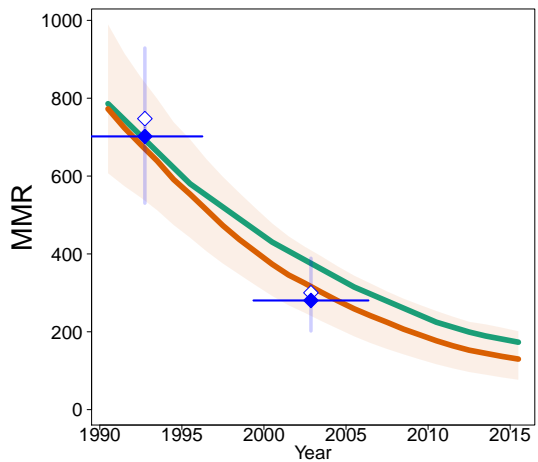

Nepal [NPL; B]
(Southern Asia)

– Estimates –
Bmat 2014
MMEIG 2014
– Data –
Observed Data
Adjusted Data
– Data type –
DHS

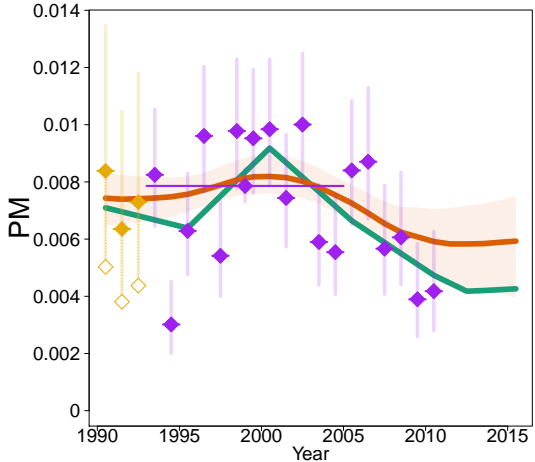

Netherlands [NLD; A]
(Developed regions)

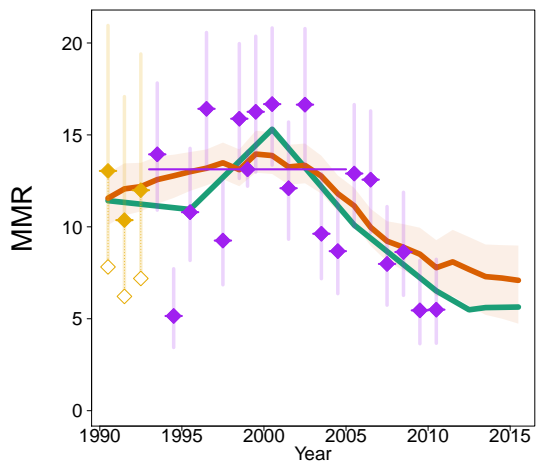

Netherlands [NLD; A]
(Developed regions)

– Estimates –
Bmat 2014
MMEIG 2014
– Data –
Observed Data
Adjusted Data
– Data type –
VR
Spec. studies

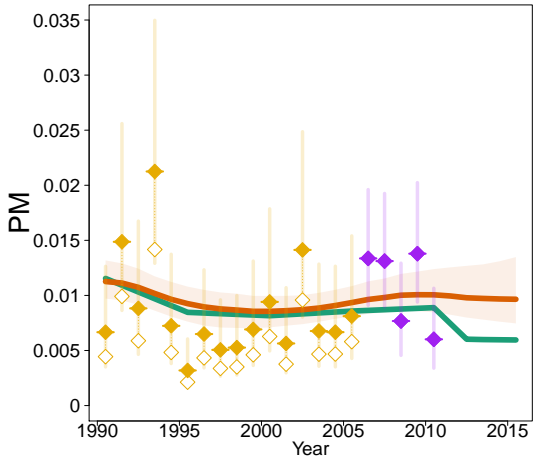

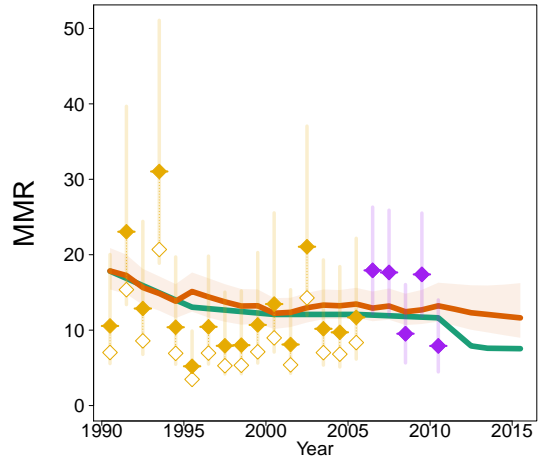

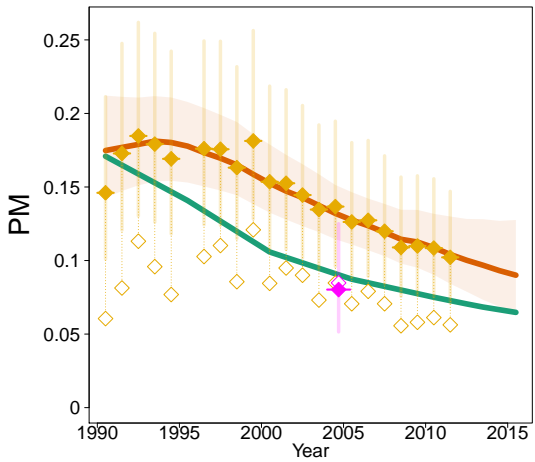

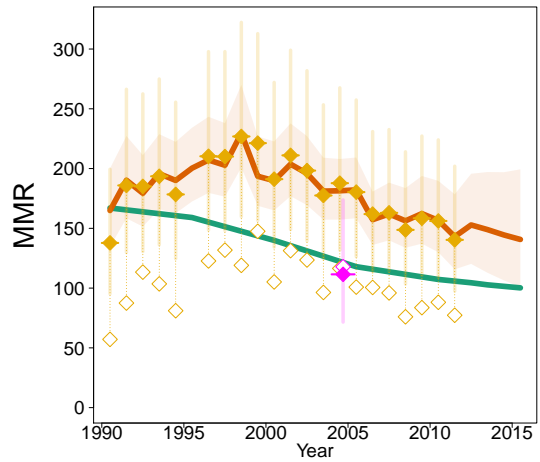

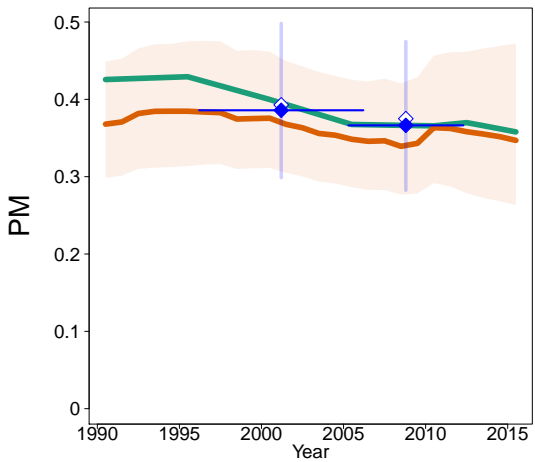

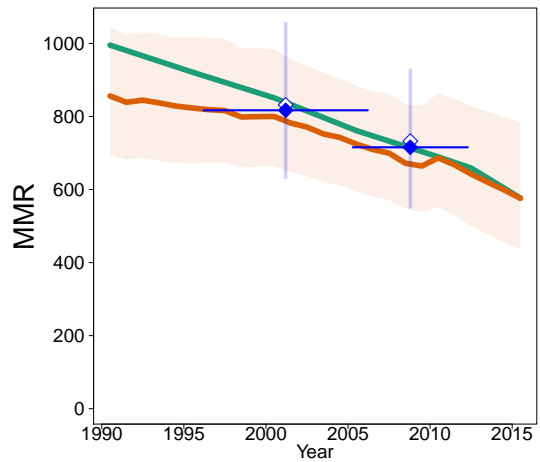

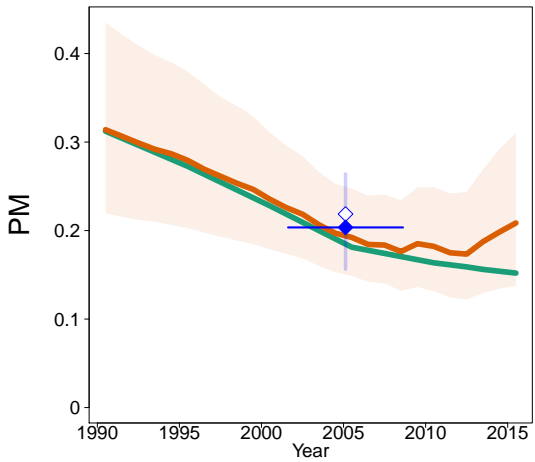

Nigeria [NGA; B]
(Western Africa)

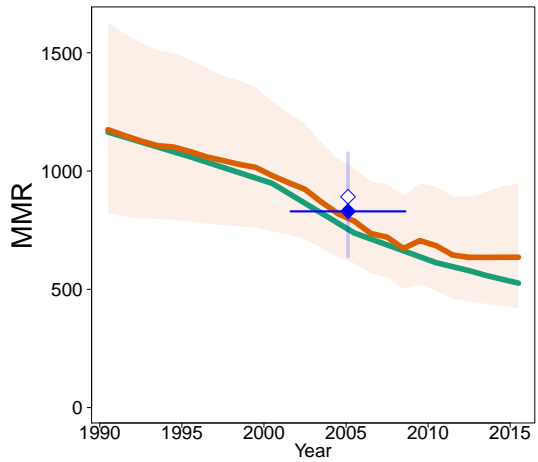

Nigeria [NGA; B]
(Western Africa)

– Estimates –
Bmat 2014
MMEIG 2014
– Data –
◇ Observed Data
◆ Adjusted Data
– Data type –
◆ DHS

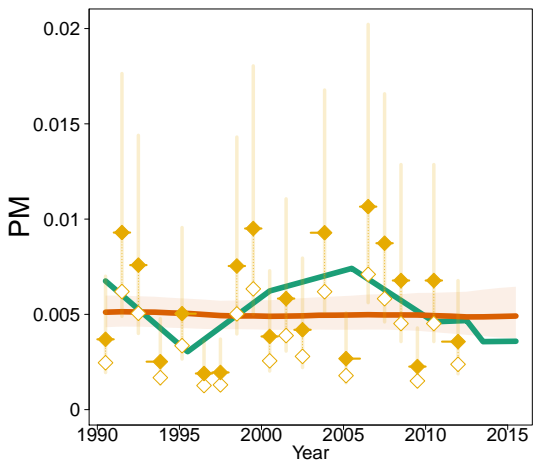

Norway [NOR; A]
(Developed regions)

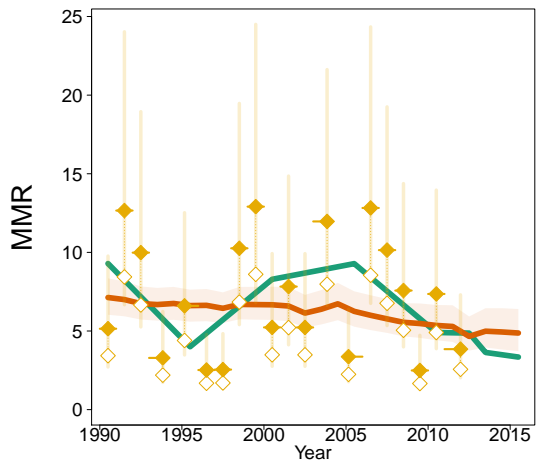

Norway [NOR; A]
(Developed regions)

– Estimates –
Bmat 2014
MMEIG 2014
– Data –
◇ Observed Data
◆ Adjusted Data
– Data type –
◆ VR

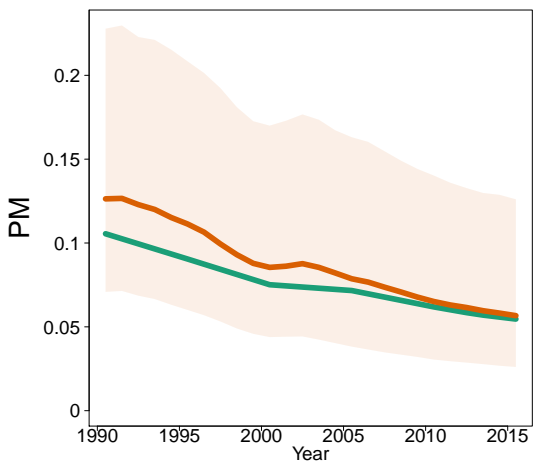

Occupied Palestinian Territory [PSE; C]
(Western Asia)

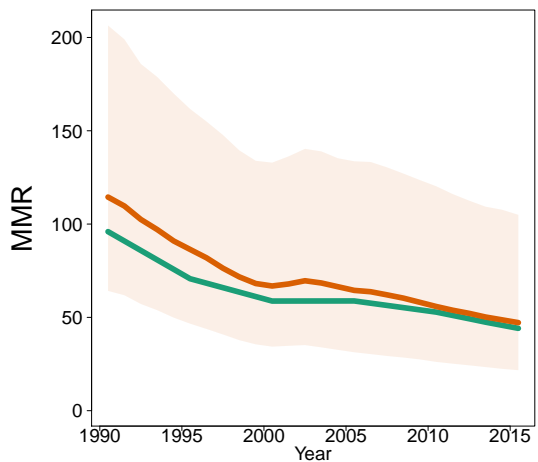

Occupied Palestinian Territory [PSE; C]
(Western Asia)

– Estimates –
Bmat 2014
MMEIG 2014
– Data –

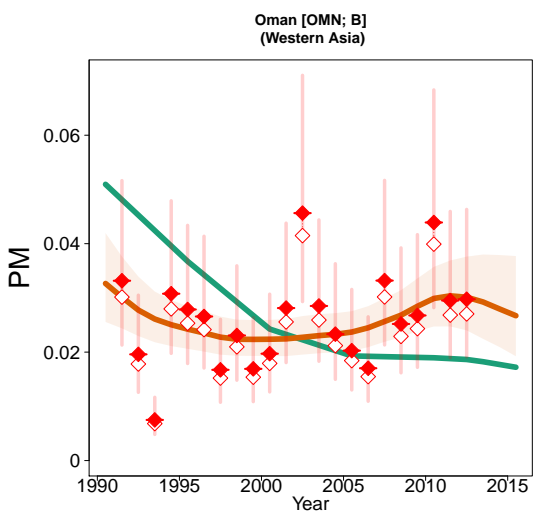

**Oman [OMN; B]**
**(Western Asia)**

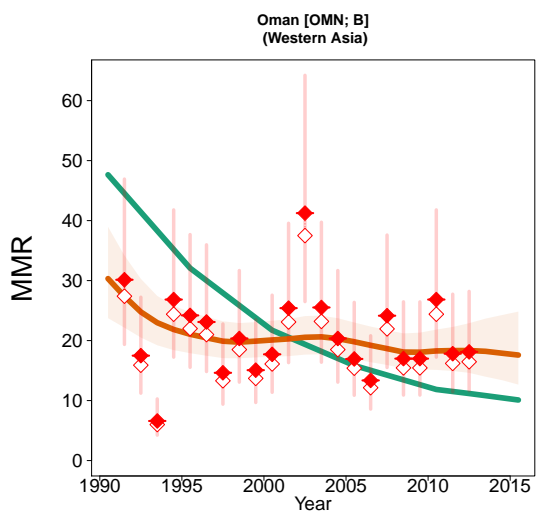

**Oman [OMN; B]**
**(Western Asia)**

– Estimates –
Bmat 2014
MMEIG 2014
– Data –
◇ Observed Data
◆ Adjusted Data
– Data type –
◆ Misc. studies

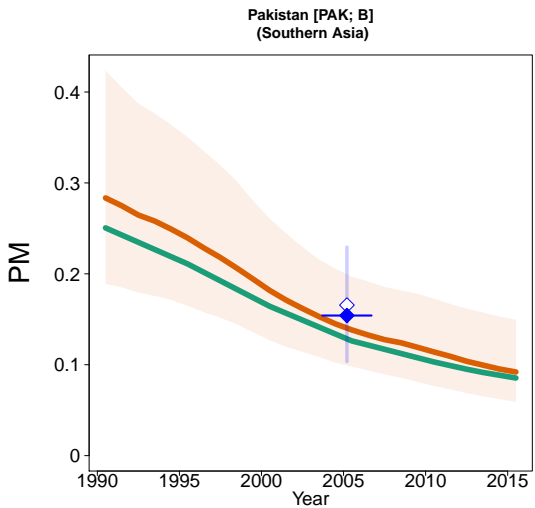

**Pakistan [PAK; B]**
**(Southern Asia)**

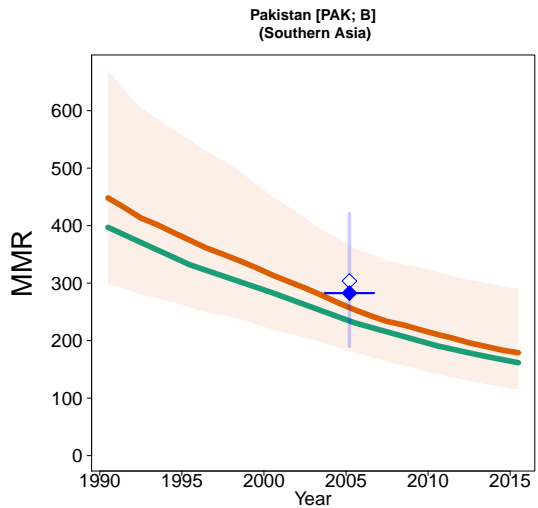

**Pakistan [PAK; B]**
**(Southern Asia)**

– Estimates –
Bmat 2014
MMEIG 2014
– Data –
◇ Observed Data
◆ Adjusted Data
– Data type –
◆ DHS

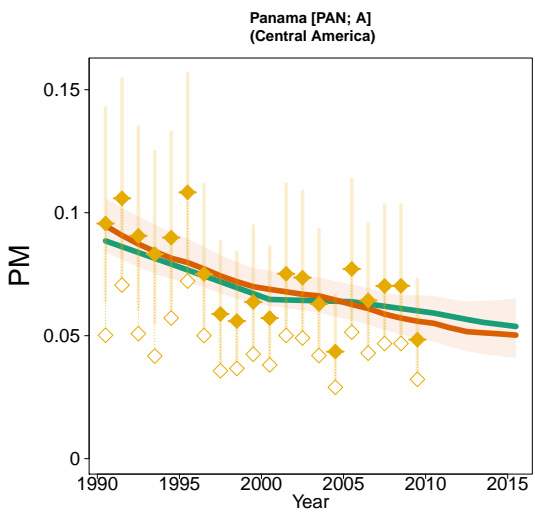

**Panama [PAN; A]**
**(Central America)**

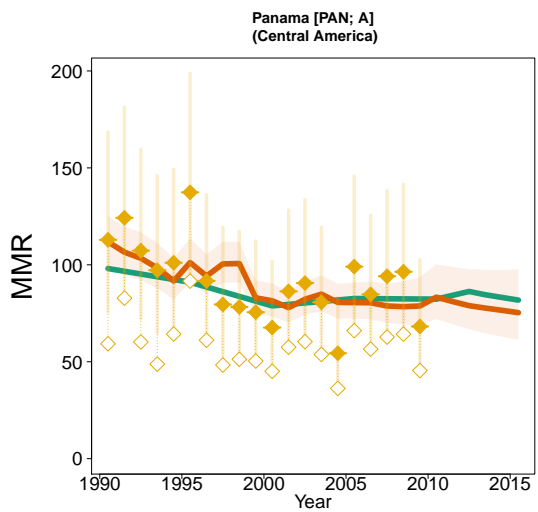

**Panama [PAN; A]**
**(Central America)**

– Estimates –
Bmat 2014
MMEIG 2014
– Data –
◇ Observed Data
◆ Adjusted Data
– Data type –
◆ VR

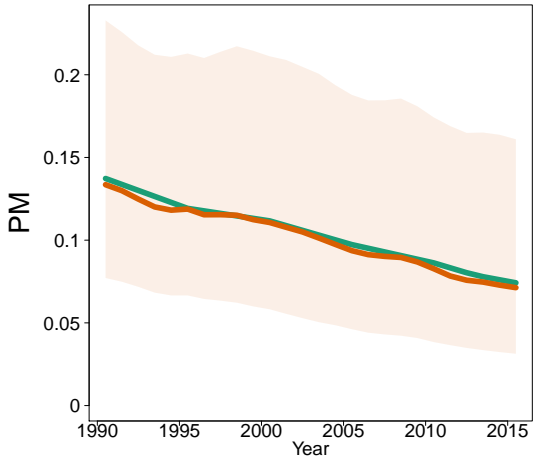

**Papua New Guinea [PNG; C]**
**(South−Eastern Asia / Oceania)**

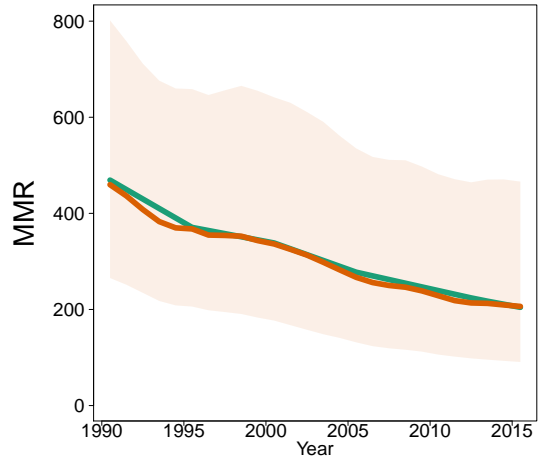

**Papua New Guinea [PNG; C]**
**(South−Eastern Asia / Oceania)**

– Estimates –
Bmat 2014
MMEIG 2014
– Data –

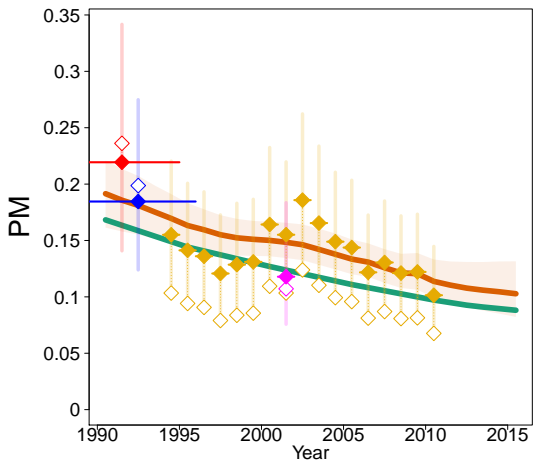

**Paraguay [PRY; B]**
**(South America)**

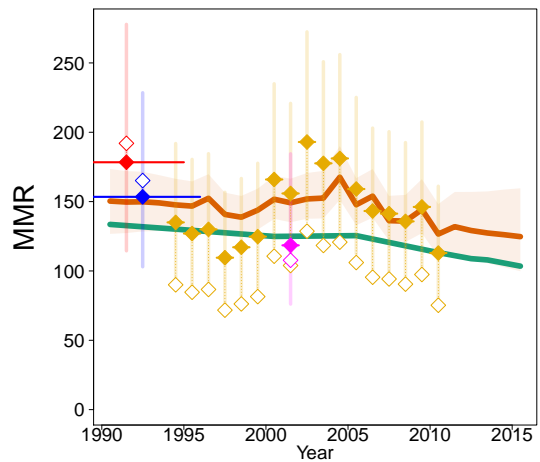

**Paraguay [PRY; B]**
**(South America)**

– Estimates –
Bmat 2014
MMEIG 2014
– Data –
Observed Data
Adjusted Data
– Data type –
VR
DHS
Census
Misc. studies

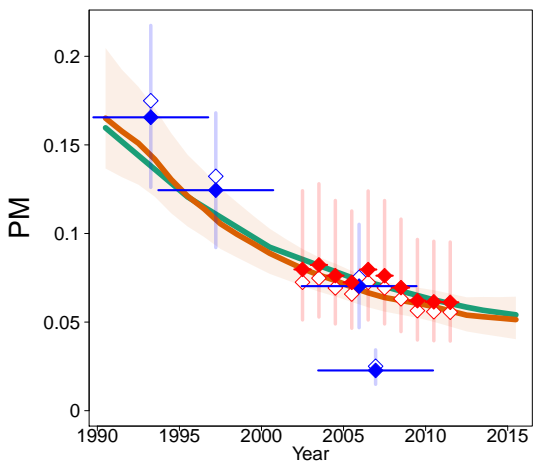

**Peru [PER; B]**
**(South America)**

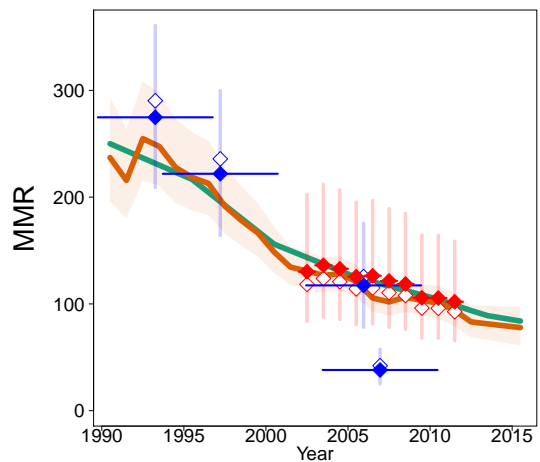

**Peru [PER; B]**
**(South America)**

– Estimates –
Bmat 2014
MMEIG 2014
– Data –
Observed Data
Adjusted Data
– Data type –
DHS
Misc. studies

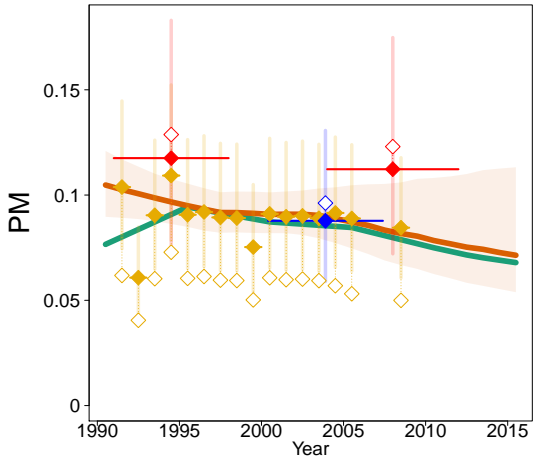

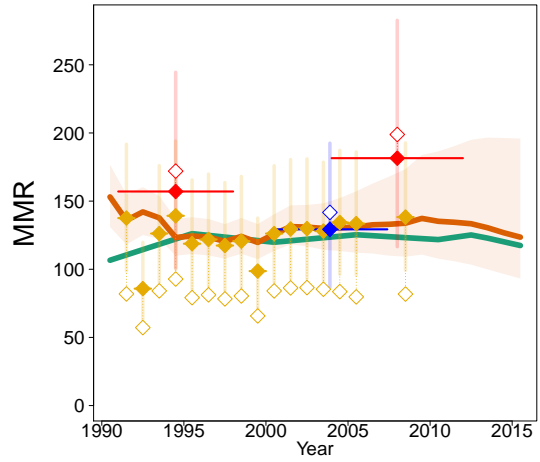

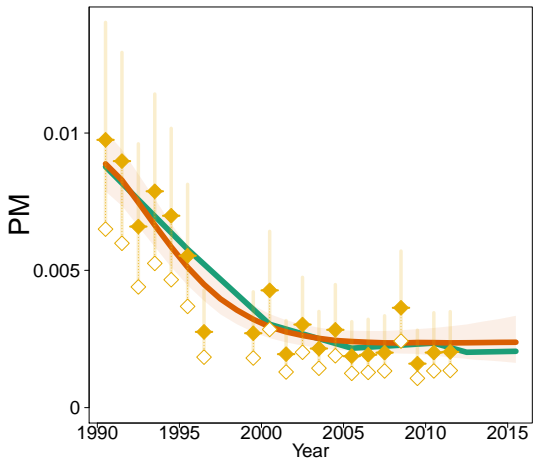

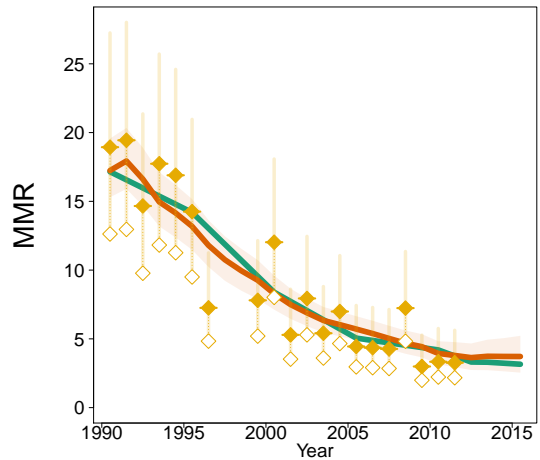

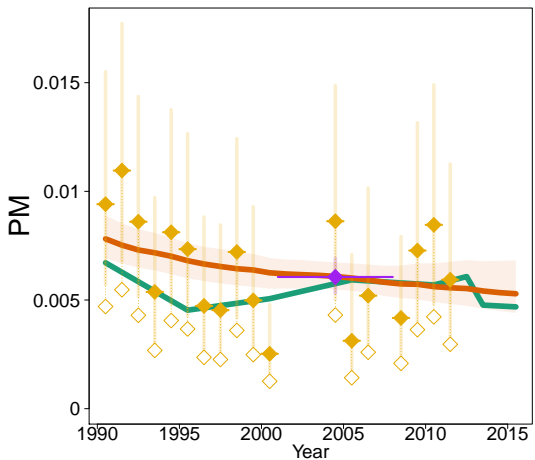

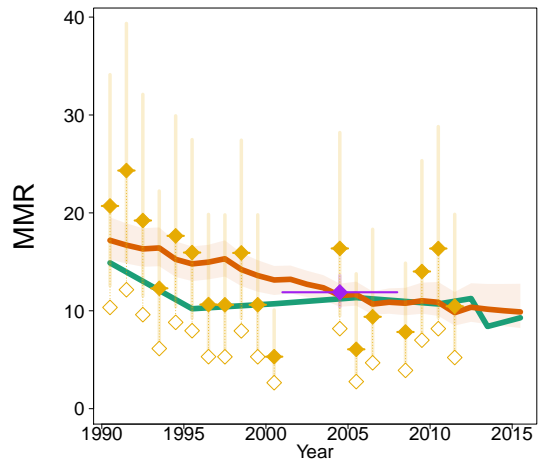

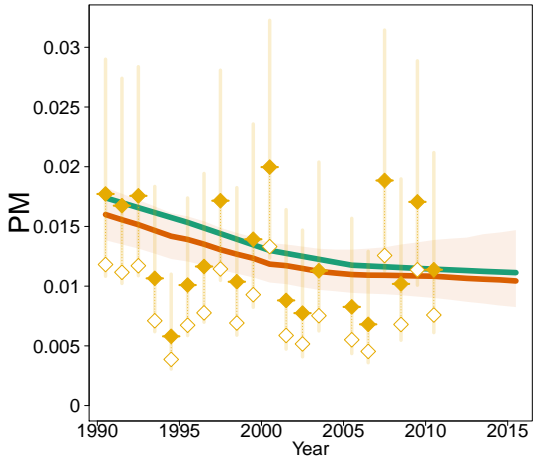

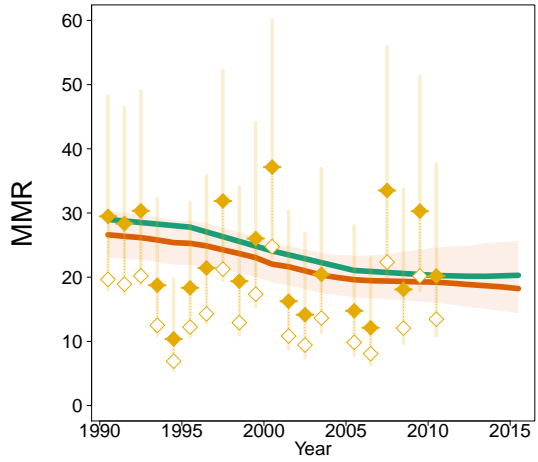

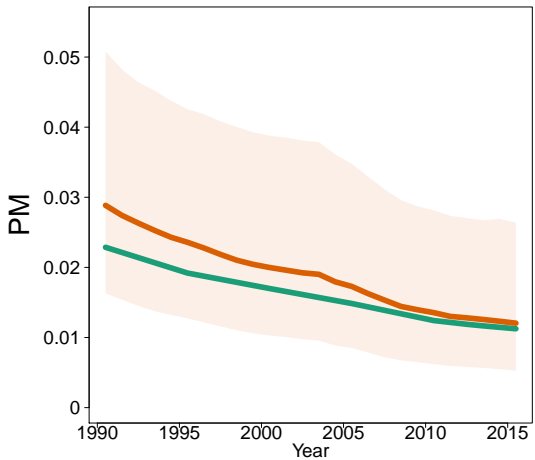

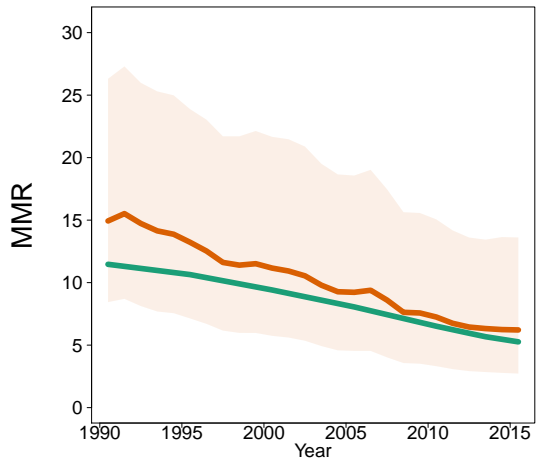

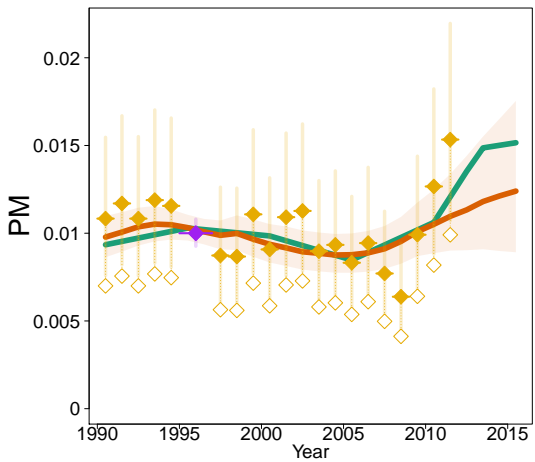

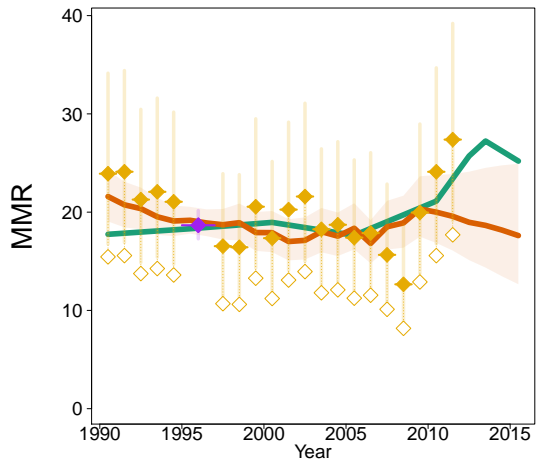

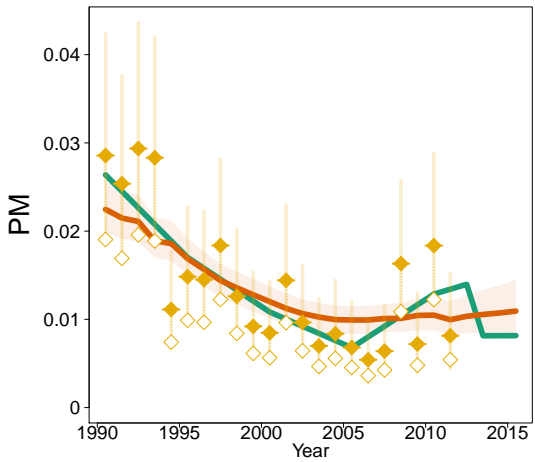

**Republic of Moldova [MDA; A]**
(Developed regions)

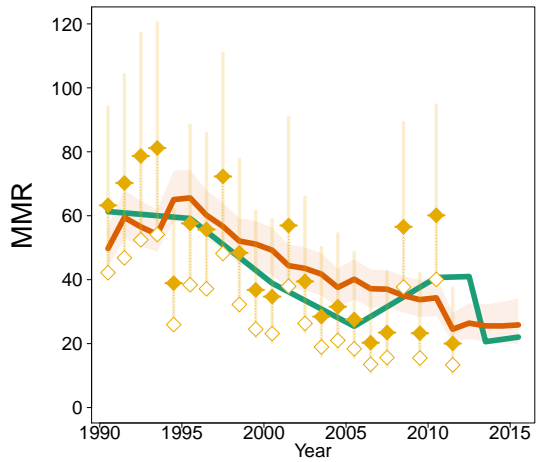

**Republic of Moldova [MDA; A]**
(Developed regions)

– Estimates –
Bmat 2014
MMEIG 2014
– Data –
◇ Observed Data
◆ Adjusted Data
– Data type –
◆ VR

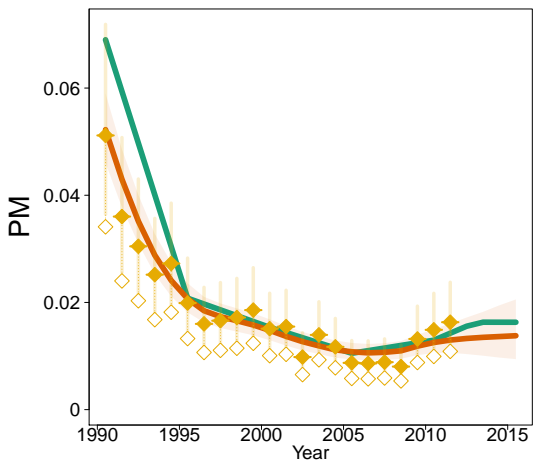

**Romania [ROU; A]**
(Developed regions)

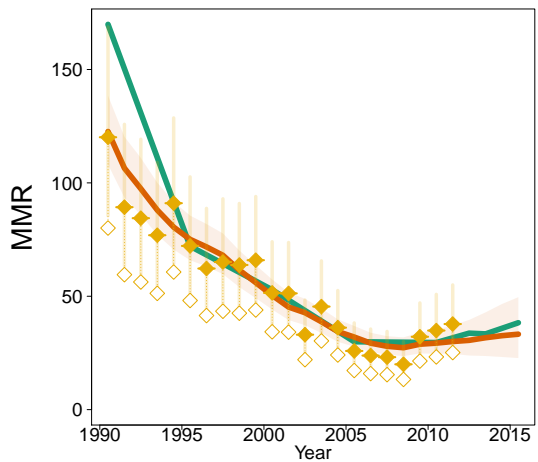

**Romania [ROU; A]**
(Developed regions)

– Estimates –
Bmat 2014
MMEIG 2014
– Data –
◇ Observed Data
◆ Adjusted Data
– Data type –
◆ VR

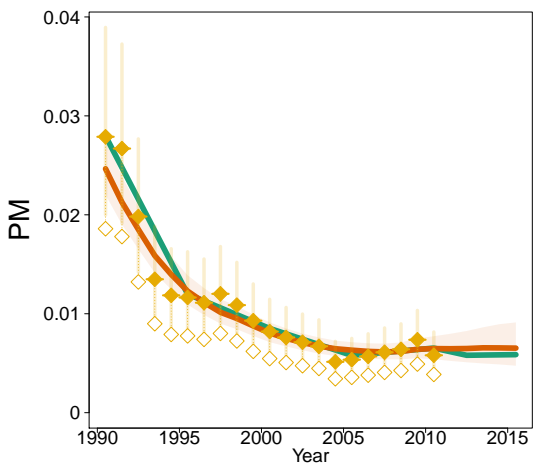

**Russian Federation [RUS; A]**
(Developed regions)

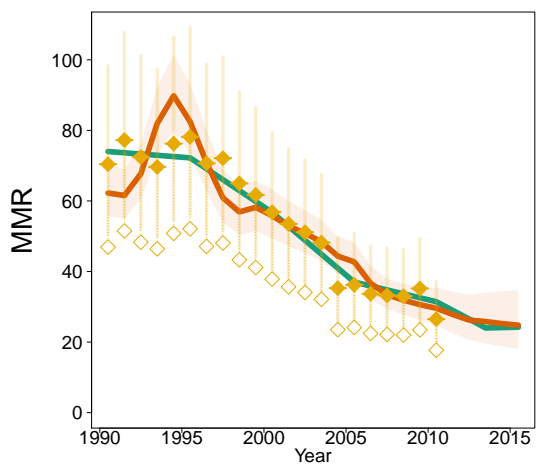

**Russian Federation [RUS; A]**
(Developed regions)

– Estimates –
Bmat 2014
MMEIG 2014
– Data –
◇ Observed Data
◆ Adjusted Data
– Data type –
◆ VR

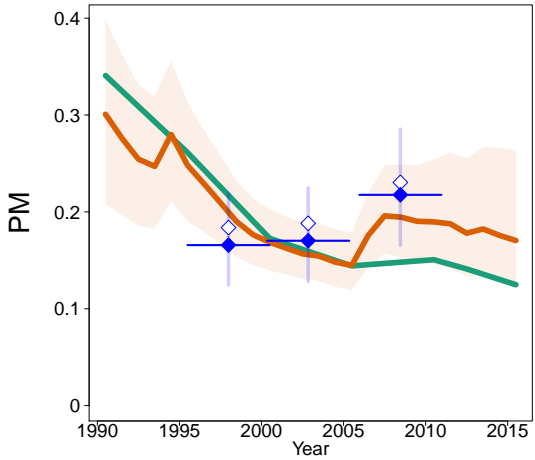

Rwanda [RWA; B]
(Eastern Africa)

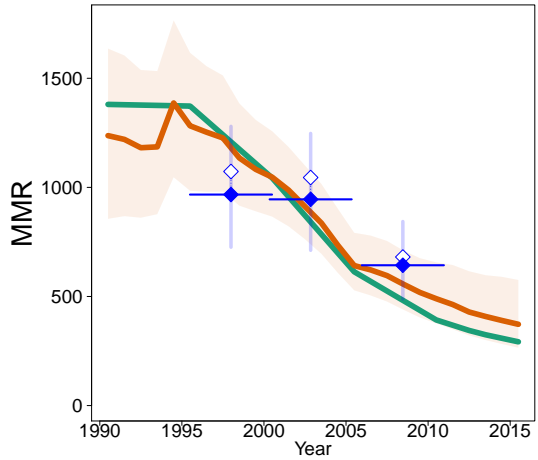

Rwanda [RWA; B]
(Eastern Africa)

– Estimates –
Bmat 2014
MMEIG 2014
– Data –
Observed Data
Adjusted Data
– Data type –
DHS

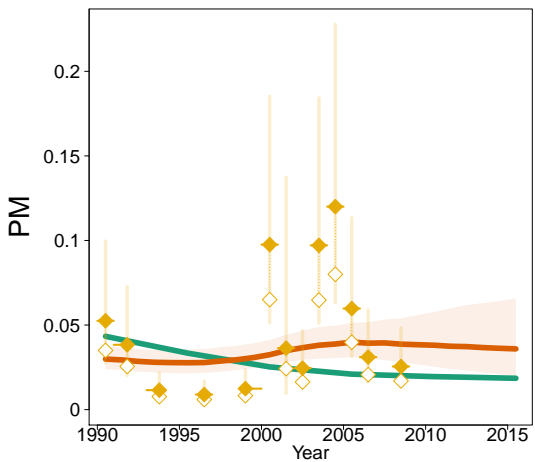

Saint Lucia [LCA; A]
(Caribbean)

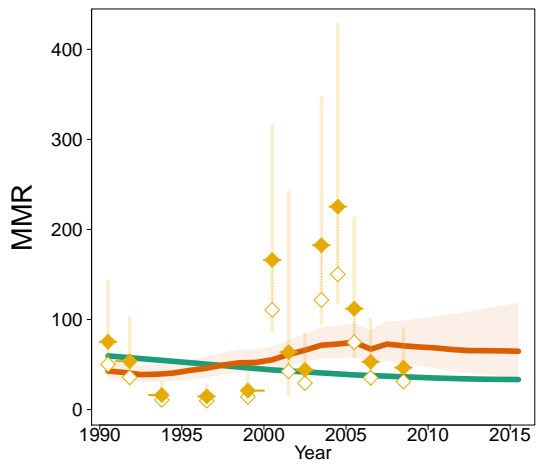

Saint Lucia [LCA; A]
(Caribbean)

– Estimates –
Bmat 2014
MMEIG 2014
– Data –
Observed Data
Adjusted Data
– Data type –
VR

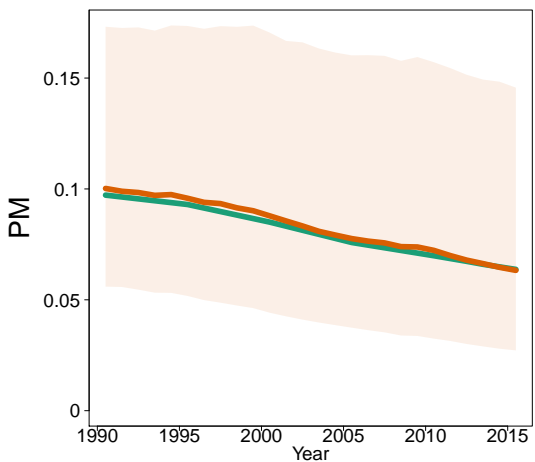

Samoa [WSM; C]
(South−Eastern Asia / Oceania)

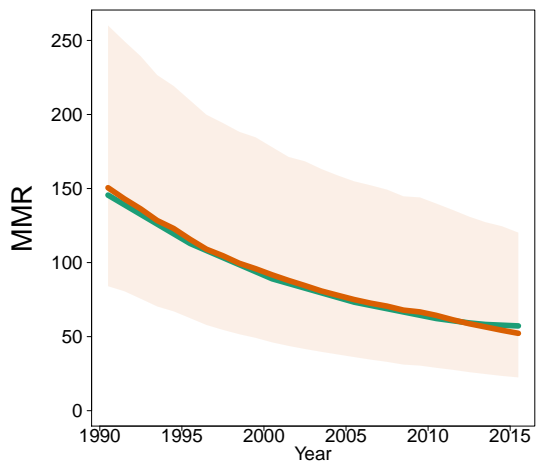

Samoa [WSM; C]
(South−Eastern Asia / Oceania)

– Estimates –
Bmat 2014
MMEIG 2014
– Data –

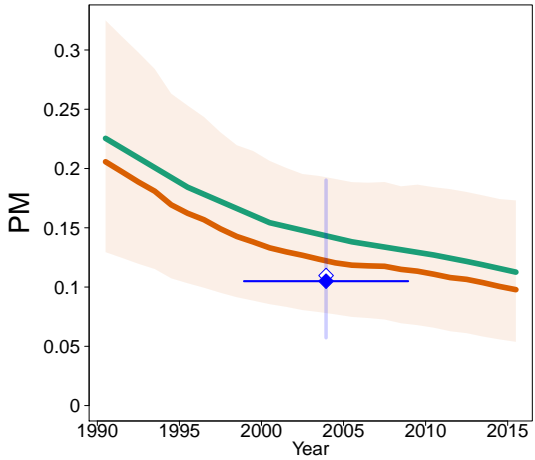

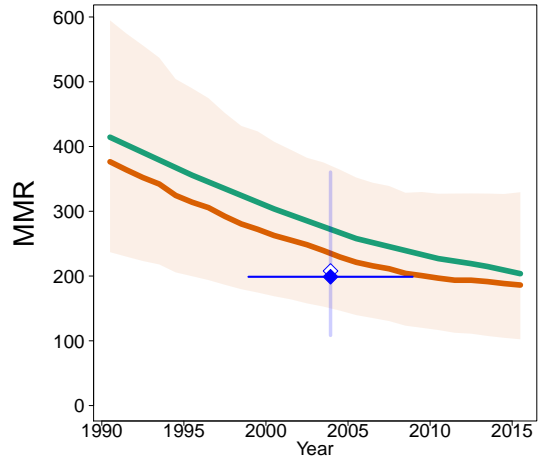

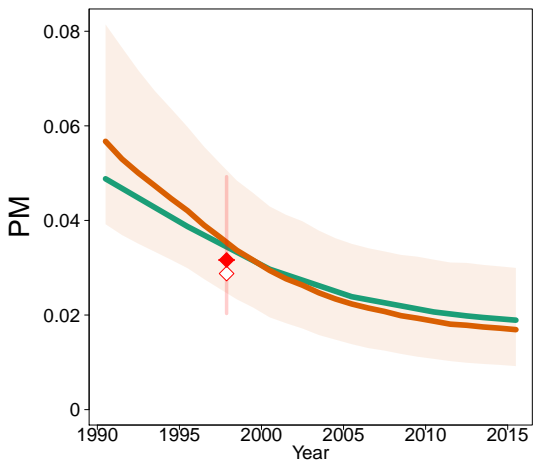

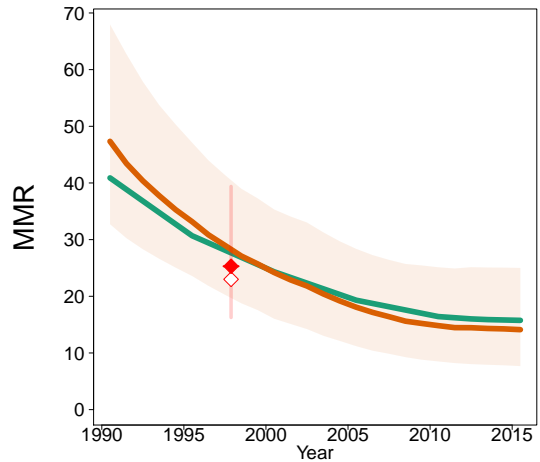

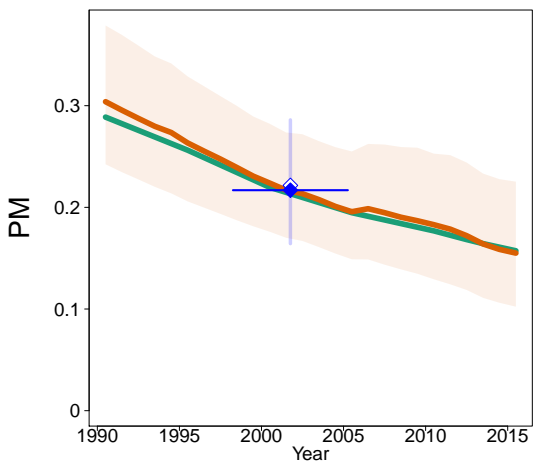

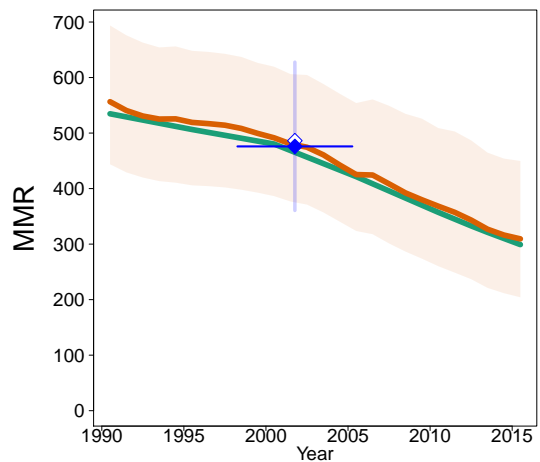

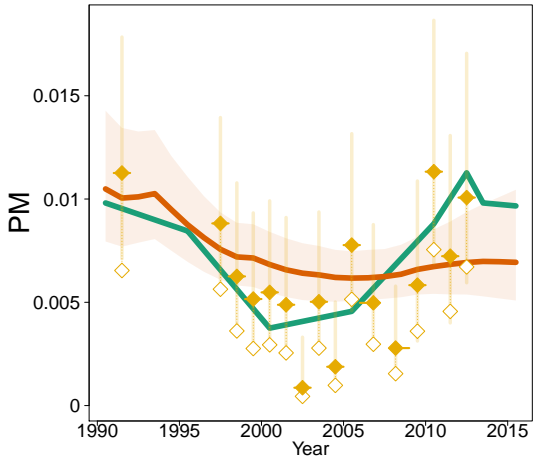

**Serbia [SRB; A]**
**(Transition countries of South-Eastern Europe)**

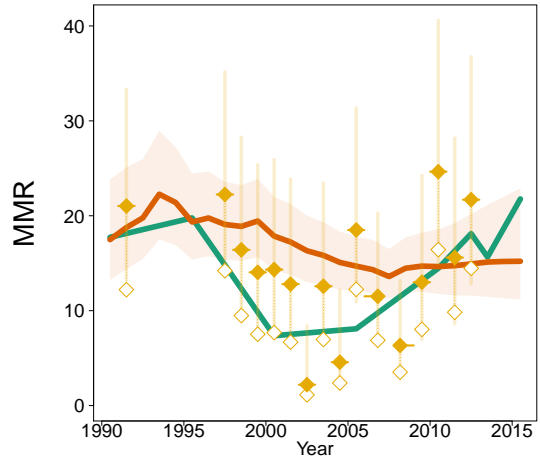

**Serbia [SRB; A]**
**(Transition countries of South-Eastern Europe)**

– Estimates –
Bmat 2014
MMEIG 2014
– Data –
◇ Observed Data
◆ Adjusted Data
– Data type –
◆ VR

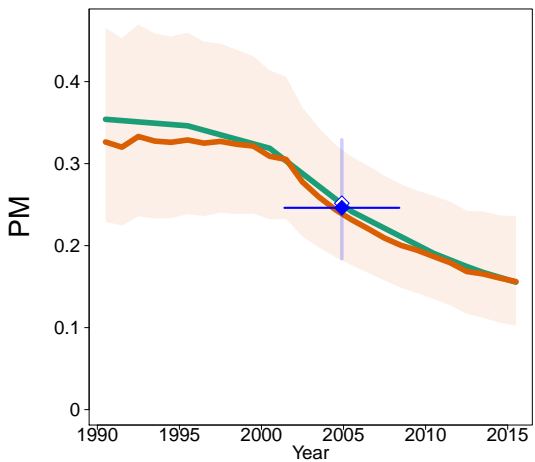

**Sierra Leone [SLE; B]**
**(Western Africa)**

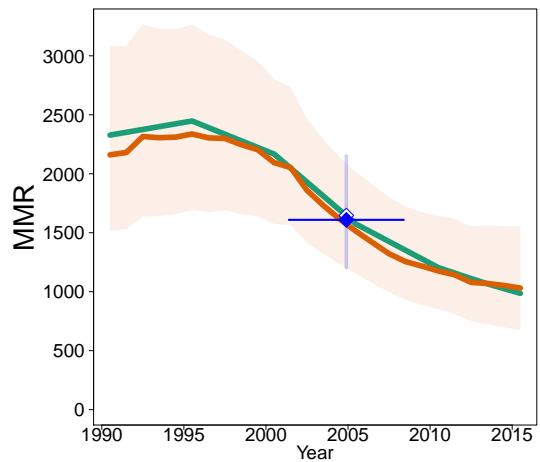

**Sierra Leone [SLE; B]**
**(Western Africa)**

– Estimates –
Bmat 2014
MMEIG 2014
– Data –
◇ Observed Data
◆ Adjusted Data
– Data type –
◆ DHS

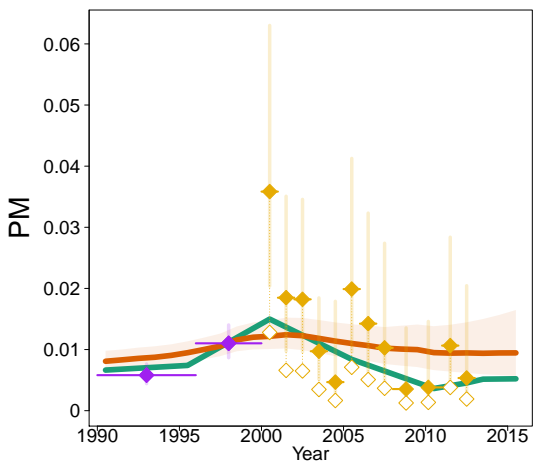

**Singapore [SGP; A]**
**(South-Eastern Asia / Oceania)**

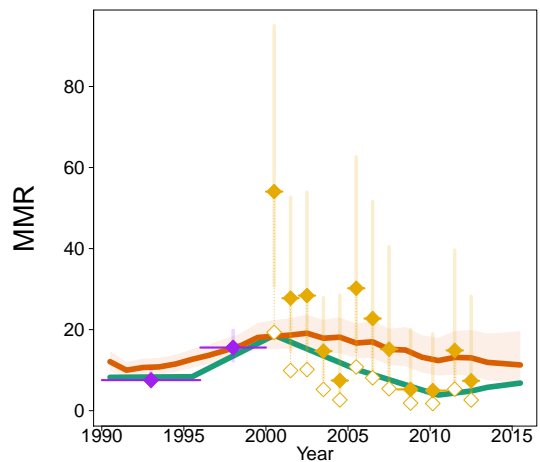

**Singapore [SGP; A]**
**(South-Eastern Asia / Oceania)**

– Estimates –
Bmat 2014
MMEIG 2014
– Data –
◇ Observed Data
◆ Adjusted Data
– Data type –
◆ VR
◆ Spec. studies

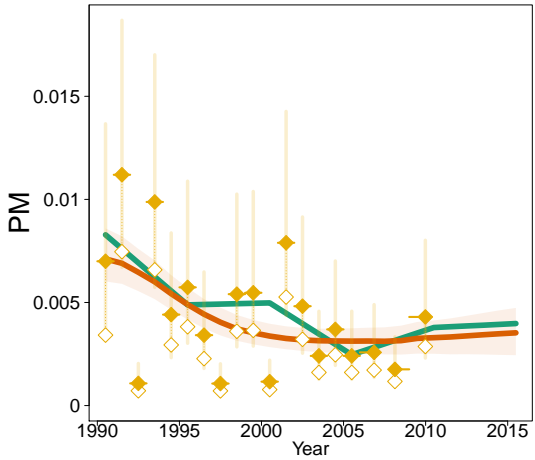

**Slovakia [SVK; A]**
**(Developed regions)**

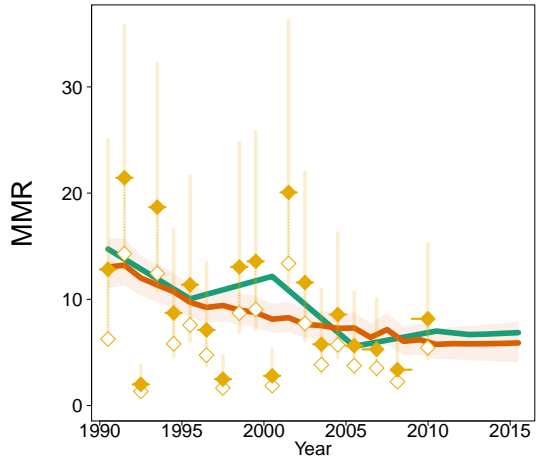

**Slovakia [SVK; A]**
**(Developed regions)**

– Estimates –
— Bmat 2014
— MMEIG 2014
– Data –
◇ Observed Data
◆ Adjusted Data
– Data type –
◆ VR

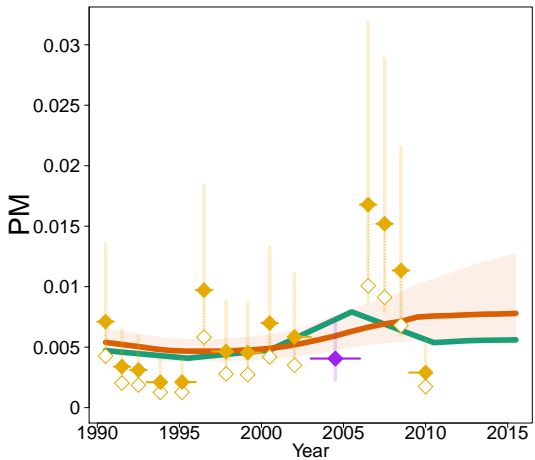

**Slovenia [SVN; A]**
**(Developed regions)**

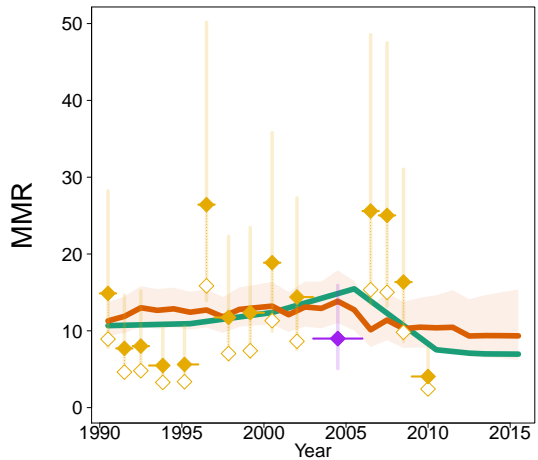

**Slovenia [SVN; A]**
**(Developed regions)**

– Estimates –
— Bmat 2014
— MMEIG 2014
– Data –
◇ Observed Data
◆ Adjusted Data
– Data type –
◆ VR
◆ Spec. studies

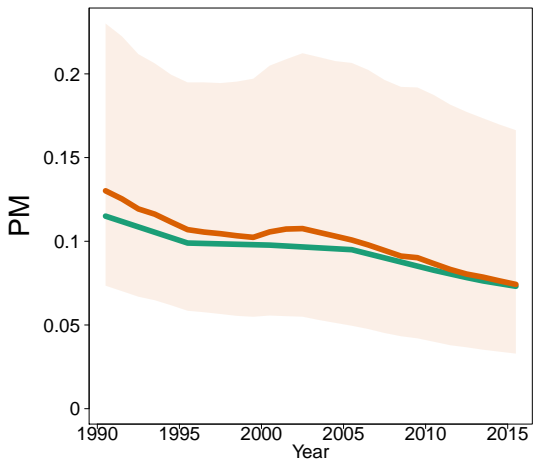

**Solomon Islands [SLB; C]**
**(South−Eastern Asia / Oceania)**

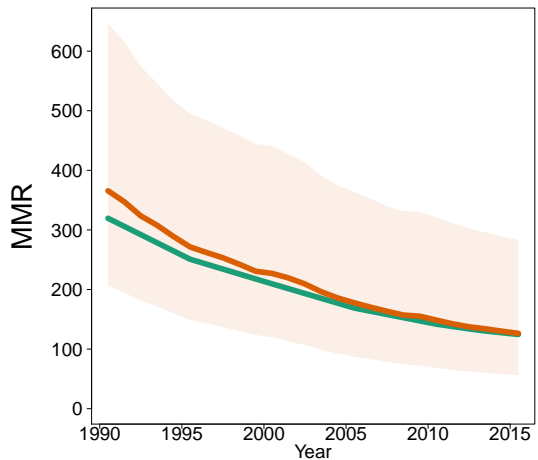

**Solomon Islands [SLB; C]**
**(South−Eastern Asia / Oceania)**

– Estimates –
— Bmat 2014
— MMEIG 2014
– Data –

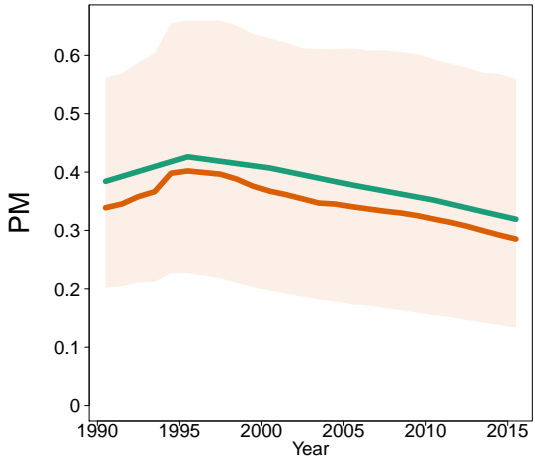

**Somalia [SOM; C]**
**(Eastern Africa)**

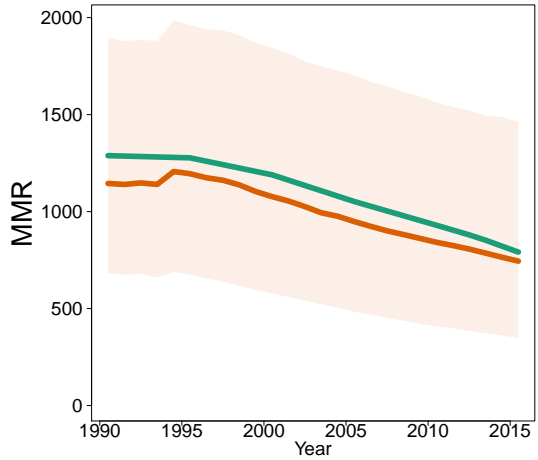

**Somalia [SOM; C]**
**(Eastern Africa)**

– Estimates –
Bmat 2014
MMEIG 2014
– Data –

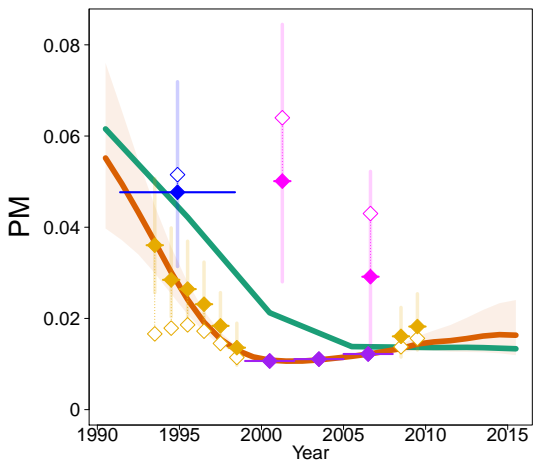

**South Africa [ZAF; B]**
**(Southern Africa)**

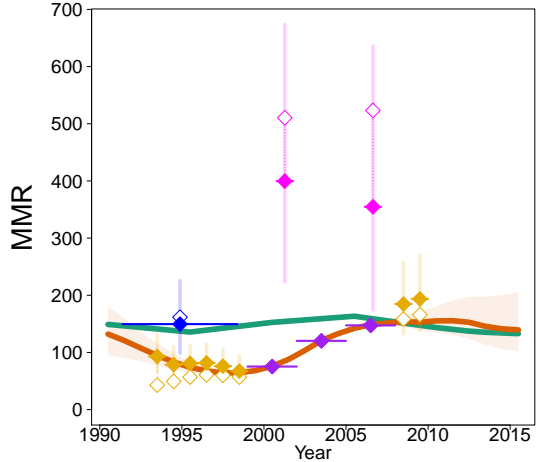

**South Africa [ZAF; B]**
**(Southern Africa)**

– Estimates –
Bmat 2014
MMEIG 2014
– Data –
Observed Data
Adjusted Data
– Data type –
VR
Spec. studies
DHS
Census

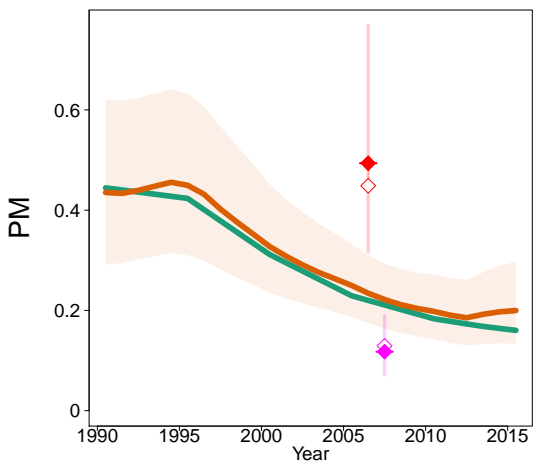

**South Sudan [SSD; B]**
**(Eastern Africa)**

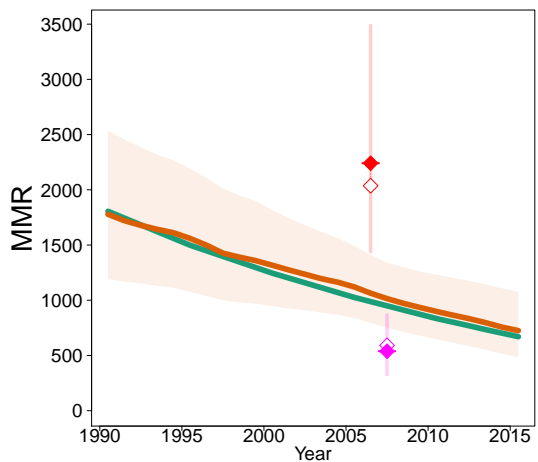

**South Sudan [SSD; B]**
**(Eastern Africa)**

– Estimates –
Bmat 2014
MMEIG 2014
– Data –
Observed Data
Adjusted Data
– Data type –
Census
Misc. studies

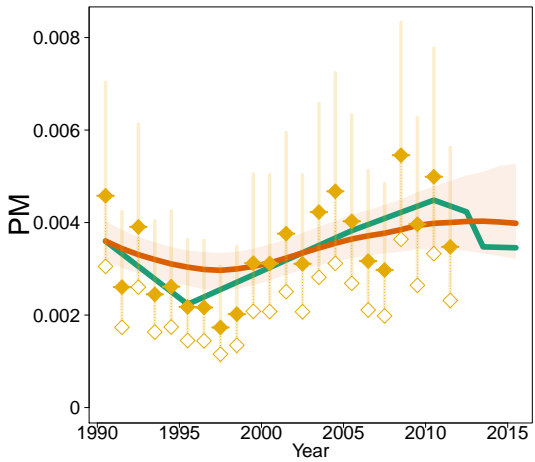

Spain [ESP; A]
(Developed regions)

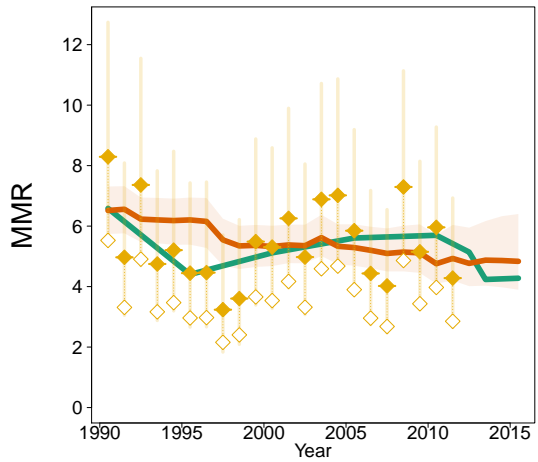

Spain [ESP; A]
(Developed regions)

– Estimates –
Bmat 2014
MMEIG 2014
– Data –
◇ Observed Data
◆ Adjusted Data
– Data type –
◆ VR

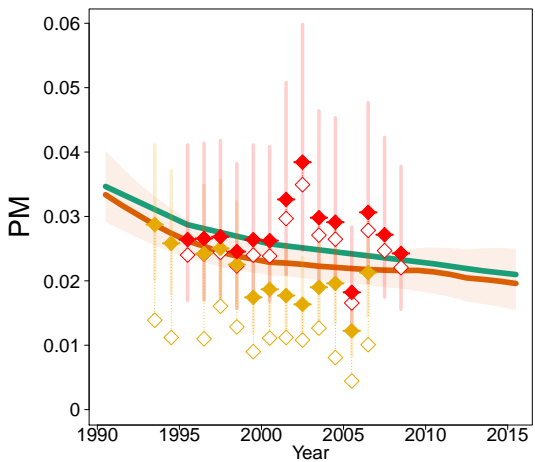

Sri Lanka [LKA; B]
(Southern Asia)

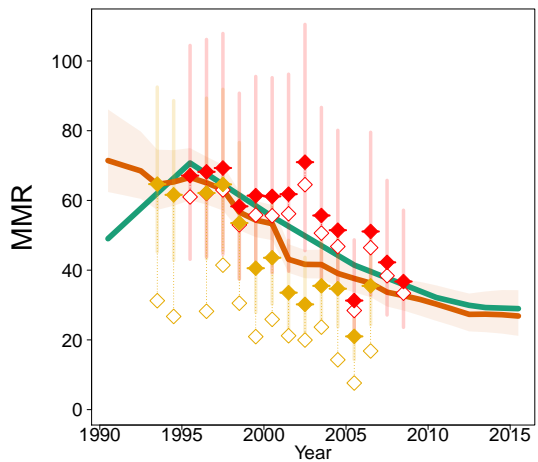

Sri Lanka [LKA; B]
(Southern Asia)

– Estimates –
Bmat 2014
MMEIG 2014
– Data –
◇ Observed Data
◆ Adjusted Data
– Data type –
◆ VR
◆ Misc. studies

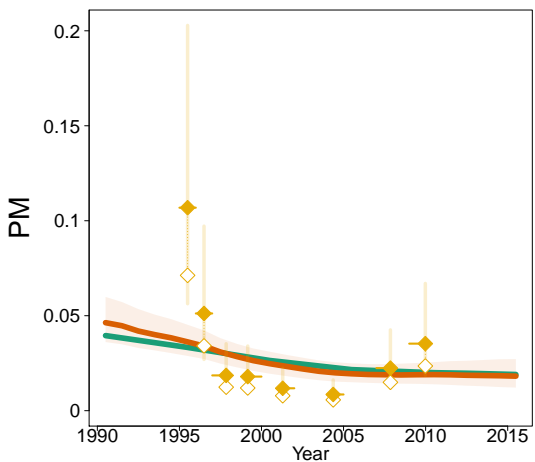

St Vincent and the Grenadines [VCT; A]
(Caribbean)

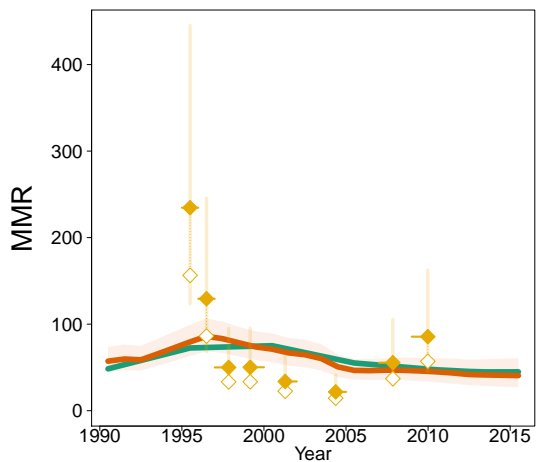

St Vincent and the Grenadines [VCT; A]
(Caribbean)

– Estimates –
Bmat 2014
MMEIG 2014
– Data –
◇ Observed Data
◆ Adjusted Data
– Data type –
◆ VR

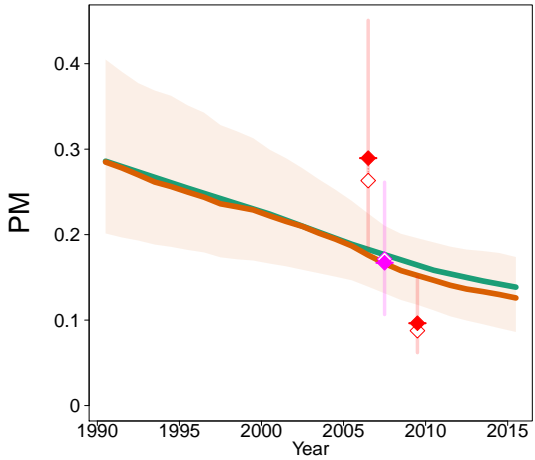

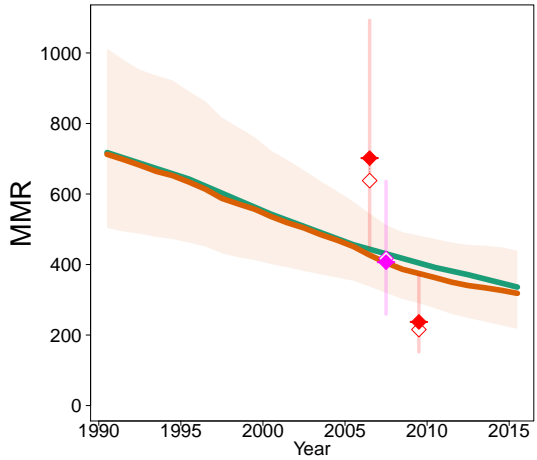

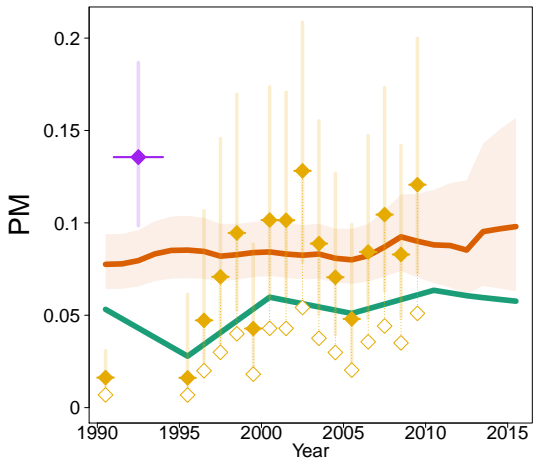

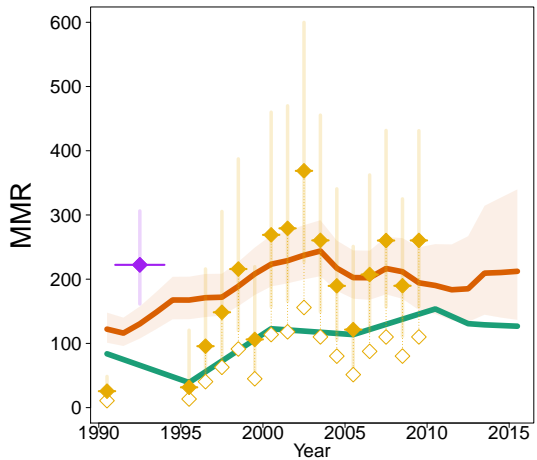

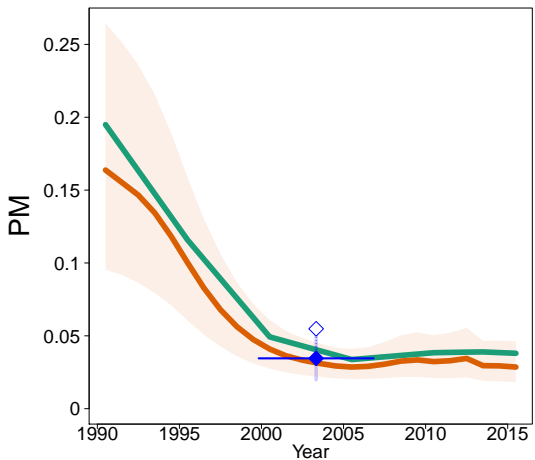

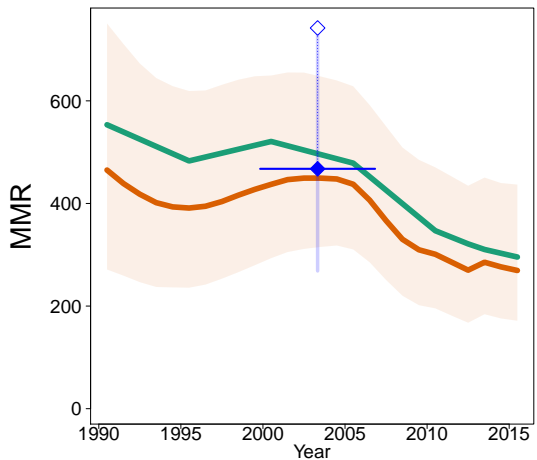

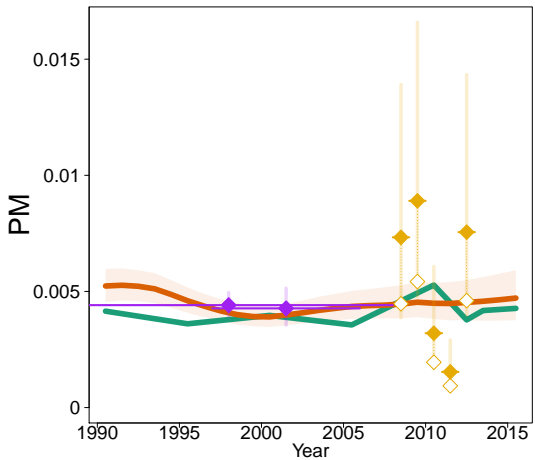

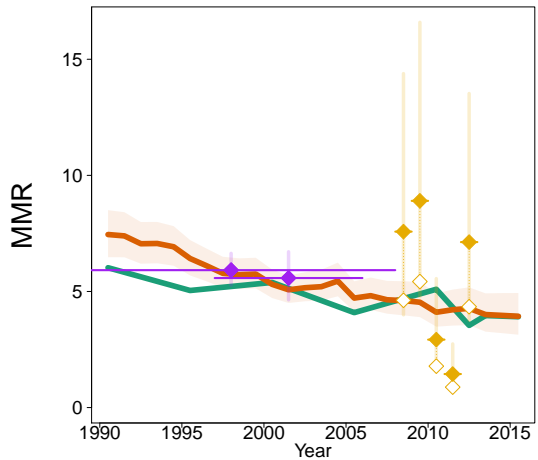

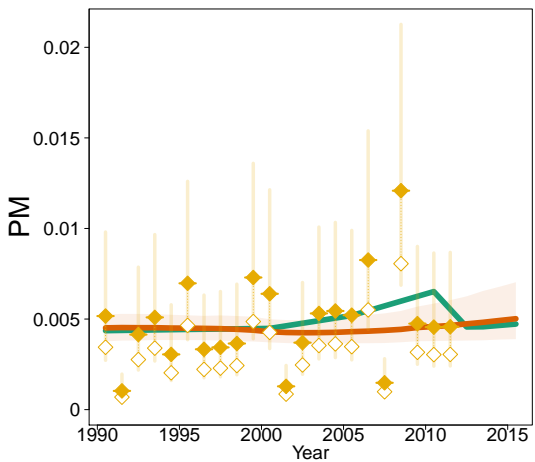

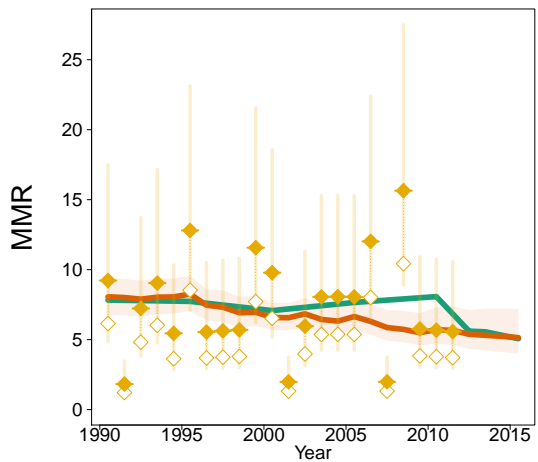

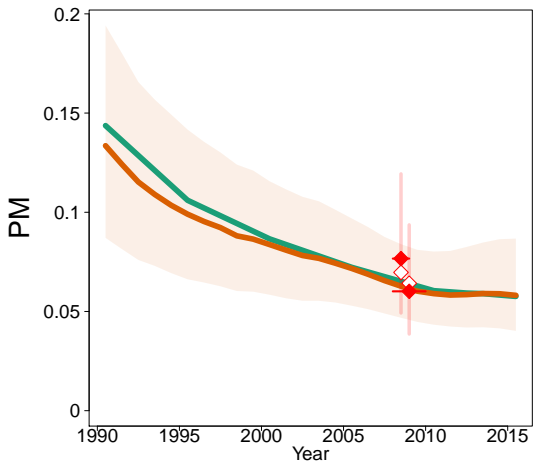

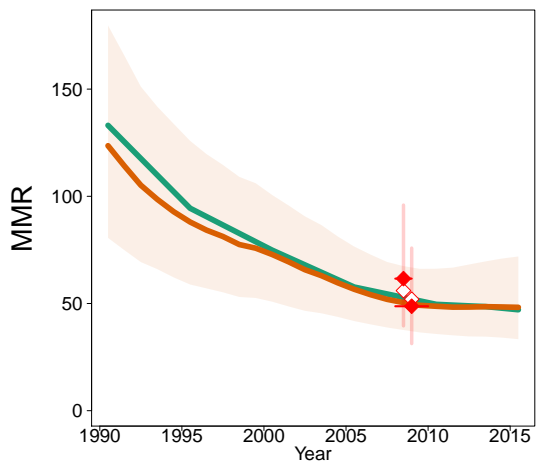

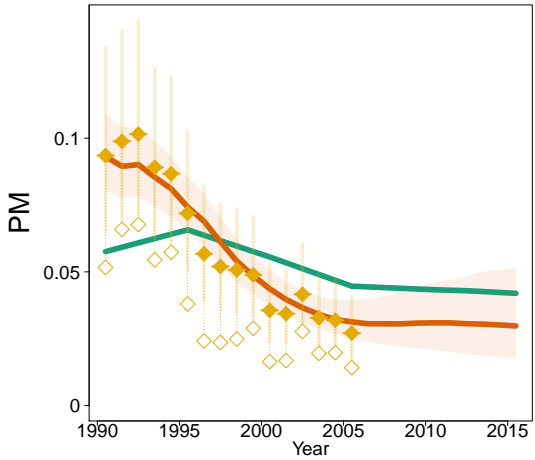

Tajikistan [TJK; B]
(Central Asia)

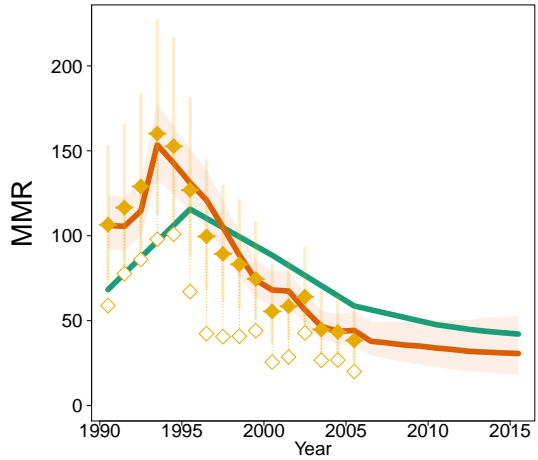

Tajikistan [TJK; B]
(Central Asia)

– Estimates –
Bmat 2014
MMEIG 2014
– Data –
Observed Data
Adjusted Data
– Data type –
VR

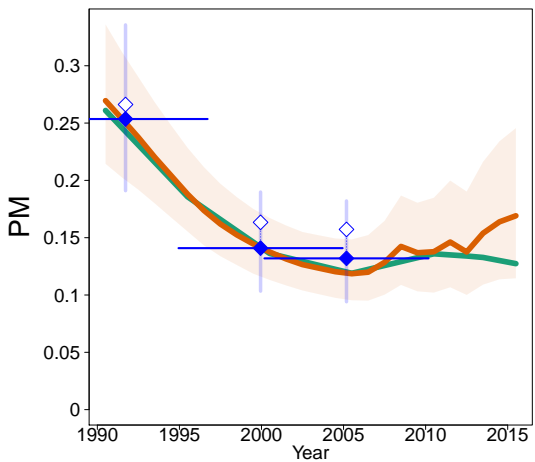

Tanzania [TZA; B]
(Eastern Africa)

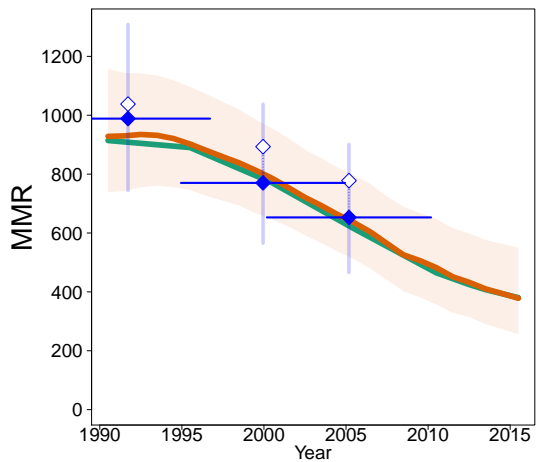

Tanzania [TZA; B]
(Eastern Africa)

– Estimates –
Bmat 2014
MMEIG 2014
– Data –
Observed Data
Adjusted Data
– Data type –
DHS

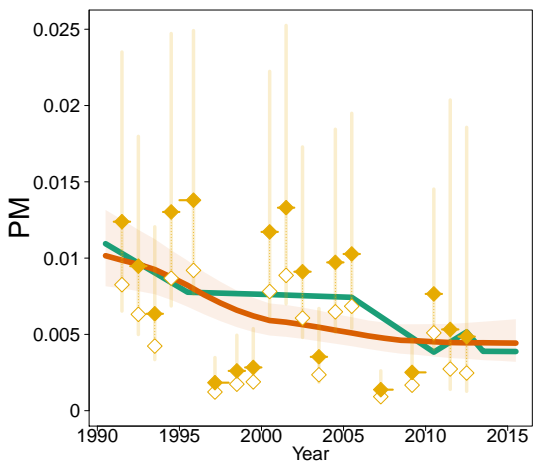

TFYR Macedonia [MKD; A]
(Transition countries of South–Eastern Europe)

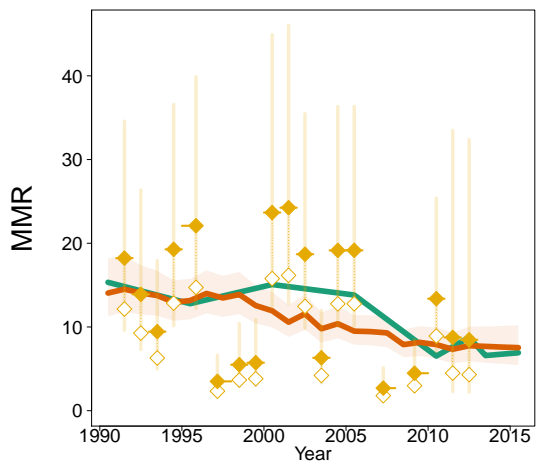

TFYR Macedonia [MKD; A]
(Transition countries of South–Eastern Europe)

– Estimates –
Bmat 2014
MMEIG 2014
– Data –
Observed Data
Adjusted Data
– Data type –
VR

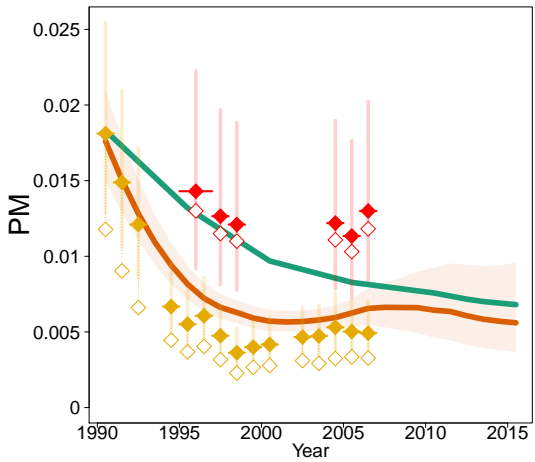

**Thailand [THA; B]**
**(South–Eastern Asia / Oceania)**

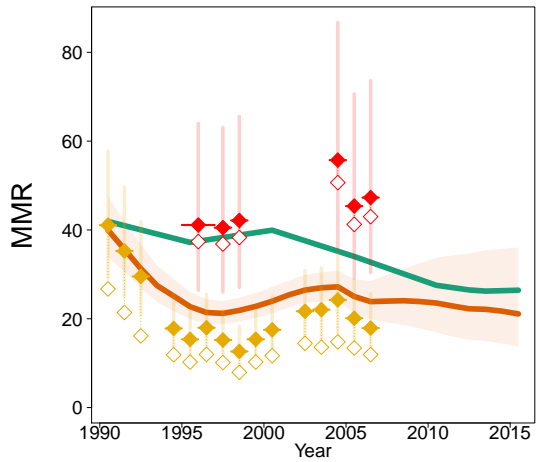

**Thailand [THA; B]**
**(South–Eastern Asia / Oceania)**

– Estimates –
Bmat 2014
MMEIG 2014
– Data –
Observed Data
Adjusted Data
– Data type –
VR
Misc. studies

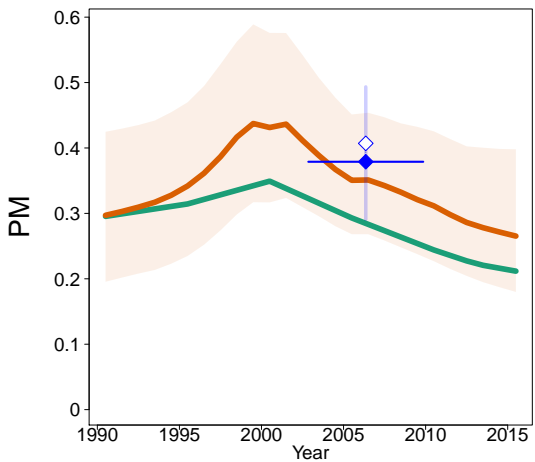

**Timor–Leste [TLS; B]**
**(South–Eastern Asia / Oceania)**

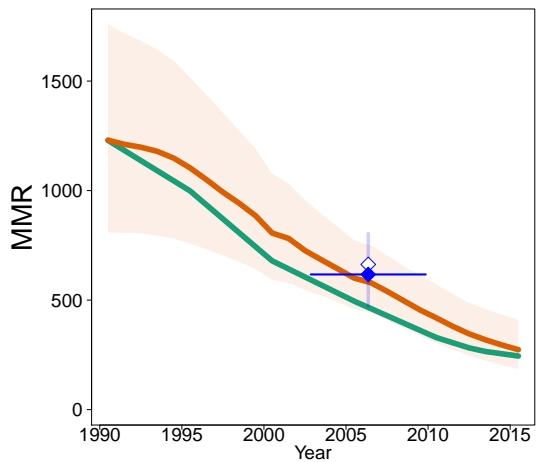

**Timor–Leste [TLS; B]**
**(South–Eastern Asia / Oceania)**

– Estimates –
Bmat 2014
MMEIG 2014
– Data –
Observed Data
Adjusted Data
– Data type –
DHS

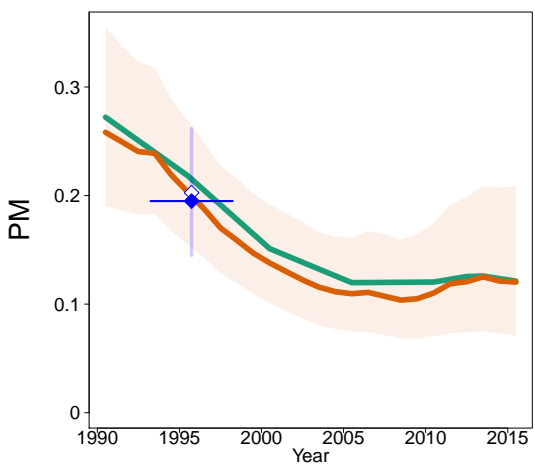

**Togo [TGO; B]**
**(Western Africa)**

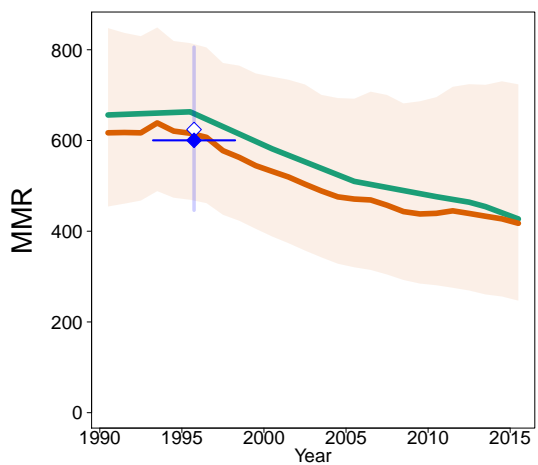

**Togo [TGO; B]**
**(Western Africa)**

– Estimates –
Bmat 2014
MMEIG 2014
– Data –
Observed Data
Adjusted Data
– Data type –
DHS

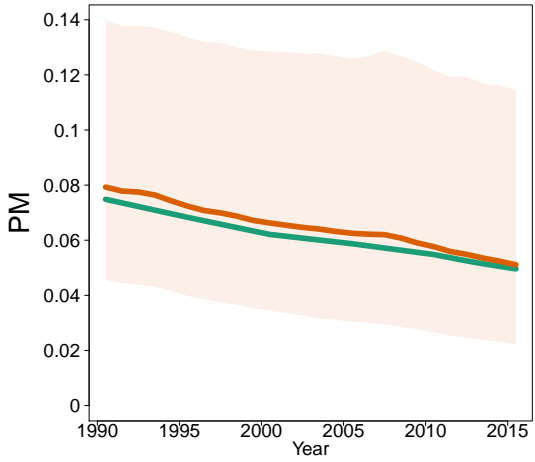

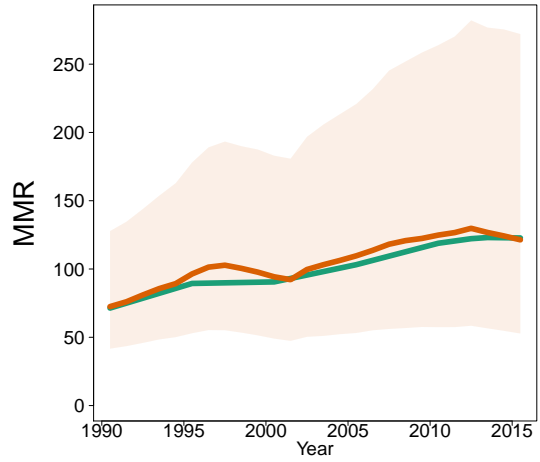

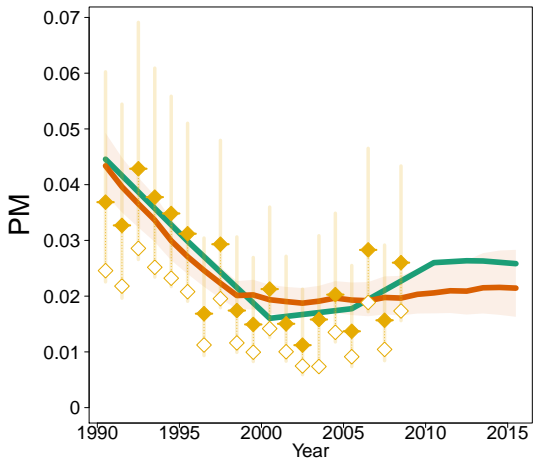

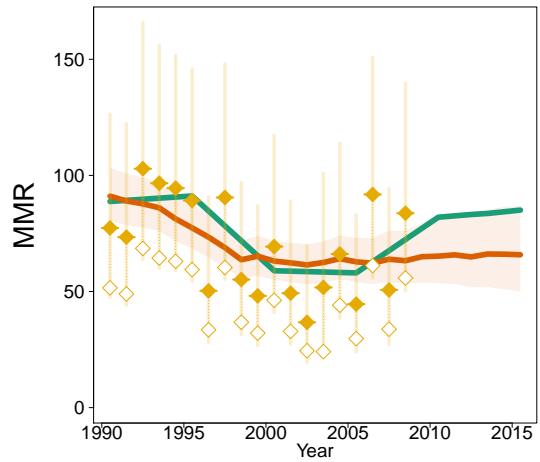

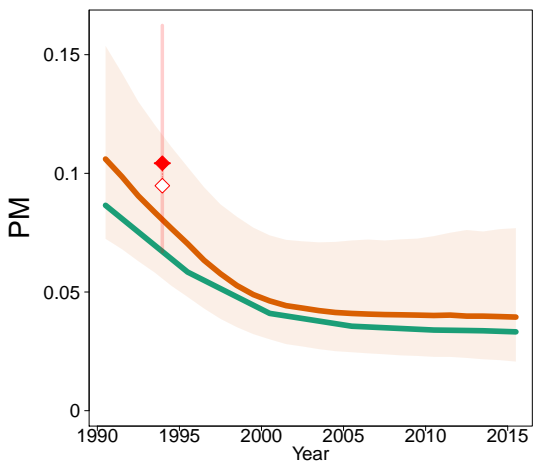

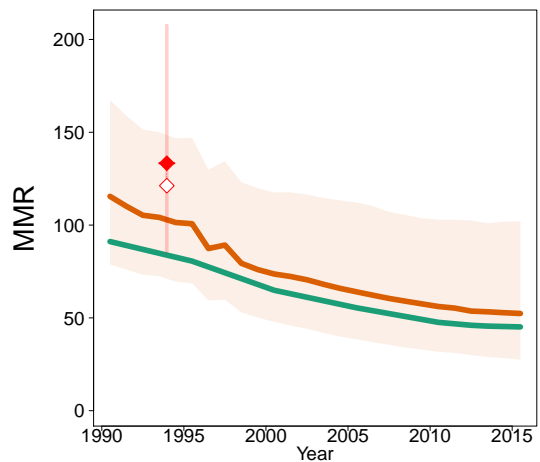

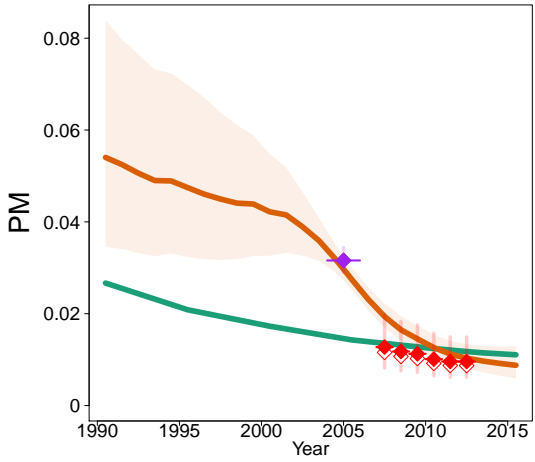

**Turkey [TUR; B]**
**(Western Asia)**

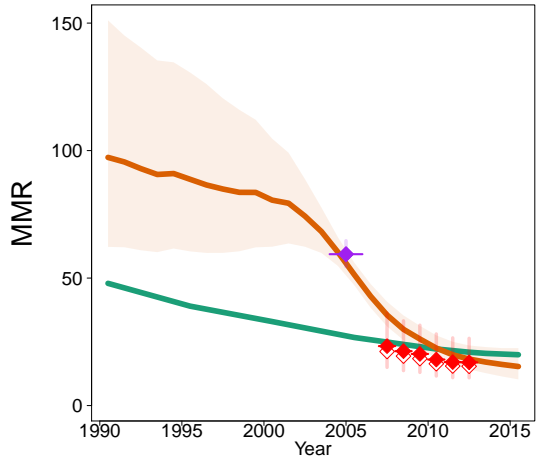

**Turkey [TUR; B]**
**(Western Asia)**

– Estimates –
Bmat 2014
MMEIG 2014
– Data –
◇ Observed Data
◆ Adjusted Data
– Data type –
◆ Spec. studies
◆ Misc. studies

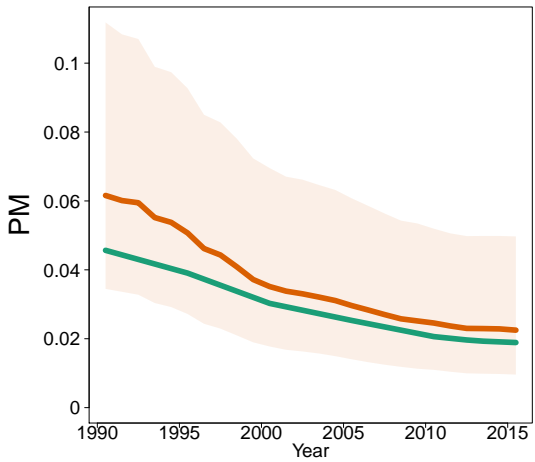

**Turkmenistan [TKM; C]**
**(Central Asia)**

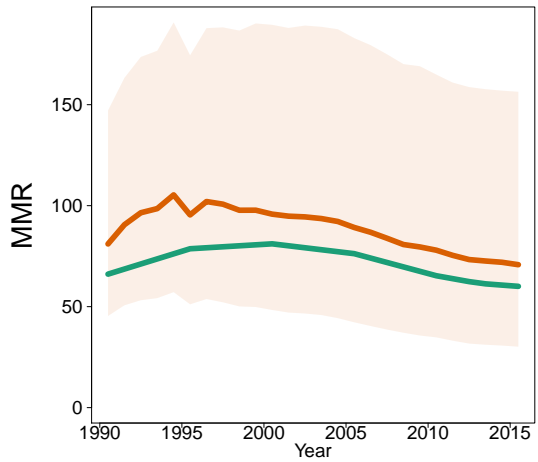

**Turkmenistan [TKM; C]**
**(Central Asia)**

– Estimates –
Bmat 2014
MMEIG 2014
– Data –
◇ Observed Data
◆ Adjusted Data
– Data type –

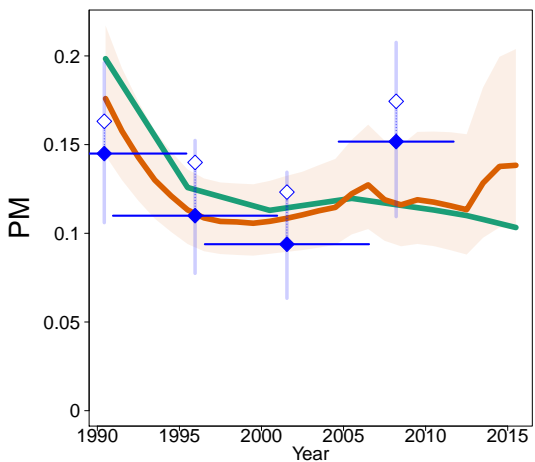

**Uganda [UGA; B]**
**(Eastern Africa)**

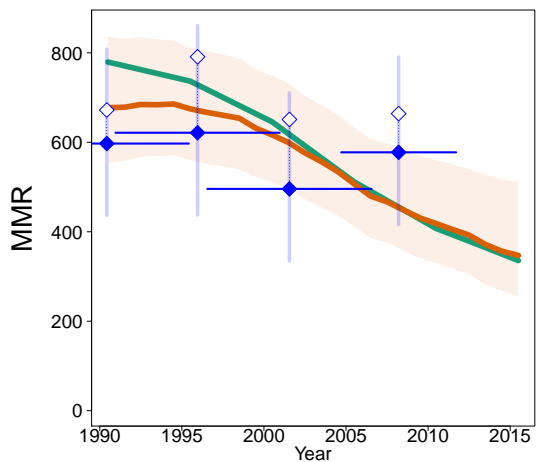

**Uganda [UGA; B]**
**(Eastern Africa)**

– Estimates –
Bmat 2014
MMEIG 2014
– Data –
◇ Observed Data
◆ Adjusted Data
– Data type –
◆ DHS

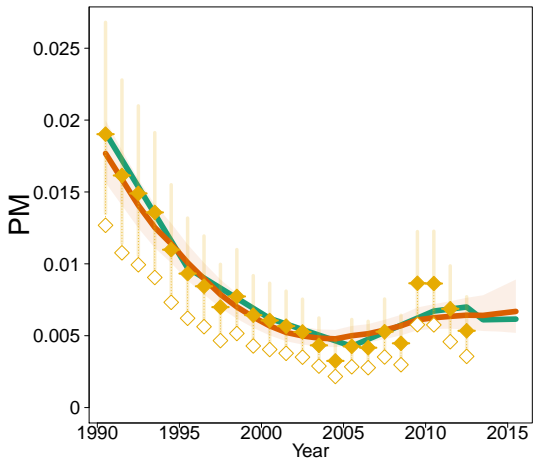

**Ukraine [UKR; A]**
**(Developed regions)**

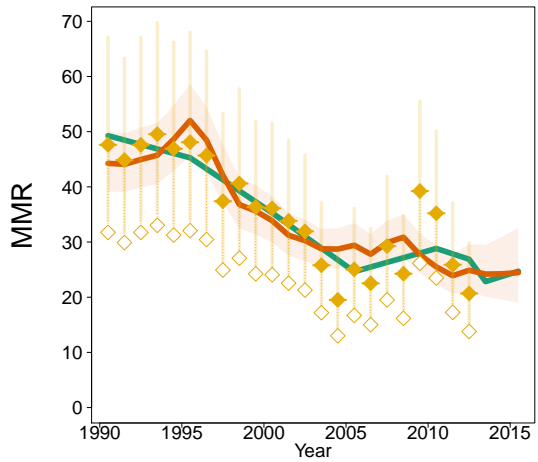

**Ukraine [UKR; A]**
**(Developed regions)**

– Estimates –
Bmat 2014
MMEIG 2014
– Data –
◇ Observed Data
◆ Adjusted Data
– Data type –
◆ VR

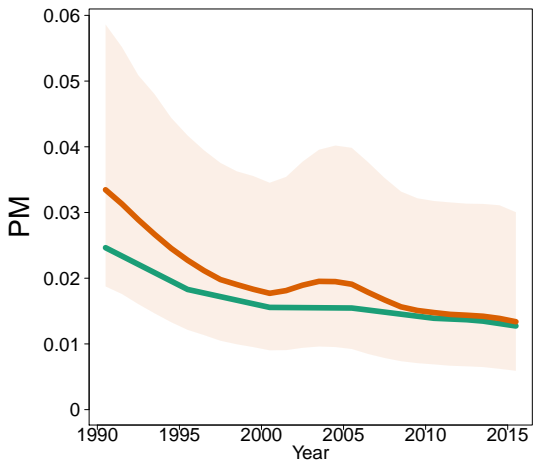

**United Arab Emirates [ARE; C]**
**(Western Asia)**

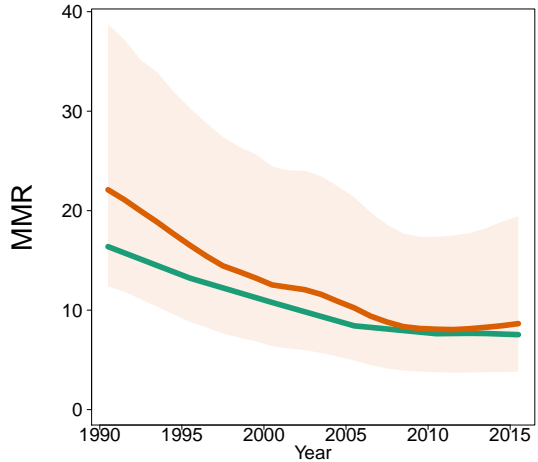

**United Arab Emirates [ARE; C]**
**(Western Asia)**

– Estimates –
Bmat 2014
MMEIG 2014
– Data –
◇ Observed Data
◆ Adjusted Data
– Data type –

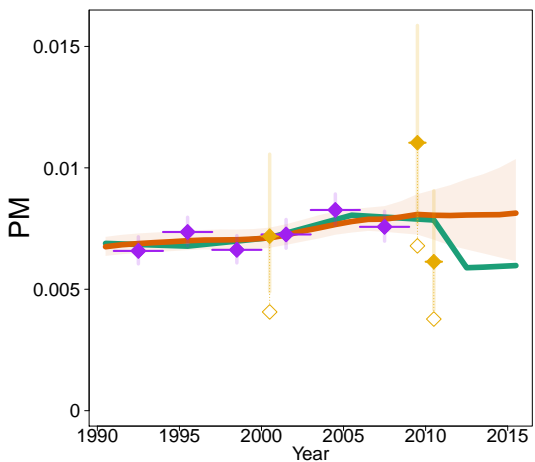

**United Kingdom [GBR; A]**
**(Developed regions)**

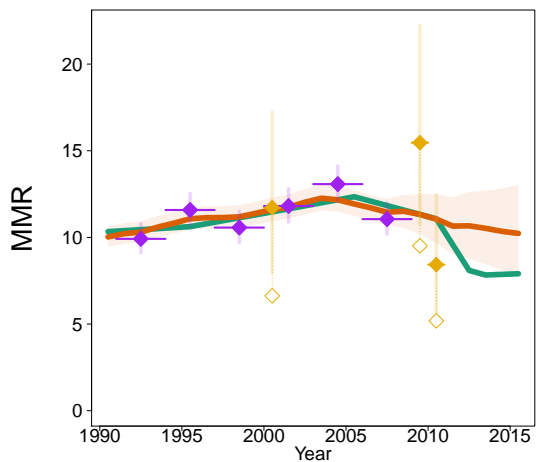

**United Kingdom [GBR; A]**
**(Developed regions)**

– Estimates –
Bmat 2014
MMEIG 2014
– Data –
◇ Observed Data
◆ Adjusted Data
– Data type –
◆ VR
◆ Spec. studies

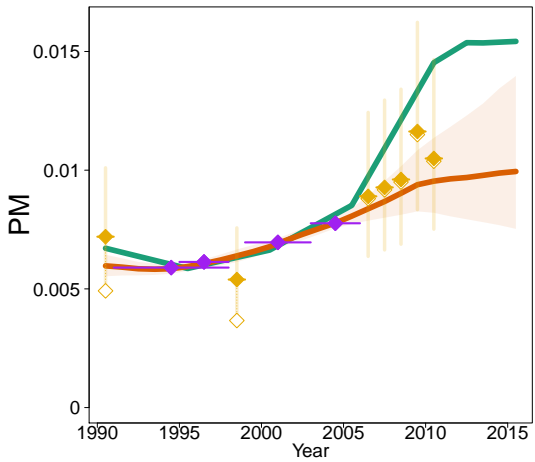

**United States of America [USA; A]**
**(Developed regions)**

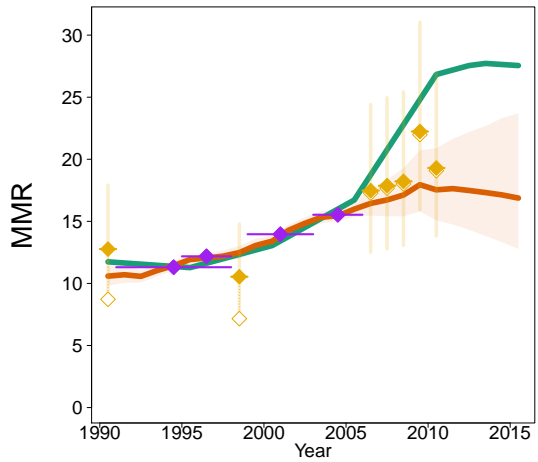

**United States of America [USA; A]**
**(Developed regions)**

– Estimates –
Bmat 2014
MMEIG 2014
– Data –
◇ Observed Data
◆ Adjusted Data
– Data type –
◆ VR
◆ Spec. studies

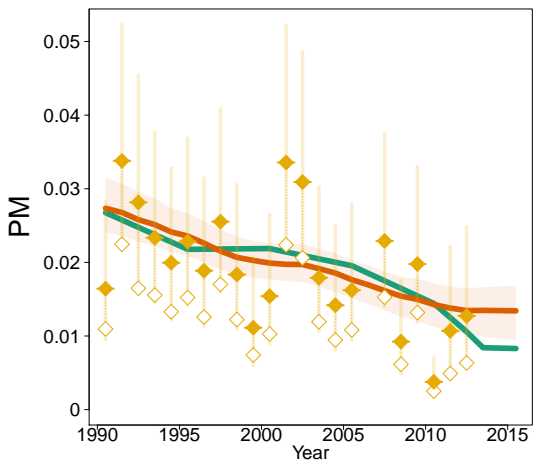

**Uruguay [URY; A]**
**(South America)**

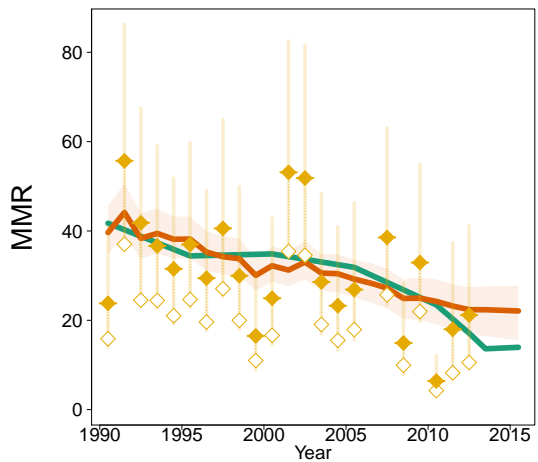

**Uruguay [URY; A]**
**(South America)**

– Estimates –
Bmat 2014
MMEIG 2014
– Data –
◇ Observed Data
◆ Adjusted Data
– Data type –
◆ VR

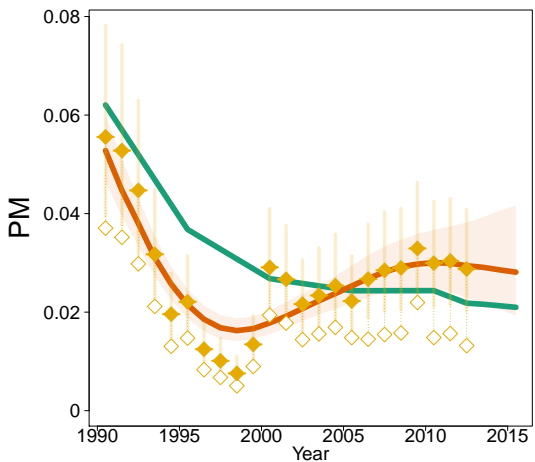

**Uzbekistan [UZB; A]**
**(Central Asia)**

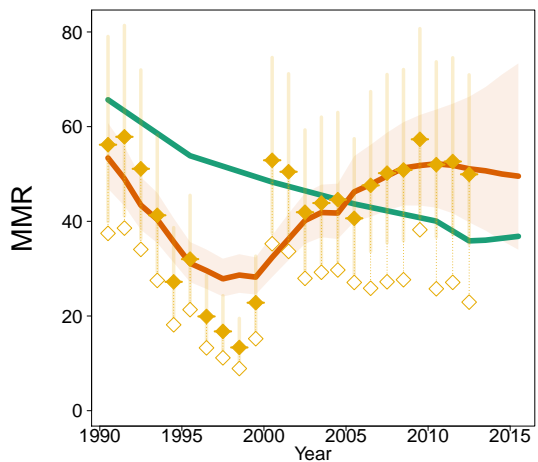

**Uzbekistan [UZB; A]**
**(Central Asia)**

– Estimates –
Bmat 2014
MMEIG 2014
– Data –
◇ Observed Data
◆ Adjusted Data
– Data type –
◆ VR

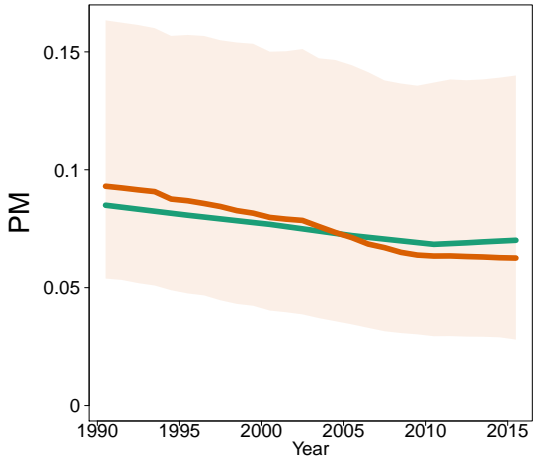

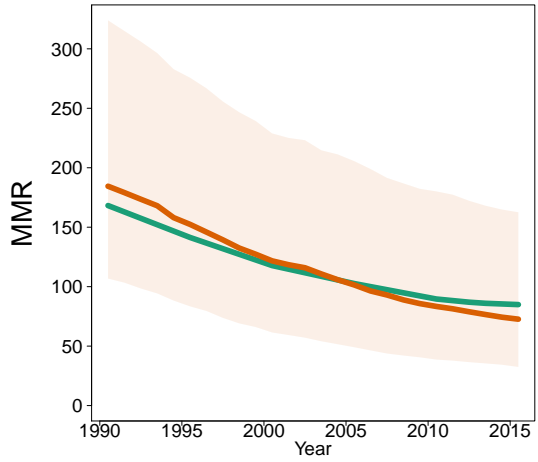

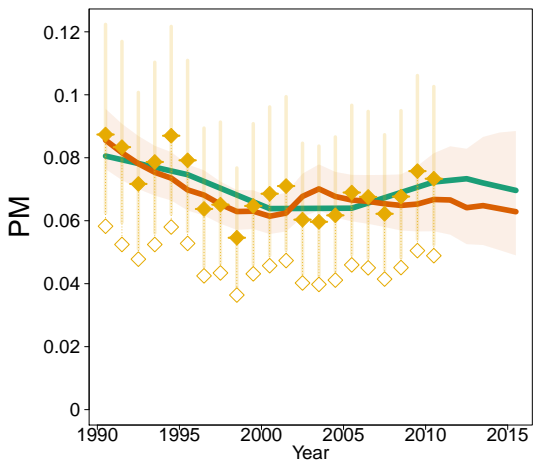

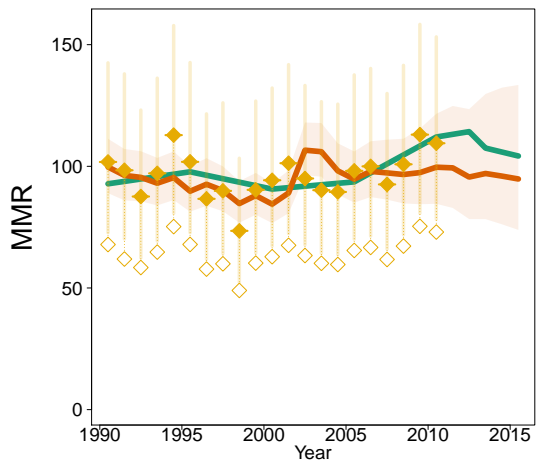

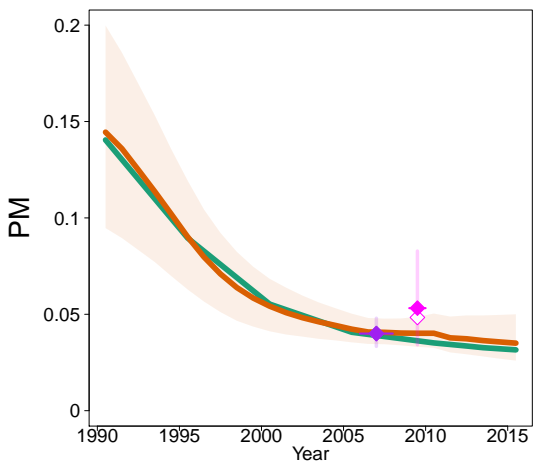

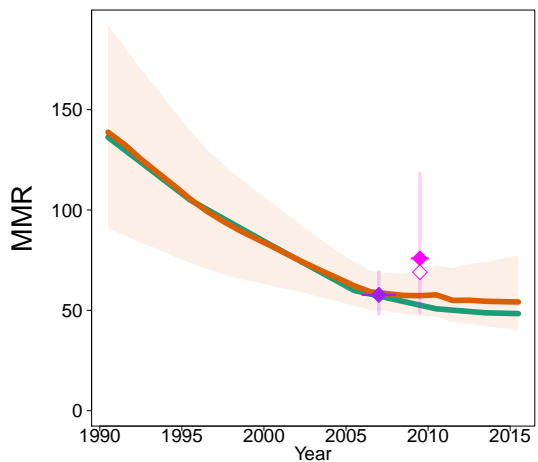

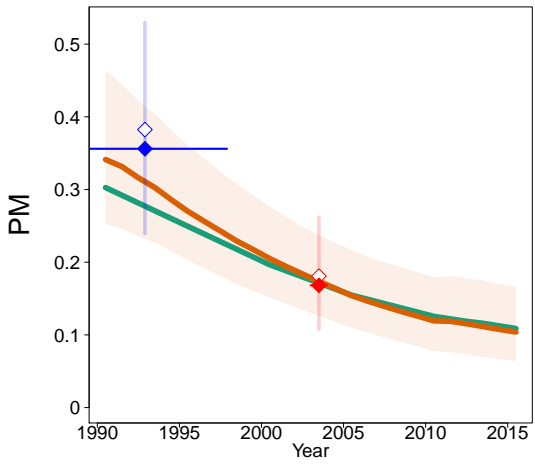

**Yemen [YEM; B]**
**(Western Asia)**

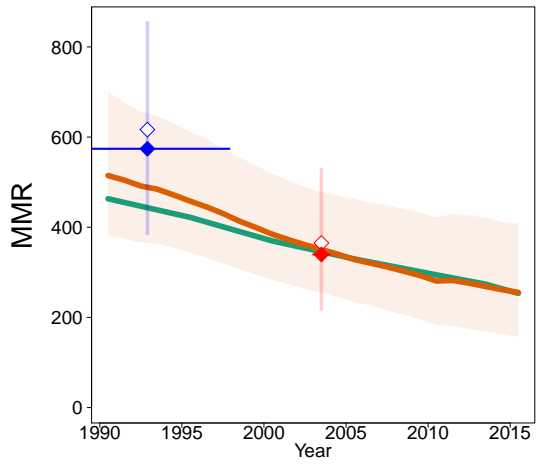

**Yemen [YEM; B]**
**(Western Asia)**

– Estimates –
Bmat 2014
MMEIG 2014
– Data –
◇ Observed Data
◆ Adjusted Data
– Data type –
◆ DHS
◆ Misc. studies

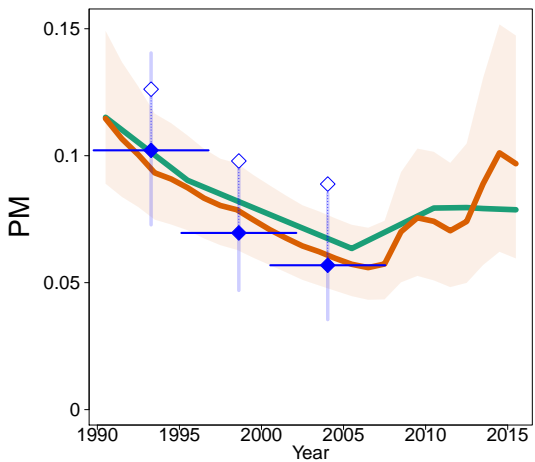

**Zambia [ZMB; B]**
**(Eastern Africa)**

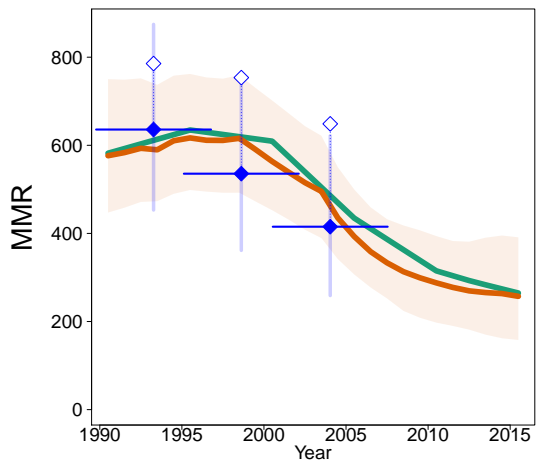

**Zambia [ZMB; B]**
**(Eastern Africa)**

– Estimates –
Bmat 2014
MMEIG 2014
– Data –
◇ Observed Data
◆ Adjusted Data
– Data type –
◆ DHS

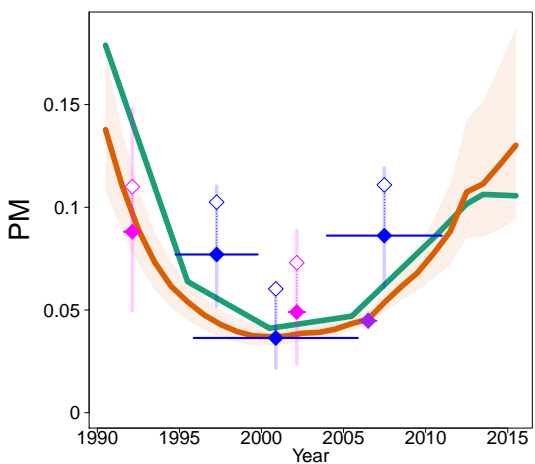

**Zimbabwe [ZWE; B]**
**(Eastern Africa)**

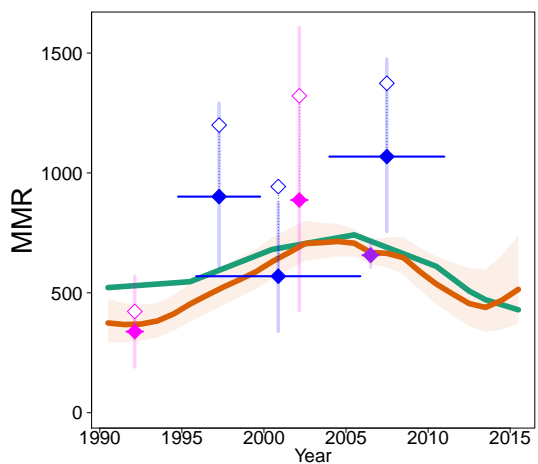

**Zimbabwe [ZWE; B]**
**(Eastern Africa)**

– Estimates –
Bmat 2014
MMEIG 2014
– Data –
◇ Observed Data
◆ Adjusted Data
– Data type –
◆ Spec. studies
◆ DHS
◆ Census



\end{document}